\documentclass[prd,twocolumn,showpacs,showkeys,preprintnumbers,floatfix,nofootinbib,superscriptaddress]{revtex4-1}
\usepackage{amsfonts} % AMS
\usepackage{amssymb} % AMS
\usepackage{amsmath} % AMS
\usepackage{graphicx} % Include figure files
\usepackage{subfigure} % Include figure files
\usepackage{array} % array
\usepackage{dcolumn} % Align table columns on decimal point
\usepackage{bm} % bold math
\usepackage{latexsym} % latex symbols
\usepackage{longtable} % long tables
\usepackage{bbold}
\usepackage{color}
\usepackage{multirow}
\usepackage{rotating}

\graphicspath{{./figs/}}

\newcommand{\qslash}[1]{\text{$\not \! #1$}}

\newcommand{\tsep}{\mathop{t_{\rm sep}}\nolimits}
\newcommand{\tskip}{\mathop{\tau_{\rm skip}}\nolimits}
\newcommand{\MeV}{\mathop{\rm MeV}\nolimits}
\newcommand{\GeV}{\mathop{\rm GeV}\nolimits}
\newcommand{\fm}{\mathop{\rm fm}\nolimits}
\newcommand{\gsim}{\raisebox{-0.7ex}{$\stackrel{\textstyle >}{\sim}$ }}
\newcommand{\lsim}{\raisebox{-0.7ex}{$\stackrel{\textstyle <}{\sim}$ }}

\DeclareMathOperator{\Tr}{Tr}

\definecolor{green}{rgb}{0.1, 0.8, 0.1}

\begin{document}

%%%

\title{Axial, Scalar and Tensor Charges of the Nucleon from 2+1+1-flavor Lattice QCD}
\author{Tanmoy Bhattacharya}
\email{tanmoy@lanl.gov}
\affiliation{Los Alamos National Laboratory, Theoretical Division T-2, Los Alamos, New Mexico 87545}

\author{Vincenzo Cirigliano}
\email{cirigliano@lanl.gov}
\affiliation{Los Alamos National Laboratory, Theoretical Division T-2, Los Alamos, New Mexico 87545}

\author{Saul D.~Cohen}
\email{saul.cohen@gmail.com}
\affiliation{Institute for Nuclear Theory, University of Washington, Seattle, Washington 98195}

\author{Rajan Gupta}
\email{rajan@lanl.gov}
\affiliation{Los Alamos National Laboratory, Theoretical Division T-2, Los Alamos, New Mexico 87545}

\author{Huey-Wen Lin}
\email{hwlin@pa.msu.edu}
\affiliation{Physics Department, University of California, Berkeley, California 94720}

\author{Boram Yoon}
\email{boram@lanl.gov}
\affiliation{Los Alamos National Laboratory, Theoretical Division T-2, Los Alamos, New Mexico 87545}

\collaboration{Precision Neutron Decay Matrix Elements (PNDME) Collaboration}
\preprint{LA-UR-16-20522}
\preprint{INT-PUB-16-008}
\pacs{11.15.Ha, % Lattice gauge theory
      12.38.Gc  % Lattice QCD calculations
}
\keywords{nucleon charges, lattice QCD, excited-state contamination, neutron EDM}
\date{\today}
\begin{abstract}
We present results for the isovector axial, scalar and tensor charges
$g^{u-d}_A$, $g^{u-d}_S$ and $g^{u-d}_T$ of the nucleon needed to probe the Standard
Model and novel physics. The axial charge is a fundamental parameter
describing the weak interactions of nucleons. The scalar and tensor
charges probe novel interactions at the TeV scale in neutron and
nuclear $\beta$-decays, and the flavor-diagonal tensor charges
$g^{u}_T$, $g^{d}_T$ and $g^{s}_T$ are needed to quantify the
contribution of the quark electric dipole moment (EDM) to the neutron
EDM.
The lattice-QCD calculations were done using nine ensembles of gauge
configurations generated by the MILC Collaboration using the highly improved staggered quarks 
action with 2+1+1 dynamical flavors. These ensembles span three
lattice spacings $a \approx 0.06, 0.09$ and $0.12 \fm$ and light-quark
masses corresponding to the pion masses $M_\pi \approx 135, 225$ and
$315 \MeV$. High-statistics estimates on five ensembles using the
all-mode-averaging method allow us to quantify all systematic
uncertainties and perform a simultaneous extrapolation in the lattice
spacing, lattice volume and light-quark masses for the connected
contributions. Our final estimates, in the $\overline{\text{MS}}$
scheme at $2\GeV$, of the isovector charges are $g_A^{u-d} =
1.195(33)(20)$, $g_S^{u-d} = 0.97(12)(6) $ and $g_T^{u-d} =
0.987(51)(20)$. The first error includes statistical and all
systematic uncertainties except that due to the extrapolation Ansatz,
which is given by the second error estimate.  Combining our estimate
for $g_S^{u-d}$ with the difference of light quarks masses
$(m_d-m_u)^{\rm QCD}=2.67(35)$~MeV given by the Flavor Lattice Average
Group, we obtain $(M_N-M_P)^{\rm QCD} = 2.59(49)$~MeV.
Estimates of the connected part of the flavor-diagonal tensor charges
of the proton are $g^{u}_T=0.792(42)$ and
$g^{d}_T=-0.194(14)$. Combining our new estimates with precision
low-energy experiments, we present updated constraints on novel scalar
and tensor interactions, $\epsilon_{S,T}$, at the TeV scale.
\end{abstract}
\maketitle
%
%
%
%\nopagebreak
%
%%%%%%%%%%%%%%%%%%%%%%%%%%%%%%%%%%%%%%%%%%%%%%%%%%%%%%%%%%%%%%%%%%%%%
%%%  SECTION                                                      %%%
%%%%%%%%%%%%%%%%%%%%%%%%%%%%%%%%%%%%%%%%%%%%%%%%%%%%%%%%%%%%%%%%%%%%%
\section{Introduction}
\label{sec:into}

The nucleon axial charge $g_A^{u-d}$ is an important parameter that
encapsulates the strength of weak interactions of nucleons. It enters
in many analyses of nucleon structure and of the Standard Model (SM)
and beyond-the-SM (BSM) physics. For example, the rate of
proton-proton fusion, which is the first step in the thermonuclear
reaction chains that power low-mass hydrogen-burning stars like the
Sun, is sensitive to it. It impacts the extraction of $V_{ud}$ and
tests of the unitarity of the Cabibbo-Kobayashi-Maskawa (CKM) matrix, as well as the analysis of
neutrinoless double-beta decay.  At present, the ratio of the axial to
the vector charge, $g_A/g_V$, is best determined from the experimental
measurement of neutron beta decay using polarized ultracold neutrons
by the UCNA Collaboration, $1.2756(30)$~\cite{Mendenhall:2012tz}, and
by PERKEO II, $1.2761{}^{+14}_{-17}$~\cite{Mund:2012fq}. Note that, in
the SM, $g_V=1$ up to second order corrections in isospin
breaking~\cite{Ademollo:1964sr,Donoghue:1990ti} as a result of the
conservation of the vector current.  Using $V_{ud}$ determined from superallowed
nuclear beta decay or pion decay in combination with the average
neutron lifetime measurement also gives a consistent value for
$g_A^{u-d}$~\cite{Agashe:2014kda,Superallowed}. Given the important role
$g_A^{u-d}$ plays in parametrizing the structure and weak interactions of
nucleons, and probing signatures of new physics, it is important to
calculate it directly with $O(1\%)$ accuracy using lattice QCD and
eventually confront the theoretical prediction with experimental
measurements.

%% We show a compilation of experimental measurements of
%% $g_A^{u-d}$ used in PDG average in the top of Fig.~\ref{fig:gA}. The
%% extraction of $g_A^{u-d}$ from experiments, however, uses some assumptions
%% about the nonexistence of new physics at the TeV
%% scale~\cite{Cirigliano:2012ab} that can contribute to the measured
%% process.  

The isovector scalar and tensor charges of the nucleon, combined with
the helicity-flip parameters $b$ and $b_\nu$ in the neutron decay
distribution, probe novel scalar and tensor interactions at the TeV
scale~\cite{Bhattacharya:2011qm}.  To optimally bound such scalar and
tensor interactions using planned measurements of these $b$ and
$b_\nu$ parameters at the $10^{-3}$ precision
level~\cite{abBA,WilburnUCNB,Pocanic:2008pu}, requires the matrix
elements of the local scalar and tensor quark bilinear operators
within the nucleon state to be calculated with a precision of
10\%--15\%~\cite{Bhattacharya:2011qm}.  Future higher-precision
measurements of $b$ and $b_\nu$ would require correspondingly
higher-precision calculations of the matrix elements to place even
more stringent bounds on TeV-scale couplings.  In a recent
work~\cite{Bhattacharya:2015wna}, we showed that lattice-QCD
calculations have reached a level of control over all sources of
systematic errors needed to yield the tensor charge with the required
precision. The data for the scalar 3-point functions is about a factor
of 5 more noisy. In this paper we show that by using the
all-mode-averaging (AMA) error-reduction technique~\cite{Bali:2009hu,Blum:2012uh}
on the same set of ensembles used in Ref.~\cite{Bhattacharya:2015wna},
we can increase the statistics significantly and extract the scalar
charge with $O(15\%)$ uncertainty. These higher-statistics results
also improve upon our previous estimates of the tensor charges.

In addition to probing novel scalar and tensor interactions at the TeV
scale, precise estimates of the matrix elements of the flavor-diagonal
tensor operators are needed to quantify the contributions of the
$u,\ d, \ s, \ c$ quark electric dipole moments (EDM) to the neutron
electric dipole moment
(nEDM)~\cite{Bhattacharya:2015wna,Pospelov:2005pr}. Most extensions of
the Standard Model designed to explain nature at the TeV scale have
new sources of $CP$ violation, and the nEDM is a very sensitive probe of
these. Thus, planned experiments aiming to reduce the current bound on
the nEDM of $2.9 \times 10^{-26}\ e$~cm~\cite{Baker:2006ts} 
to around $ 10^{-28}\ e$~cm will put stringent constraints on many BSM
theories, provided the matrix elements of novel $CP$-violating
interactions, of which the quark EDM is one, are calculated with the
required precision.

The tensor charges are also given by the zeroth moment of the
transversity distributions that are measured in many experiments
including Drell-Yan and semi-inclusive deep inelastic scattering
(SIDIS). Transversity distributions describe the net transverse
polarization of quarks in a transversely polarized nucleon, and there
exists an active program at Jefferson Lab (JLab) to measure
them~\cite{Dudek:2012vr}. The extraction of the transversity
distributions from the data taken over a limited range of $Q^2$ and
Bjorken $x$, however, is not straightforward and requires additional
phenomenological modeling. As discussed in
Sec.~\ref{sec:comparison}, lattice-QCD estimates of
$g_T^{u-d}$ are the most accurate at present. Future experiments will significantly
improve the extraction of the transversity distributions.  Thus,
accurate calculations of the tensor charges using lattice QCD will
continue to help elucidate the structure of the nucleon in terms of
quarks and gluons and provide a benchmark against which
phenomenological estimates utilizing measurements at JLab and other
experimental facilities worldwide can be compared.

The methodology for calculating the isovector charges in an isospin
symmetric theory, that is, measuring the contribution to the matrix
elements of the insertion of the zero-momentum bilinear quark
operators in one of the three valence quarks in the nucleon, is well
developed~\cite{Lin:2012ev,Syritsyn:2014saa,Constantinou:2014tga}.
Calculation of the flavor-diagonal charges is similar except that it
gets additional contributions from contractions of the operator as a
vacuum quark loop that interacts with the nucleon propagator through
the exchange of gluons.
%% as shown in Fig.~\ref{fig:con_disc}.  
Our estimates of disconnected contributions to $g_T^{u, d, s}$ were
given in Ref.~\cite{Bhattacharya:2015wna}, where we showed that these
contributions are small, $O(0.01)$, and in most cases consistent with
zero within errors.\footnote{The five ensembles analyzed were 
$a12m310$, $a12m220$, $a09m310$, $a09m220$ and $a06m310$. Analysis of the 
physical mass ensemble $a09m130$ is ongoing.}
For the disconnected contribution of the strange
quark, also needed for the neutron EDM analysis, we were able to
extrapolate the data to the continuum limit and find $g_T^s =
0.008(9)$~\cite{Bhattacharya:2015wna,Bhattacharya:2015esa}.  We do not
have new results for these disconnected contributions. In this paper, we
report on improvements in the estimate of the isovector charges
$g_A^{u-d} $, $g_S^{u-d} $ and $g_T^{u-d} $ and in the connected parts
of the flavor-diagonal charges $g_{A,S,T}^{u} $ and $g_{A,S,T}^{d} $,
and the isoscalar combination $g_T^{u+d} $ through a high-statistics
study using the AMA method on five ensembles.

%
%%  \begin{figure*}[tb]
%%    \subfigure{
%%      \includegraphics[width=0.4\linewidth]{connected}
%%    }
%%    \hspace{0.04\linewidth}
%%    \subfigure{
%%      \includegraphics[width=0.4\linewidth]{disconnected}
%%    }
%%  \caption{The connected (left) and disconnected (right) three-point
%%    diagrams needed to calculate the matrix elements of bilinear quark
%%    operators, labeled by the symbol $\otimes$, in the nucleon state.
%%    \label{fig:con_disc}}
%%  \end{figure*}
%

Overall, we analyze nine ensembles of $2+1+1$ flavors of highly
improved staggered quarks (HISQ)~\cite{Follana:2006rc} generated by
the MILC Collaboration~\cite{Bazavov:2012xda}.  The high-statistics
study using the AMA method~\cite{Bali:2009hu,Blum:2012uh} allows us to
demonstrate control over various sources of systematic errors and
obtain reliable error estimates.  Using these data, we perform a
combined extrapolation to infinite volume, the continuum limit and the
physical light-quark masses and obtain $g_A^{u-d} =1.195(33)(20)$,
$g_S^{u-d} =0.97(12)(6)$ and $g_T^{u-d} = 0.987(51)(20)$. The first
error includes statistical and all systematic uncertainties except
that due to the extrapolation Ansatz, which is given by the second
error estimate.  Throughout the paper, we present results for the
charges of the proton, which by convention are called nucleon charges
in the literature.  From these, results for the neutron are obtained
by $u \leftrightarrow d$ interchange. A preliminary version of these
results was presented in Ref.~\cite{Gupta:2016rli}.

This paper is organized as follows. In Sec.~\ref{sec:Methodology}, we
describe the parameters of the gauge ensembles analyzed and the
lattice methodology. The fits used to isolate excited-state
contamination are described in Sec.~\ref{sec:excited}.  The
renormalization of the operators is discussed in
Sec.~\ref{sec:renorm}. Our final results for the isovector charges and
the connected parts of the flavor-diagonal charges are presented in
Sec.~\ref{sec:results}.  Results from additional simulations to
validate our analysis of excited-state contamination are presented in
Sec.~\ref{sec:confirmation}, and the estimation of errors is revisited
in Sec.~\ref{sec:errors}.  A comparison with previous works is given
in Sec.~\ref{sec:comparison}. In Sec.~\ref{sec:est}, we provide
constraints on novel scalar and tensor interactions at the TeV scale
using our new estimates of the charges and precision beta decay experiments and
compare them to those from the LHC.  Our final conclusions are
presented in Sec.~\ref{sec:conclusions}.

%%%%%%%%%%%%%%%%%%%%%%%%%%%%%%%%%%%%%%%%%%%%%%%%%%%%%%%%%%%%%%%%%%%%%
%%%  SECTION                                                      %%%
%%%%%%%%%%%%%%%%%%%%%%%%%%%%%%%%%%%%%%%%%%%%%%%%%%%%%%%%%%%%%%%%%%%%%
\section{Lattice Methodology}
\label{sec:Methodology}

\begin{table*}
\begin{center}
\renewcommand{\arraystretch}{1.2} % Change horizontal spacing
\begin{ruledtabular}
\begin{tabular}{lc<{ }|ccc|cc|cccc}
\multicolumn{2}{c}{Ensemble ID} & $a$ (fm) & $M_\pi^{\rm sea}$ (MeV) & $M_\pi^{\rm val}$ (MeV) & $L^3\times T$    & $M_\pi^{\rm val} L$ & $t_\text{sep}/a$ & $N_\text{conf}$  & $N_{\rm meas}^{\rm HP}$  & $N_{\rm meas}^{\rm AMA}$  \\
\hline
a12m310 & \includegraphics[viewport=16 13 22 19]{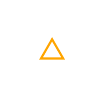}   & 0.1207(11) & 305.3(4) & 310.2(2.8) & $24^3\times 64$ & 4.55 & $\{8,9,10,11,12\}$ $\{8,10,12\}$& 1013 & 8104  &   64832   \\
a12m220S& \includegraphics[viewport=16 13 22 19]{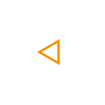}  & 0.1202(12) & 218.1(4) & 225.0(2.3) & $24^3\times 64$ & 3.29 & $\{8, 10, 12\}$   & 1000 & 24000 &           \\
a12m220 & \includegraphics[viewport=16 13 22 19]{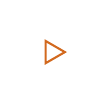}   & 0.1184(10) & 216.9(2) & 227.9(1.9) & $32^3\times 64$ & 4.38 & $\{8, 10, 12\}$   & 958  & 7664  &           \\
a12m220L& \includegraphics[viewport=16 13 22 19]{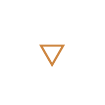}  & 0.1189(09) & 217.0(2) & 227.6(1.7) & $40^3\times 64$ & 5.49 & $\{10\}$ $\{8, 10, 12, 14\}$         & 1010 & 8080  &  68680   \\
\hline                                                                                                                                                                    
a09m310 & \includegraphics[viewport=16 13 22 19]{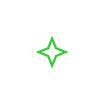}   & 0.0888(08) & 312.7(6) & 313.0(2.8) & $32^3\times 96$ & 4.51 & $\{10,12,14\}$    & 881  & 7048  &           \\
a09m220 & \includegraphics[viewport=16 13 22 19]{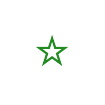}   & 0.0872(07) & 220.3(2) & 225.9(1.8) & $48^3\times 96$ & 4.79 & $\{10,12,14\}$    & 890  & 7120  &           \\
a09m130 & \includegraphics[viewport=16 13 22 19]{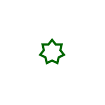}   & 0.0871(06) & 128.2(1) & 138.1(1.0) & $64^3\times 96$ & 3.90 & $\{10,12,14\}$    & 883  & 7064  &  84768   \\
\hline                                                                                                                                                                    
a06m310 & \includegraphics[viewport=16 13 22 19]{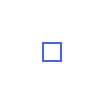}   & 0.0582(04) & 319.3(5) & 319.6(2.2) & $48^3\times 144$& 4.52 & $\{16,20,22,24\}$ & 1000 & 8000  &  64000   \\
a06m220 & \includegraphics[viewport=16 13 22 19]{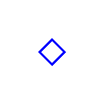}   & 0.0578(04) & 229.2(4) & 235.2(1.7) & $64^3\times 144$& 4.41 & $\{16,20,22,24\}$ & 650  & 2600  &  41600   \\
\end{tabular}
\end{ruledtabular}
\caption{Parameters, including the Goldstone pion mass $M_\pi^{\rm
    sea}$, of the 2+1+1-flavor HISQ lattices generated by the MILC
  Collaboration and analyzed in this study are quoted from
  Ref.~\cite{Bazavov:2012xda}.  The lattice scale is determined using
  $r_1$~\cite{Bazavov:2012xda}.  Symbols used in the plots are defined
  along with the ensemble ID. All fits are made versus $M_\pi^{\rm
    val}$ and finite-size effects are analyzed in terms of $M_\pi^{\rm
    val} L$.  Estimates of $M_\pi^{\rm val}$, the clover-on-HISQ pion
  mass, are the same as given in Ref.~\cite{Bhattacharya:2015wna} and
  the error is governed mainly by the uncertainty in the lattice
  scale. For each ensemble, we also give the values of the source-sink
  separation $t_{\rm sep}$ simulated, the number of configurations
  analyzed, and the number of measurements made using the HP and AMA
  methods.  The HP calculation on the $a12m220L$ ensemble has been
  done with a single $t_{\rm sep}=10$ while the LP analysis has been
  done with $t_{\rm sep}=\{8,10,12,14\}$.  }
\label{tab:ens}
\end{center}
\end{table*}

The nine ensembles used in the analysis cover a range of lattice
spacings ($0.06 \, \lsim a \, \lsim 0.12$~fm), pion masses ($135 \,
\lsim M_\pi \, \lsim 320$~MeV) and lattice volumes ($3.3\, \lsim M_\pi
L\, \lsim5.5$) and were generated using 2+1+1-flavors of 
HISQ~\cite{Follana:2006rc} by the MILC
Collaboration~\cite{Bazavov:2012xda}.  These are the same ensembles as
used in Ref.~\cite{Bhattacharya:2015wna}, where we presented an
analysis of the tensor charge. Their parameters are summarized in
Table~\ref{tab:ens}.

The correlation functions needed to calculate the matrix elements are
constructed using Wilson-clover fermions on these HISQ ensembles. This
mixed-action, clover-on-HISQ approach, leads to a nonunitary
formulation that at small, but {\it a priori} unknown, quark masses
suffers from the problem of exceptional configurations. In
Ref.~\cite{Bhattacharya:2015wna}, we described the tests performed to
show the absence of any such exceptional configurations in our
statistical samples.  The mixed-action approach also introduces
additional corrections to the leading chiral and continuum
extrapolation Ansatz.  In Sec.~\ref{sec:results}, we analyze the
observed dependence of the charges on $M_\pi$ over the range $135\,
\lsim M_\pi \, \lsim 320\ {\rm MeV}$ with and without the chiral
logarithm corrections in the fit Ansatz. It turns out that with our
current data, the observed dependence on the lattice spacing $a$ and
the quark masses is accounted for by the lowest-order correction
terms.

The parameters used in the analysis of 2- and 3-point functions with
clover fermions are given in Table~\ref{tab:cloverparams}. The
Sheikholeslami-Wohlert coefficient~\cite{Sheikholeslami:1985ij} used
in the clover action is fixed to its tree-level value with tadpole
improvement, $c_\text{sw} = 1/u_0^3$, where $u_0$ is the fourth root of
the plaquette expectation value calculated on the hypercubic (HYP)
smeared~\cite{Hasenfratz:2001hp} HISQ lattices.

The masses of light clover quarks were tuned so that the
clover-on-HISQ pion masses, $M^{\rm val}_\pi$, match the HISQ-on-HISQ
Goldstone ones, $M_\pi^{\rm sea}$. Both estimates are given in
Table~\ref{tab:ens}. All fits in $M_\pi^2$ to study the chiral
behavior are made using the clover-on-HISQ $M^{\rm val}_{\pi}$ since
the correlation functions, and thus the chiral behavior of the
charges, have a greater sensitivity to it. Henceforth, for
brevity, we drop the superscript and denote the clover-on-HISQ pion
mass as $M_\pi$. Performing fits using the HISQ-on-HISQ values,
${M_\pi^{\rm sea}}$, did not change the estimates significantly.

Most of the details of the methodology, the calculation strategy and
the analysis are the same as described in
Ref.~\cite{Bhattacharya:2015wna}.  The new feature in the current work
is the use of the AMA method~\cite{Bali:2009hu,Blum:2012uh} to recalculate all
quantities on the five ensembles that had the largest uncertainty:
$a12m310$, $a12m220L$, $a09m130$, $a06m310$ and $a06m220$. These new
estimates have significantly smaller statistical errors, and this
improvement allows us to better understand and quantify the
excited-state contamination. Using these more precise estimates
improves our final combined extrapolation in the lattice volume, $M_\pi
L \to \infty$, lattice spacing, $a \to 0$, and the light-quark mass, 
$M_{\pi^0} \to 135$~MeV.

\begin{table}[b]
\centering
\begin{ruledtabular}
\begin{tabular}{l|lc|c}
\multicolumn1c{ID}       & \multicolumn1c{$m_l$} &  $c_{\text{SW}}$ & Smearing    \\
         &       &                  & Parameters  \\
\hline
a12m310  & $-0.0695$  & 1.05094 & \{5.5, 70\}  \\
a12m220S & $-0.075$   & 1.05091 & \{5.5, 70\}  \\
a12m220  & $-0.075$   & 1.05091 & \{5.5, 70\}  \\
a12m220L & $-0.075$   & 1.05091 & \{5.5, 70\}  \\ 
\hline
a09m310  & $-0.05138$ & 1.04243 & \{5.5, 70\}  \\
a09m220  & $-0.0554$  & 1.04239 & \{5.5, 70\}  \\
a09m130  & $-0.058$   & 1.04239 & \{5.5, 70\}  \\
\hline
a06m310  & $-0.0398$  & 1.03493 & \{6.5, 70\}  \\
a06m220  & $-0.04222$ & 1.03493 & \{5.5, 70\}  \\
\end{tabular}
\end{ruledtabular}
\caption{The parameters used in the calculation of clover propagators.
  The hopping parameter $\kappa$ in the clover action is given by
  $2\kappa_{l} = 1/(m_{l}+4)$.  The Gaussian smearing parameters are
  defined by $\{\sigma, N_{\text{GS}}\}$, both in Chroma
  convention~\cite{Edwards:2004sx}. The parameter $N_{\text{GS}}$ is
  the number of applications of the Laplacian operator and the width
  of the smearing is controlled by $\sigma$.  $m_l$ is tuned to
  achieve $M_\pi^{\rm val} \approx M_\pi^\text{sea}$. }
  \label{tab:cloverparams}
\end{table}

%%%%%%%%%%%%%%%%%%%%%%%%%%%%%%%%%
\subsection{Correlation Functions}
\label{sec:CorrelationFunctions}
%%%%%%%%%%%%%%%%%%%%%%%%%%%%%%%%%

The interpolating operator $\chi$ used to create$/$annihilate the nucleon
state is 
\begin{align}
 \chi(x) = \epsilon^{abc} \left[ {q_1^a}^T(x) C \gamma_5 
            \frac{(1 \pm \gamma_4)}{2} q_2^b(x) \right] q_1^c(x)
\label{eq:nucl_op}
\end{align}
with color indices $\{a, b, c\}$, charge conjugation matrix
$C=\gamma_0 \gamma_2$, and $q_1$ and $q_2$ denote the two different
flavors of light quarks. At zero momentum, this operator couples only
to the spin-$\frac{1}{2}$ state.
The nonrelativistic projection $(1 \pm \gamma_4)/2$ is inserted to
improve the signal, with the plus and minus signs applied to the
forward and backward propagations in Euclidean time, respectively~\cite{Gockeler:1995wg}.

The 2-point and 3-point nucleon correlation functions at zero momentum 
are defined as 
\begin{align}
{\mathbf C}_{\alpha \beta}^{\text{2pt}}(t)
  &= \sum_{\mathbf{x}} 
   \langle 0 \vert \chi_\alpha(t, \mathbf{x}) \overline{\chi}_\beta(0, \mathbf{0}) 
   \vert 0 \rangle \,, 
\label{eq:corr_fun2} \\
{\mathbf C}_{\Gamma; \alpha \beta}^{\text{3pt}}(t, \tau)
  &= \sum_{\mathbf{x}, \mathbf{x'}} 
  \langle 0 \vert \chi_\alpha(t, \mathbf{x}) \mathcal{O}_\Gamma(\tau, \mathbf{x'})
  \overline{\chi}_\beta(0, \mathbf{0}) 
   \vert 0 \rangle \,,
\label{eq:corr_fun3}
\end{align}
where $\alpha$ and $\beta$ are spinor indices. The source is
placed at time slice $0$, $t$ is the sink time slice, and $\tau$ is an
intermediate time slice at which the local quark bilinear operator
$\mathcal{O}_\Gamma^q(x) = \bar{q}(x) \Gamma q(x)$ is inserted. The
Dirac matrix $\Gamma$ is $1$, $\gamma_4$, $\gamma_i \gamma_5$ and
$\gamma_i \gamma_j$ for scalar (S), vector (V), axial (A) and tensor
(T) operators, respectively.
In this work, subscripts $i$ and $j$ on gamma matrices run over $\{1,2,3\}$, 
with $i<j$. 

The nucleon charges $g_\Gamma^q$ are obtained from the matrix element
\begin{align}
 \langle N(p, s) \vert \mathcal{O}_\Gamma^q \vert N(p, s) \rangle
 = g_\Gamma^q \bar{u}_s(p) \Gamma u_s(p)
\end{align}
with spinors satisfying
\begin{align}
 \sum_s u_s(\mathbf{p}) \bar{u}_s(\mathbf{p})  = {\qslash{p} + m_N} \,.
%  = \frac{\qslash{p} + m_N} {2 E(\mathbf{p})}\,.
\end{align}

To extract the charges, we first construct the projected 2- and 3-point correlation functions
\begin{align}
C^{\text{2pt}}(t) & = {\langle \Tr [ \mathcal{P}_\text{2pt} {\mathbf C}^{\text{2pt}}(t) ] \rangle} 
\label{eq:2pt_proj}  \\
C_{\Gamma}^{\text{3pt}}(t, \tau)  & = \langle \Tr [ \mathcal{P}_{\rm 3pt} {\mathbf C}_{\Gamma}^{\text{3pt}}(t, \tau) ]\rangle \, .
 \label{eq:3pt_proj}
\end{align}
The operator $\mathcal{P}_\text{2pt} = (1+\gamma_4)/2$ is
used to project on to the positive parity contribution for the nucleon
propagating in the forward direction. For the connected 3-point
contributions, $\mathcal{P}_{\rm 3pt} =  \mathcal{P}_\text{2pt}(1+i\gamma_5\gamma_3)$ is used.  Note that the
$C_{\Gamma}^{\text{3pt}}(t, \tau)$ defined in
Eq.~\eqref{eq:3pt_proj} becomes zero if $\Gamma$ anticommutes
with $\gamma_4$, so only $\Gamma = 1$, $\gamma_4$, $\gamma_i \gamma_5$
and $\gamma_i \gamma_j$ elements of the Clifford algebra survive.
The fits used 
to extract the charges from the 2- and 3-point functions 
defined in Eqs.~\eqref{eq:2pt_proj} and~\eqref{eq:3pt_proj} are  discussed in
Sec.~\ref{sec:excited}.
%

%%%%%%%%%%%%%%%%%%%%%%%%%%%%%%%%
\subsection{The AMA Method}
\label{sec:AMA}
%%%%%%%%%%%%%%%%%%%%%%%%%%%%%%%%

The high-statistics calculation using the AMA 
technique~\cite{Bali:2009hu,Blum:2012uh} was carried out on five
ensembles.  To implement the AMA method, we choose four different
source time slices separated by $T/4$ on each configuration. Starting
from each of these time slices we calculate the 2- and 3-point
correlators by choosing $N_{\rm LP}=16$ source locations from which
low-precision (LP) evaluation of the quark propagator is carried
out. The resulting LP estimates for 2- and 3-point functions from
these $4\times 16 = 64$ sources may be biased due to incomplete
inversion of the Dirac matrix. To remove this bias, we place an
additional high-precision (HP) source on each of the four time slices
from which we calculate both LP and HP correlation functions. Thus, in
our implementation of the AMA method, $64+4$ LP and 4 HP calculations
are done on each configuration. These 4 HP calculations are the same
as used in the full HP study presented in
Ref.~\cite{Bhattacharya:2015wna} and, therefore, needed no additional
calculations. In total, the new simulations generated $4 \times 16 +
4=68$ LP 2- and 3-point correlation functions per configuration.

Using four HP and $64+4$ LP correlators on each configuration, the
bias corrected 2- and 3-point functions are given by
\begin{align}
 C^\text{imp}& 
 = \frac{1}{N_\text{LP}} \sum_{i=1}^{N_\text{LP}} 
    C_\text{LP}(\mathbf{x}_i^\text{LP}) \nonumber \\
  +& \frac{1}{N_\text{HP}} \sum_{i=1}^{N_\text{HP}} \left[
    C_\text{HP}(\mathbf{x}_i^\text{HP})
    - C_\text{LP}(\mathbf{x}_i^\text{HP})
    \right] \,,
  \label{eq:2-3pt_AMA}
\end{align}
where $C_\text{LP}$ and $C_\text{HP}$ are the 2- and 
3-point correlation functions calculated in LP and HP, respectively,
and $\mathbf{x}_i^\text{LP}$ and $\mathbf{x}_i^\text{HP}$ are the two
kinds of source positions. 

\begin{figure*}[tbp]
  \vspace{-0.05in}
  \subfigure{
    \includegraphics[height=1.64in,trim={0.1cm 0.67cm 0 0},clip]{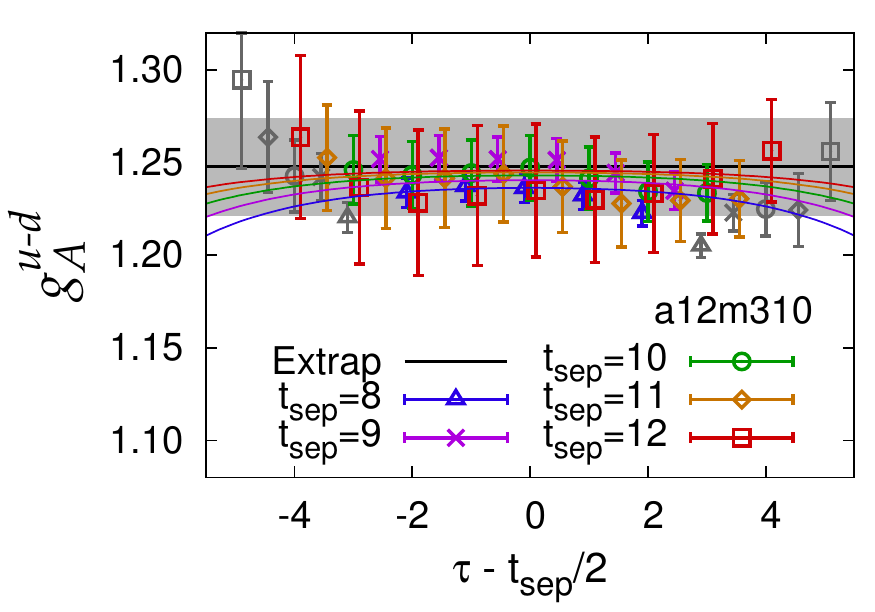}
    \includegraphics[height=1.64in,trim={0.9cm 0.67cm 0 0},clip]{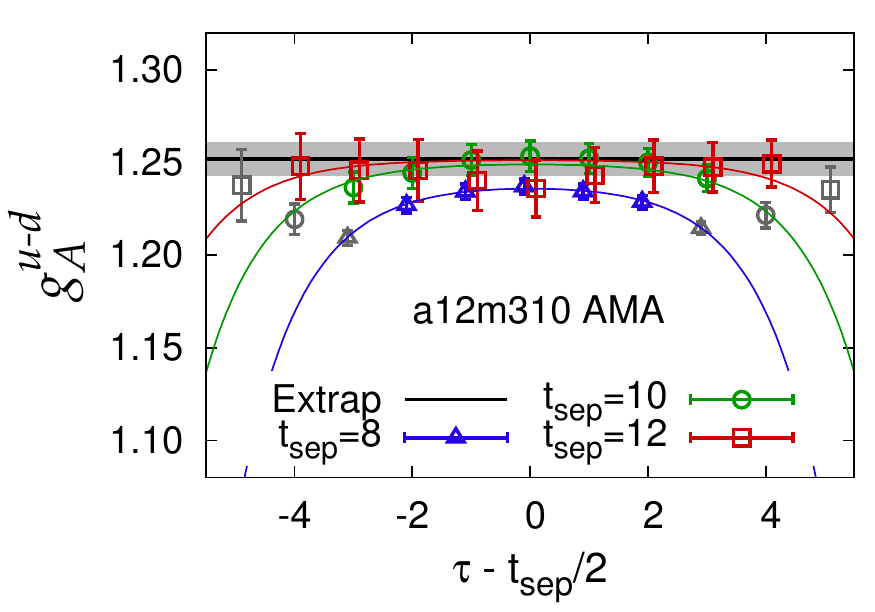}
}\\
  \vspace{-0.1in}
  \subfigure{
    \includegraphics[height=1.64in,trim={0.1cm 0.67cm 0 0},clip]{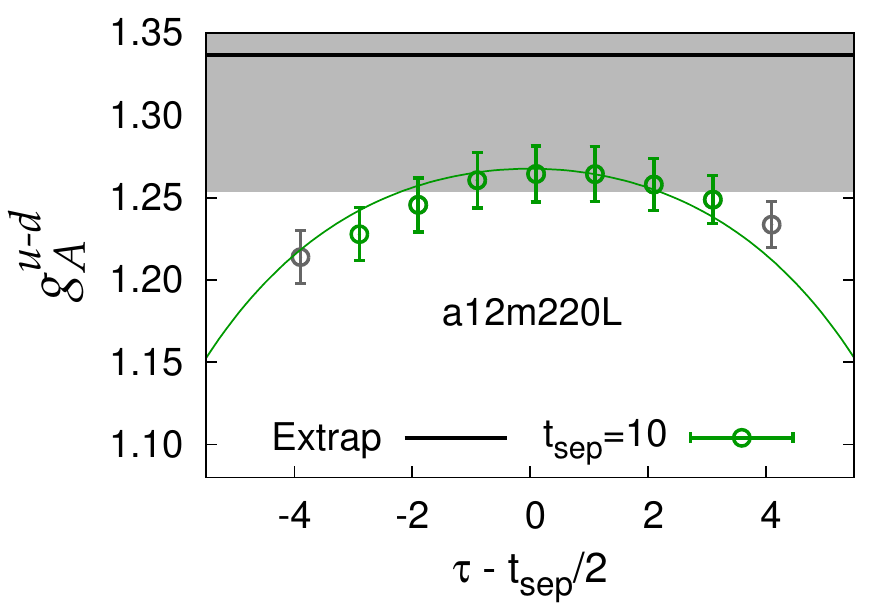}
    \includegraphics[height=1.64in,trim={0.9cm 0.67cm 0 0},clip]{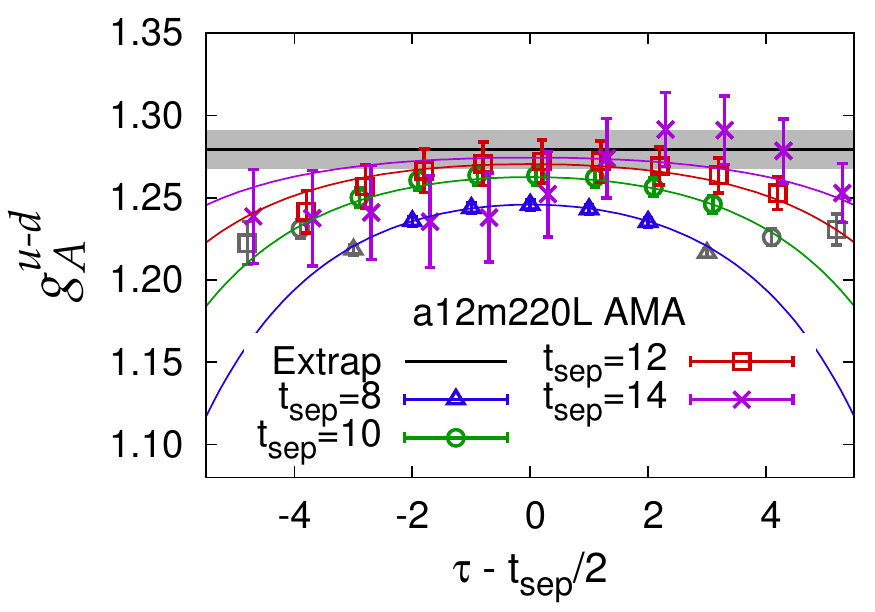}
}\\
  \vspace{-0.1in}
  \subfigure{
    \includegraphics[height=1.64in,trim={0.1cm 0.67cm 0 0},clip]{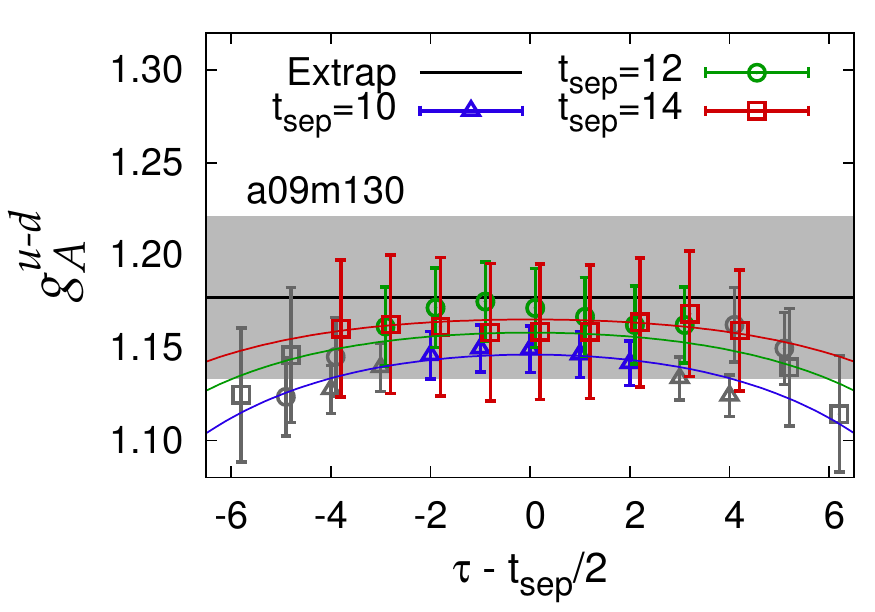}
    \includegraphics[height=1.64in,trim={0.9cm 0.67cm 0 0},clip]{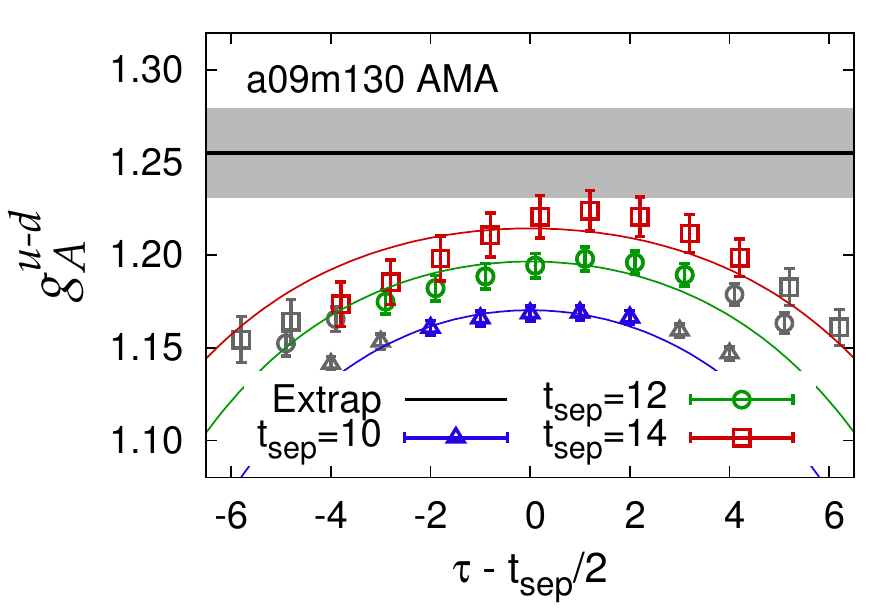}
}\\
  \vspace{-0.1in}
  \subfigure{
    \includegraphics[height=1.64in,trim={0.1cm 0.67cm 0 0},clip]{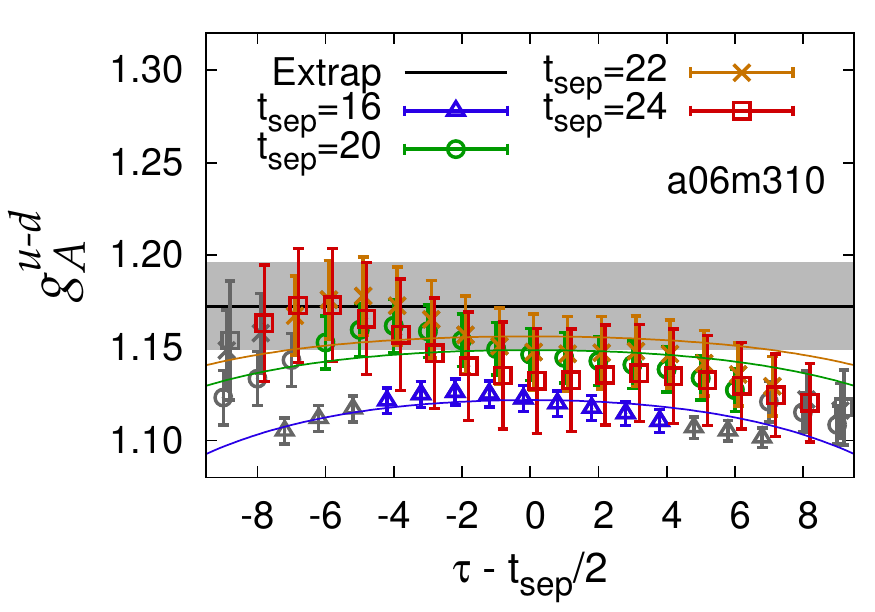}
    \includegraphics[height=1.64in,trim={0.9cm 0.67cm 0 0},clip]{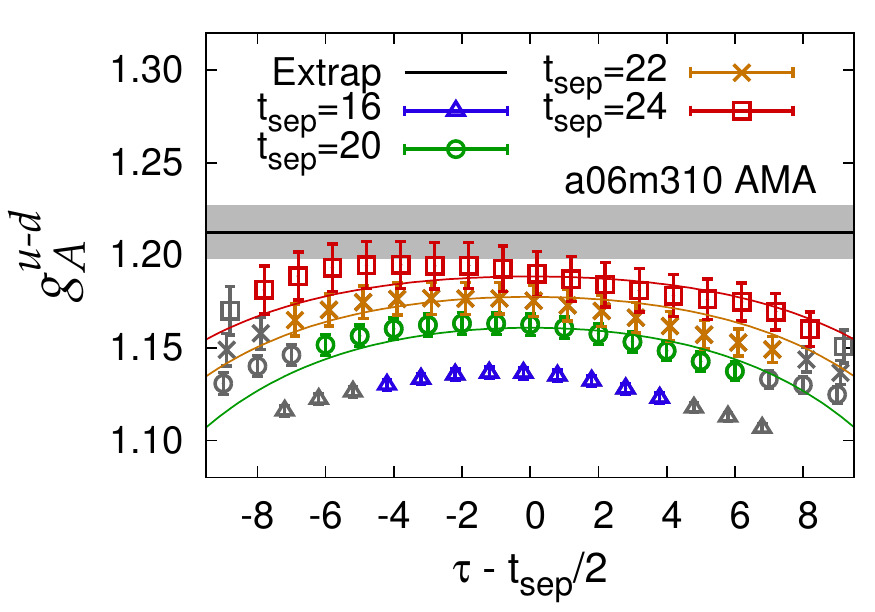}
}\\
  \vspace{-0.1in}
  \subfigure{
    \includegraphics[height=1.81in,trim={0.1cm 0.1cm 0 0},clip]{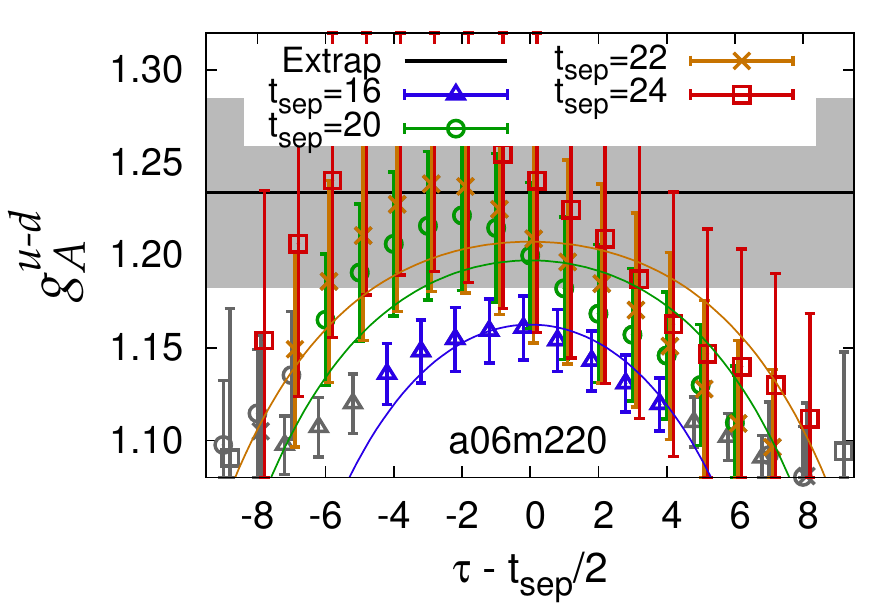}
    \includegraphics[height=1.81in,trim={0.9cm 0.1cm 0 0},clip]{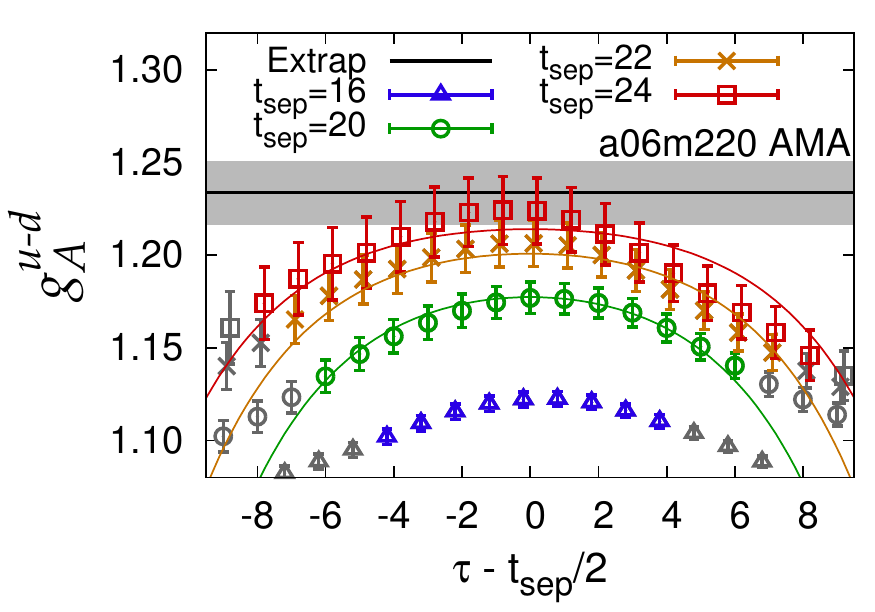}
}
  \vspace{-0.1in}
\caption{Comparison of the unrenormalized $g_A^{u-d}$ data obtained using
  all HP measurements (left) with the AMA method (right). The
  parameters for these five ensembles are given in
  Table~\protect\ref{tab:ens}. The increase in statistics with the
  AMA method significantly improves the resolution of the data at 
  the various source-sink separations, $t_{\rm sep}$. In each case, the solid line within the grey
  error band is the $t_{\rm sep} \to \infty$ result given by the 2-state fit using
  Eqs.~\eqref{eq:2pt} and~\eqref{eq:3pt}, and the colored lines are its values 
  for different $t_{\rm sep}$ plotted in the same color as the
  data. On the $a12m220L$ ensemble, the HP data was generated only with 
  $t_{\rm sep} = 10$. 
  \label{fig:gA_HPversusAMA}}
\end{figure*}

\begin{figure*}[tbp]
  \subfigure{
    \includegraphics[height=1.64in,trim={0.1cm 0.67cm 0 0},clip]{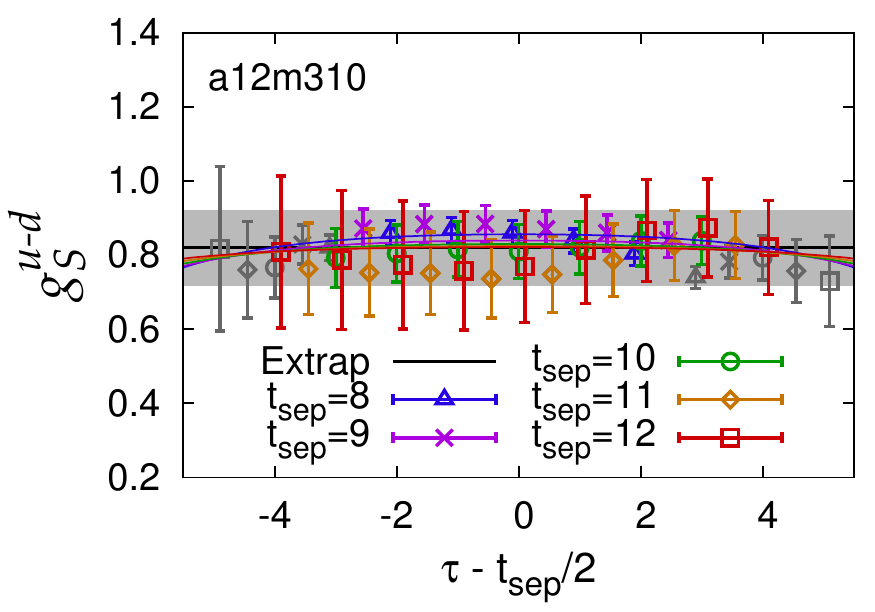}
    \includegraphics[height=1.64in,trim={0.9cm 0.67cm 0 0},clip]{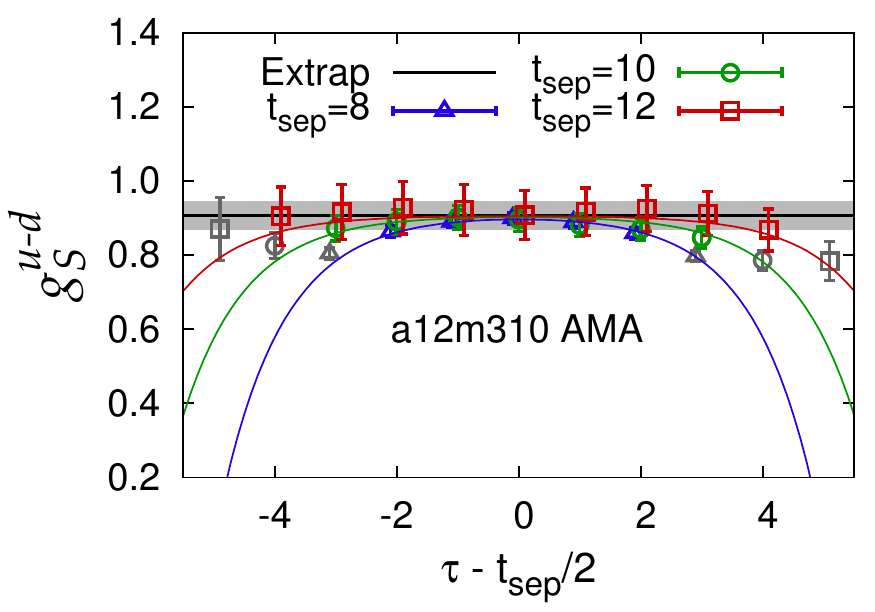}
  \vspace{-0.03\linewidth}
  \vspace{-0.03\linewidth}
}
  \vspace{-0.03\linewidth}
  \subfigure{
    \includegraphics[height=1.64in,trim={0.1cm 0.67cm 0 0},clip]{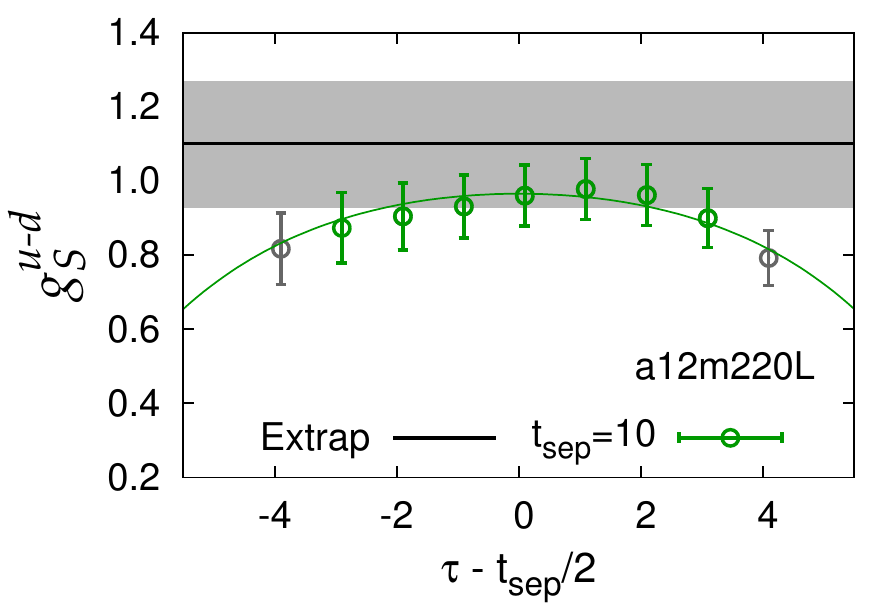}
    \includegraphics[height=1.64in,trim={0.9cm 0.67cm 0 0},clip]{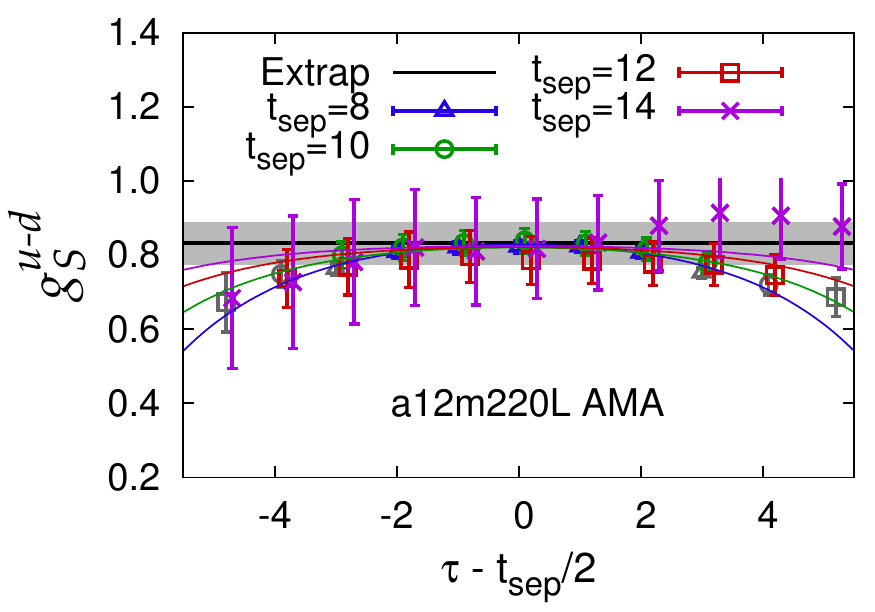}
}
  \vspace{-0.03\linewidth}
  \subfigure{
    \includegraphics[height=1.64in,trim={0.1cm 0.67cm 0 0},clip]{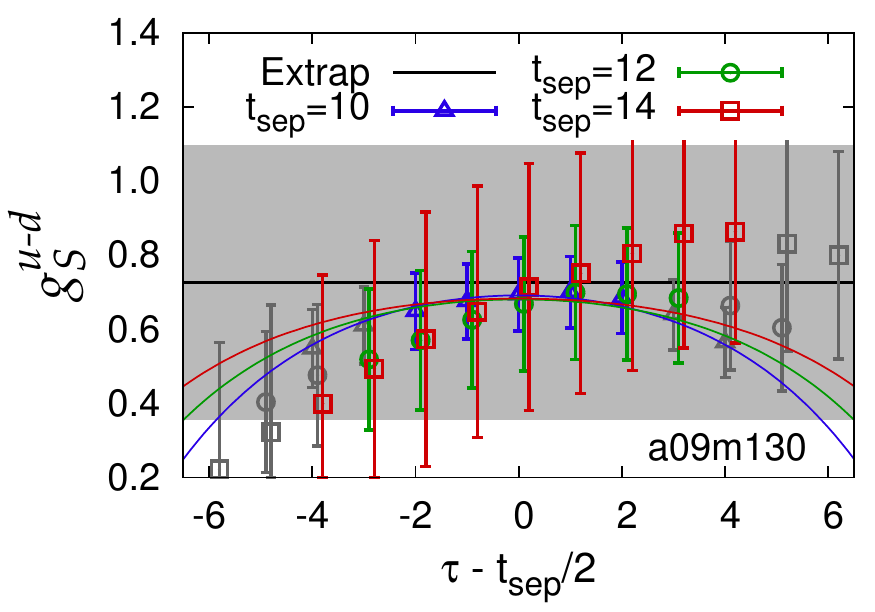}
    \includegraphics[height=1.64in,trim={0.9cm 0.67cm 0 0},clip]{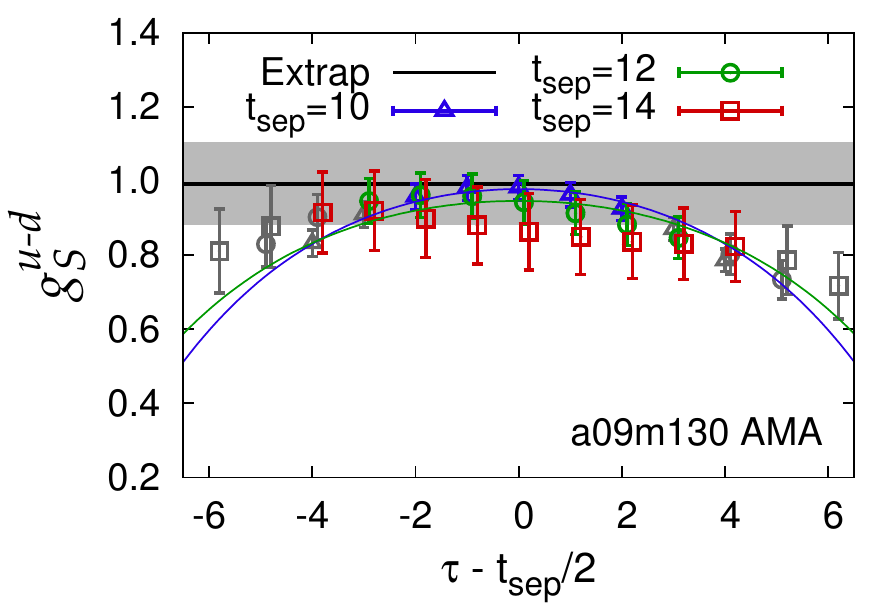}
}
  \vspace{-0.03\linewidth}
  \subfigure{
    \includegraphics[height=1.64in,trim={0.1cm 0.67cm 0 0},clip]{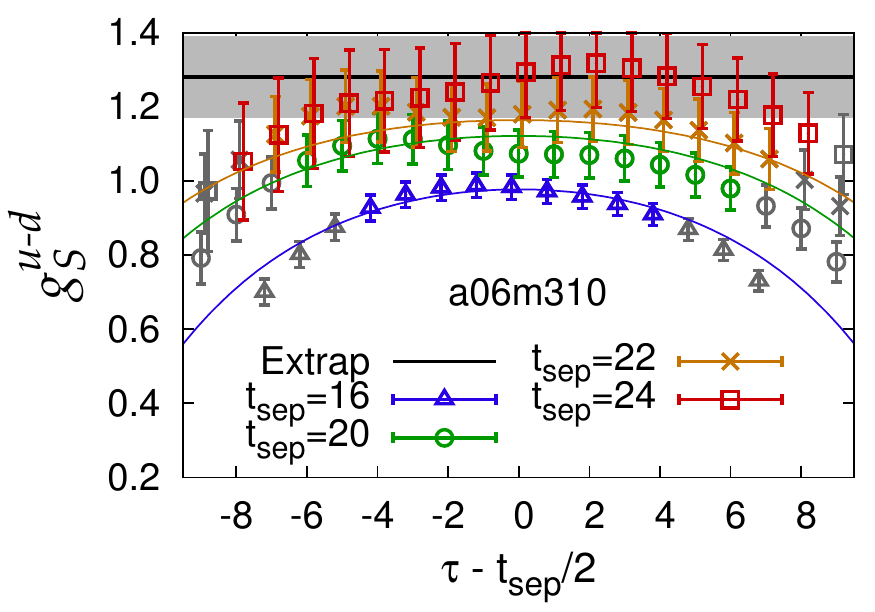}
    \includegraphics[height=1.64in,trim={0.9cm 0.67cm 0 0},clip]{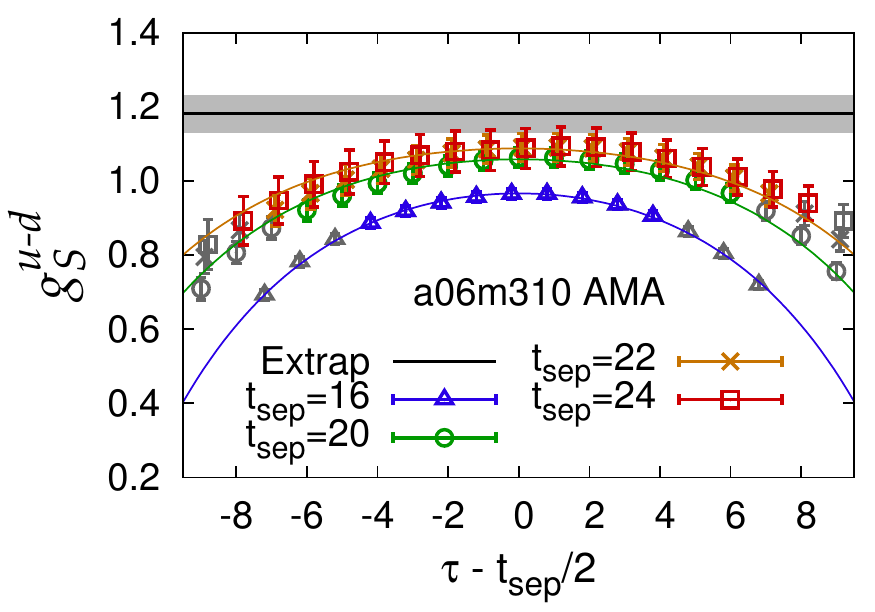}
}
  \vspace{-0.02\linewidth}
  \subfigure{
    \includegraphics[height=1.81in,trim={0.1cm 0.1cm 0 0},clip]{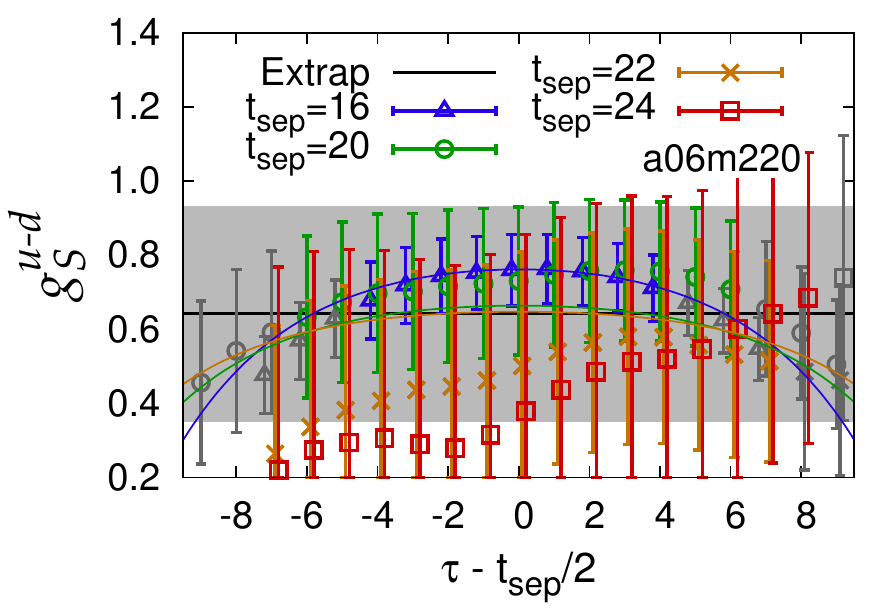}
    \includegraphics[height=1.81in,trim={0.9cm 0.1cm 0 0},clip]{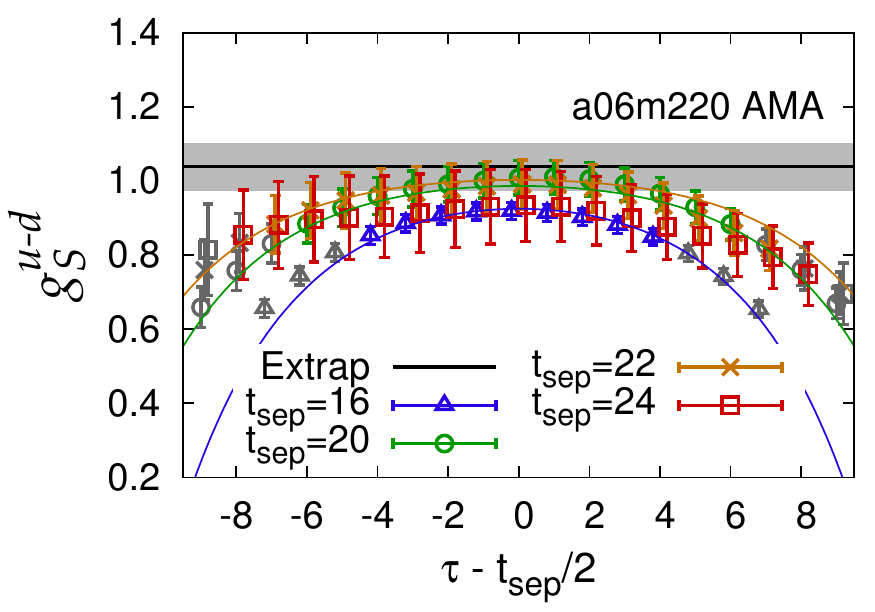}
}
\caption{Comparison of the unrenormalized $g_S^{u-d}$ data obtained using
  all HP measurements (left) with the AMA method (right). The
  rest is the same as in Fig.~\protect\ref{fig:gA_HPversusAMA}. 
  \label{fig:gS_HPversusAMA}}
\end{figure*}

\begin{figure*}[tbp]
  \subfigure{
    \includegraphics[height=1.64in,trim={0.1cm 0.67cm 0 0},clip]{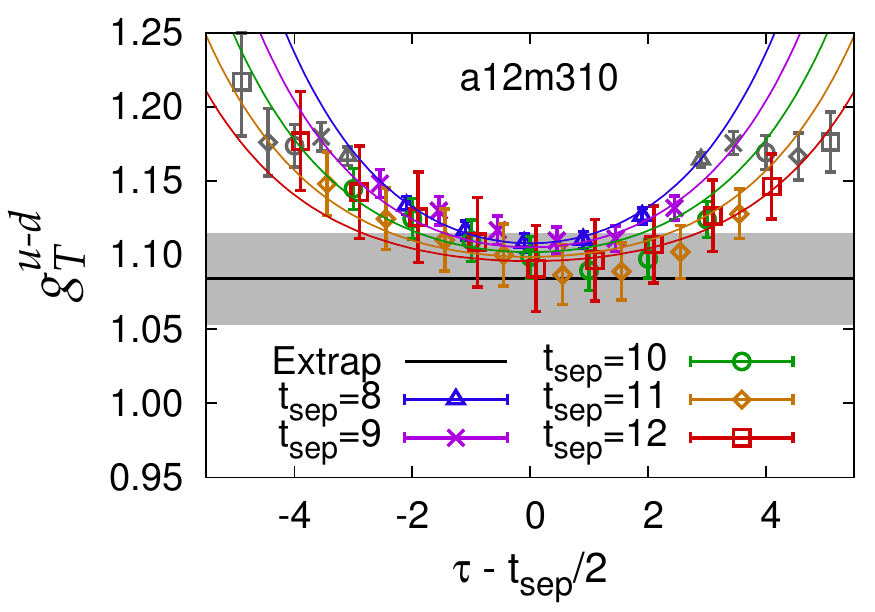}
    \includegraphics[height=1.64in,trim={0.9cm 0.67cm 0 0},clip]{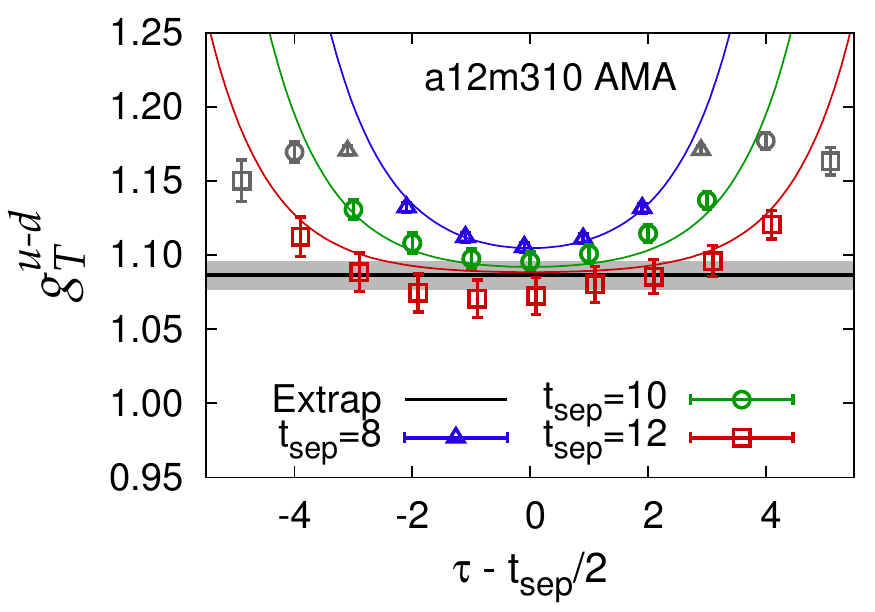}
  \vspace{-0.03\linewidth}
}
  \vspace{-0.03\linewidth}
  \subfigure{
    \includegraphics[height=1.64in,trim={0.1cm 0.67cm 0 0},clip]{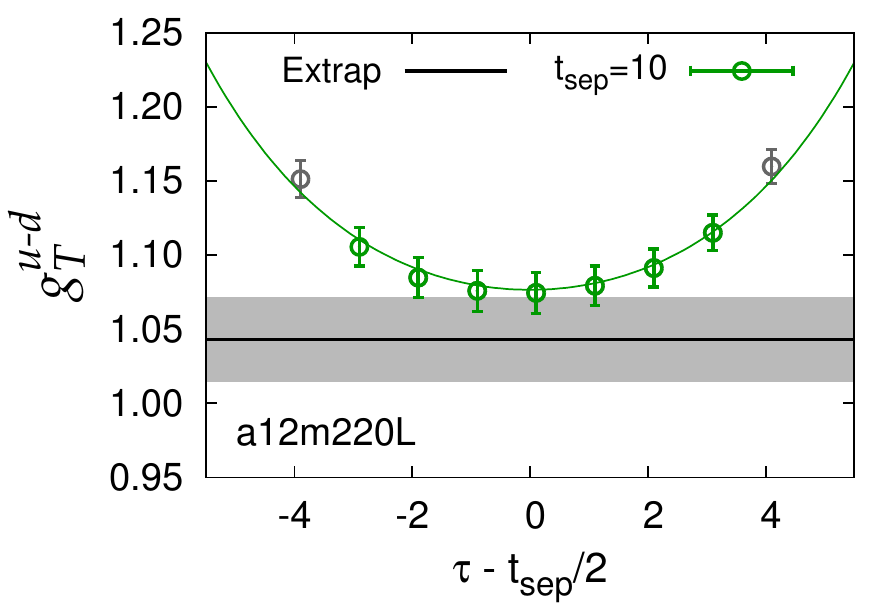}
    \includegraphics[height=1.64in,trim={0.9cm 0.67cm 0 0},clip]{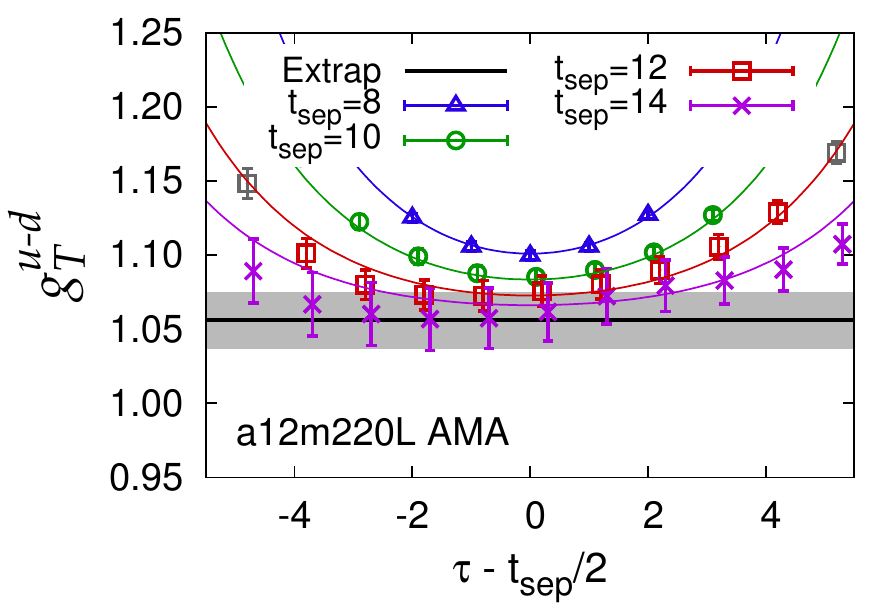}
}
  \vspace{-0.03\linewidth}
  \subfigure{
    \includegraphics[height=1.64in,trim={0.1cm 0.67cm 0 0},clip]{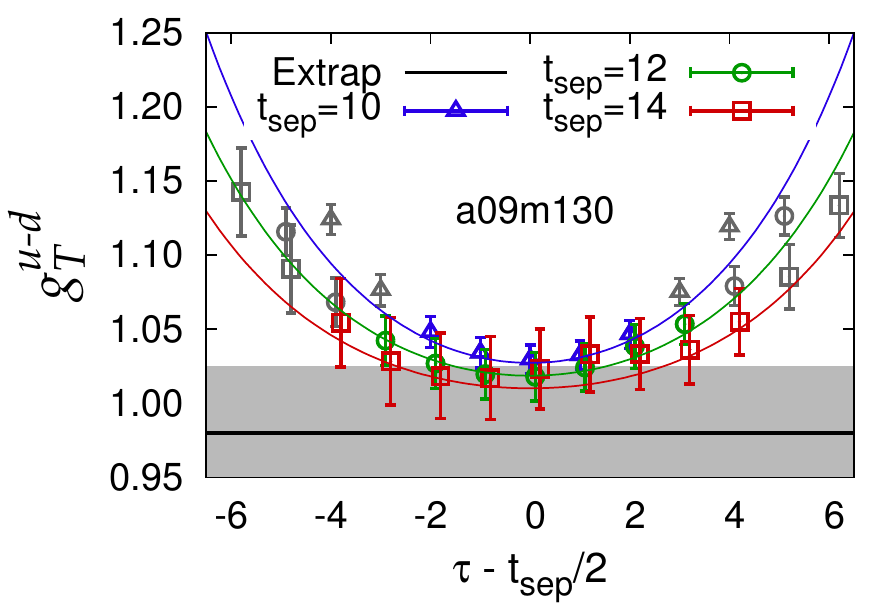}
    \includegraphics[height=1.64in,trim={0.9cm 0.67cm 0 0},clip]{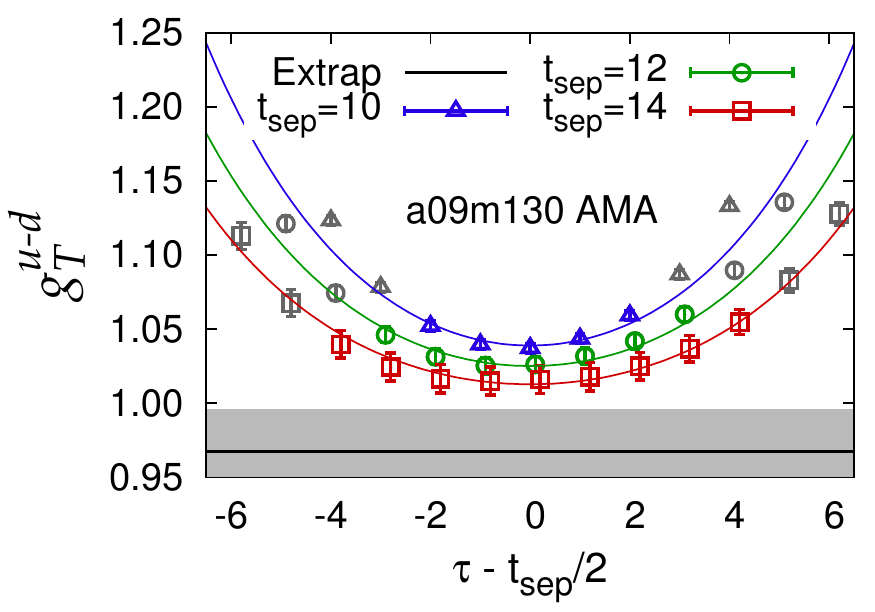}
}
  \vspace{-0.03\linewidth}
  \subfigure{
    \includegraphics[height=1.64in,trim={0.1cm 0.67cm 0 0},clip]{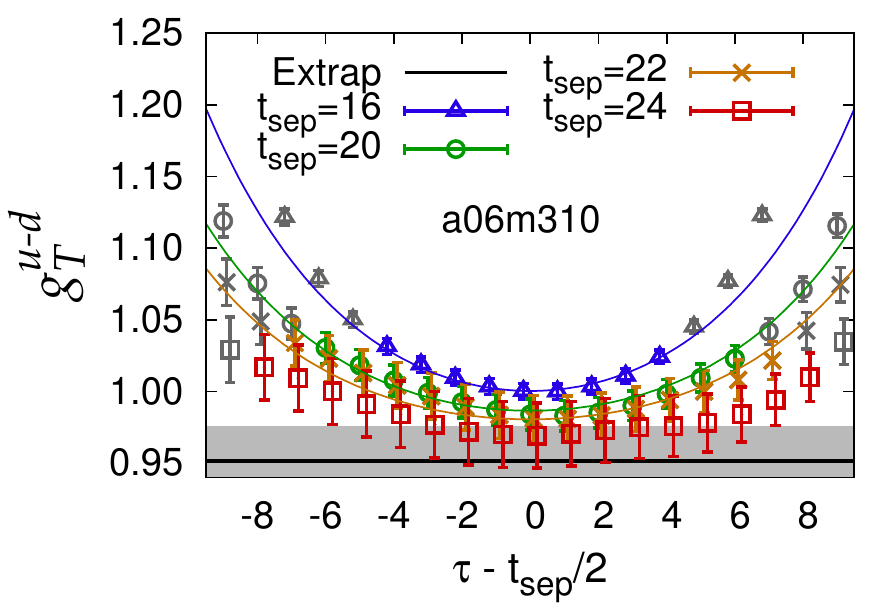}
    \includegraphics[height=1.64in,trim={0.9cm 0.67cm 0 0},clip]{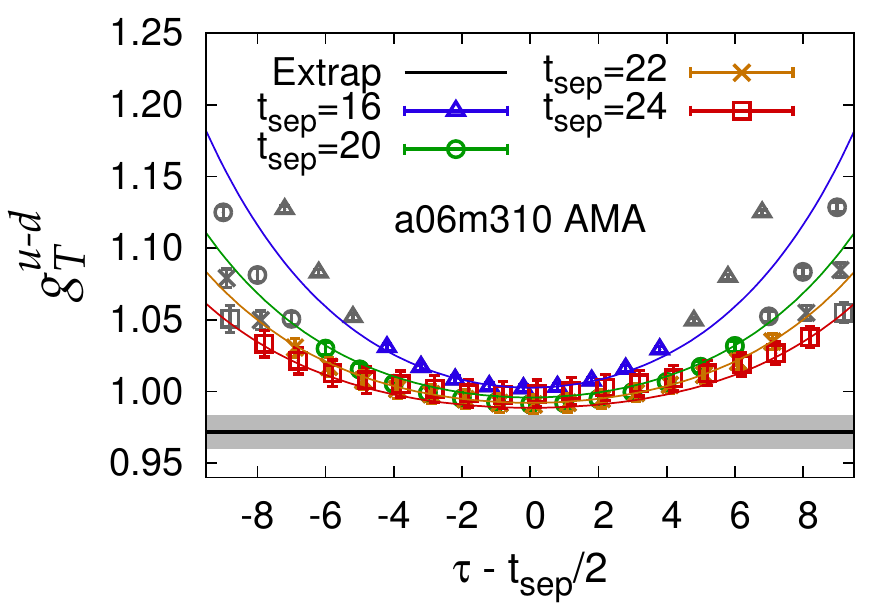}
}
  \vspace{-0.02\linewidth}
  \subfigure{
    \includegraphics[height=1.81in,trim={0.1cm 0.1cm 0 0},clip]{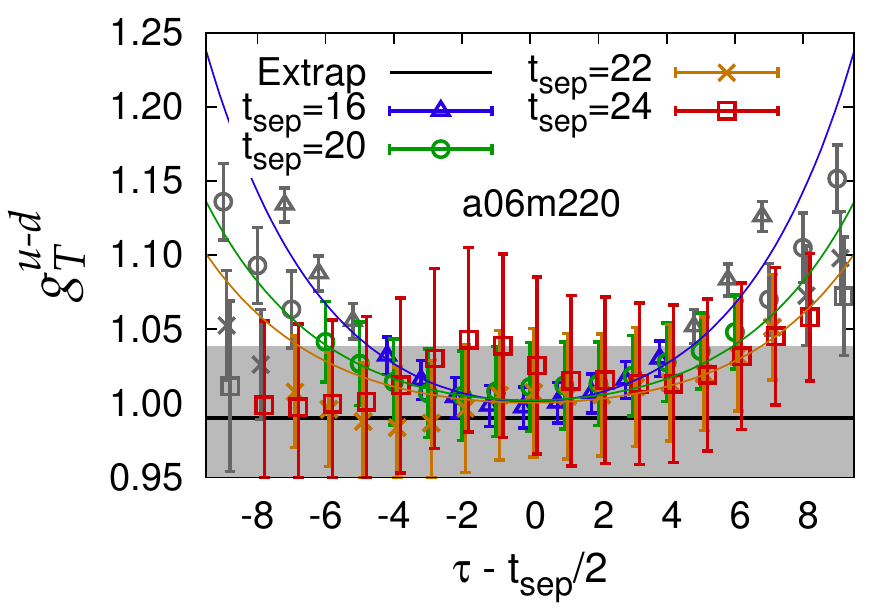}
    \includegraphics[height=1.81in,trim={0.9cm 0.1cm 0 0},clip]{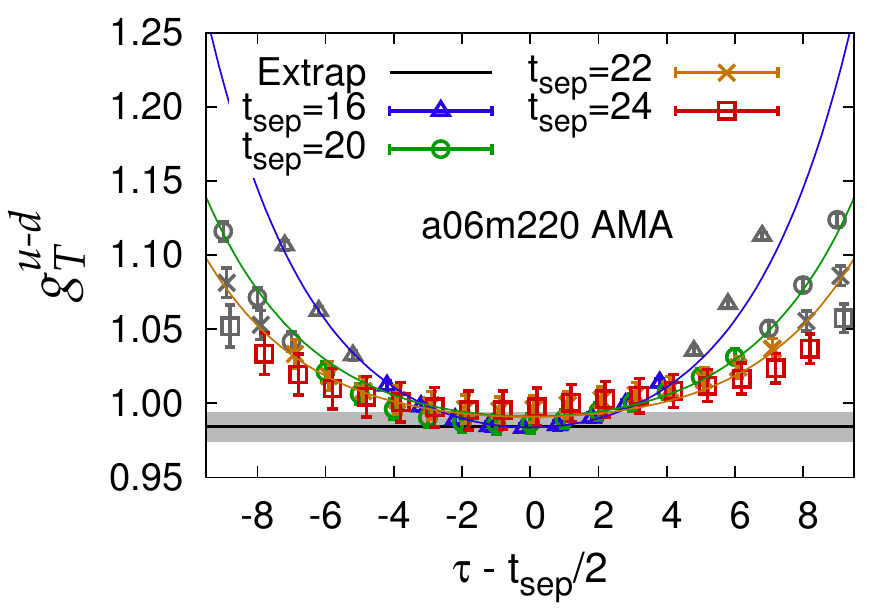}
}
\caption{Comparison of the unrenormalized $g_T^{u-d}$ data obtained using
  all HP measurements (left) with the AMA method (right). The
  rest is the same as in Fig.~\protect\ref{fig:gA_HPversusAMA}. 
  \label{fig:gT_HPversusAMA}}
\end{figure*}

\begin{figure*}[tbp]
  \subfigure{
    \includegraphics[height=1.64in,trim={0.1cm 0.67cm 0 0},clip]{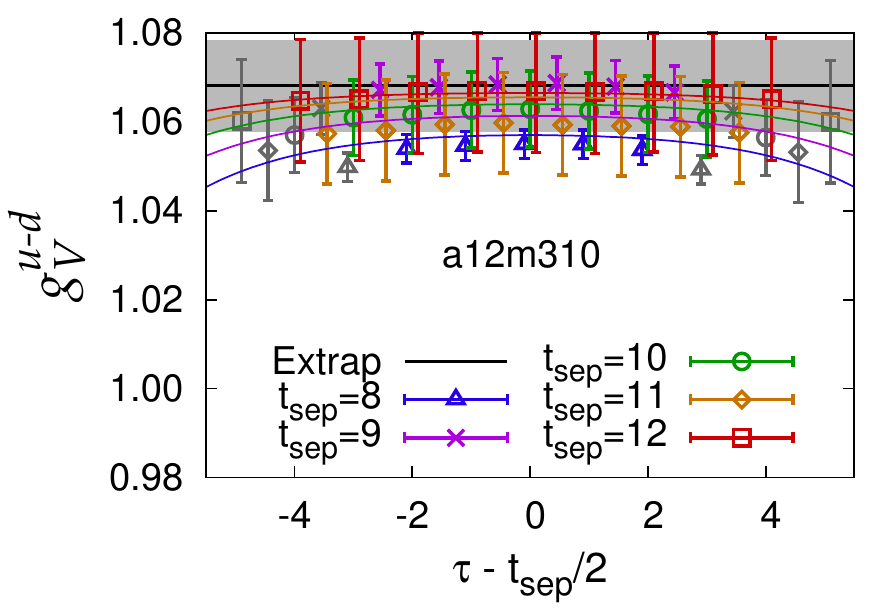}
    \includegraphics[height=1.64in,trim={0.9cm 0.67cm 0 0},clip]{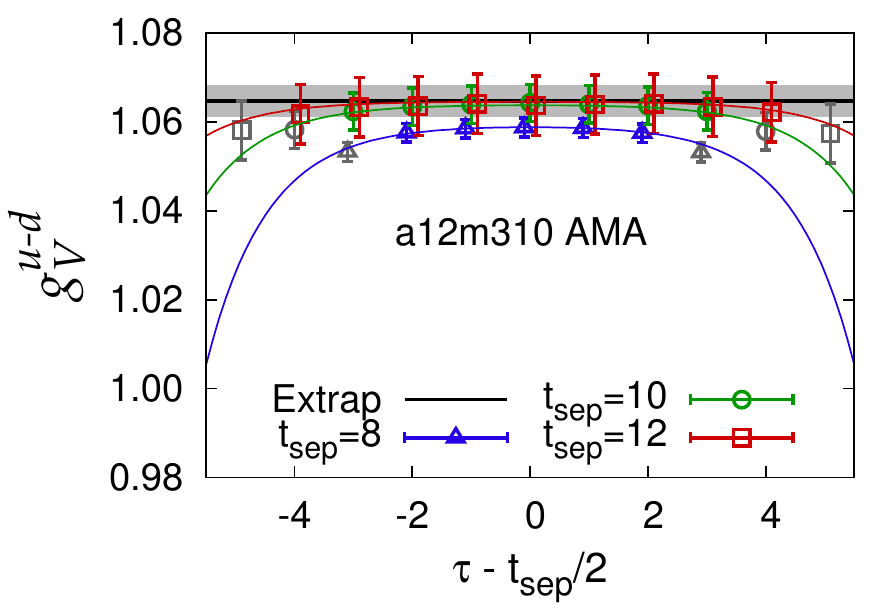}
  \vspace{-0.03\linewidth}
}
  \vspace{-0.03\linewidth}
  \subfigure{
    \includegraphics[height=1.64in,trim={0.1cm 0.67cm 0 0},clip]{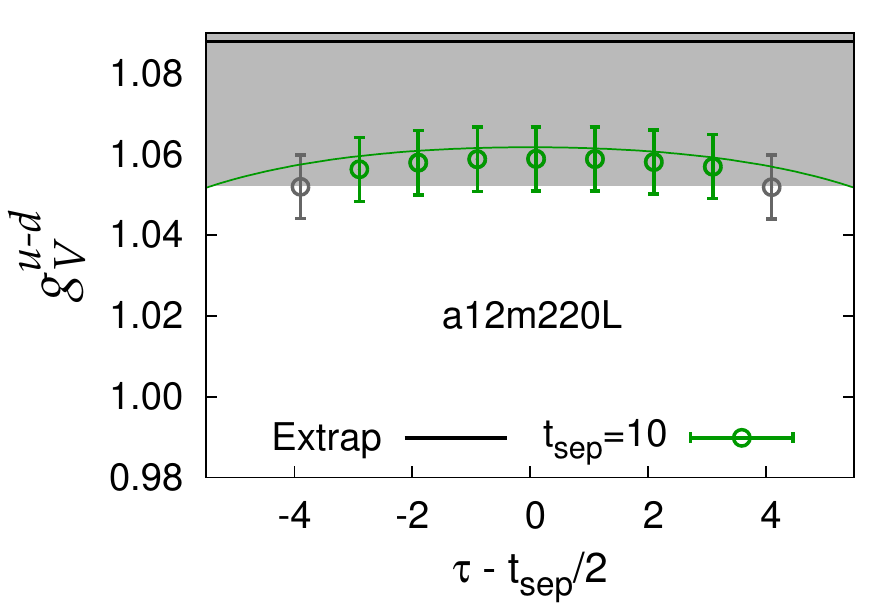}
    \includegraphics[height=1.64in,trim={0.9cm 0.67cm 0 0},clip]{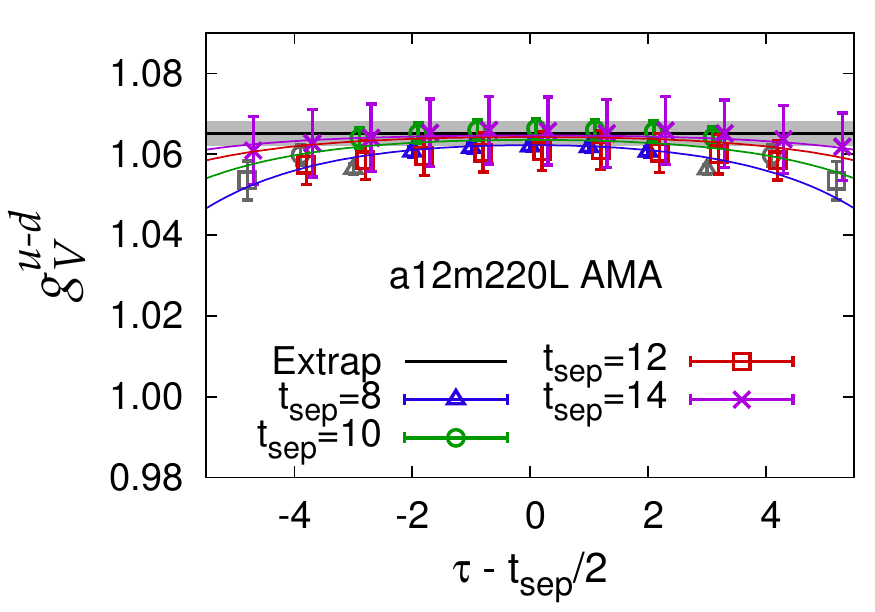}
}
  \vspace{-0.03\linewidth}
  \subfigure{
    \includegraphics[height=1.64in,trim={0.1cm 0.67cm 0 0},clip]{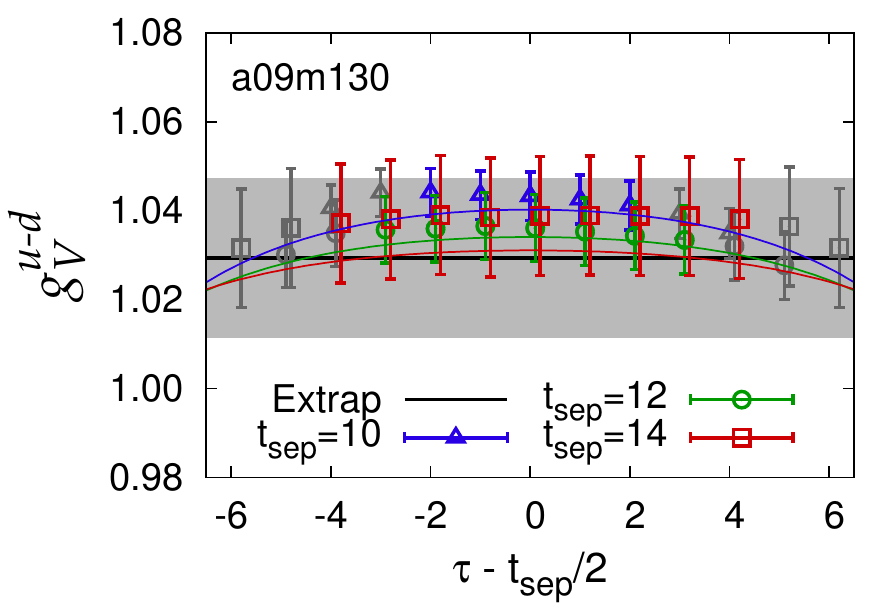}
    \includegraphics[height=1.64in,trim={0.9cm 0.67cm 0 0},clip]{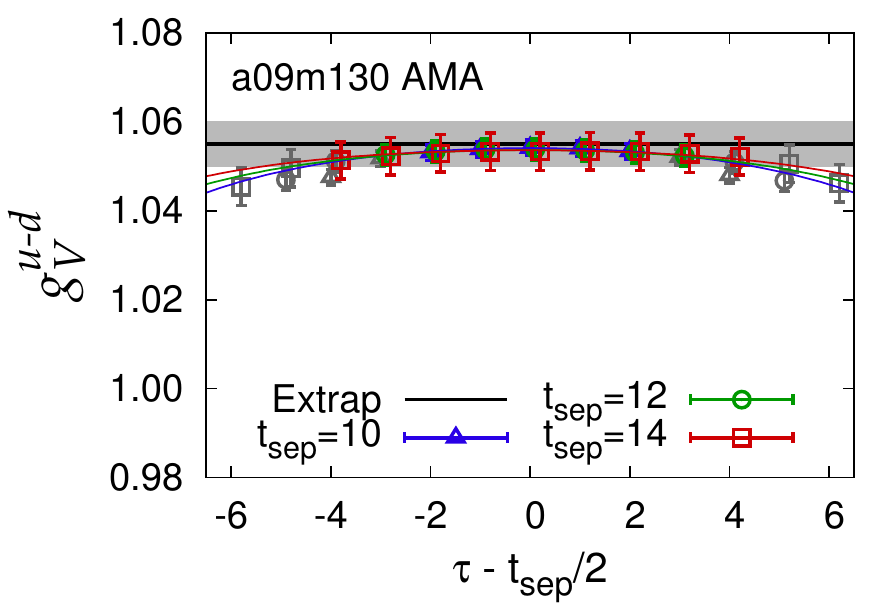}
}
  \vspace{-0.03\linewidth}
  \subfigure{
    \includegraphics[height=1.64in,trim={0.1cm 0.67cm 0 0},clip]{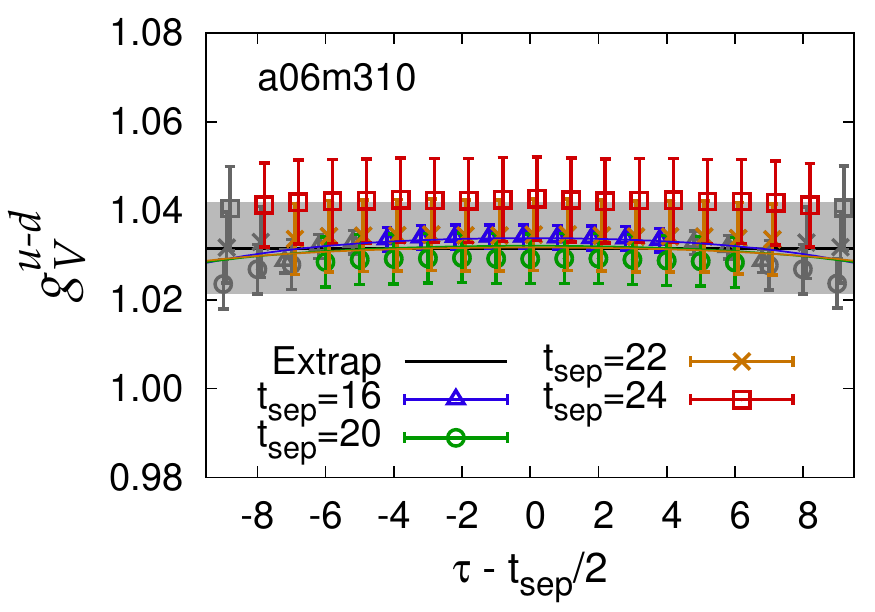}
    \includegraphics[height=1.64in,trim={0.9cm 0.67cm 0 0},clip]{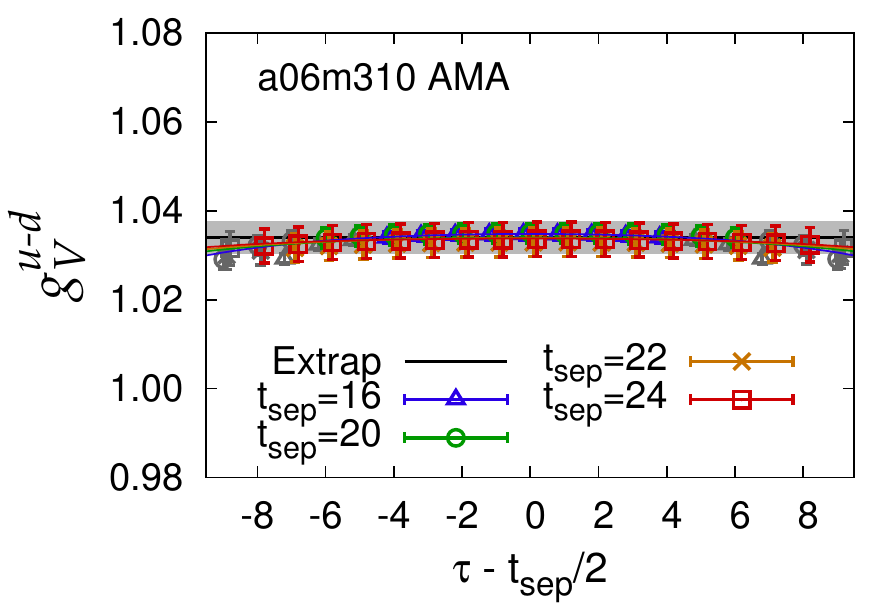}
}
  \vspace{-0.02\linewidth}
  \subfigure{
    \includegraphics[height=1.81in,trim={0.1cm 0.1cm 0 0},clip]{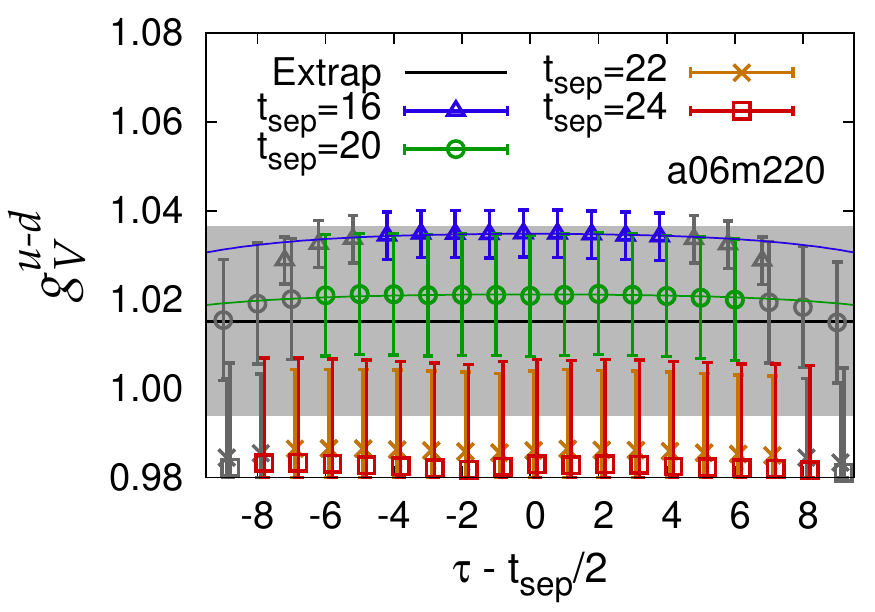}
    \includegraphics[height=1.81in,trim={0.9cm 0.1cm 0 0},clip]{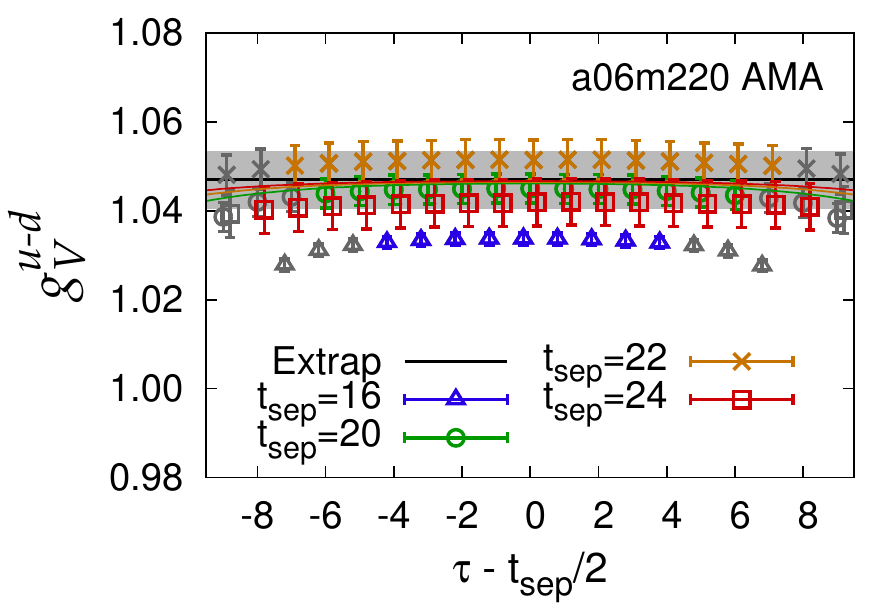}
}
\caption{Comparison of the unrenormalized $g_V^{u-d}$ data obtained using
  all HP measurements (left) with the AMA method (right). The
  rest is the same as in Fig.~\protect\ref{fig:gA_HPversusAMA}. 
  \label{fig:gV_HPversusAMA}}
\end{figure*}

The basic idea of AMA is that, in the low-precision evaluation, the LP
average, first term in Eq.~\eqref{eq:2-3pt_AMA}, may be biased.  The
bias is corrected by the second low-statistics term without
significantly increasing the overall statistical errors. To determine
the LP stopping criteria, we compared LP and AMA results using 50
configurations from the $a12m310$ ensemble with $r_{\rm LP} \equiv
|{\rm residue}|_{\rm LP}/|{\rm source}| = 10^{-2}$, $5 \times
10^{-3}$, $10^{-3}$ and $ 10^{-4}$. All four LP estimates agreed with
the AMA estimate within $1\sigma$.  To reduce computational cost and yet be
conservative, we selected $r_{\rm LP} = 10^{-3}$ for all calculations
presented in this work.

In our current implementation,
17 LP measurements cost the same as one HP when using the multigrid
algorithm for inverting the Dirac matrix~\cite{Babich:2010qb}. Adding
64+4 LP measurements doubled the cost compared to 4 HP measurements,
whereas the increase in statistics is by a factor of 16 in the AMA
analysis with bias correction. Note that only four of the
existing eight HP source positions were used for bias correction as 
these were found to be sufficient. Also, on the $a06m220$ ensemble,
only four HP measurements were made on each configuration.

To test how the errors change with the bias correction, we compare the
results for the charges on the five ensembles analyzed using both HP
and AMA methods.  We compared both HP versus AMA and AMA versus LP
estimates.  A comparison of the HP versus the AMA data for the masses
and amplitudes extracted from the nucleon 2-point function is shown in
Tables~\ref{tab:2ptfits1} and~\ref{tab:2ptfits2}.  Comparison of the
isovector charges $g_A^{u-d}$, $g_S^{u-d}$, $g_T^{u-d}$ and
$g_V^{u-d}$ is shown in Figs.~\ref{fig:gA_HPversusAMA},
~\ref{fig:gS_HPversusAMA}, ~\ref{fig:gT_HPversusAMA}
and~\ref{fig:gV_HPversusAMA}, respectively.  The increase in
statistics with the AMA method (see Table~\protect\ref{tab:ens})
significantly improves the precision of the data at the various
source-sink separations, $t_{\rm sep}$, we have used in the analysis
of the 3-point functions. In each panel of Figs. 1--4, the grey
error band and the solid line within it give the $t_\text{sep} \to
\infty$ value obtained from the 2-state fit using Eqs.~\eqref{eq:2pt}
and~\eqref{eq:3pt}. The two estimates are, in most cases, consistent
and in the rest the difference is less than $2\sigma$.  The AMA data
at each $t_{\rm sep}$ has much smaller errors; consequently, the
$t_{\rm sep} \to \infty$ estimates from the 2-state fit to the AMA
data are more precise.

Comparing the AMA with the LP data, we find that in each case the
difference is a tiny fraction of the statistical error in the AMA
calculation; that is, the bias correction term is negligible. This is
shown in Tables~\ref{tab:2ptfits1} and~\ref{tab:2ptfits2} for the
2-point data and in
Tables~\ref{tab:3ptgAcompare},~\ref{tab:3ptgScompare}
and~\ref{tab:3ptgTcompare} for the matrix elements extracted from the
3-point data.  Thus, our conclusion is that using the multigrid
solver~\cite{Babich:2010qb, Osborn:2010mb} with a low-accuracy
stopping criterion $r_{\rm LP} \equiv |{\rm residue}|_{\rm LP}/|{\rm
  source}| \approx 10^{-3}$, is as good as the HP inversion with
$r_{\rm HP} = 10^{-8}$--$10^{-12}$ for the calculation of the nucleon
charges at the level of statistical precision achieved in this work. 

\begin{table*}
\centering
\begin{ruledtabular}
\begin{tabular}{c|c|ccccc|c}
ID       & Type & Fit Range  & $aM_0$      & $aM_1$     & ${\cal A}_0^2 \times 10^{11}$  & ${\cal A}_1^2 \times 10^{11}$   & ${\cal A}_1^2/{\cal A}_0^2$   \\
\hline
a12m310  & HP   & 2--15     & 0.6669(53) & 1.36(11)  & 6.57(27)  & 6.28(61)  & 0.96(7)     \\
a12m310  & AMA  & 3--15     & 0.6722(22) & 1.64(16)  & 6.95(12)  & 9.5(3.4)  & 1.36(47)    \\
a12m220S & HP   & 2--15     & 0.6233(55) & 1.42(13)  & 6.58(26)  & 6.94(93)  & 1.05(11)    \\
a12m220  & HP   & 2--15     & 0.6232(49) & 1.45(15)  & 6.58(24)  & 6.8(1.1)  & 1.03(14)    \\
a12m220L & HP   & 2--15     & 0.6046(71) & 1.16(12)  & 5.68(37)  & 5.63(51)  & 0.99(6)     \\
a12m220L & LP   & 3--15     & 0.6118(26) & 1.18(7)   & 5.99(15)  & 4.64(50)  & 0.78(7)     \\
\hline                                                                                       
a09m310  & HP   & 4--20     & 0.4943(62) & 0.87(9)   & 13.6(1.1) & 14.4(1.7) & 1.05(8)     \\
a09m220  & HP   & 4--20     & 0.4535(58) & 0.86(8)   & 11.8(9)   & 15.2(2.1) & 1.29(11)    \\
a09m130  & HP   & 3--20     & 0.4186(76) & 0.83(6)   & 9.74(89)  & 17.2(1.0) & 1.76(10)    \\
a09m130  & AMA  & 5--20     & 0.4150(45) & 0.73(5)   & 8.70(59)  & 11.7(8)   & 1.34(6)     \\
\hline                                                                                       
a06m310  & HP   & 5--30     & 0.3219(37) & 0.58(2)   & 0.53(4)   & 1.26(6)   & 2.35(12)    \\
a06m310  & AMA  & 7--30     & 0.3277(18) & 0.59(2)   & 0.59(2)   & 1.10(8)   & 1.86(8)     \\
a06m220  & HP   & 5--30     & 0.3166(66) & 0.64(5)   & 13.0(1.5) & 38.5(5.4) & 2.96(20)    \\
a06m220  & AMA  & 7--30     & 0.3068(17) & 0.63(2)   & 11.3(3)   & 38.5(3.0) & 3.41(18)    \\
\end{tabular}
\end{ruledtabular}
\caption{Estimates of the masses $M_0$ and $M_1$ and the amplitudes
  ${\cal A}_0$ and ${\cal A}_1$ extracted from the fits to the 2-point
  correlation functions using the 2-state Ansatz given in
  Eqs.~\eqref{eq:2pt} and~\eqref{eq:3pt}. We give the results
  of fits to both the HP and the AMA data for the five ensembles $a12m310$,
  $a12m220L$, $a09m130$, $a06m310$ and $a06m220$.  The Gaussian
  smearing parameters, $\{\sigma, N_{\text{GS}}\}$, used in the
  calculation of the 2- and 3-point connected correlation functions
  are given in Table~\protect\ref{tab:cloverparams}.  The fit range
  used is listed as Case 1 in Table~\protect\ref{tab:4cases}.}
  \label{tab:2ptfits1}
\end{table*}

\begin{table*}
\centering
\begin{ruledtabular}
\begin{tabular}{c|c|ccccc|c}
ID       & Type & Fit Range  & $aM_0$      & $aM_1$     & ${\cal A}_0^2 \times 10^{11}$  & ${\cal A}_1^2 \times 10^{11}$   & ${\cal A}_1^2/{\cal A}_0^2$   \\
\hline
a12m310  & HP   & 3--15     & 0.6641(76) & 1.20(17) & 6.38(46) & 4.5(9)    & 0.70(11)    \\
a12m310  & AMA  & 3--15     & 0.6722(22) & 1.64(16) & 6.95(12) & 9.5(3.4)  & 1.36(47)    \\
a12m220S & HP   & 3--15     & 0.6202(94) & 1.20(27) & 6.37(54) & 4.3(1.5)  & 0.67(18)    \\
a12m220  & HP   & 3--15     & 0.6216(73) & 1.27(30) & 6.47(42) & 4.5(2.2)  & 0.69(30)    \\
a12m220L & HP   & 3--15     & 0.597(14)  & 0.96(17) & 5.16(87) & 4.19(33)  & 0.81(14)    \\
a12m220L & LP   & 4--15     & 0.6109(35) & 1.11(13) & 5.92(22) & 3.78(98)  & 0.64(15)    \\
\hline                                                                                     
a09m310  & HP   & 5--20     & 0.4933(78) & 0.84(12) & 13.4(1.5)& 13.0(2.9) & 0.97(15)    \\
a09m220  & HP   & 5--20     & 0.4529(68) & 0.84(12) & 11.7(1.1)& 14.1(3.8) & 1.21(24)    \\
a09m130  & HP   & 4--20     & 0.413(12)  & 0.75(10) & 8.9(1.6) & 14.0(1.4) & 1.56(18)    \\
a09m130  & AMA  & 6--20     & 0.4137(60) & 0.71(7)  & 8.50(85) & 10.8(1.5) & 1.27(9)     \\
\hline                                                                                     
a06m310  & HP   & 6--30     & 0.3190(47) & 0.54(3)  & 0.50(5)  & 1.08(6)   & 2.17(16)    \\
a06m310  & AMA  & 8--30     & 0.3268(23) & 0.56(3)  & 0.58(3)  & 0.95(10)  & 1.66(11)    \\
a06m220  & HP   & 6--30     & 0.3149(84) & 0.61(8)  & 12.5(2.0)& 32.3(7.3) & 2.59(28)    \\
a06m220  & AMA  & 8--30     & 0.3069(18) & 0.63(3)  & 11.3(4)  & 39.2(5.0) & 3.47(35)    \\
\end{tabular}
\end{ruledtabular}
\caption{Same as Table~\protect\ref{tab:2ptfits1} except that the fit
  range used is listed as Case 3 in Table~\protect\ref{tab:4cases}. }
  \label{tab:2ptfits2}
\end{table*}

\begin{table}
\centering
\begin{ruledtabular}
\begin{tabular}{c|c|c|c}
Fit     & $t_{\rm min}$ (2pt HP)  & $t_{\rm min}$ (2pt AMA)  & $\tskip$ (3pt)   \\
\hline
Case 1  & $\{2,4,5\}$ &  $\{3,5,7\}$     & $\{2,3,4\}$ \\
Case 2  & $\{2,4,5\}$ &  $\{3,5,7\}$     & $\{3,4,6\}$ \\
Case 3  & $\{3,5,6\}$ &  $\{4,6,8\}$     & $\{2,3,4\}$ \\
Case 4  & $\{3,5,6\}$ &  $\{4,6,8\}$     & $\{3,4,6\}$ \\
\end{tabular}
\end{ruledtabular}
\caption{The fit parameters $t_{\rm min}$ and $\tskip$ defining the
  four cases used to analyze the 2- and 3-point functions data. The
  parameter $t_{\rm min}$ is the starting value of time used in the
  2-point fit and $\tskip$ is the number of points skipped adjacent to
  the source and sink in the fit to the 3-point data.  The triplets of
  numbers correspond to the three lattice spacings $a=\{0.12,
  0.09,0.06\}$~fm, respectively.  The three exceptions to these fit
  ranges are specified in Tables~\ref{tab:2ptfits1}
  and~\ref{tab:2ptfits2}. The ending time slices in the 2-point fits,
  were chosen to be $t_{\rm max} = \{15, 20, 30\}$ in all
  cases. }
\label{tab:4cases}
\end{table}

\begin{figure}
\centering
    \includegraphics[width=1.05\linewidth]{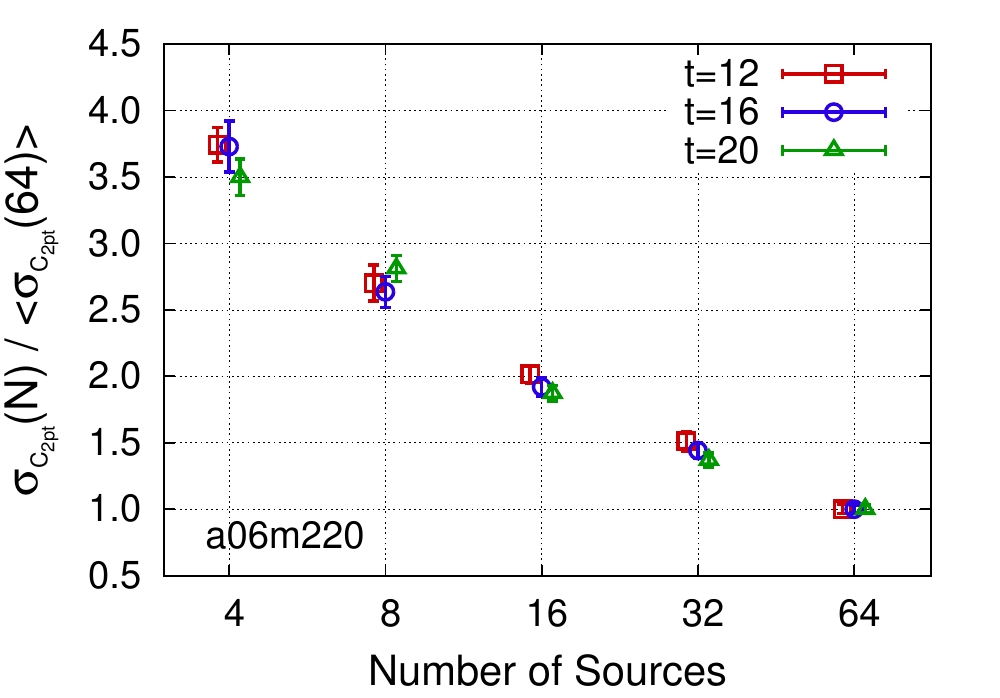}
    \includegraphics[width=1.05\linewidth]{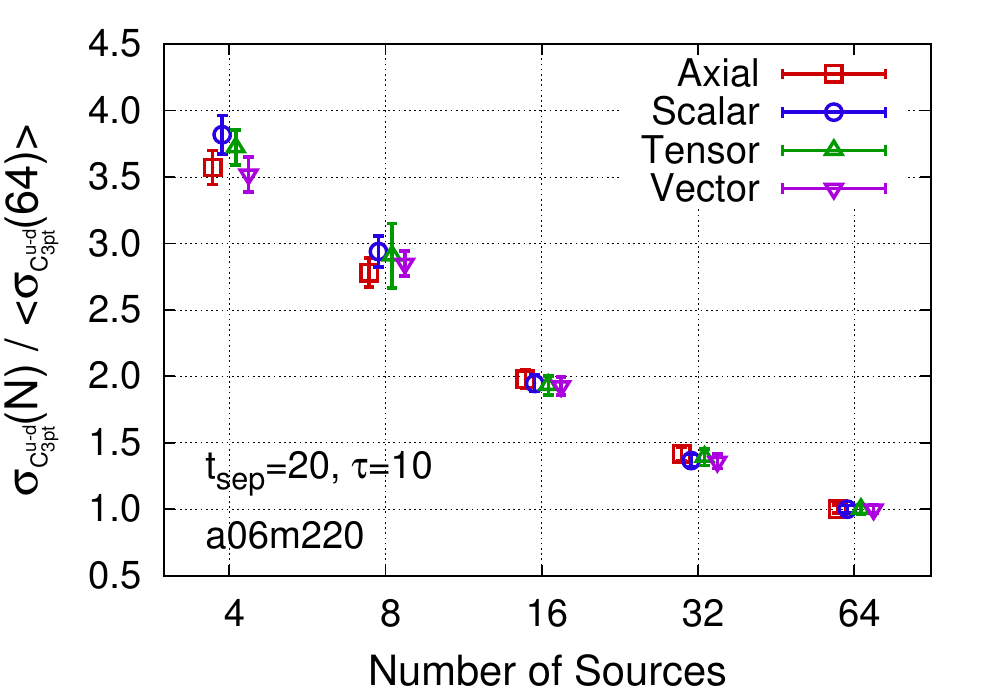}
\caption{Illustration of the decrease in the errors in the 2- and
  3-point functions calculated on the $a06m220$ lattices with the
  number $N$ of LP measurements made per configuration. (Top) The
  2-point nucleon correlation function at time separations $12,
  \ 16,\ 20$. (Bottom) The data for the four charges at the midpoint
  $\tau=10$ of the $\tsep=20$ calculation. The errors are normalized
  using the $N = 64$ estimates and decrease by a factor of about $
  3.7$ between $N = 4$ and 64 measurements compared to the expected
  factor of 4 for uncorrelated data.
  \label{fig:ErrorScaling}}
\end{figure}

Our final errors are obtained using a single elimination jacknife
analysis over the configurations; that is, we first construct the average 
defined in Eq.~\eqref{eq:2-3pt_AMA} on each configuration. Because of this ``binning'' of
the data, we do not need to correct the jacknife estimate of the
error for correlations between the 64 LP measurements per
configuration. It is, however, useful to determine the number of source
positions one can place on each configuration beyond which 
the additional computational cost offsets the gain in 
statistics. The following tests indicate that the
correlations between estimates from various sources are reasonably
small with up to $O(100)$ sources per configuration on lattices with
$M_\pi L\, \gsim 4$ and $M_\pi T\, \gsim 8$.
\begin{itemize}
\item
Comparing the errors in the estimates for masses and amplitudes given
in Tables~\ref{tab:2ptfits1} and~\ref{tab:2ptfits2}, we find that the
AMA errors are a factor of 2--4 smaller than those in the all HP
analysis. (In the limit of no correlations, the scaling factor 
should be $\sqrt{64/8} = 2.83$.) 
A similar improvement is seen in the matrix elements as
shown in Tables~\ref{tab:3ptgAcompare},~\ref{tab:3ptgScompare}
and~\ref{tab:3ptgTcompare}. In most cases, the improvement is observed
to become better with decreasing quark mass and lattice spacing.
\item
Figure~\ref{fig:ErrorScaling} illustrates that the errors decrease by
a factor of about $3.7$ when the number of LP sources are increased
from 4 to 64 on the $a06m220$ lattices.  The gain is similar in both
the 2- and 3-point functions.
\item
On the {\it a06m310} ensemble, we compared estimates of the 2- and
3-point correlation functions using 64 and 128 LP sources during the
study of the disconnected diagram contribution reported in
Ref.~\cite{Bhattacharya:2015wna}.  We found that the errors are
reduced by only a factor of $ 1.2$. A similar comparison of 2- and
3-point functions obtained using 64 and 128 LP sources in the
$a06m310$ AMA2 study discussed in Sec.~\ref{sec:confirmation} shows
that errors reduce by only a factor of $1.15$. Thus, both tests
indicate that the gain in statistics becomes small when more the 64 LP
sources per configuration are used on lattices with $M_\pi L \sim 4$.
\end{itemize}

A second technique used to reduce correlations in the AMA calculations
is to choose the 64 LP source points randomly within and between
configurations.  On each configuration, the four source time slices are
chosen randomly to be $r + \alpha T/4$ with $r$ a random integer $ \in
\{1,T/4\}$ and $\alpha \in \{0,1,2,3\}$. On the $a06m220$ ensemble,
each of these four time slices is then divided into 16 boxes of size
$L/2 \times L/2 \times L/4$ and a random source point is chosen
within each of these boxes. On the rest of the ensembles, the 16
source points on each of the 4 time slices are chosen randomly. In the
HP calculation, the 8 source points were taken to be the same on all
the configurations. A statistical analysis comparing the HP and AMA
data shows that choosing the source points randomly reduces the final
errors. For example, the standard error in the mean constructed by
choosing a single source point per configuration is smaller for the LP
data with randomly selected source point versus the HP data in which
the source point was fixed to be the same on all the configurations.

To reduce computational cost, we have used the coherent sequential
source method~\cite{Bratt:2010jn}.  Four sequential sources on the
four sink time slices were calculated and held in computer memory. These 
were added and a single inversion performed to construct the coherent
sequential propagator. Details of the method are given in
Ref.~\cite{Yoon:2016dij}, where we also showed that this method does
not increase the bias or the errors. 

To estimate errors, we performed both correlated and uncorrelated fits
to the nucleon 2- and 3-point functions data.  The final statistical
errors were calculated using a single elimination jackknife method
with uncorrelated fits to both the 2- and 3-point functions since
correlated fits were not stable for the 3-point functions in all
cases. In all cases in which the correlated fits to the 3-point
functions were stable under changes in the fit ranges and had
$\chi^2 \le 1$, the correlated and uncorrelated fits gave
overlapping estimates.

Having high-statistics data with the AMA method significantly improves
the analysis.  For example, the plateaus in the effective-mass plots
extend to larger Euclidean time as discussed in
Sec.~\ref{sec:excited}. This allowed fits to the 2-point AMA data to
be made using a later starting time slice, $t_{\rm min}$, thus reducing
contributions from excited states in the extraction of the masses and
amplitudes as shown in Figs.~\ref{fig:Meffa09m130} and
~\ref{fig:Meffa12m220}. Similarly, the improvement in the estimates of
the matrix elements from the 3-point functions is shown in
Figs.~\ref{fig:gA_HPversusAMA}, ~\ref{fig:gS_HPversusAMA},
~\ref{fig:gT_HPversusAMA} and~\ref{fig:gV_HPversusAMA}.

To summarize, we find that the AMA method with $O(100)$ randomly
selected source positions on each configuration is a very
cost-effective way to increase the statistics significantly and
consequently improve the quality of the fits used to extract the
estimates for the charges on each ensemble.  Having these estimates
with small errors improves the quality of the final fits made to
obtain results in the continuum limit and at the physical light-quark
mass as discussed in Sec.~\ref{sec:results}.

\section{Excited-State Contamination}
\label{sec:excited}

Nucleon charges are given by the matrix elements of the bilinear quark
operators between ground-state nucleons. The lattice operator $\chi$
given in Eq.~\eqref{eq:nucl_op}, however, couples to the nucleon, its
excitations and multiparticle states with the same quantum numbers. We
incorporate three strategies to reduce excited-state contamination:
\begin{itemize}
\item
The overlap between the nucleon operator and the excited states is
reduced by using tuned smeared sources when calculating the quark
propagators. We construct gauge-invariant Gaussian smeared sources by
applying the three-dimensional Laplacian operator $\nabla^2$ a
fixed number of times $N_{\rm GS}$, $(1 -
\sigma^2\nabla^2/(4N_{\rm GS}))^{N_{\rm GS}}$.  The smearing
parameters $\{\sigma, N_{\rm GS}\}$ for each ensemble are given in
Table~\ref{tab:cloverparams} and are the same as in
Ref.~\cite{Bhattacharya:2015wna} in order to avoid repeating the
expensive HP calculation needed for the AMA analysis. Also, the same smearing
is used at the source and sink points to construct the smeared-smeared
2- and 3-point correlation functions.
\item
We calculate the 3-point correlation functions for a number of
values of the source-sink separation $t_{\rm sep}$ given in
Table~\ref{tab:ens}. We fit the data at all $t_{\rm sep}$
simultaneously using the 2-state Ansatz given in Eq.~\eqref{eq:3pt} to 
estimate the $t_{\rm sep} \to \infty$ value. 
\item
We include one excited state in the analysis of the 2-point and
3-point functions as given in Eqs.~\eqref{eq:2pt} and~\eqref{eq:3pt}. We find
that this Ansatz has enough free parameters to fit the data and the
additional five parameters needed to include a second excited state
would be very poorly determined.
\end{itemize}

The 2-state Ansatz used to fit the 2- and 3-point functions is, 
\begin{align}
C^\text{2pt}
  &(t_f,t_i) = \nonumber \\
  &{|{\cal A}_0|}^2 e^{-M_0 (t_f-t_i)} + {|{\cal A}_1|}^2 e^{-M_1 (t_f-t_i)}\,, 
\label{eq:2pt}  \\
C^\text{3pt}_{\Gamma}&(t_f,\tau,t_i) = \nonumber\\
  & |{\cal A}_0|^2 \langle 0 | \mathcal{O}_\Gamma | 0 \rangle  e^{-M_0 (t_f-t_i)} +{}\nonumber\\
  & |{\cal A}_1|^2 \langle 1 | \mathcal{O}_\Gamma | 1 \rangle  e^{-M_1 (t_f-t_i)} +{}\nonumber\\
  & {\cal A}_0{\cal A}_1^* \langle 0 | \mathcal{O}_\Gamma | 1 \rangle  e^{-M_0 (\tau-t_i)} e^{-M_1 (t_f-\tau)} +{}\nonumber\\
  & {\cal A}_0^*{\cal A}_1 \langle 1 | \mathcal{O}_\Gamma | 0 \rangle  e^{-M_1 (\tau-t_i)} e^{-M_0 (t_f-\tau)} ,
\label{eq:3pt}
\end{align}
where $\tau$ is the time at which the operator is inserted and $t_f
-t_i = t_\text{sep}$ in the 3-point function calculation.  The states
$|0\rangle$ and $|1\rangle$ represent the ground and ``first excited''
nucleon states, respectively. To extract the charges $g_A^{u-d}$,
$g_S^{u-d}$ and $g_T^{u-d}$, we only need to examine the real part of
the correlation functions with the operator insertion at zero momentum,
in which case ${\cal A}_0$ and ${\cal A}_1$ are real and the matrix
element $\langle 0 | \mathcal{O}_\Gamma | 1 \rangle = \langle 1 |
\mathcal{O}_\Gamma | 0 \rangle$.  Thus, we need to extract seven
parameters from fits to the 2- and 3-point functions.  The four
parameters, $M_0$, $M_1$, ${\cal A}_0$ and ${\cal A}_1$ are estimated
first from the 2-point data and are then used as inputs in the
extraction of matrix elements from fits to the 3-point data.  Fits to
both 2- and 3-point data are done within the same jacknife process to
take into account correlations.

Five of the seven parameters, $M_0$, $M_1$ and the three matrix
elements $\langle 0 | \mathcal{O}_\Gamma | 0 \rangle$, $\langle 0 |
\mathcal{O}_\Gamma | 1 \rangle $ and $ \langle 1 | \mathcal{O}_\Gamma
| 1 \rangle$ are physical once the discretization errors and higher
excited-state contaminations have been removed from them.  The
amplitudes ${\cal A}_0$ and ${\cal A}_1$ depend on the choice of the
interpolating nucleon operator and the smearing parameters used to
generate the smeared sources.  It is clear from Eqs.~\eqref{eq:2pt} and~\eqref{eq:3pt}
that the ratio of the amplitudes, ${\cal A}_1/{\cal A}_0$, is the
quantity to minimize in order to reduce excited-state contamination
since it determines the relative strength of the overlap of the
nucleon operator with the first excited state.  Estimates of $\langle
0 | \mathcal{O}_\Gamma | 1 \rangle $, $ \langle 1 | \mathcal{O}_\Gamma
| 1 \rangle$, the mass gap $M_1-M_0$, and the ratio ${\cal A}_1/{\cal
  A}_0$ can be used, {\it post facto}, to bound the size of the
excited-state contamination for a given source-sink separation $t_{\rm
  sep}$.

Results of the 2-state fits to the 2-point functions needed to extract
$M_0$, $M_1$, ${\cal A}_0$ and ${\cal A}_1$ are consistent within
errors for two sets of starting time slices $t_{\rm min}$ as shown in
Tables~\ref{tab:2ptfits1} and~\ref{tab:2ptfits2}.  The $\chi^2/$d.o.f. of
the correlated fits are also similar for the two fit ranges.  These
fit ranges are specified in Table~\ref{tab:4cases}.

Higher statistics with the AMA method allow us to fit the data with
larger $t_{\rm min}$. The HP and AMA estimates for $M_0$ and ${\cal
  A}_0$ with different fit ranges are consistent for all the ensembles
and are independent of $t_{\rm min}$. This is expected since these
ground-state estimates are sensitive only to the data at large $t$ and that
is the same in both choices of the fit ranges.  We also note that the
scaling of the errors is consistent with the na\"{\i}ve expectation,
$\sqrt{N_{HP}/N_{LP}}$.

\begin{figure}
\centering
    \includegraphics[width=1.04\linewidth]{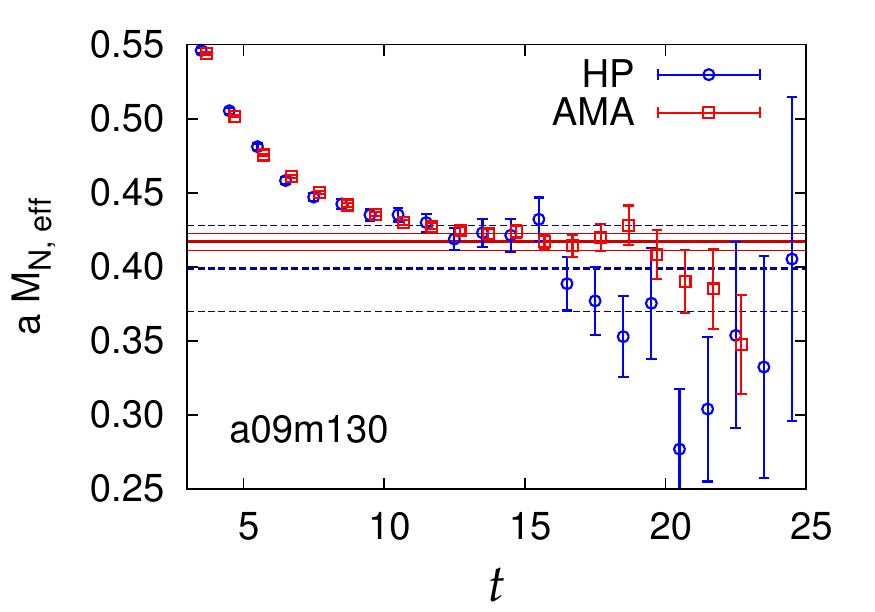}
\caption{Effective-mass plot for the nucleon on the $a09m130$
  ensemble.  The solid line and the error band are the estimate of
  $M_0$ from the fits to the AMA (red squares) data. The dashed blue line and
  the error band give the HP (blue circles) estimate.  The fit ranges used
  correspond to Case 3 defined in Table~\protect\ref{tab:4cases}.
  Results for all the nine ensembles are given in
  Table~\ref{tab:2ptfits2}.
  \label{fig:Meffa09m130}}
\end{figure}

We illustrate the improvement with the AMA method compared to just the
HP analysis in Fig.~\ref{fig:Meffa09m130} using the $a09m130$ data fit
with $t_{\rm min}=6$.  The plateau in the effective-mass $M_{\rm
  eff}(t+0.5) = - \log \{C^{\rm 2pt}(t+1)/C^{\rm 2pt}(t)\}$, where
$C^{\rm 2pt}(t)$ is the 2-point nucleon correlation function at time
separation $t$, extends about four time slices further with the AMA
data.  In the limit $t \to \infty$, $M_{\rm eff}(t)$ converges to the
ground-state mass $M_0$. A comparison of the error estimates in both
the data points and in the result of the fit shows that the extraction
of the masses and the amplitudes improves very significantly with the
AMA method.

\begin{figure}
\centering
    \includegraphics[width=1.05\linewidth]{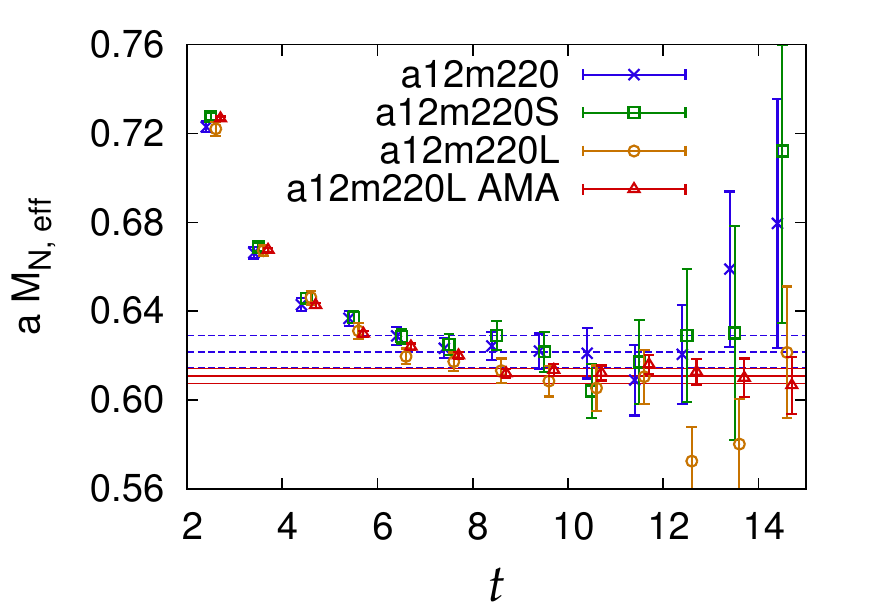}
\caption{Effective-mass plots for the nucleon on the $a12m220S$,
  $a12m220$, $a12m220L$ and $a12m220L$ with AMA ensembles.  The two
  bands with errors are the results for $M_0$ from the 2-state fit for
  the $a12m220$ (blue dashed line) and $a12m220L$ with AMA (red solid line)
  ensembles.  These two estimates differ by a combined $1\sigma$. The
  estimates of the masses and the amplitudes for all four calculations
  are summarized in Table~\ref{tab:2ptfits2}.  The fit ranges used
  correspond to Case 3 defined in Table~\protect\ref{tab:4cases}.
  \label{fig:Meffa12m220}}
\end{figure}

The data in Tables~\ref{tab:2ptfits1} and~\ref{tab:2ptfits2} show a
$2\sigma$ difference in the extraction of $M_0$ from the four
ensembles $a12m220S$, $a12m220$, $a12m220L$ (HP) and $a12m220L$ with
AMA\footnote{We checked for bias in the $a12m220L$ data using the
  $t_{\rm sep}=10$ HP data and found it to be negligible as in the
  other four AMA studies.  Assuming that this is true at other values of 
$\tsep$, we include the $a12m220L$ LP data as part of the AMA discussion.}. The
effective-mass data for the four calculations are shown in
Fig.~\ref{fig:Meffa12m220}.  The errors in the $a12m220S$, $a12m220$
data are large and the data show a noisy plateau over the range $7 \le
t \le 10$. As a result, the 2-state fit is sensitive to the choice of
$t_{\rm min}$.  The higher statistics and larger volume $a12m220L$ AMA
data show a plateau over the interval $9 \le t \le 15$ and the fit is
stable under changes in $t_{\rm min}$.  Because of this difference in
the quality of the data, it is difficult to assess if the $t \to
\infty$ estimates have been obtained in the $a12m220S$ and $a12m220$
data.  As a result, we are not able to determine what fraction of the
difference in $M_0$ is due to finite-volume effects and how much is
due to statistics.  

The quantities that are harder to extract and are sensitive to the
choice of $t_{\rm min}$ are $M_1$ and ${\cal A}_1$. In most cases,
results of fits with larger $t_{\rm min}$ (Case 3 versus Case 1 in
Table~\ref{tab:4cases}) yield smaller values for $M_1$ and ${\cal
  A}_1$. This is expected since, within a 2-state fit Ansatz, these
parameters effectively capture the contributions of all the higher
states and fitting with larger $t_{\rm min}$ uses data with relatively
less higher-states contamination. In practice, one has to compromise
between using data with smaller errors at smaller $t_{\rm min}$ and
picking a large $t_{\rm min}$ to reduce higher excited-state
contamination. We have picked the second set with the larger $t_{\rm
  min}$ for presenting our final results. Note that the error
estimates are slightly larger with this choice.

\begin{figure*}[tb]
\centering
%  \begin{flushleft}                                                                                                                                        
  \subfigure{
    \includegraphics[height=1.7in,trim={0.095cm 0.11cm 0 0},clip]{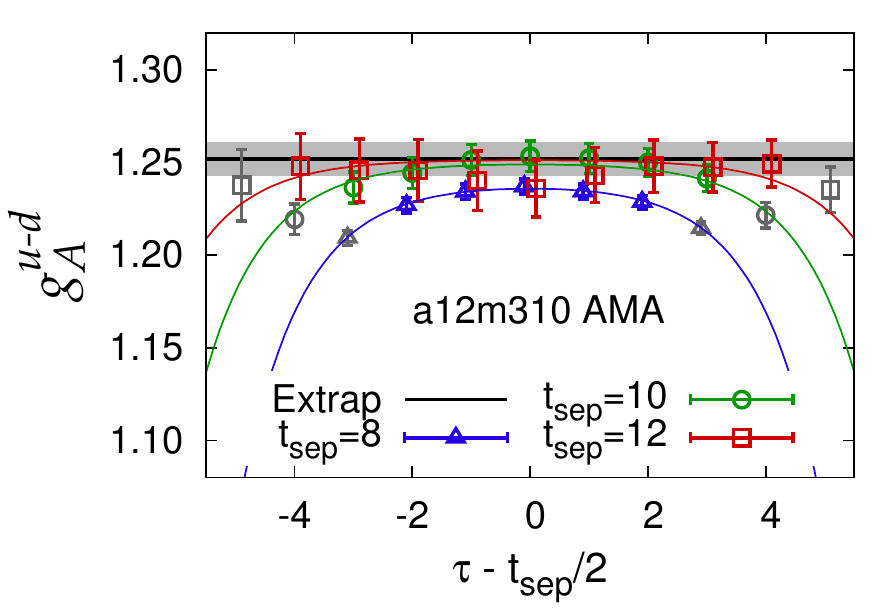}
    \includegraphics[height=1.7in,trim={0.9cm 0.11cm 0 0},clip]{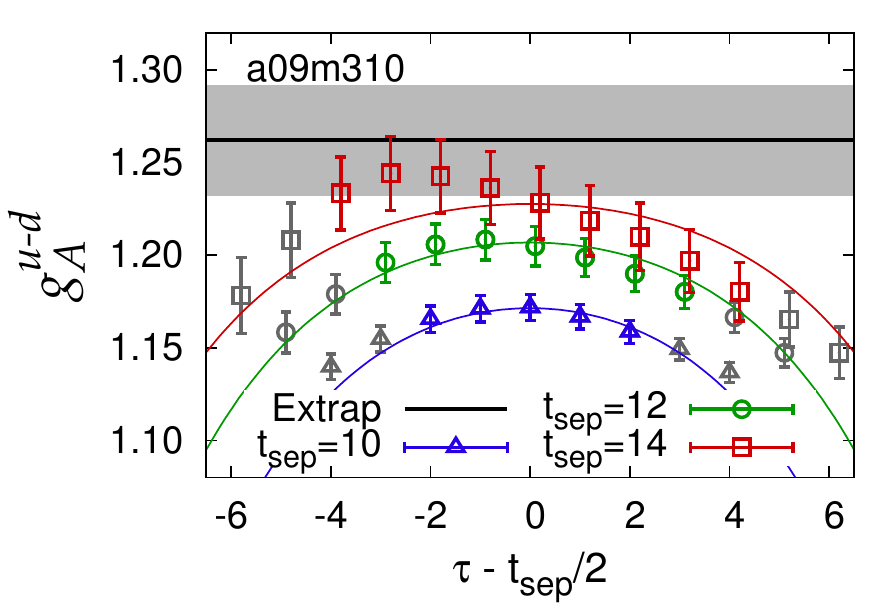}
    \includegraphics[height=1.7in,trim={0.9cm 0.11cm 0 0},clip]{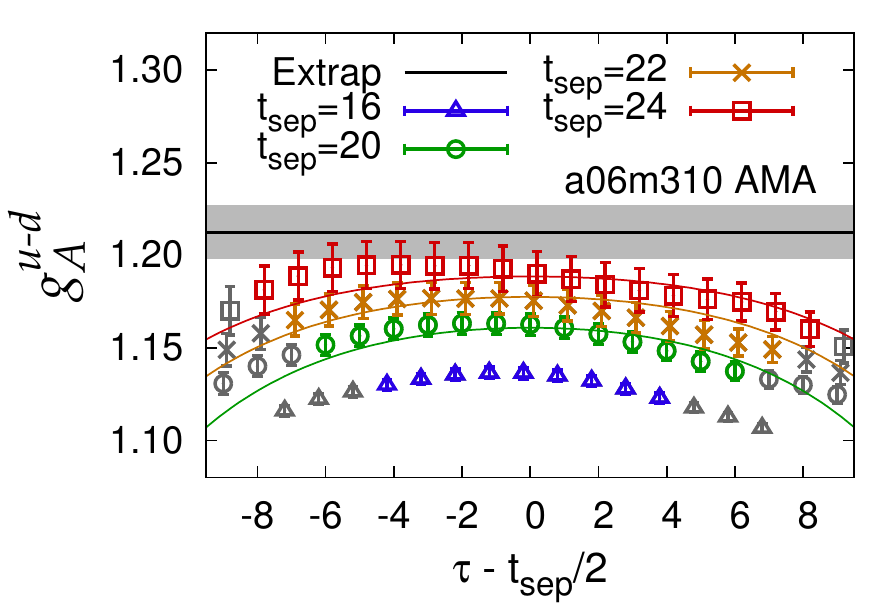}
  }
  \hspace{0.04\linewidth}
  \subfigure{
    \includegraphics[height=1.7in,trim={0.095cm 0.11cm 0 0},clip]{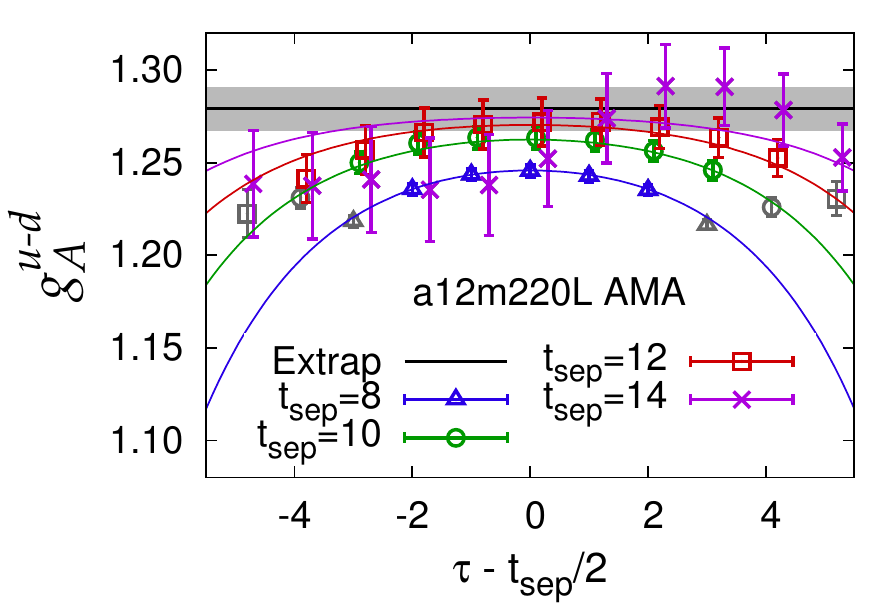}
    \includegraphics[height=1.7in,trim={0.9cm 0.11cm 0 0},clip]{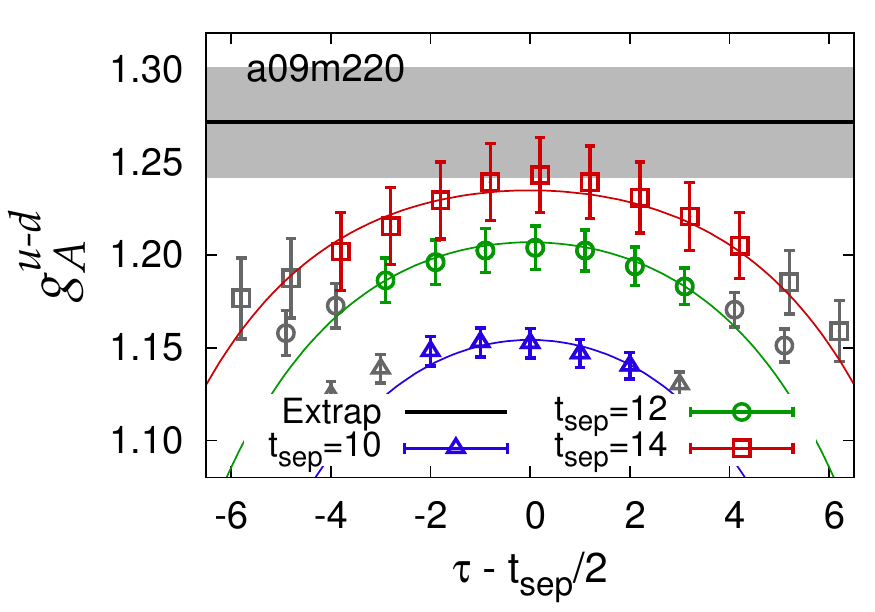}
    \includegraphics[height=1.7in,trim={0.9cm 0.11cm 0 0},clip]{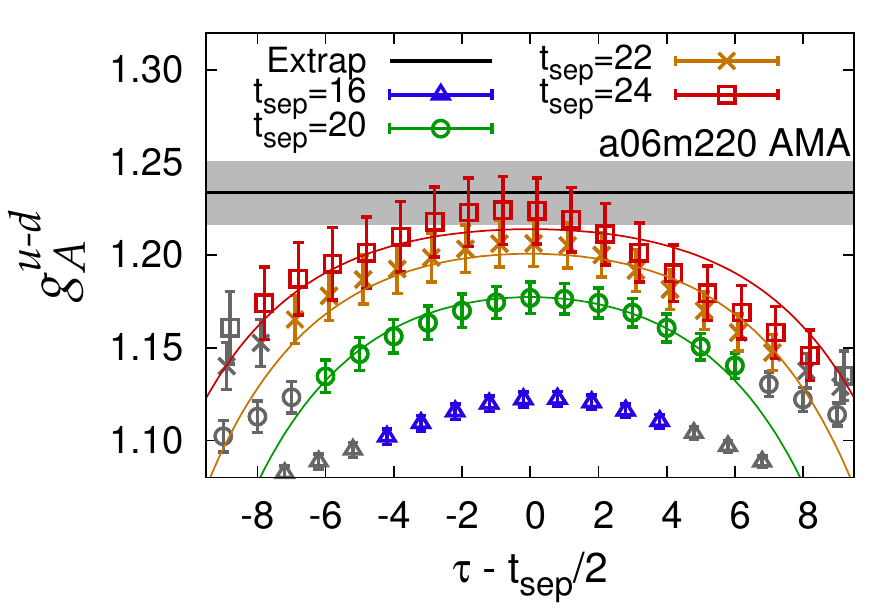}
  }
  \hspace{0.04\linewidth}
  \subfigure{
    \includegraphics[height=1.7in,trim={0.095cm 0.11cm 0 0},clip]{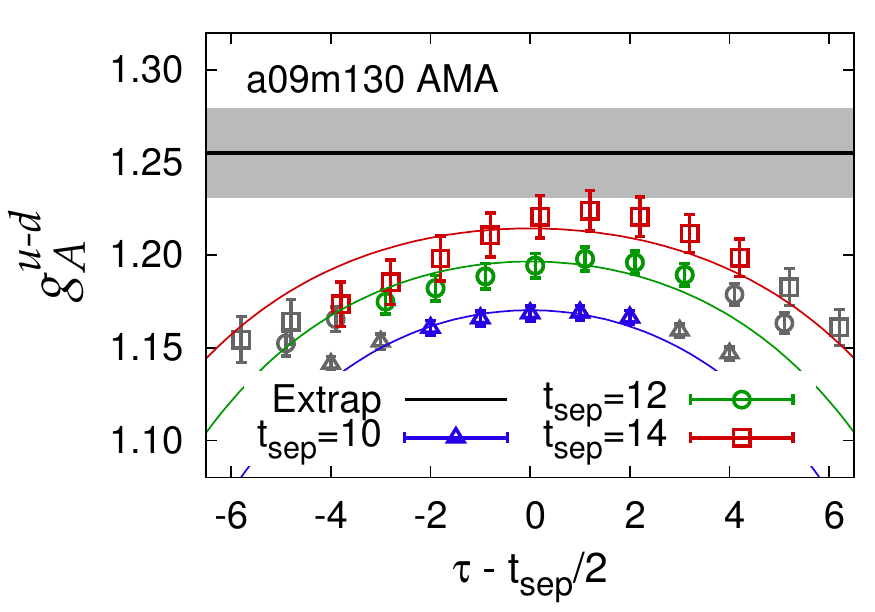}
  }
%  \end{flushleft}
%%    \includegraphics[viewport=60 380 570 740,clip,width=0.98\linewidth]{figs/Excited_states_gA}
\caption{The 2-state fit to the unrenormalized axial charge
  $g_A^{u-d}$ data for the seven ensembles at different values of the
  lattice spacing and pion mass. The grey
  error band and the solid line within it is the $t_{\rm sep} \to
  \infty$ estimate obtained using the 2-state fit. The result of the
  fit for each individual $t_{\rm sep}$ is shown by a solid
  line with the same color as the data points. Note that the data
  with $t_{\rm sep}=16$ in the two $a06$ ensembles are not used in the
  fit.
  \label{fig:gA7}}
\end{figure*}

\begin{figure*}
\centering
  \subfigure{
    \includegraphics[height=1.7in,trim={0.095cm 0.11cm 0 0},clip]{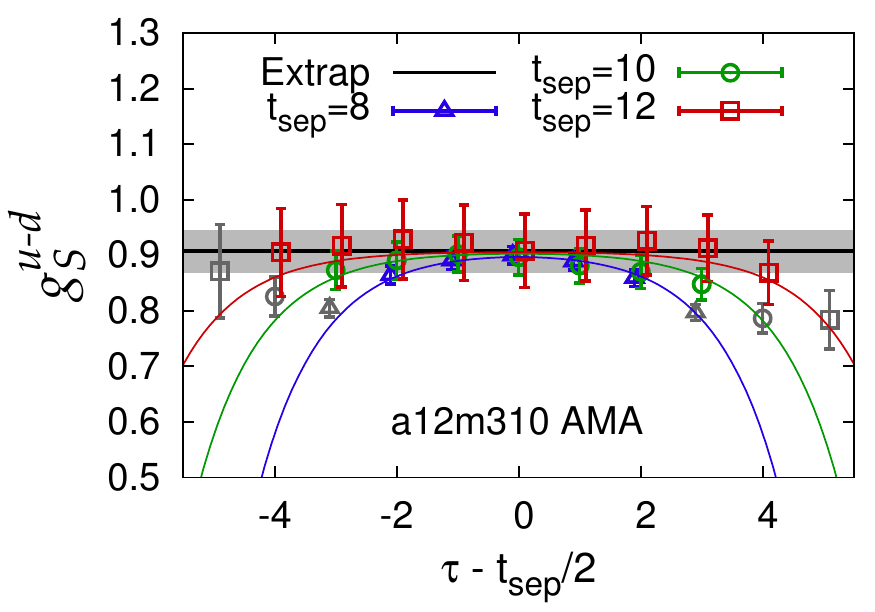}
    \includegraphics[height=1.7in,trim={0.9cm 0.11cm 0 0},clip]{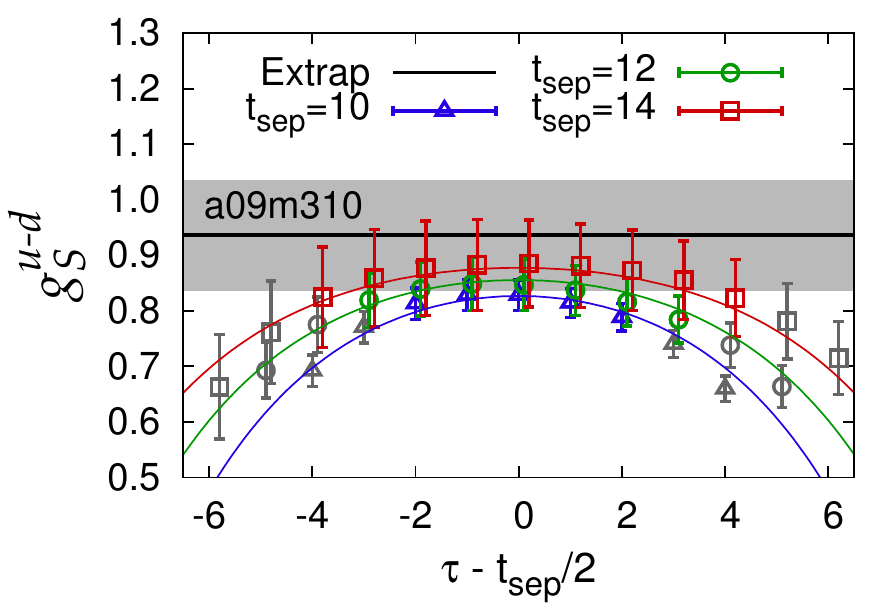}
    \includegraphics[height=1.7in,trim={0.9cm 0.11cm 0 0},clip]{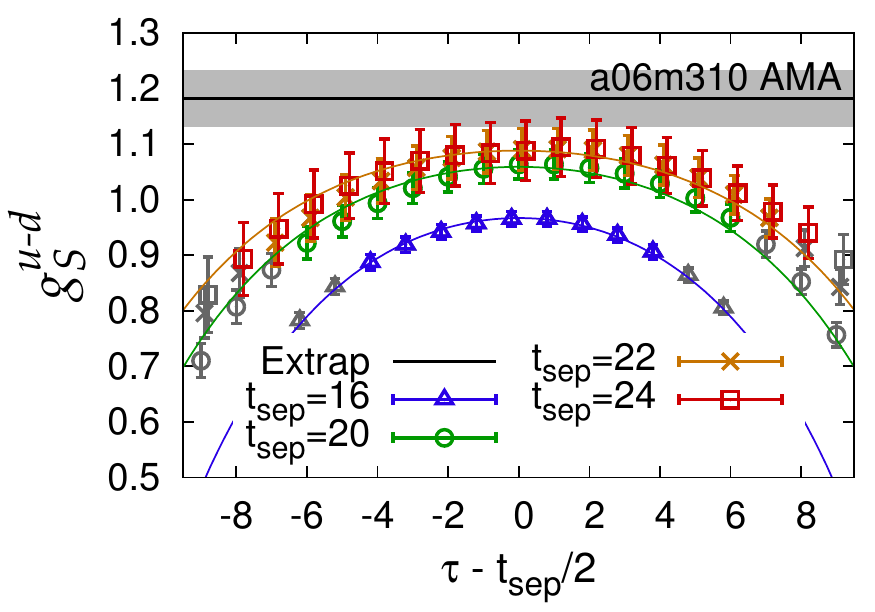}
  }
  \hspace{0.04\linewidth}
  \subfigure{
    \includegraphics[height=1.7in,trim={0.095cm 0.11cm 0 0},clip]{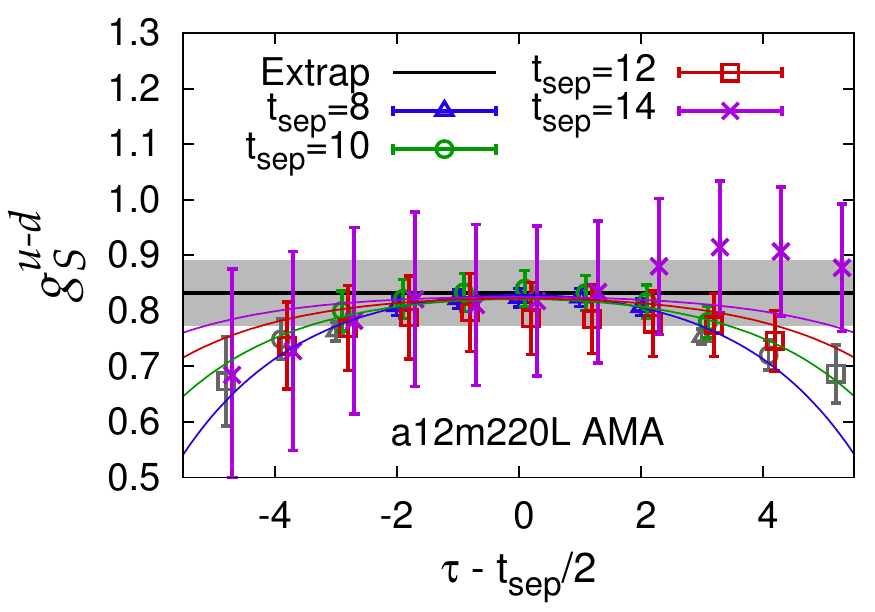}
    \includegraphics[height=1.7in,trim={0.9cm 0.11cm 0 0},clip]{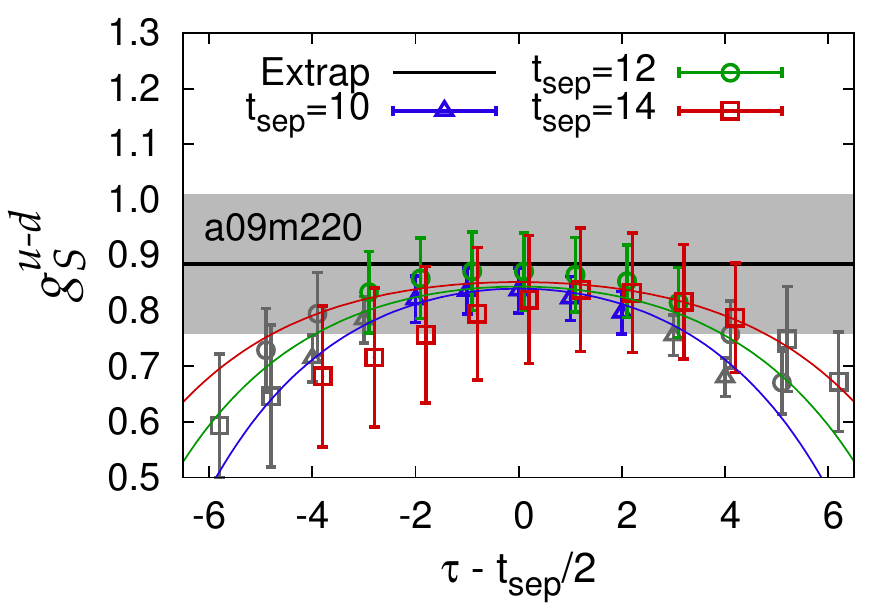}
    \includegraphics[height=1.7in,trim={0.9cm 0.11cm 0 0},clip]{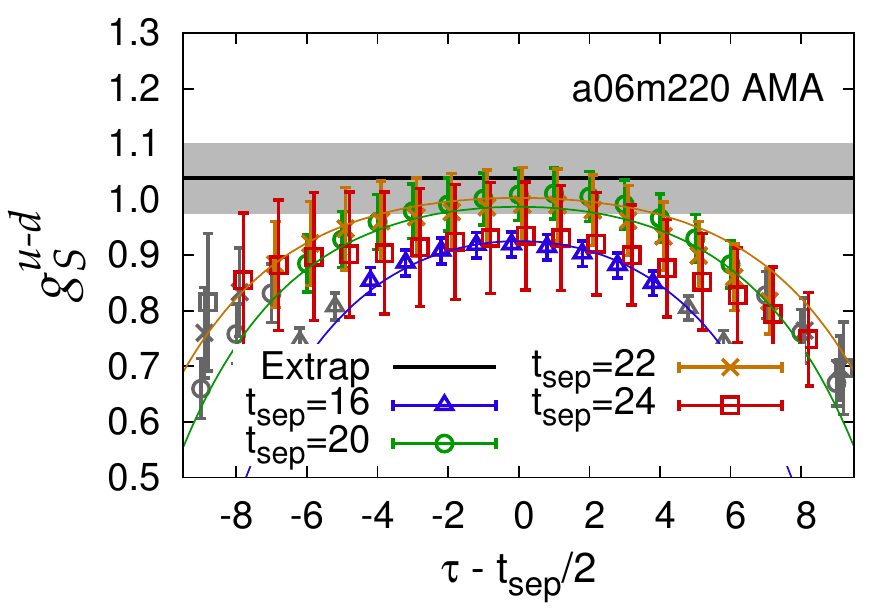}
  }
  \hspace{0.04\linewidth}
  \subfigure{
    \includegraphics[height=1.7in,trim={0.095cm 0.11cm 0 0},clip]{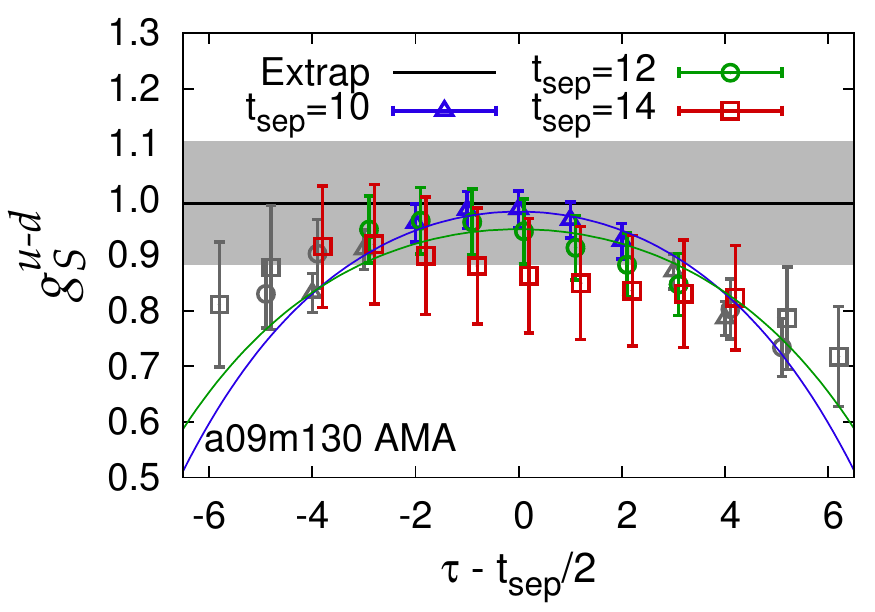}
  }
\caption{The 2-state fit results for the unrenormalized scalar charge
  $g_S^{u-d}$ data. The rest is the same as in
  Fig.~\protect\ref{fig:gA7}.
  \label{fig:gS7}}
\end{figure*}

\begin{figure*}
\centering
  \subfigure{
    \includegraphics[height=1.7in,trim={0.095cm 0.11cm 0 0},clip]{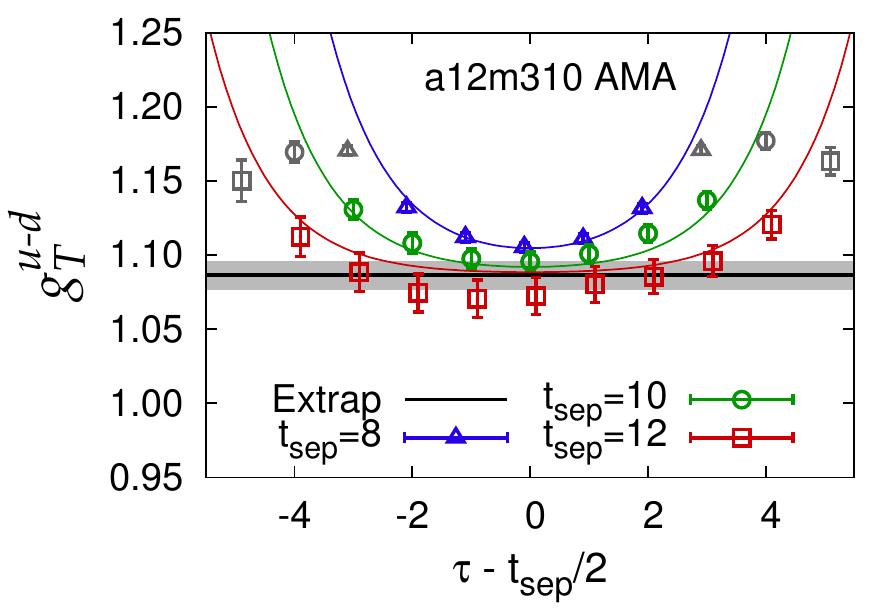}
    \includegraphics[height=1.7in,trim={0.9cm 0.11cm 0 0},clip]{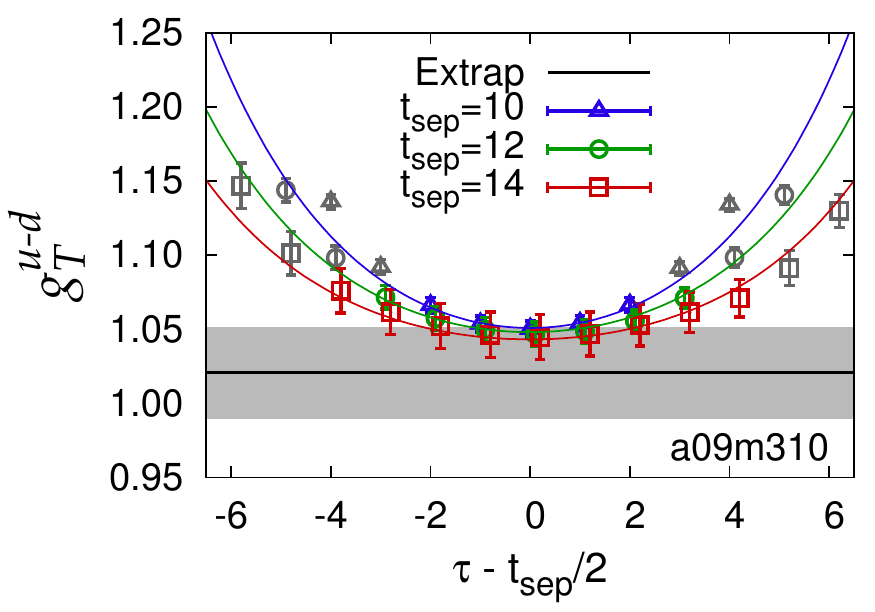}
    \includegraphics[height=1.7in,trim={0.9cm 0.11cm 0 0},clip]{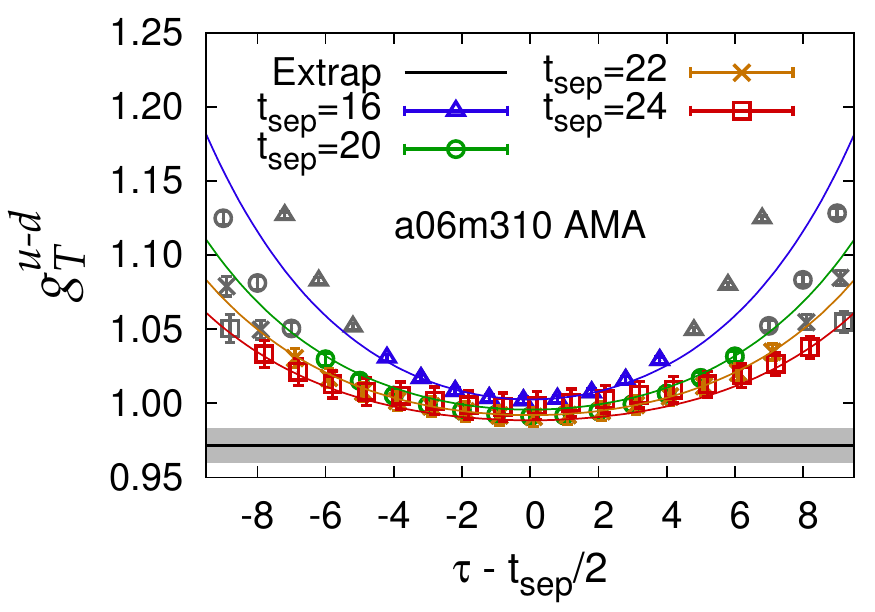}
  }
  \hspace{0.04\linewidth}
  \subfigure{
    \includegraphics[height=1.7in,trim={0.095cm 0.11cm 0 0},clip]{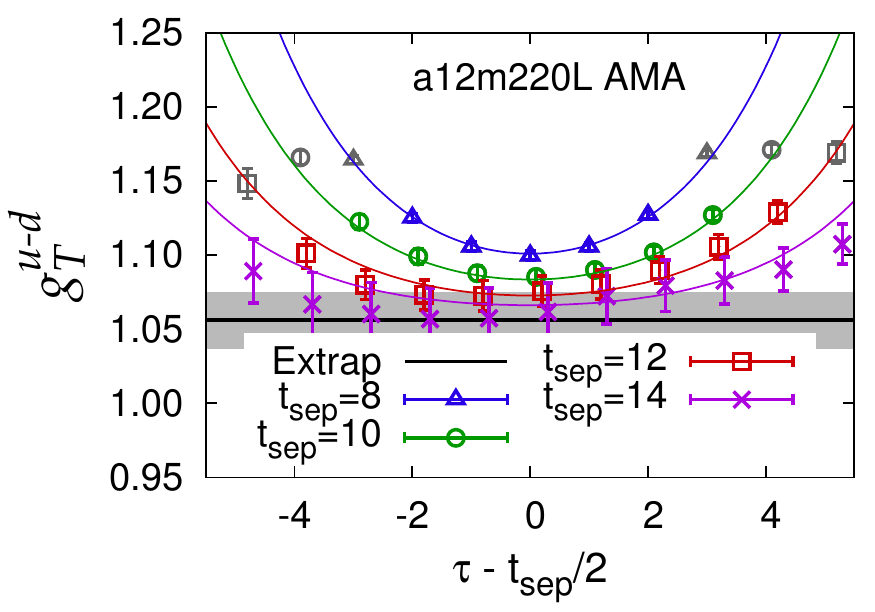}
    \includegraphics[height=1.7in,trim={0.9cm 0.11cm 0 0},clip]{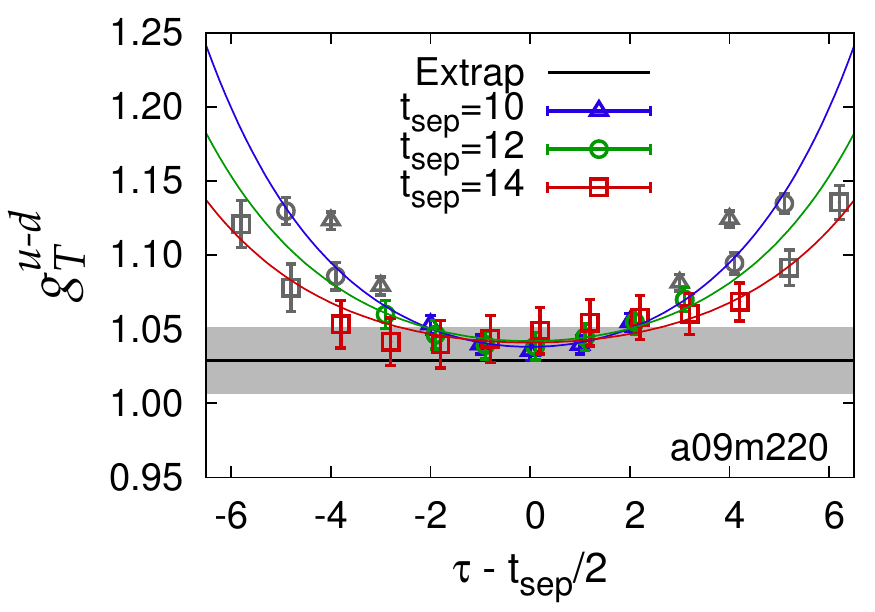}
    \includegraphics[height=1.7in,trim={0.9cm 0.11cm 0 0},clip]{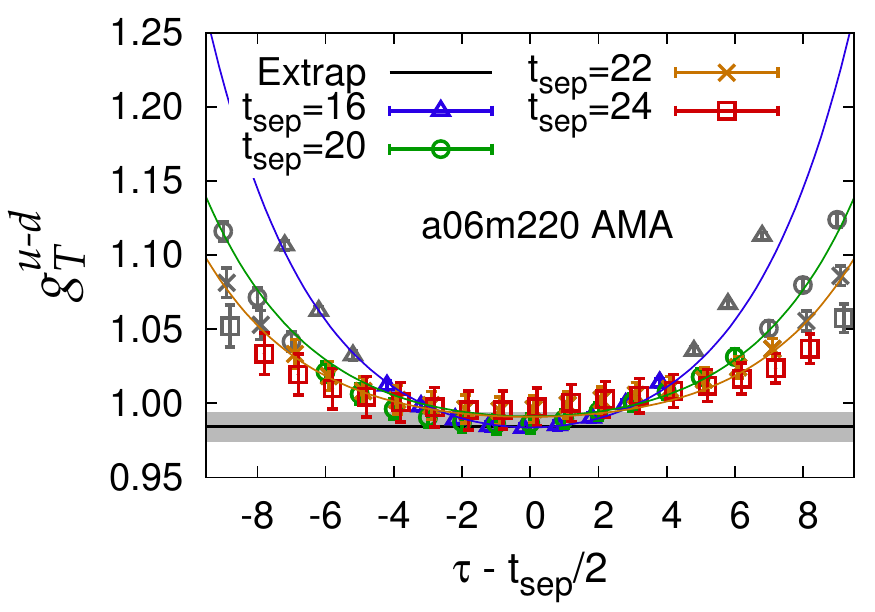}
  }
  \hspace{0.04\linewidth}
  \subfigure{
    \includegraphics[height=1.7in,trim={0.095cm 0.11cm 0 0},clip]{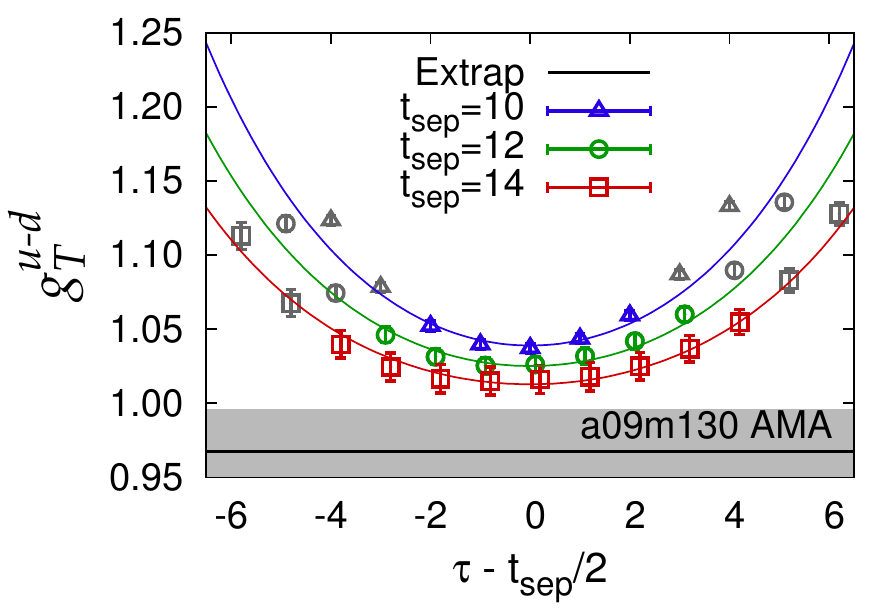}
  }
\caption{The 2-state fit results for the unrenormalized tensor charge $g_T^{u-d}$ data. 
  The rest is the same as in Fig.~\protect\ref{fig:gA7}.
  \label{fig:gT7}}
\end{figure*}

As discussed above, we need to minimize the ratio of the amplitudes,
${\cal A}_1/{\cal A}_0$, to reduce the overlap of the nucleon operator
with the first excited state by tuning the smearing parameters. (In
general, all the ${\cal A}_n/{\cal A}_0$ when up to $n$ excited states
are included in the fit Ansatz.)  Our additional tests on the $a06$
ensembles discussed in Sec.~\ref{sec:confirmation} show that
increasing the smearing size $\sigma$ over the range simulated reduces
${\cal A}_1/{\cal A}_0$ and the excited-state contamination, most
notably in the axial and scalar charges. On the other hand, beyond a
certain size $\sigma$, the statistical errors based on a given number
of gauge configurations start to increase. Also, when calculating the
form factors, one expects the optimal $\sigma$ to decrease with
increasing momentum.  Thus, one has to compromise between obtaining a
good statistical signal and reducing excited-state contamination in
both the charges and the form factors, when all these quantities are
being calculated with a single choice of the smearing parameters.

The data in Tables~\ref{tab:2ptfits1} and~\ref{tab:2ptfits2} show an
increase in the ratio ${\cal A}_1/{\cal A}_0$ as the lattice spacing
is decreased. This suggests that the smearing parameter $\sigma$ (see
Table~\ref{tab:cloverparams}) should have been scaled with the lattice
spacing $a$.  The dependence of the ratio on the two choices of
$t_{\rm min}$ used in the fits (estimates in Table~\ref{tab:2ptfits1}
versus Table~\ref{tab:2ptfits2}) and between the HP and AMA estimates
for each choice is much smaller.  Based on these trends and additional
tests discussed in Sec.~\ref{sec:confirmation}, a better choice for
the smearing parameters when calculating the matrix elements at
zero-momentum transfer is estimated to be $\{5,70\}$, $\{7,120\}$ and
$\{9,200\}$ for the $a=0.12,\ 0.09$ and $0.06$~fm ensembles,
respectively. In physical units, a rule-of-thumb estimate for tuning
the smearing size is $\sigma a \approx 0.55$~fm.

To extract the three matrix elements $\langle 0 | \mathcal{O}_\Gamma |
0 \rangle$, $ \langle 1 | \mathcal{O}_\Gamma | 0 \rangle$ and $
\langle 1 | \mathcal{O}_\Gamma | 1 \rangle$, for each operator
$\mathcal{O}_\Gamma = \mathcal{O}_{A,S,T,V}$, from the 3-point
functions, we make one overall fit using the data at all values of the
operator insertion time $\tau$ and the various source-sink separations
$t_{\rm sep}$ using Eq~\eqref{eq:3pt}.  From such fits we extract
the $t_{\rm sep} \to \infty$ estimates under the assumption that the
contribution of all higher states is isolated by the 2-state
Ansatz given in Eqs.~\eqref{eq:2pt} and~\eqref{eq:3pt}.

In Figs.~\ref{fig:gA7}, \ref{fig:gS7} and~\ref{fig:gT7}, we show the
data for the unrenormalized isovector charges $g_A^{u-d}$, $g_S^{u-d}$
and $g_T^{u-d}$ for the seven ensembles at different values of $a$ and
$M_\pi$ used in the final analysis. The other two ensembles,
$a12m220S$ and $a12m220$ at $a=0.114$~fm and $M_\pi=227$~MeV, allow us
to study the finite-volume effect. The data at central values of $\tau
\approx t_{\rm sep}/2$ show significant, about $15\%$, dependence of
$g_A^{u-d}$ on $t_{\rm sep}$ between 1 and 1.5~fm for the $a=0.09$ and
$0.06$~fm ensembles.  In $g_T^{u-d}$, this effect is less than
$5\%$. The size of the effect in $g_S^{u-d}$ is not clear due to the
much larger errors. Based on these data and additional tests discussed
in Sec.~\ref{sec:confirmation}, the two key observations are the following: (i) the
2-state Ansatz given in Eqs.~\eqref{eq:2pt} and~\eqref{eq:3pt} fits
the data at multiple values of $\tsep$ and gives a reliable estimate
of the $\tsep \to \infty$ value, and (ii) the behavior of the
unrenormalized data versus $\tau$ and $\tsep$ are less sensitive to
the value of $M_\pi$ for fixed $a$ compared to versus $a$ for fixed
$M_\pi$. The behavior of the renormalized charges is discussed in
Sec.~\ref{sec:results}.

The estimates of the matrix elements obtained from the fits to the
3-point correlation functions for the unrenormalized charges are
given in Tables~\ref{tab:3ptgAcompare},~\ref{tab:3ptgScompare}
and~\ref{tab:3ptgTcompare} for the nine ensembles.  Again, for the
case of the five ensembles analyzed using the AMA method, we give
the HP and the AMA estimates.  For each of
the two choices of fit ranges for the 2-point functions, listed in
Tables~\ref{tab:2ptfits1} and~\ref{tab:2ptfits2}, we determined the
three matrix elements for two different choices of the number of
time-slices, $\tskip$,  skipped on either end of the 3-point
correlation functions. These two choices are $\tskip =$
$\{2,3,4\}$ and $\{3,4,6\}$ for the three lattice spacings $a=\{0.12,
0.09, 0.06\}$~fm, respectively. The four cases of fit parameters used in 
the analysis are summarized in Table~\ref{tab:4cases}.

The results for $\langle 0 | \mathcal{O}_\Gamma | 0 \rangle$ for all
four cases (two fit ranges for the 2-point functions and two choices
of $\tskip$ for the 3-point functions) are consistent as shown in
Tables~\ref{tab:3ptgAcompare}, ~\ref{tab:3ptgScompare}
and~\ref{tab:3ptgTcompare}. Estimates of $\langle 0 |
\mathcal{O}_\Gamma | 1 \rangle$ and $\langle 1 | \mathcal{O}_\Gamma |
1 \rangle$ are also similar between Case 1 and Case 3, especially for
the data with the AMA method.  We choose Case 3 for our final
estimates as it has (i) the larger $t_{\rm min}$ corresponding to a
larger suppression of excited-states in the 2-point fits; and (ii) the
smaller value for $\tskip$ whereby more points at each $t_{\rm sep}$
are included in the 3-point fits to give a better determination of
$\langle 0 | \mathcal{O}_\Gamma | 1 \rangle$. Last, note that the
estimates from the HP data presented here are consistent but slightly
different from those published in Ref.~\cite{Bhattacharya:2015wna}
because of the different choices of the fit ranges.
\begin{table*}
\centering
\begin{ruledtabular}
\begin{tabular}{c|c|cccc|cc|cc}
         &      &  \multicolumn{4}{c|}{$\langle 0 | \mathcal{O}_A | 0 \rangle$ } &  \multicolumn{2}{c|}{Case 1} &  \multicolumn{2}{c}{Case 3}  \\
ID       & Type & Case 1      & Case 2      &   Case 3    &   Case 4   
                & $\langle 0 | \mathcal{O}_A | 1 \rangle$ & $\langle 1 | \mathcal{O}_A | 1 \rangle$
                & $\langle 0 | \mathcal{O}_A | 1 \rangle$ & $\langle 1 | \mathcal{O}_A | 1 \rangle$   \\ 
\hline
a12m310  & HP   & 1.249(22)   & 1.246(26)   &  1.248(26)   & 1.243(33)  &  $-$0.019(42)  &  $-$2.4(62)    &  $-$0.016(46)  &  0.4(2.8)    \\
a12m310  & AMA  & 1.252(9)    & 1.251(11)   &  1.252(9)    & 1.251(11)  &  $-$0.060(24)  &  $-$22(28)     &  $-$0.060(24)  &  $-$22(28)   \\
a12m310  & LP   & 1.252(9)    & 1.251(11)   &  1.252(9)    & 1.251(11)  &  $-$0.060(24)  &  $-$22(28)     &  $-$0.060(24)  &  $-$22(28)   \\
a12m220S & HP   & 1.275(32)   & 1.287(39)   &  1.283(39)   & 1.295(48)  &  $-$0.187(64)  &  $-$7.(18)     &  $-$0.199(69)  &  0.6(6.2)    \\
a12m220  & HP   & 1.271(27)   & 1.265(33)   &  1.273(30)   & 1.263(39)  &  $-$0.068(58)  &  $-$16.0(2.5)  &  $-$0.063(63)  &  $-$5.(15)     \\
a12m220L & HP   & 1.285(22)   & 1.291(25)   &  1.337(83)   & 1.346(89)  &  $-$0.125(47)  &                &  $-$0.145(60)  &              \\
a12m220L & LP   & 1.276(10)   & 1.275(14)   &  1.279(12)   & 1.276(16)  &  $-$0.084(22)  &  $-$0.8(18)    &  $-$0.087(25)  &  $-$0.0(1.7)   \\
\hline                                                                                                                            
a09m310  & HP   & 1.255(21)   & 1.255(25)   &  1.262(30)   & 1.262(30)  &  $-$0.118(35)  &  $-$0.7(18)    &  $-$0.123(38)  &  $-$0.3(1.9)   \\
a09m220  & HP   & 1.267(22)   & 1.277(25)   &  1.272(30)   & 1.283(33)  &  $-$0.135(34)  &  $-$2.0(25)    &  $-$0.138(37)  &  $-$1.7(2.9)   \\
a09m130  & HP   & 1.172(39)   & 1.164(46)   &  1.177(44)   & 1.163(55)  &  $-$0.029(44)  &  0.4(13)       &  $-$0.028(49)  &  0.84(77)    \\
a09m130  & AMA  & 1.247(16)   & 1.258(19)   &  1.255(24)   & 1.266(27)  &  $-$0.096(19)  &  0.64(35)      &  $-$0.102(24)  &  0.72(38)    \\
a09m130  & LP   & 1.247(16)   & 1.257(19)   &  1.255(24)   & 1.265(26)  &  $-$0.094(19)  &  0.60(36)      &  $-$0.100(23)  &  0.69(38)    \\
\hline                                                                                                                            
a06m310  & HP   & 1.167(22)   & 1.156(26)   &  1.172(24)   & 1.157(29)  &  $-$0.020(22)  &  0.03(70)      &  $-$0.019(24)  &  0.45(50)    \\
a06m310  & AMA  & 1.209(14)   & 1.209(15)   &  1.212(14)   & 1.212(16)  &  $-$0.058(16)  &  $-$2.3(16)    &  $-$0.060(17)  &  $-$1.2(1.4)   \\
a06m310  & LP   & 1.210(13)   & 1.210(15)   &  1.213(14)   & 1.213(16)  &  $-$0.058(16)  &  $-$2.4(17)    &  $-$0.061(17)  &  $-$1.3(1.4)   \\
a06m220  & HP   & 1.227(50)   & 1.238(57)   &  1.234(51)   & 1.244(60)  &  $-$0.182(61)  &  $-$0.2(3.1)   &  $-$0.186(62)  &  0.6(2.3)    \\
a06m220  & AMA  & 1.234(17)   & 1.241(19)   &  1.234(17)   & 1.241(19)  &  $-$0.122(18)  &  $-$6.0(3.2)   &  $-$0.121(18)  &  $-$6.3(3.9)   \\
a06m220  & LP   & 1.234(17)   & 1.241(18)   &  1.234(17)   & 1.241(18)  &  $-$0.117(18)  &  $-$7.0(3.4)   &  $-$0.116(18)  &  $-$7.5(3.9)   \\
\end{tabular}
\end{ruledtabular}
\caption{Estimates of the matrix element $\langle 0 | \mathcal{O}_A |
  0 \rangle$ for the isovector axial operators for four cases of the
  fit ranges defined in Table~\protect\ref{tab:4cases}. We also give
  estimates of the matrix elements $\langle 0 | \mathcal{O}_A | 1
  \rangle$ and $\langle 1 | \mathcal{O}_A | 1 \rangle$ for Case 1
  and Case 3.  For the four ensembles, $a12m310$, $a09m130$, $a06m310$ and
  $a06m22$, we give both the AMA and LP estimates. We find that the
  bias correction term is negligible in all cases. }
\label{tab:3ptgAcompare}
\end{table*}

\begin{table*}
\centering
\begin{ruledtabular}
\begin{tabular}{c|c|cccc|cc|cc}
         &      &  \multicolumn{4}{c|}{$\langle 0 | \mathcal{O}_S | 0 \rangle$ } &  \multicolumn{2}{c|}{Case 1} &  \multicolumn{2}{c}{Case 3}  \\
ID       & Type & Case 1      & Case 2      &   Case 3    &   Case 4   
                & $\langle 0 | \mathcal{O}_S | 1 \rangle$ & $\langle 1 | \mathcal{O}_S | 1 \rangle$
                & $\langle 0 | \mathcal{O}_S | 1 \rangle$ & $\langle 1 | \mathcal{O}_S | 1 \rangle$   \\ 
\hline
a12m310  & HP   & 0.83(9)     & 0.79(10)    &  0.82(10)   & 0.77(12)   &  $-$0.06(11) &  10(23)       &  $-$0.05(12)  &  6(11)        \\
a12m310  & AMA  & 0.91(4)     & 0.91(4)     &  0.91(4)    & 0.91(4)    &  $-$0.29(5)  &  6(56)        &  $-$0.29(5)   &  6(56)        \\
a12m310  & LP   & 0.91(4)     & 0.91(4)     &  0.91(4)    & 0.91(4)    &  $-$0.29(5)  &  6(56)        &  $-$0.29(5)   &  6(56)        \\
a12m220S & HP   & 1.01(27)    & 1.03(30)    &  1.03(32)   & 1.07(37)   &  $-$0.16(22) &  $-$40(13)    &  $-$0.17(26)  &  $-$11(38)    \\
a12m220  & HP   & 0.69(16)    & 0.64(18)    &  0.67(18)   & 0.61(20)   &  $-$0.07(16) &  80(14)       &  $-$0.06(18)  &  35(77)       \\
a12m220L & HP   & 1.01(10)    & 1.00(10)    &  1.10(17)   & 1.08(17)   &  $-$0.34(15) &               &  $-$0.39(18)  &               \\
a12m220L & LP   & 0.83(5)     & 0.83(7)     &  0.83(6)    & 0.83(7)    &  $-$0.18(6)  &  4.9(6.2)     &  $-$0.20(7)   &  4.1(4.8)     \\
\hline                                                                                                                          
a09m310  & HP   & 0.92(9)     & 0.93(10)    &  0.94(10)   & 0.94(11)   &  $-$0.33(9)  &  1.1(2.5)     &  $-$0.34(10)  &  1.2(2.1)     \\
a09m220  & HP   & 0.88(12)    & 0.86(13)    &  0.88(13)   & 0.87(14)   &  $-$0.28(10) &  2.7(4.2)     &  $-$0.28(11)  &  2.7(3.8)     \\
a09m130  & HP   & 0.70(32)    & 0.72(36)    &  0.73(37)   & 0.75(42)   &  $-$0.27(17) &  3.4(9.6)     &  $-$0.29(19)  &  2.5(5.5)     \\
a09m130  & AMA  & 0.98(10)    & 0.97(11)    &  0.99(11)   & 0.97(12)   &  $-$0.37(6)  &  4.0(1.6)     &  $-$0.40(9)   &  3.8(1.5)     \\
a09m130  & LP   & 0.99(10)    & 0.97(11)    &  1.00(11)   & 0.98(11)   &  $-$0.36(6)  &  3.9(1.6)     &  $-$0.39(8)   &  3.6(1.4)     \\

\hline                                                                                                                        
a06m310  & HP   & 1.24(10)    & 1.22(11)    &  1.28(11)   & 1.24(12)   &  $-$0.26(6)  &  $-$3.4(2.1)  &  $-$0.28(7)   &  $-$1.8(1.6)  \\
a06m310  & AMA  & 1.16(4)     & 1.17(5)     &  1.18(5)    & 1.19(5)    &  $-$0.38(3)  &  $-$1.1(1.1)  &  $-$0.40(4)   &  $-$0.5(1.0)  \\
a06m310  & LP   & 1.16(4)     & 1.17(5)     &  1.18(5)    & 1.19(5)    &  $-$0.38(3)  &  $-$1.1(1.1)  &  $-$0.40(4)   &  $-$0.5(1.0)  \\
a06m220  & HP   & 0.65(28)    & 0.60(29)    &  0.64(29)   & 0.59(31)   &  $-$0.21(14) &  11(17)       &  $-$0.21(15)  &  8.(13)       \\
a06m220  & AMA  & 1.04(6)     & 1.05(7)     &  1.04(6)    & 1.05(7)    &  $-$0.30(4)  &  $-$0.6(2.5)  &  $-$0.30(4)   &  $-$0.7(2.6)  \\
a06m220  & LP   & 1.04(6)     & 1.05(7)     &  1.04(6)    & 1.05(7)    &  $-$0.30(3)  &  $-$1.1(2.5)  &  $-$0.30(3)   &  $-$1.2(2.6)  \\
\end{tabular}
\end{ruledtabular}
\caption{Estimates of the matrix elements for the isovector scalar operator. The rest is the same as in Table~\protect\ref{tab:3ptgAcompare}.
}
\label{tab:3ptgScompare}
\end{table*}

\begin{table*}
\centering
\begin{ruledtabular}
\begin{tabular}{c|c|cccc|cc|cc}
         &      &  \multicolumn{4}{c|}{$\langle 0 | \mathcal{O}_T | 0 \rangle$ } &  \multicolumn{2}{c|}{Case 1} &  \multicolumn{2}{c}{Case 3}  \\
ID       & Type & Case 1      & Case 2      &   Case 3    &   Case 4   
                & $\langle 0 | \mathcal{O}_T | 1 \rangle$ & $\langle 1 | \mathcal{O}_T | 1 \rangle$
                & $\langle 0 | \mathcal{O}_T | 1 \rangle$ & $\langle 1 | \mathcal{O}_T | 1 \rangle$   \\ 
\hline
a12m310  & HP   & 1.096(21)  &  1.092(24)  &  1.084(31)  &  1.081(33)  &  0.187(32)  &  $-$2.5(4.9)   &  0.200(35)  &  $-$0.5(2.4)   \\
a12m310  & AMA  & 1.087(10)  &  1.082(10)  &  1.087(10)  &  1.082(10)  &  0.243(20)  &  12(14)        &  0.243(20)  &  12(14)        \\
a12m310  & LP   & 1.087(10)  &  1.082(10)  &  1.087(10)  &  1.082(10)  &  0.243(20)  &  13(14)        &  0.243(20)  &  13(14)        \\
a12m220S & HP   & 1.086(26)  &  1.079(31)  &  1.067(46)  &  1.060(51)  &  0.264(55)  &  $-$1(11)      &  0.280(57)  &  0.3(3.3)      \\
a12m220  & HP   & 1.111(24)  &  1.107(28)  &  1.105(32)  &  1.102(36)  &  0.202(47)  &  $-$17(22)     &  0.209(49)  &  $-$6(14)      \\
a12m220L & HP   & 1.058(19)  &  1.059(20)  &  1.043(29)  &  1.050(29)  &  0.168(33)  &                &  0.193(43)  &                \\
a12m220L & LP   & 1.063(12)  &  1.063(13)  &  1.056(19)  &  1.058(18)  &  0.200(17)  &  1.31(74)      &  0.213(25)  &  1.03(60)      \\
\hline                                                                                                                           
a09m310  & HP   & 1.025(24)  &  1.027(26)  &  1.021(31)  &  1.023(31)  &  0.157(24)  &  0.10(87)      &  0.164(30)  &  0.15(80)      \\
a09m220  & HP   & 1.030(20)  &  1.039(21)  &  1.029(22)  &  1.039(22)  &  0.124(25)  &  $-$0.3(1.1)   &  0.127(26)  &  $-$0.2(1.1)   \\
a09m130  & HP   & 0.993(33)  &  1.001(36)  &  0.980(45)  &  0.993(47)  &  0.136(36)  &  0.48(88)      &  0.147(40)  &  0.63(53)      \\
a09m130  & AMA  & 0.974(20)  &  0.981(20)  &  0.967(28)  &  0.975(26)  &  0.164(16)  &  0.79(14)      &  0.174(25)  &  0.75(12)      \\
a09m130  & LP   & 0.973(19)  &  0.979(19)  &  0.966(28)  &  0.974(26)  &  0.167(16)  &  0.77(14)      &  0.177(25)  &  0.72(12)      \\
\hline                                                                                                                           
a06m310  & HP   & 0.961(20)  &  0.958(22)  &  0.951(24)  &  0.950(26)  &  0.124(15)  &  0.68(40)      &  0.131(16)  &  0.75(27)      \\
a06m310  & AMA  & 0.976(10)  &  0.981(10)  &  0.972(12)  &  0.978(12)  &  0.122(9)   &  0.46(26)      &  0.128(10)  &  0.50(22)      \\
a06m310  & LP   & 0.976(10)  &  0.981(10)  &  0.972(12)  &  0.978(12)  &  0.122(9)   &  0.45(26)      &  0.128(10)  &  0.49(22)      \\
a06m220  & HP   & 0.995(43)  &  1.002(47)  &  0.990(48)  &  1.000(51)  &  0.109(38)  &  $-$0.5(2.5)   &  0.110(39)  &  0.0(1.9)      \\
a06m220  & AMA  & 0.984(10)  &  0.986(10)  &  0.984(10)  &  0.986(10)  &  0.103(8)   &  $-$0.50(52)   &  0.103(8)   &  $-$0.53(59)   \\
a06m220  & LP   & 0.984(9)   &  0.985(10)  &  0.984(9)   &  0.986(10)  &  0.105(8)   &  $-$0.59(54)   &  0.105(8)   &  $-$0.63(62)   \\
\end{tabular}
\end{ruledtabular}
\caption{Estimates of the matrix elements for the isovector tensor operator. The rest is the same as in Table~\protect\ref{tab:3ptgAcompare}.
}
\label{tab:3ptgTcompare}
\end{table*}

The consistency between the four estimates of the matrix elements
allows us to draw the following qualitative conclusions
regarding excited-state contamination in $g_A^{u-d}$, $g_S^{u-d}$ and $g_T^{u-d}$.
\begin{itemize}
\item
Estimates of $\langle 1 | \mathcal{O}_\Gamma | 1 \rangle$ given in
Tables~\ref{tab:3ptgAcompare}, ~\ref{tab:3ptgScompare}
and~\ref{tab:3ptgTcompare} are poorly determined.  Within the 2-state
approximation, the effect of a nonzero $\langle 1 | \mathcal{O}_\Gamma
| 1 \rangle$ is to change the data at all $\tau$, for fixed $\tsep$,
by a constant amount. For $g_A^{u-d}$, $\langle 1 | \mathcal{O}_\Gamma | 1
\rangle$ is large on the $a=0.09$ and $0.06$~fm ensembles and has the
same sign as $\langle 0 | \mathcal{O}_\Gamma | 1 \rangle$. The two
contributions, therefore, add to give a large excited-state
contribution.  For $g_T^{u-d}$, the data in Fig.~\ref{fig:gT7} show less
than $ 5\%$ dependence on $t_{\rm sep}$ in the central regions (values
of operator insertion time $\tau$ at which the contribution of a
nonzero $\langle 0 | \mathcal{O}_\Gamma |1 \rangle$ is the smallest)
and the contribution of $\langle 1 | \mathcal{O}_\Gamma | 1 \rangle$
is small.  For $g_S^{u-d}$, the data in Fig.~\ref{fig:gS7} 
show a significant excited-state effect only for the $a=0.06$~fm
ensembles that is largely accounted for by the contribution of the
$\langle 0 | \mathcal{O}_S | 1 \rangle$ term.
\item
Estimates of $\langle 0 | \mathcal{O}_S | 1 \rangle$ vary between
$-0.05$ and $-0.41$ for $g_S^{u-d}$. This matrix element gives the
negative curvature observed in Fig.~\ref{fig:gS7}, 
and the ground-state estimate of $g_S^{u-d}$ converges from below.  
%% Using
%% Eq~\eqref{eq:2pt} and~\eqref{eq:_3pt} with ${\cal A}_1^2/{\cal A}_0^2 = \{1, 1.5,
%% 3\}$, $M_1-M_0=\{0.5,0.4,0.3\}$ and $\langle 0 | \mathcal{O}_S | 1
%% \rangle = -0.3$, rough estimates of the resulting excited-state
%% contamination in $g_S^{u-d}$ at $t_{\rm sep} = \{8, 10,16\}$, the smallest
%% $t_{\rm sep}$ analyzed on the $a=\{0.12,  0.09, 0.06\}$~fm ensembles,
%% are {0.6\%, 0.9\%, 0.8\%}, respectively.\FIXME{check numbers}
\item
Data for $g_S^{u-d}$ show excited-state contamination on the two
$a=0.06$~fm ensembles $a06m310$ and $a06m220$.  The 2-state fit gives
reliable $t_{\rm sep} \to \infty$ estimates as 
shown in Fig.~\ref{fig:gS7}. In
Sec.~\ref{sec:confirmation}, we present data from additional
simulations on the $a06m310$ and $a06m220$ ensembles and discuss the
limitations of the 2-state fit when the errors are large and the data
at different $\tsep$ overlap as a result.
\item
Estimates of $\langle 0 | \mathcal{O}_T | 1 \rangle$ vary between
0.1 and 0.26 for $g_T^{u-d}$ and give rise to the positive
curvature evident in Fig.~\ref{fig:gT7}. The 
ground-state estimate of $g_T^{u-d}$ converges from above.  
%% Again, using $\langle 0 | \mathcal{O}_T | 1 \rangle = 0.3$ with the same
%% ${\cal A}_1^2/{\cal A}_0^2$ and $M_1-M_0$ as for $g_S^{u-d}$, the corresponding
%% estimates of contamination in $g_T^{u-d}$ are the same in magnitude, 
%% {0.6\%, 0.9\%, 0.8\%}, respectively, at $t_{\rm sep} = \{8, 10,16\}$.
\item
There is evidence for excited-state contamination in $g_T^{u-d}$
on the $a12m310$, $a12m220L$ and $a09m130$ ensembles with AMA as shown
in Fig.~\ref{fig:gT7}.  This $ 5\%$ effect can be accounted for
by the larger value of $\langle 1 | \mathcal{O}_\Gamma | 1 \rangle$
that has the same sign as the $\langle 0 | \mathcal{O}_\Gamma | 1
\rangle$ contribution as shown in Table~\ref{tab:3ptgTcompare}. In all
three cases, the 2-state Ansatz fits the data at multiple values of
$t_{\rm sep}$ covering the range 1.0--1.4~fm, and gives stable  
$t_{\rm sep} \to \infty$ estimates under changes in $\tskip $ and 
$\tsep$.
\item
In all cases, the data with the largest $t_{\rm sep}$ are noisy and do
not contribute significantly to the fit. Much higher statistics are 
needed for fits to be sensitive to data with $t_{\rm sep}\, \gsim 1.5$~fm.
\end{itemize}

Our conclusion is that the 2-state fit reduces the uncertainty due to
the excited-state contamination in $g_A^{u-d}$, $g_S^{u-d}$ and $g_T^{u-d}$ to within a
few percent (size of the statistical errors) with the smearing
parameters and values of $t_{\rm sep}$ we have used. Also, the data in
Tables~\ref{tab:3ptgAcompare},~\ref{tab:3ptgScompare}
and~\ref{tab:3ptgTcompare} indicate that the variation in the
determination of ${\cal A}_1/{\cal A}_0$ and $M_1-M_0$ with different
choices of the fit ranges does not have a significant effect on the
final $t_{\rm sep} \to \infty$ estimates of the charges obtained from
the 2-state fits.

%%%%%%%%%%%%%%%%%%%%%%%%%%%%%%%%%%%%%%%%%%%%%%%%%%%%%%%%%%%%%%%%%%%%%
%%%  SECTION                                                      %%%
%%%%%%%%%%%%%%%%%%%%%%%%%%%%%%%%%%%%%%%%%%%%%%%%%%%%%%%%%%%%%%%%%%%%%
\section{Renormalization of Operators}
\label{sec:renorm}

\begin{figure*}[tb]
  \subfigure{
    \includegraphics[height=1.45in,trim={0.1cm   0.65cm 0 0},clip]{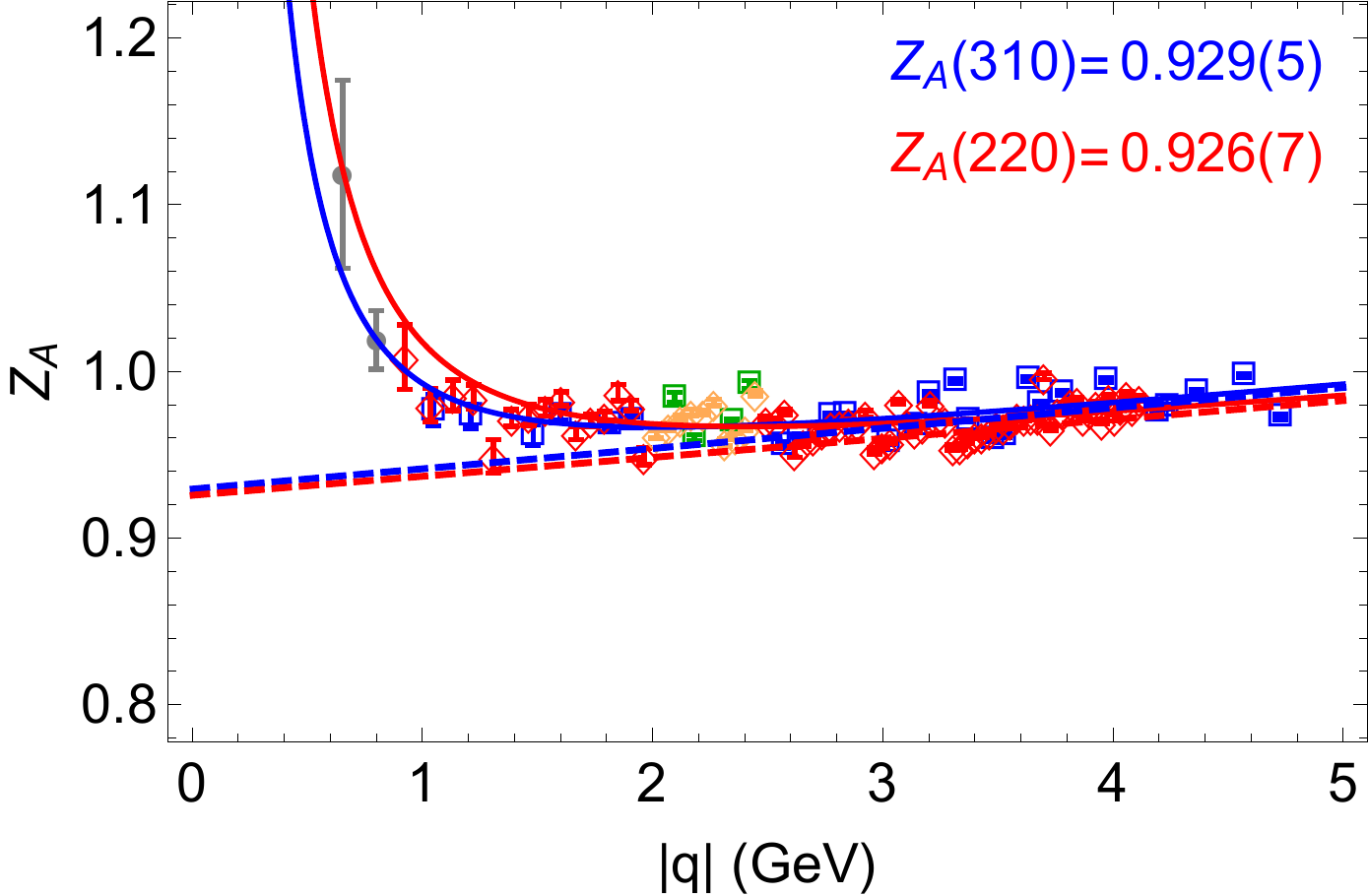}
    \includegraphics[height=1.45in,trim={0.6cm   0.65cm 0 0},clip]{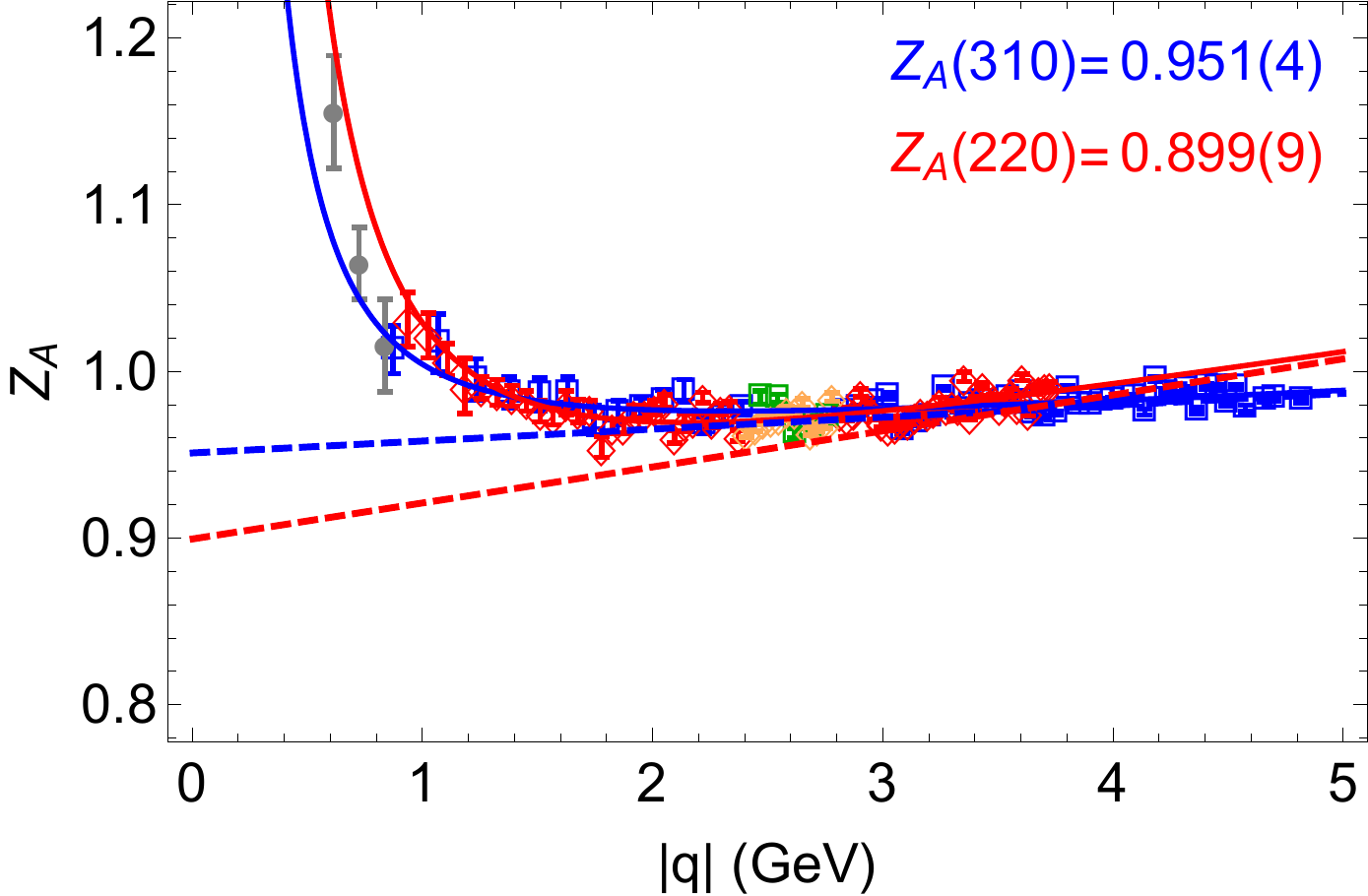}
    \includegraphics[height=1.45in,trim={0.6cm   0.65cm 0 0},clip]{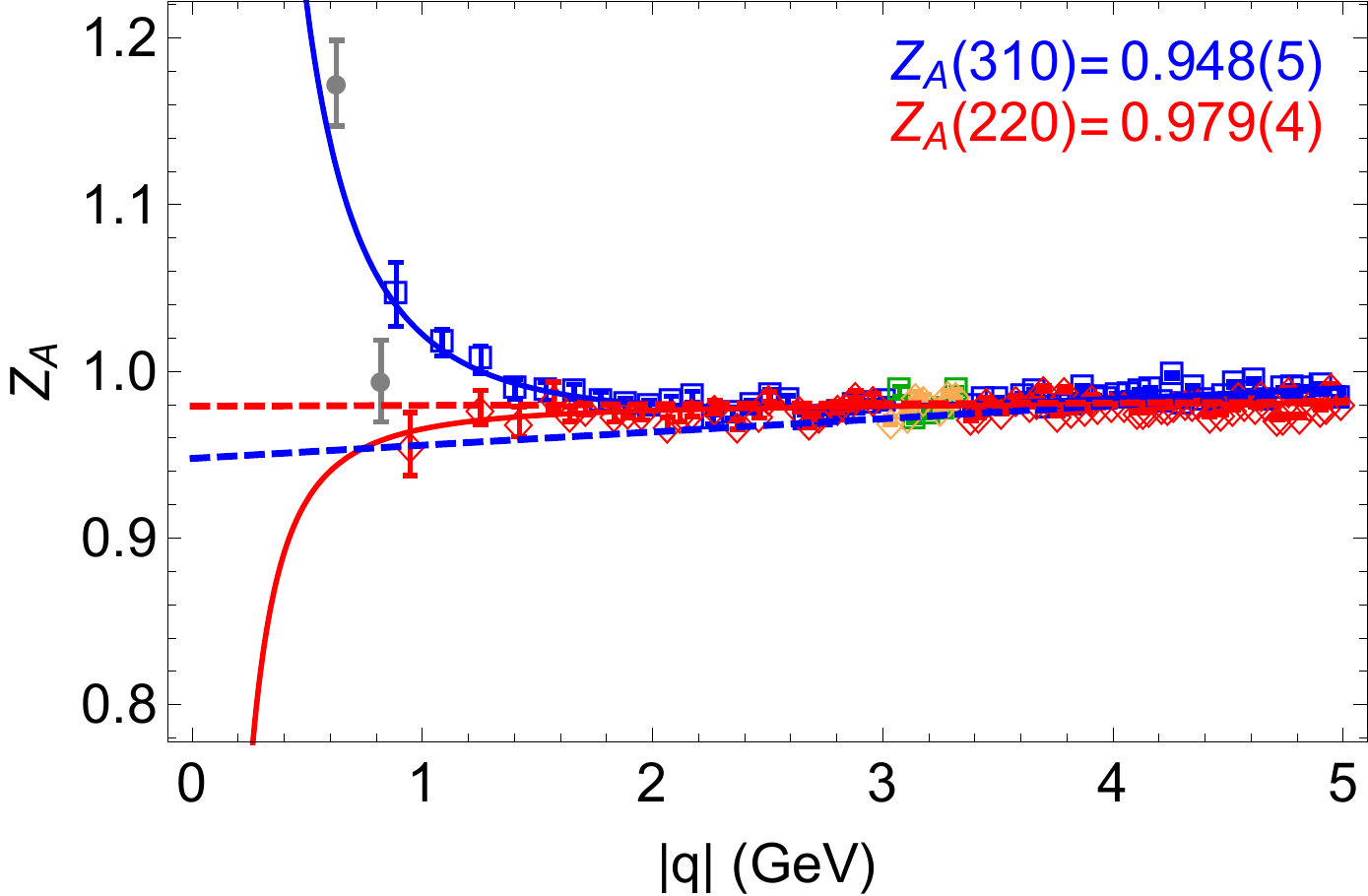}
}
  \subfigure{
    \includegraphics[height=1.45in,trim={0.1cm   0.65cm 0 0},clip]{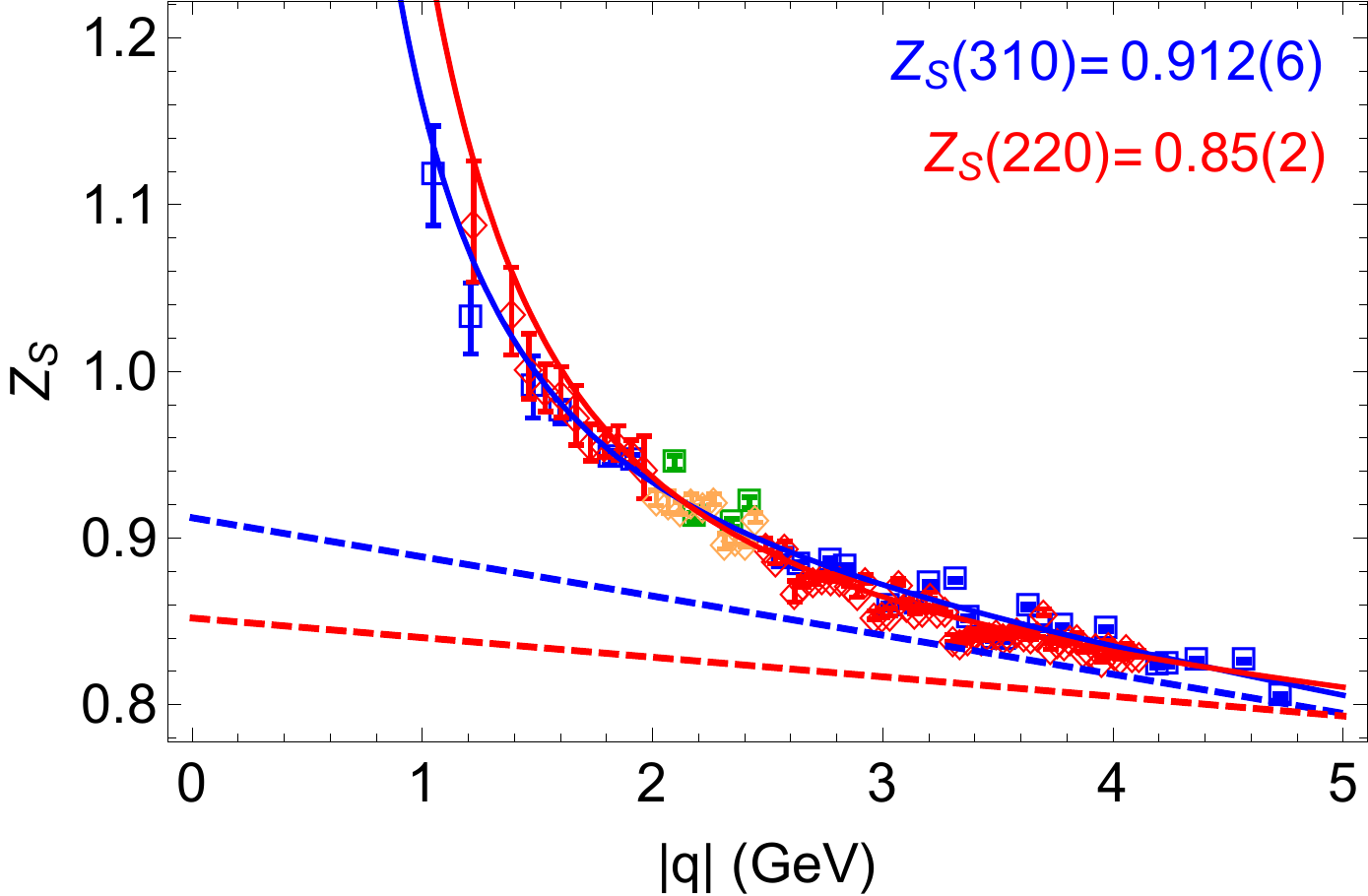}
    \includegraphics[height=1.45in,trim={0.6cm   0.65cm 0 0},clip]{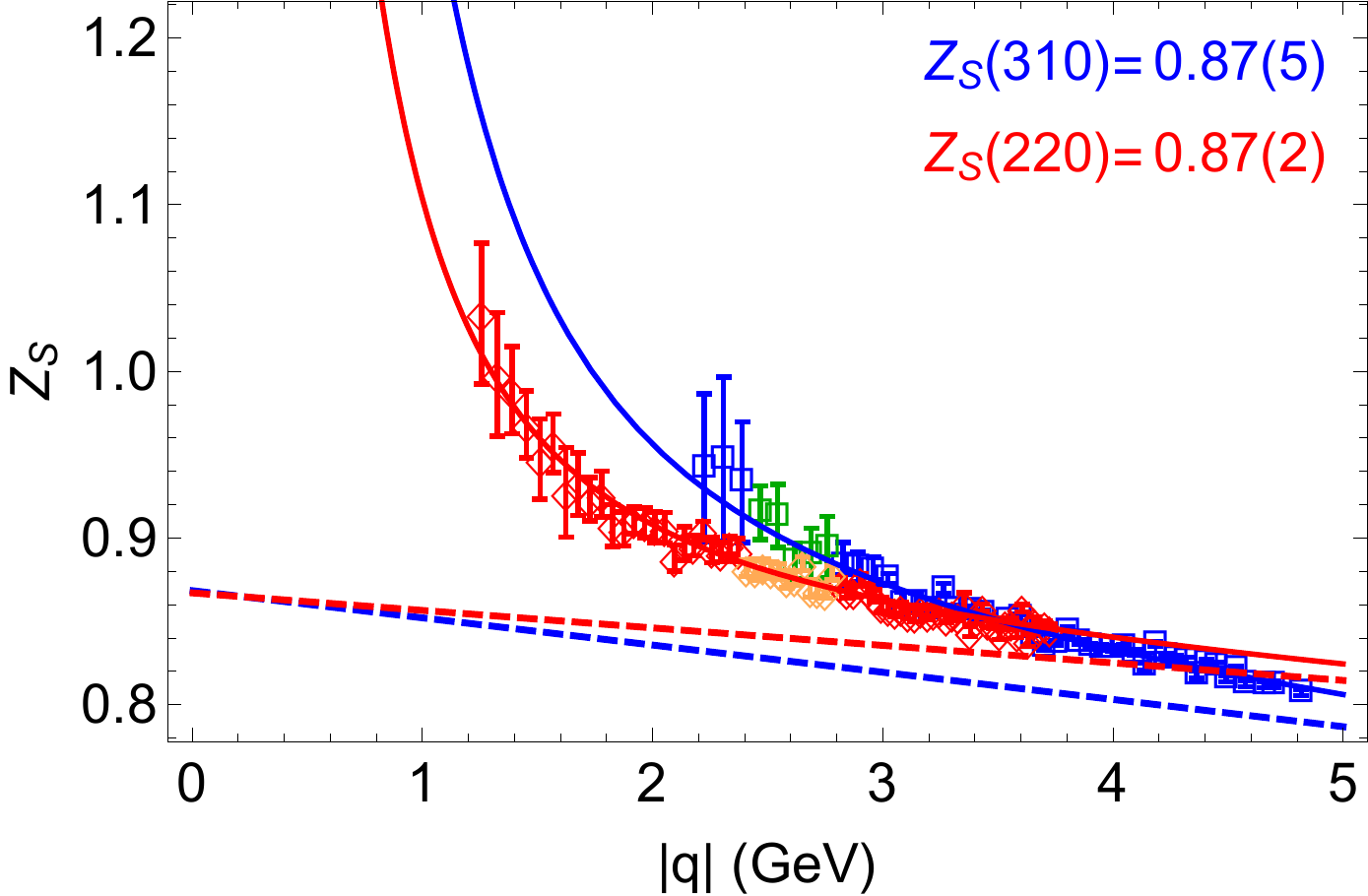}
    \includegraphics[height=1.45in,trim={0.6cm   0.65cm 0 0},clip]{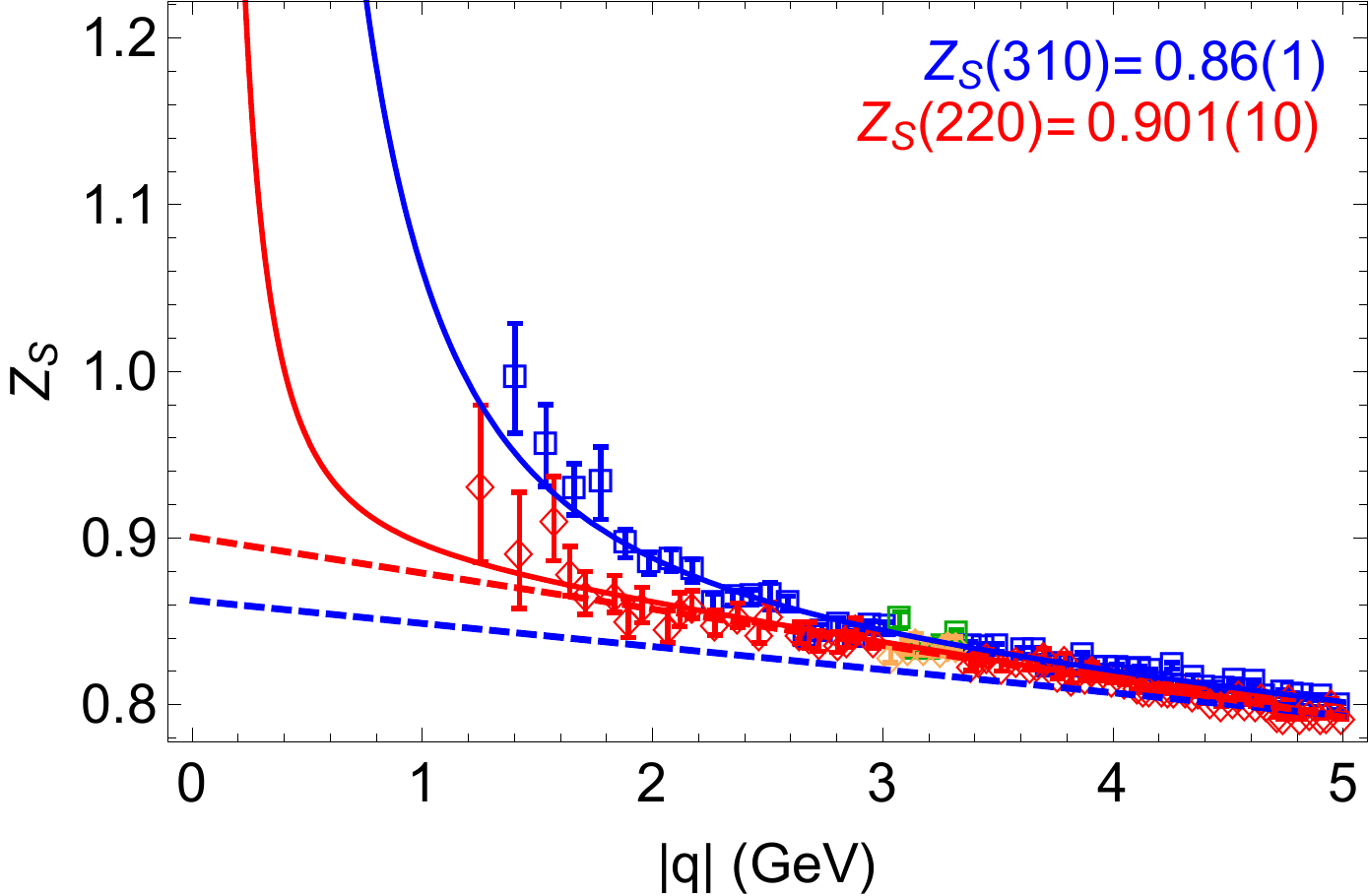}
}
  \subfigure{
    \includegraphics[height=1.43in,trim={0.0cm   0.65cm 0 0},clip]{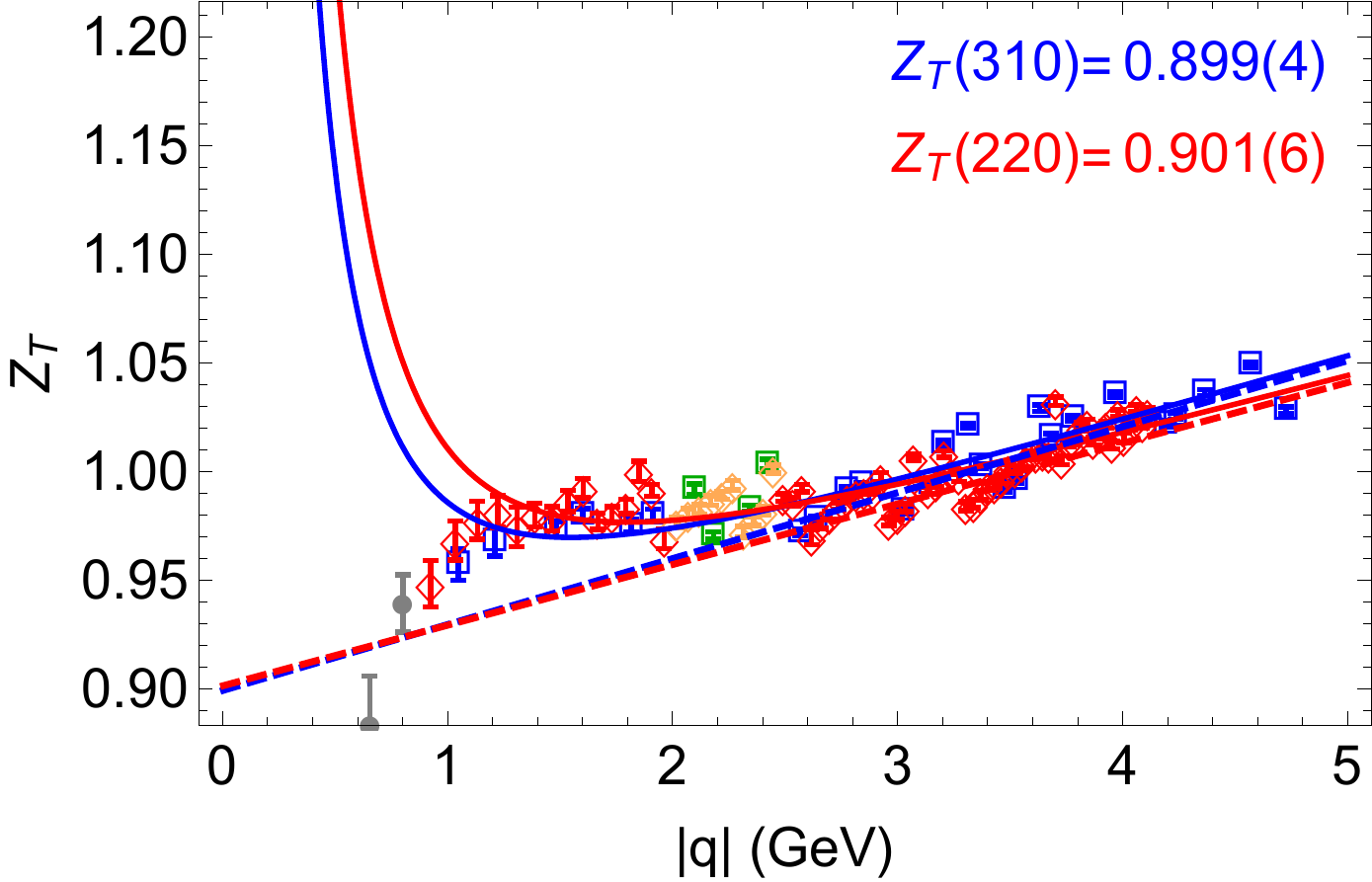}
    \includegraphics[height=1.43in,trim={0.6cm   0.65cm 0 0},clip]{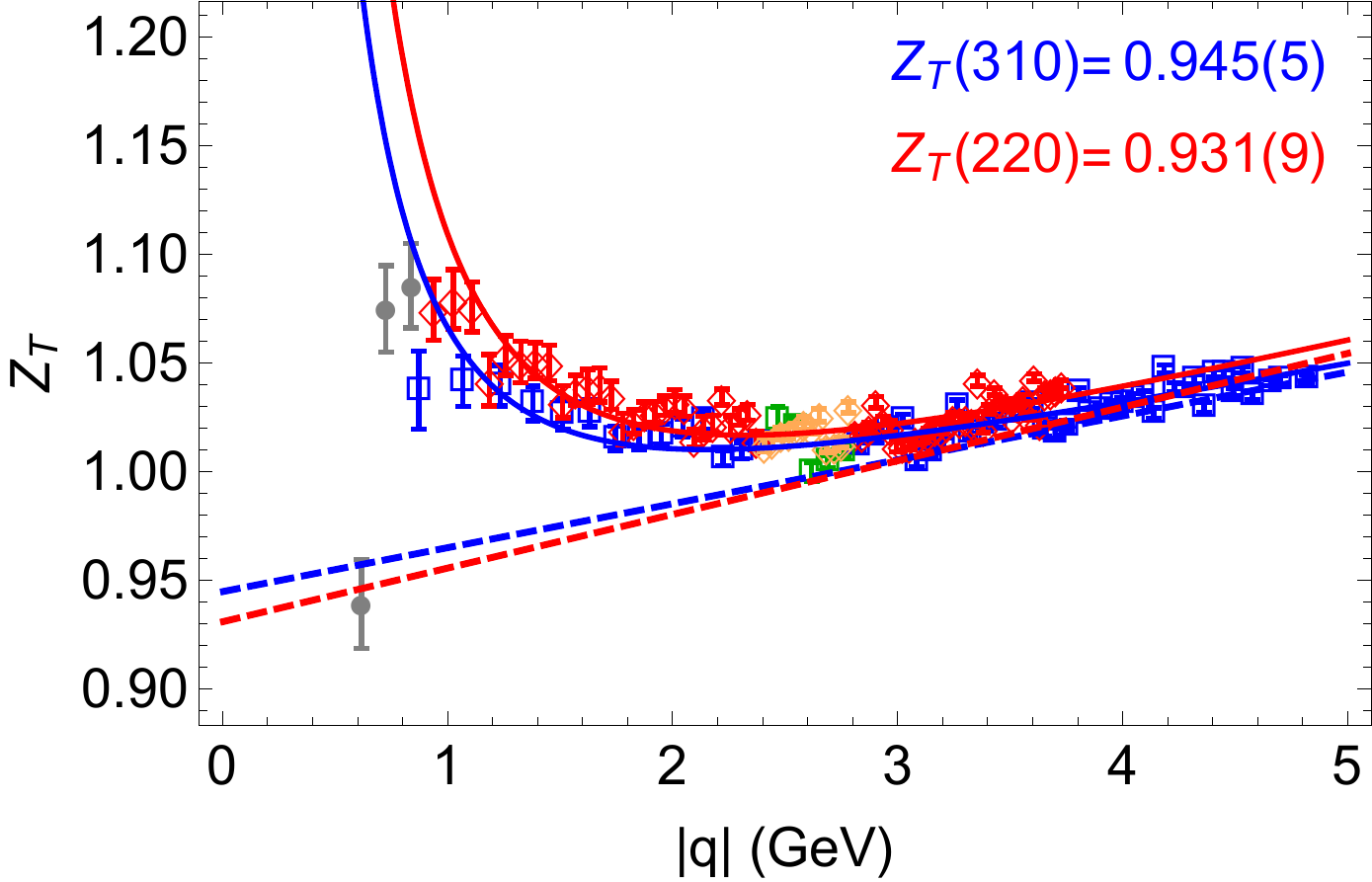}
    \includegraphics[height=1.43in,trim={0.6cm   0.65cm 0 0},clip]{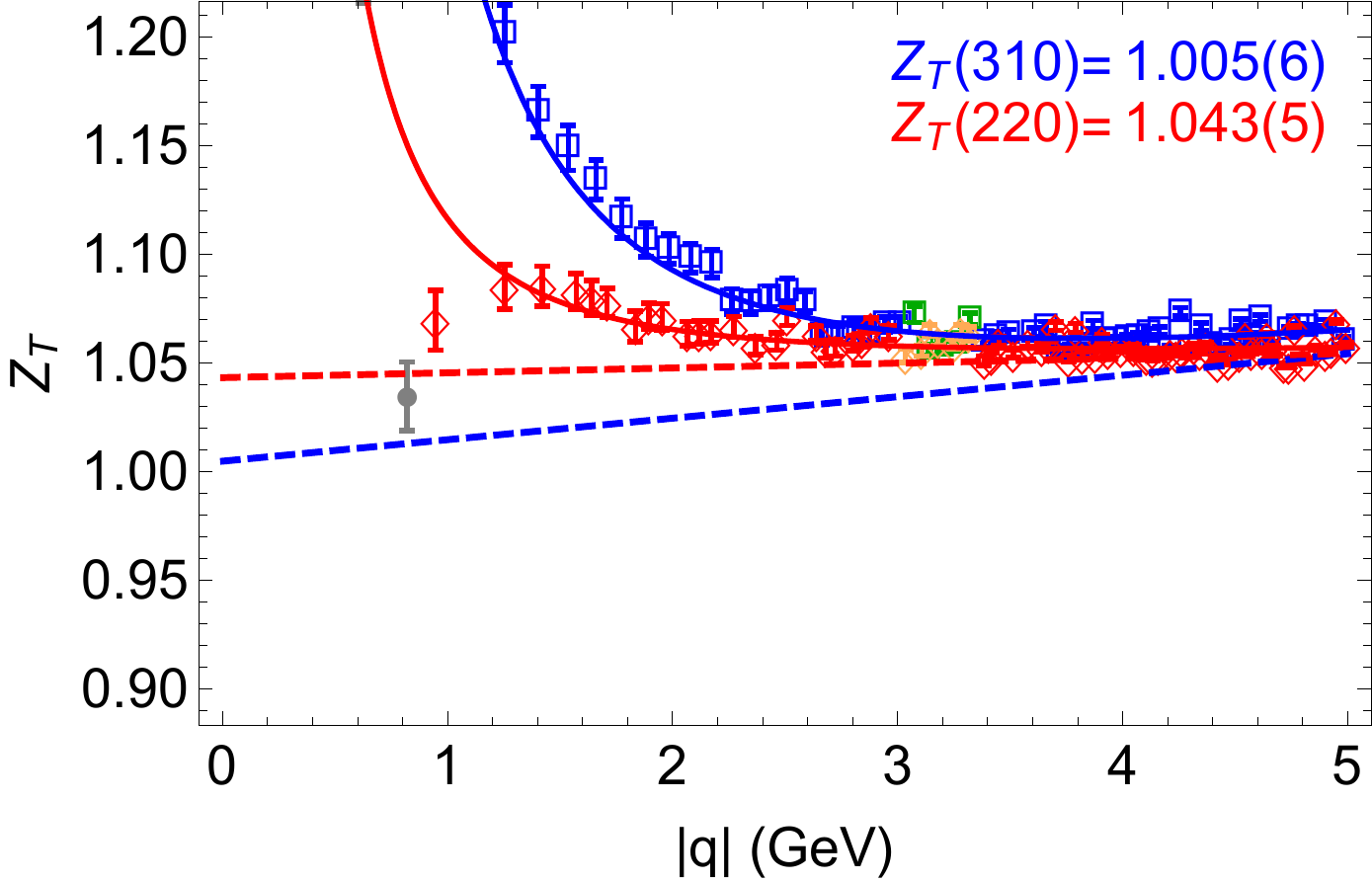}
}
  \subfigure{
    \includegraphics[height=1.56in,trim={0.1cm   0.03cm 0 0},clip]{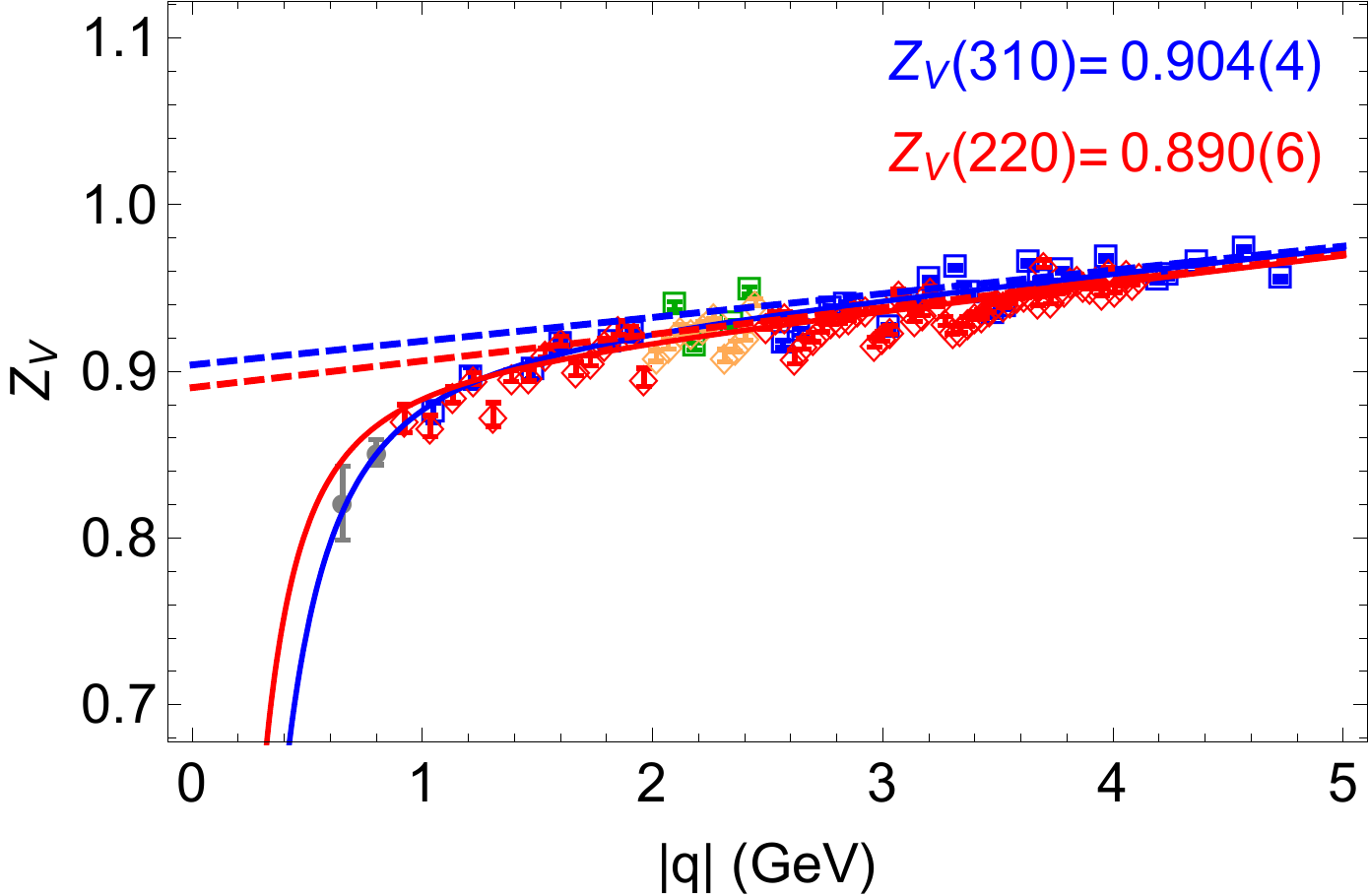}
    \includegraphics[height=1.56in,trim={0.6cm   0.03cm 0 0},clip]{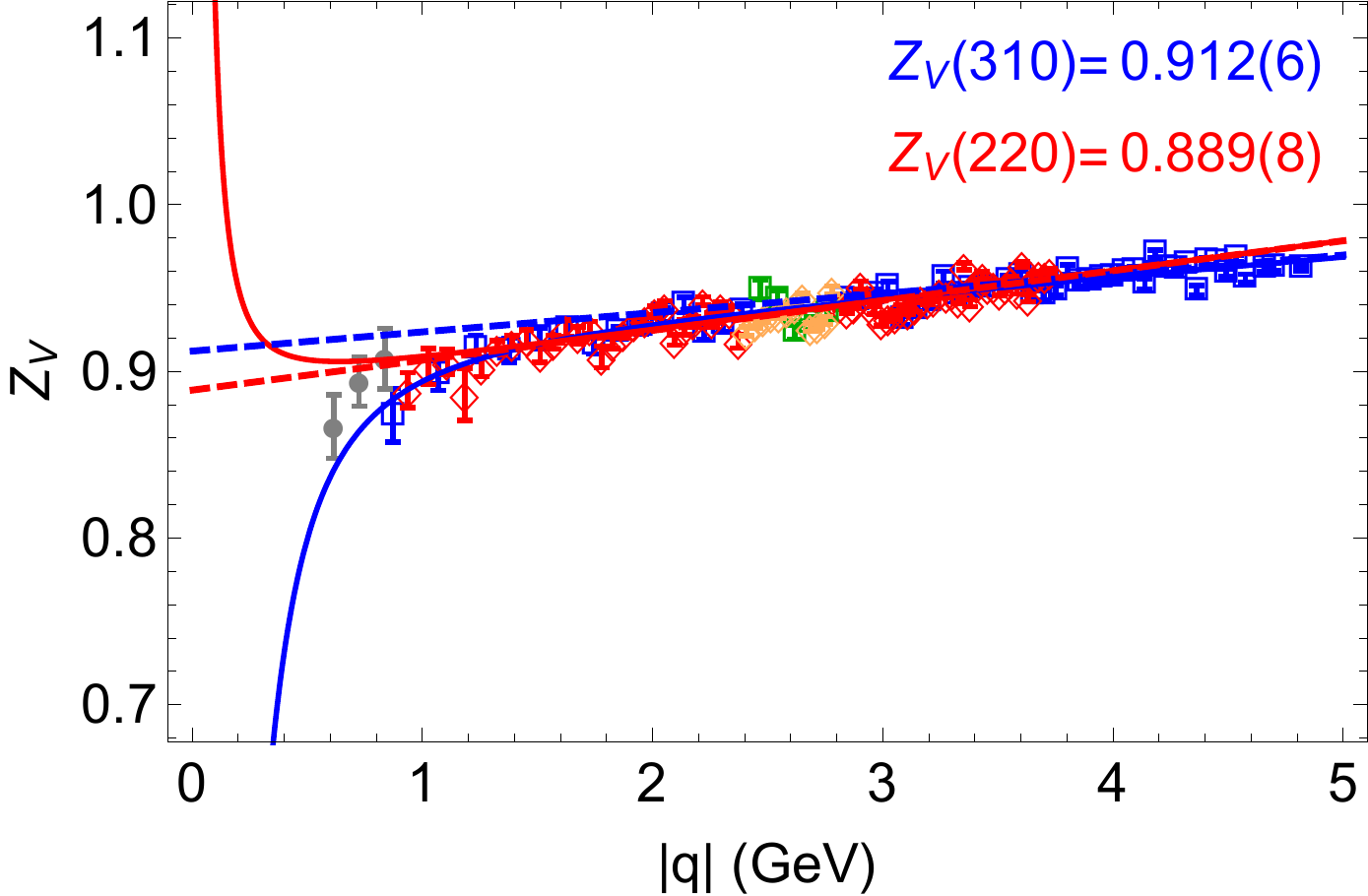}
    \includegraphics[height=1.56in,trim={0.6cm   0.03cm 0 0},clip]{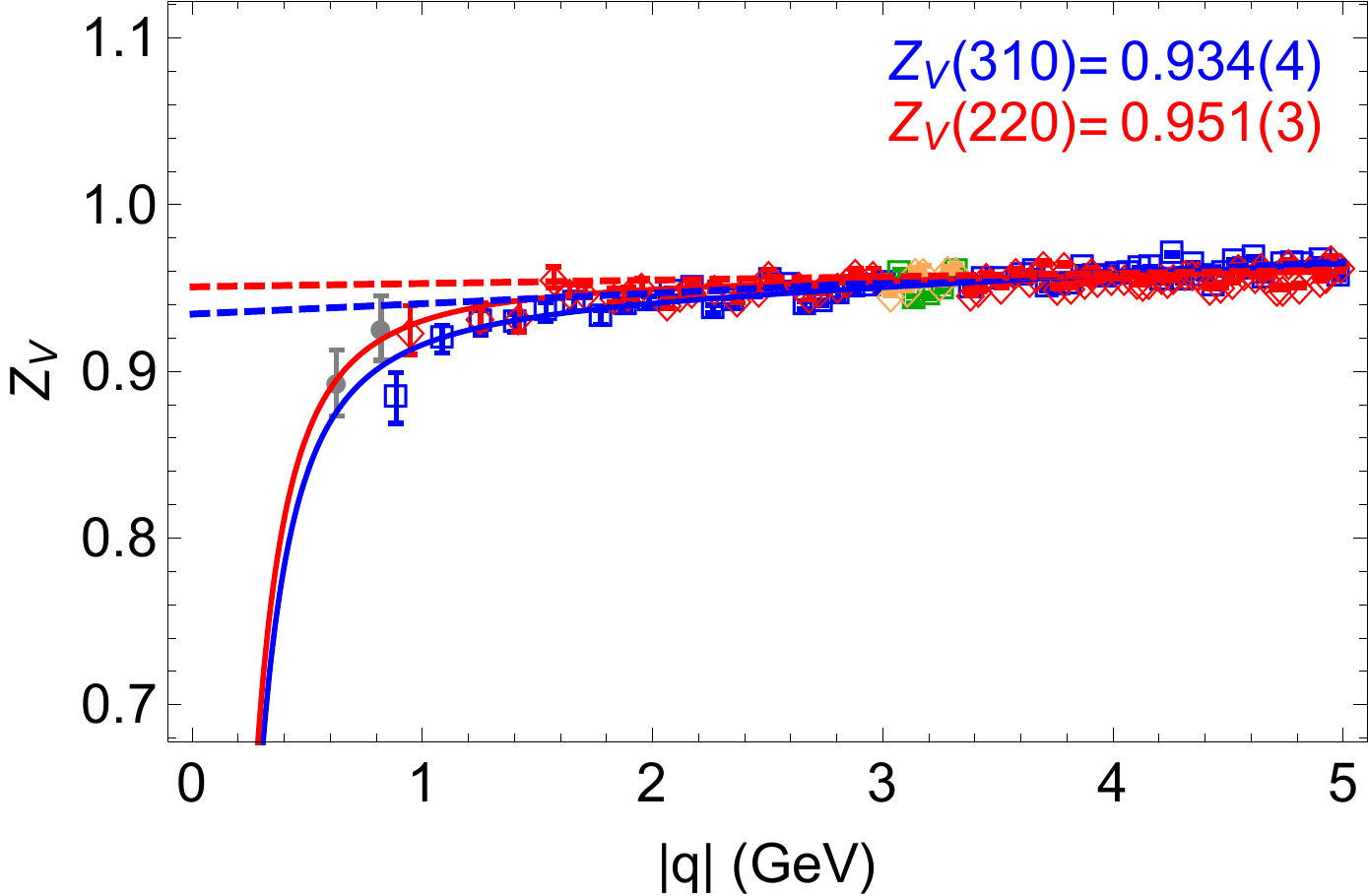}
}
\caption{Data for the four renormalization constants $Z_A$, $Z_S$,
  $Z_T$ and $Z_V$ in the $\overline{\text{MS}}$ scheme at $\mu =
  2\GeV$ as a function of the lattice momenta $|q|$ used 
  in the RI-sMOM scheme. The data are organized as
  follows: (left column) $a=0.12$~fm, (middle column) $a=0.09$~fm and
  (right column) $a=0.06$~fm ensembles. In each panel, we show the
  data at the two values of $M_\pi$ analyzed and use blue squares
  to label the $M_\pi \approx 310$~MeV and red diamonds for the $M_\pi \approx
  220$~MeV ensembles.  The dashed line is the linear part of the fit
  $c/q^2 + Z + d q $ used in method A to extract the $Z$'s. Data in
  the region of $q^2$ used to extract the $Z$'s using method B are
  shown using green ($M_\pi \approx 310$~MeV) and orange ($M_\pi
  \approx 220$~MeV) symbols.}
  \label{fig:Zplots}
\end{figure*}

\begin{figure*}[tb]
  \subfigure{
    \includegraphics[height=1.42in,trim={0.1cm   0.65cm 0 0},clip]{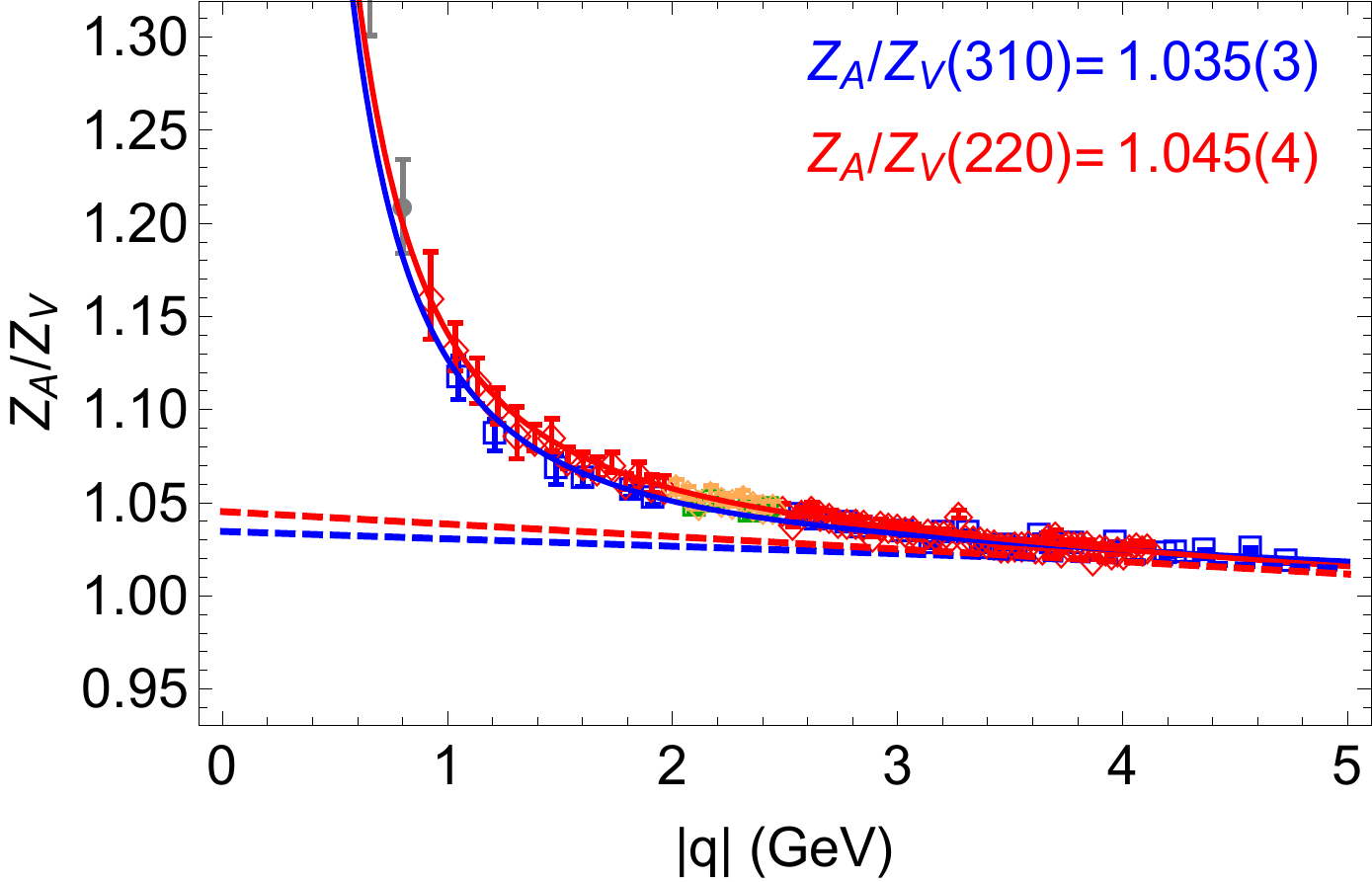}
    \includegraphics[height=1.42in,trim={0.6cm   0.65cm 0 0},clip]{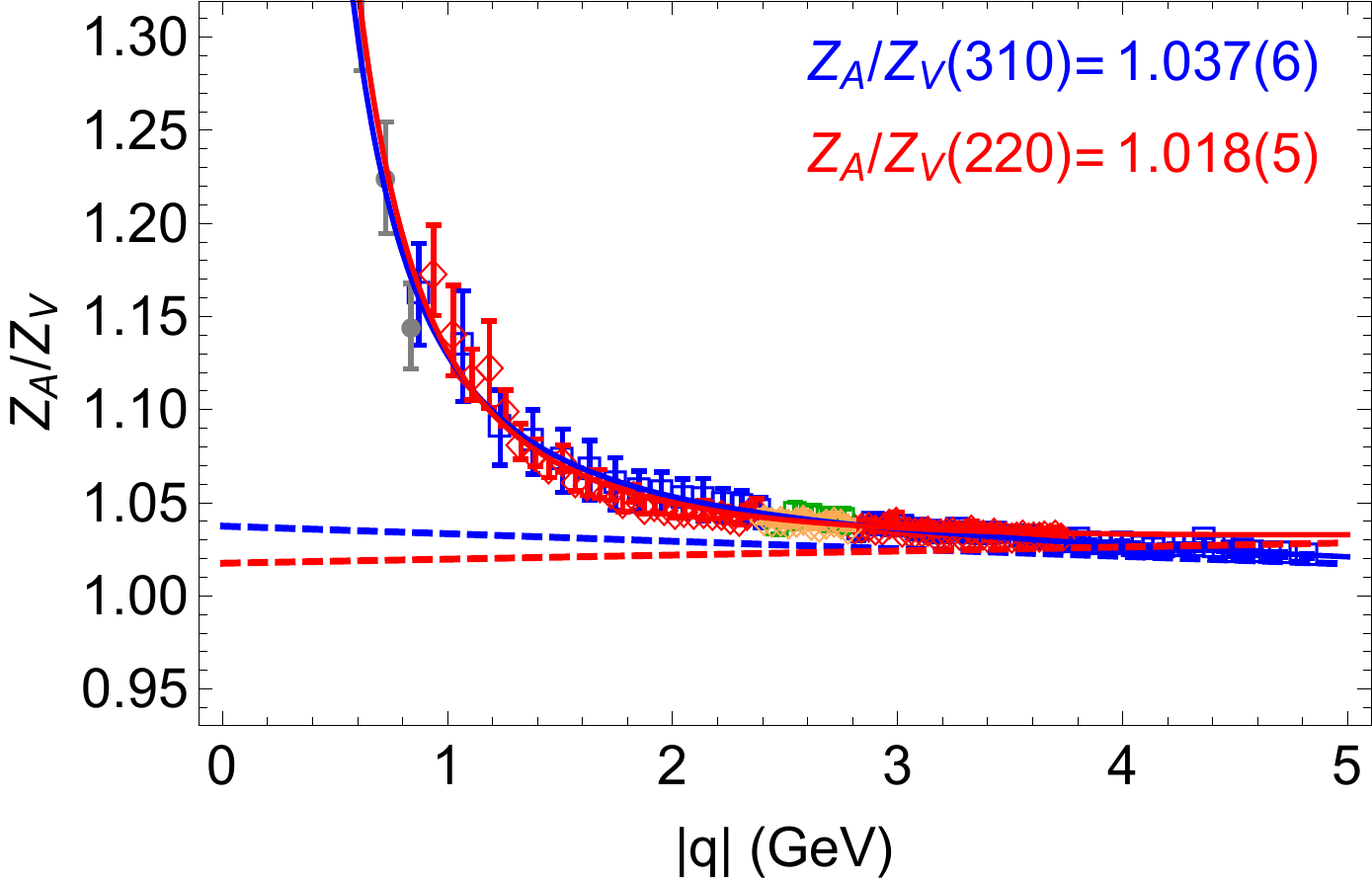}
    \includegraphics[height=1.42in,trim={0.6cm   0.65cm 0 0},clip]{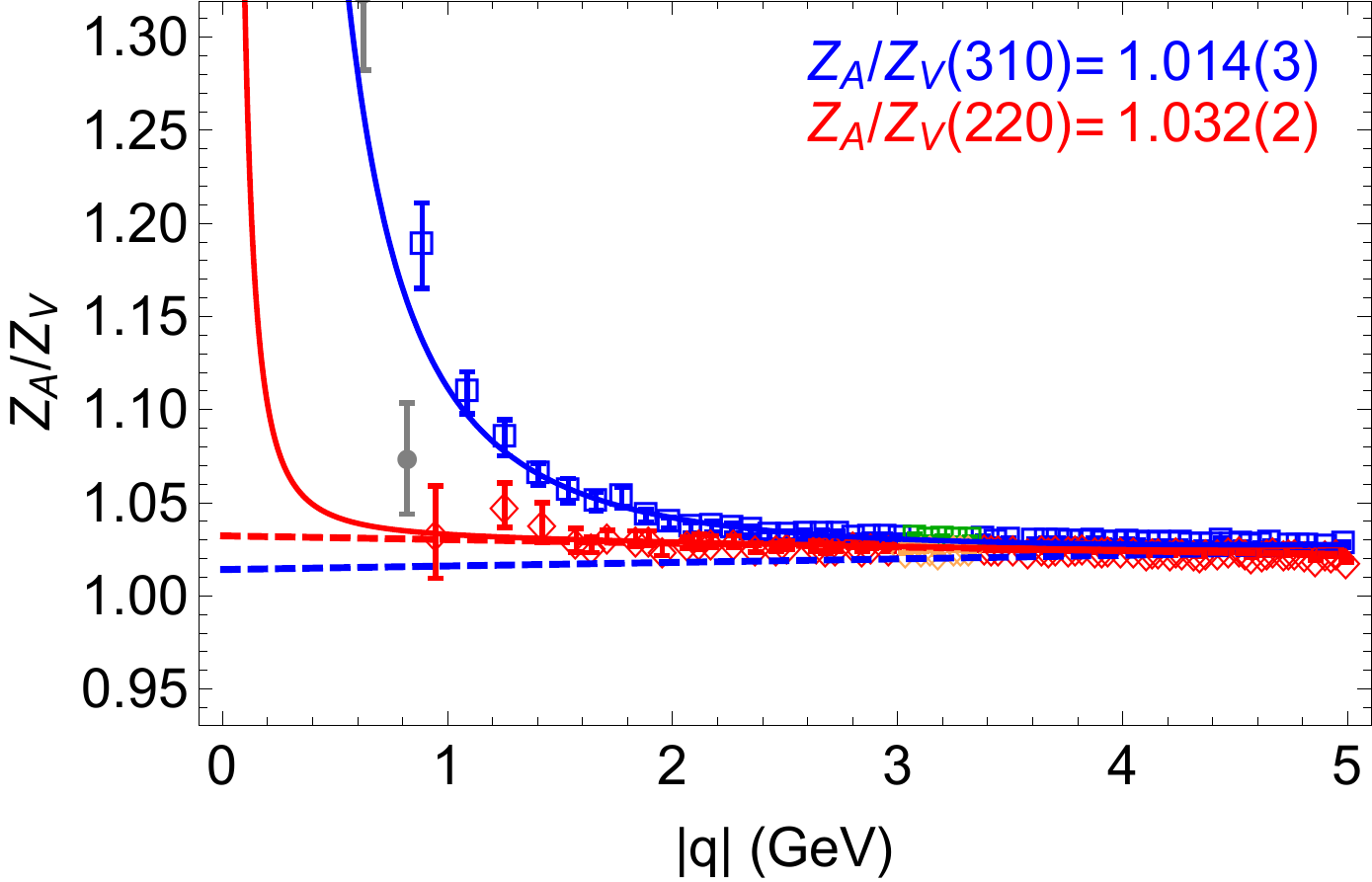}
}
  \subfigure{
    \includegraphics[height=1.45in,trim={0.1cm   0.65cm 0 0},clip]{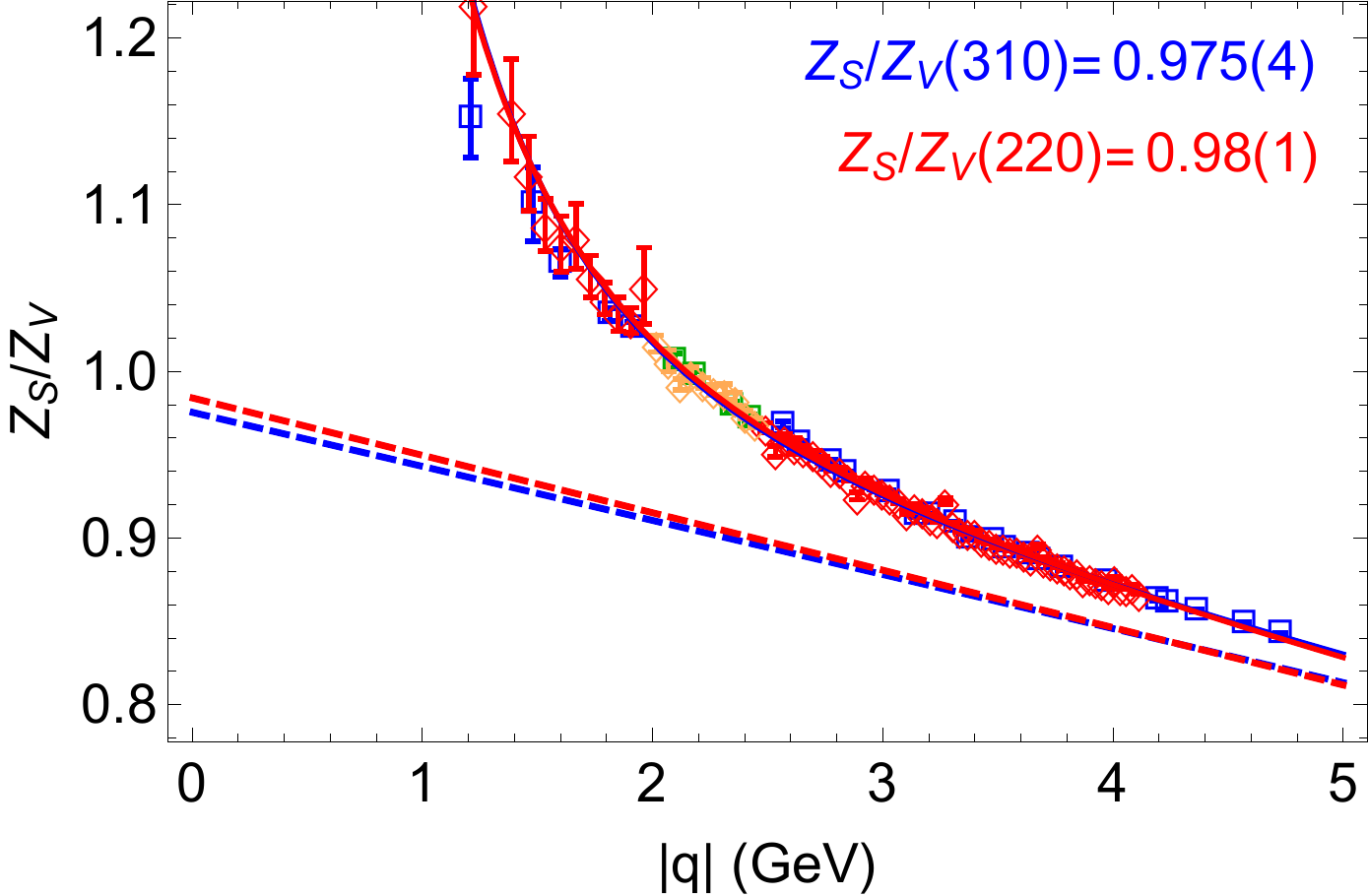}
    \includegraphics[height=1.45in,trim={0.6cm   0.65cm 0 0},clip]{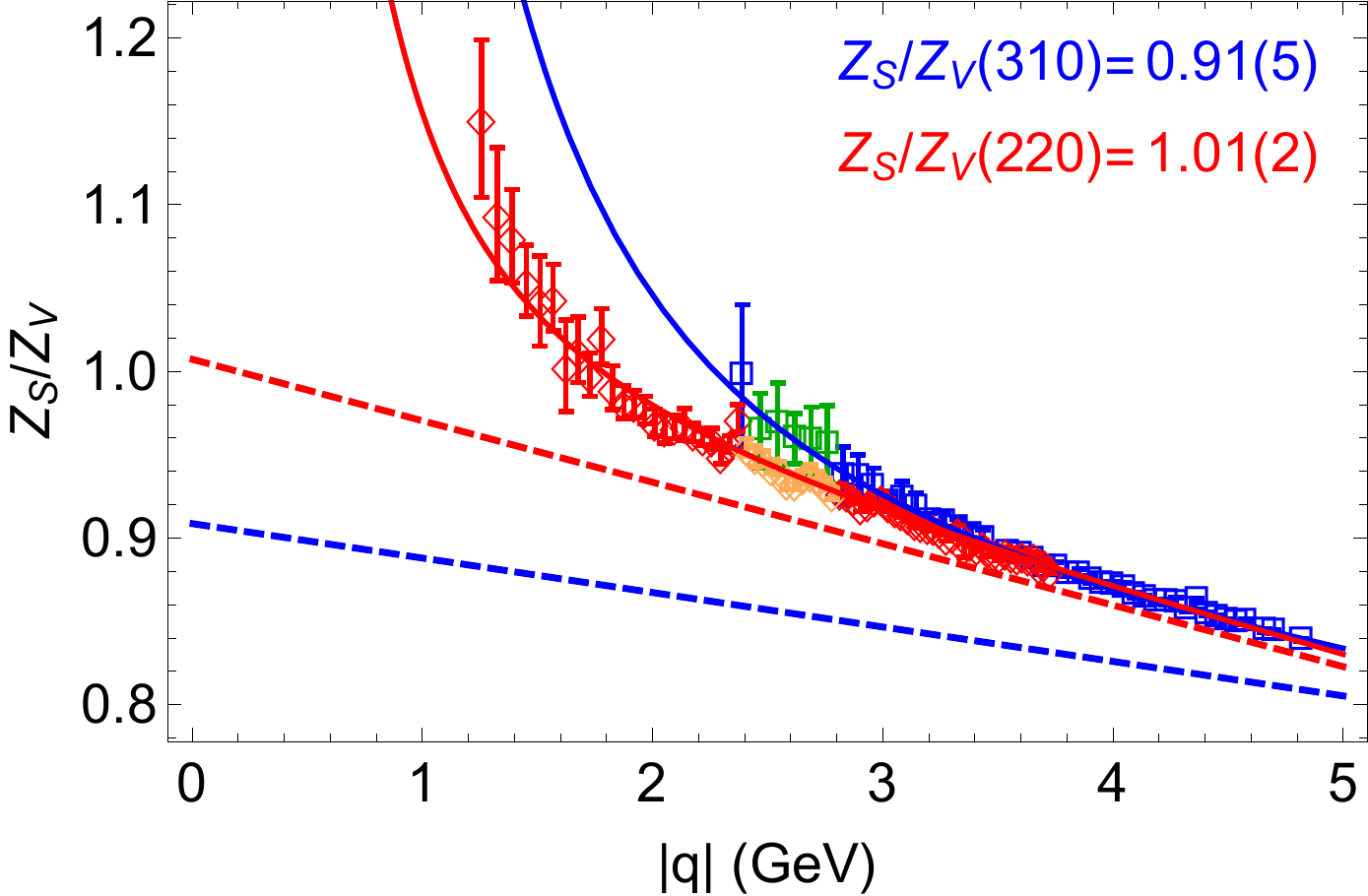}
    \includegraphics[height=1.45in,trim={0.6cm   0.65cm 0 0},clip]{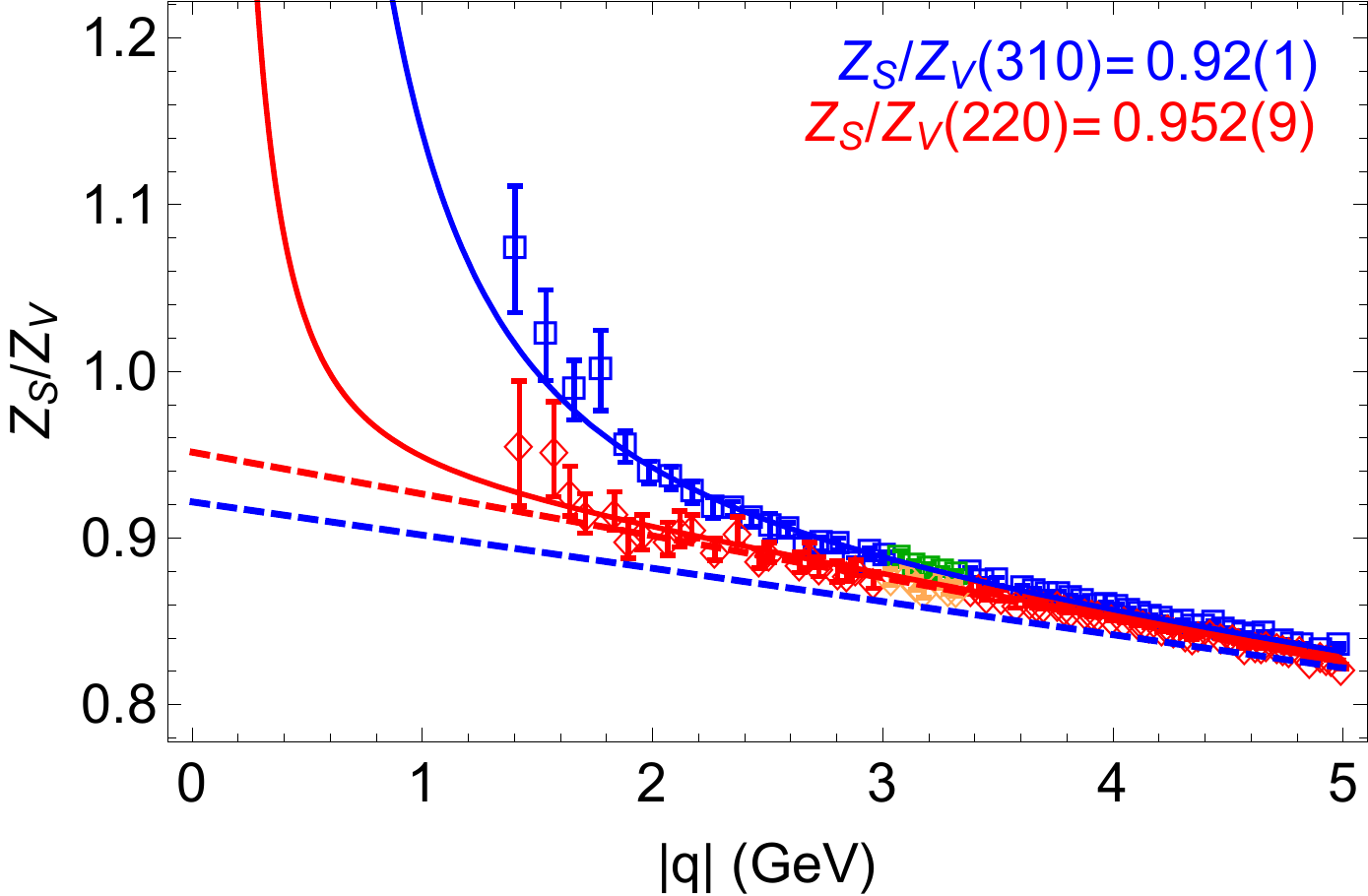}
}
  \subfigure{
    \includegraphics[height=1.53in,trim={0.1cm   0.03cm 0 0},clip]{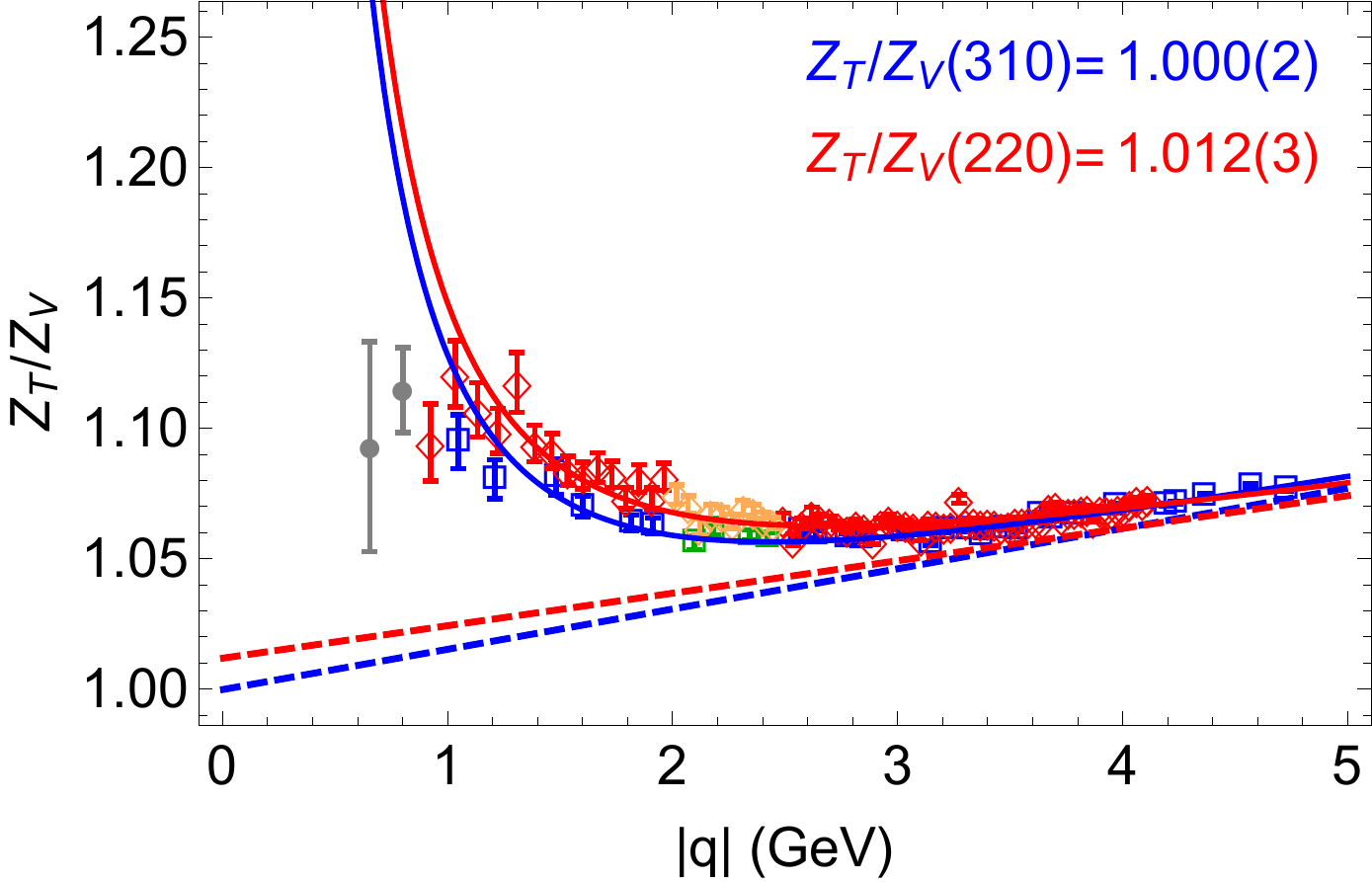}
    \includegraphics[height=1.53in,trim={0.6cm   0.03cm 0 0},clip]{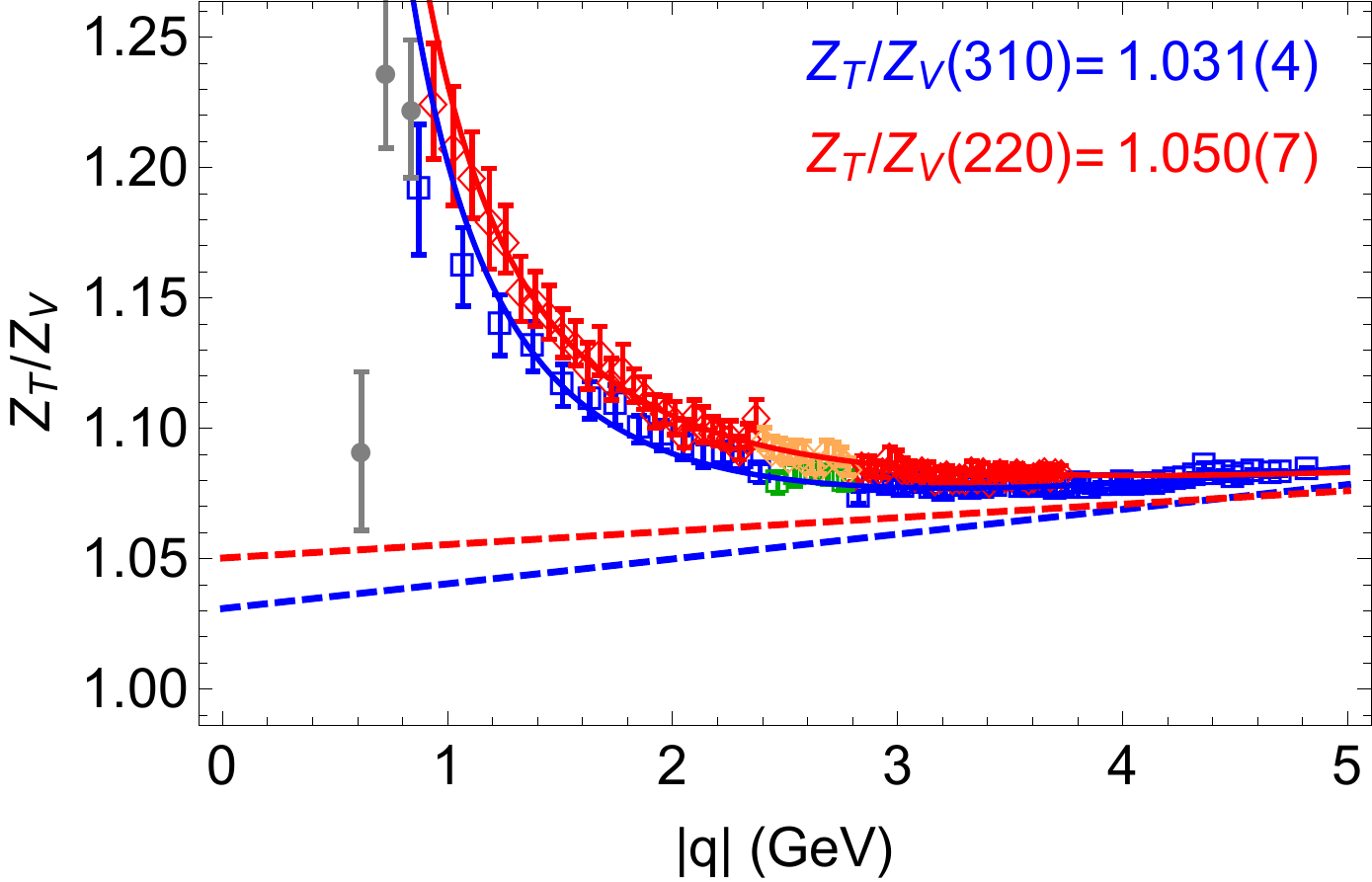}
    \includegraphics[height=1.53in,trim={0.6cm   0.03cm 0 0},clip]{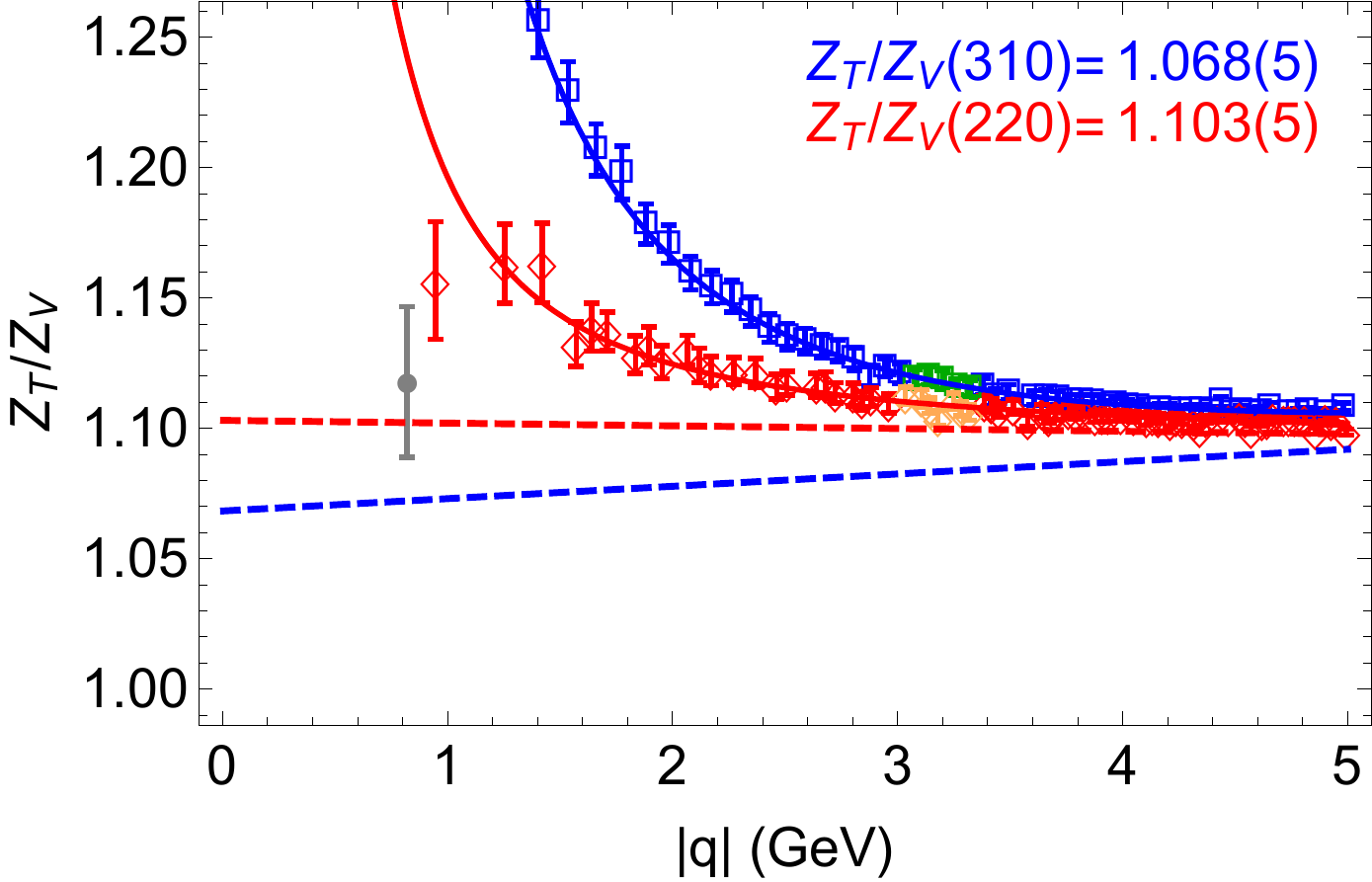}
}
\caption{Data for the ratios of renormalization constants $Z_A/Z_V$,
  $Z_S/Z_V$, $Z_T/Z_V$ in the $\overline{\text{MS}}$ scheme at $\mu =
  2\GeV$ as a function of the lattice momenta $|q|$ used 
  in the RI-sMOM scheme. The rest is the same as in
  Fig.~\protect\ref{fig:Zplots}. }
  \label{fig:Zratios}
\end{figure*}
\begin{figure*}[tb]
  \subfigure{
    \includegraphics[height=1.43in,trim={0.03cm   0.65cm 0 0},clip]{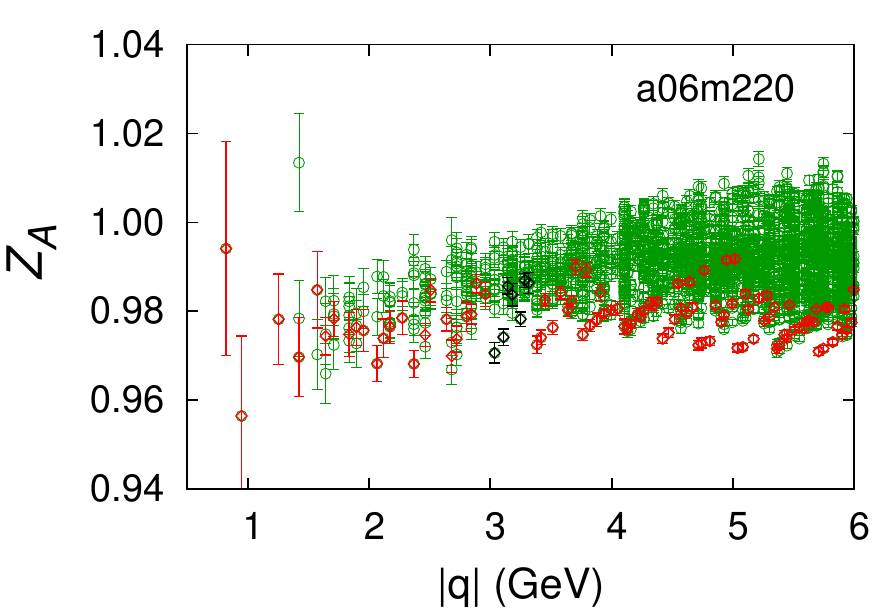}
    \includegraphics[height=1.43in,trim={0.03cm   0.65cm 0 0},clip]{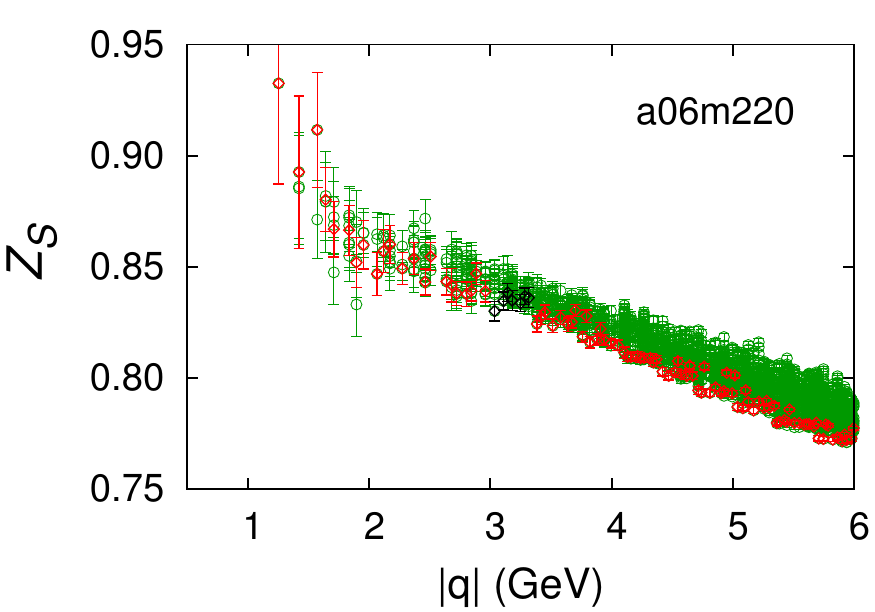}
    \includegraphics[height=1.43in,trim={0.03cm   0.65cm 0 0},clip]{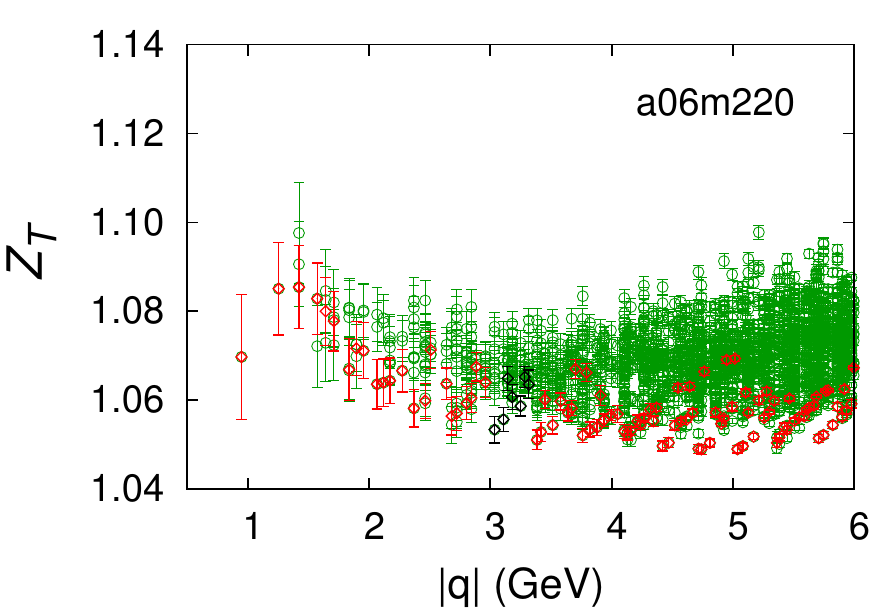}
}
  \subfigure{
    \includegraphics[height=1.60in,trim={0.03cm   0.03cm 0 0},clip]{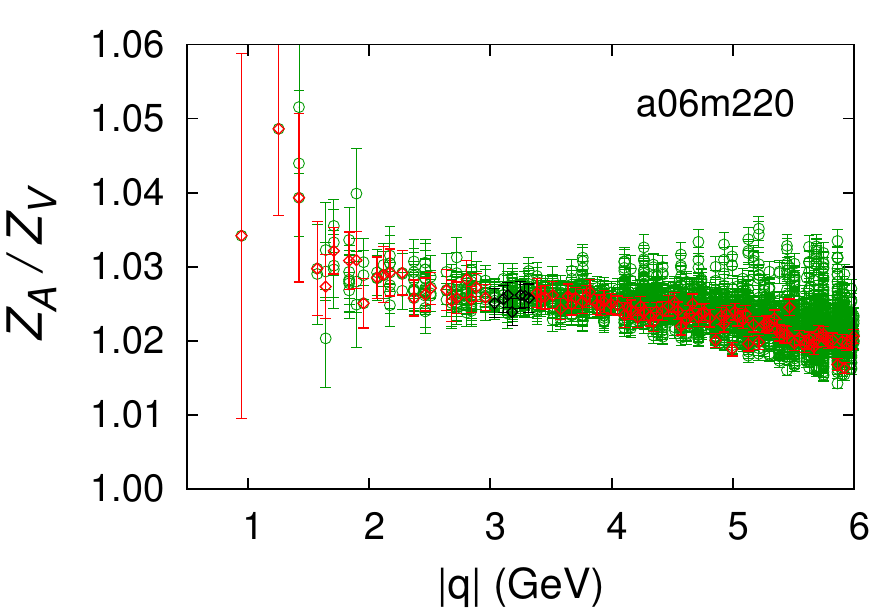}
    \includegraphics[height=1.60in,trim={0.03cm   0.03cm 0 0},clip]{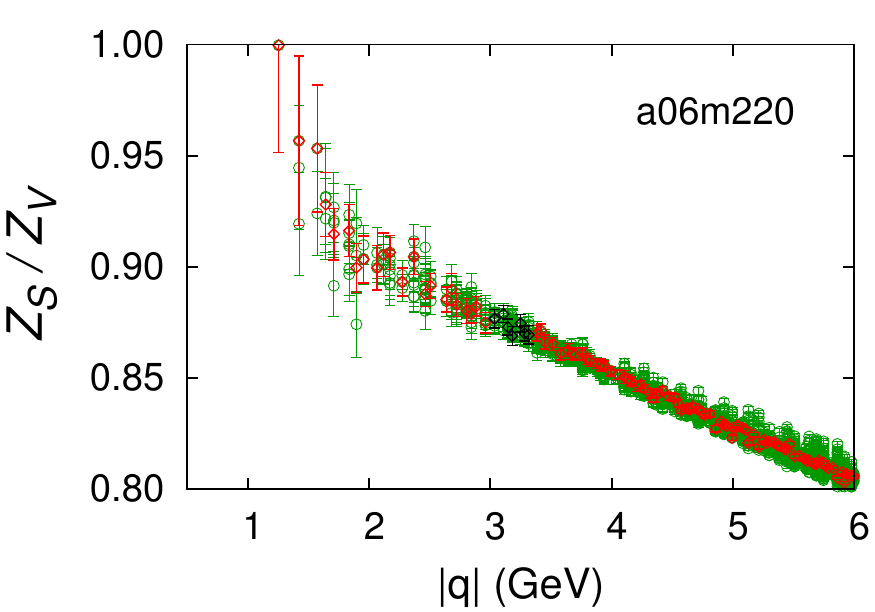}
    \includegraphics[height=1.60in,trim={0.03cm   0.03cm 0 0},clip]{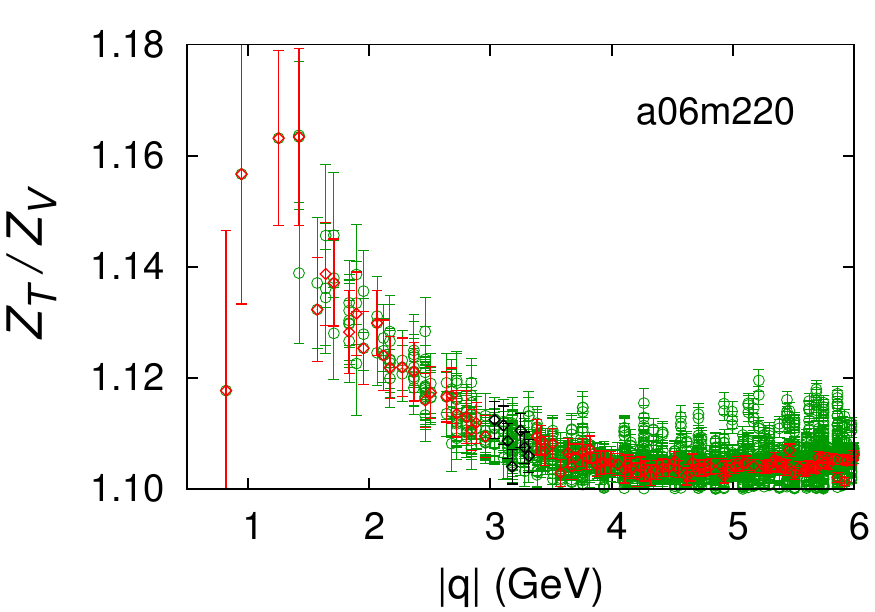}
}
\caption{Data for the renormalization constants in the
  $\overline{\text{MS}}$ scheme at $\mu = 2\GeV$ as a function of the
  lattice momenta $|q|$ used in the RI-sMOM scheme. (Top) The data for
  $Z_A$, $Z_S$ and $Z_T$ for all combinations of $(p_1)_\mu$ and
  $(p_2)_\mu$ calculated are plotted using green circles. The subset
  of points that minimize $\sum_\mu (p_1)_\mu^4/((p_1)^2)^2 + \sum_\mu
  (p_2)_\mu^4/((p_2)^2)^2$ for each $q^2$ are shown using red
  diamonds.  The value and error using method B discussed in the text
  is given by the average and variance of the points shown in black
  over the range 9.2--11.2~GeV${}^2$.  (Bottom) The data for the
  ratios $Z_A/Z_V$, $Z_S/Z_V$ and $Z_T/Z_V$ using the same symbols. }
  \label{fig:Zprune}
\end{figure*}

The calculation of the renormalization constants $Z_A$, $Z_V$, $Z_S$
and $Z_T$ of the quark bilinear operators in the 
regularization-independent symmetric momentum-subtraction (RI-sMOM)
scheme~\cite{Martinelli:1994ty,Sturm:2009kb} has been done on six 
ensembles: {\it a12m310, a12m220, a09m310, a09m220, a06m310} and {\it a06m220}.
The method and the details of this calculation have been given in
Ref.~\cite{Bhattacharya:2015wna} and here we briefly describe new 
aspects of the calculations and state the
results. 

To derive $Z_A$, $Z_S$, $Z_S$ and $Z_V$ in the continuum
$\overline{\text{MS}}$ scheme at $\mu = 2\GeV$, starting with the
lattice results obtained in the RI-sMOM scheme, we proceed as
follows. The RI-sMOM estimate obtained at a given lattice squared
four-momentum $q^2$ is first converted to the $\overline{\text{MS}}$
scheme at the same scale (horizontal matching) using two-loop
perturbative relations expressed in terms of the coupling constant
$\alpha_{\overline{\text{MS}}}(q^2)$~\cite{Gracey:2011fb}. This
estimate at $\mu=q^2$, is then run in the continuum in the
$\overline{\text{MS}}$ scheme to $2\GeV$ using the 3-loop anomalous
dimension relations for the scalar and tensor
bilinears~\cite{Gracey:2000am,Agashe:2014kda}.  The result is a set of
data points as a function of $q^2$ that are shown in
Figs.~\ref{fig:Zplots} and~\ref{fig:Zratios}. The mass-independent
renormalization factors are extracted from these as discussed below.

In calculations at sufficiently small values of the lattice spacing
$a$, one expects a window, $\eta \Lambda_{\rm QCD} \ll q \ll \xi \pi/a$,
in which the data for the $Z$'s, shown in Figs.~\ref{fig:Zplots}
and~\ref{fig:Zratios}, are independent of $q$; that is, the data should
show a plateau versus $q$. The lower cutoff $\eta \Lambda_{\rm QCD}$
is dictated by nonperturbative effects and the upper cutoff $\xi \pi/a$ by
discretization effects. Here $\eta $ and $\xi$ are, {\it a priori},
unknown dimensionless numbers of $O(1)$ that depend on the lattice action and
the link smearing procedure.  The main systematics contributing to the
lack of such a window and the resulting uncertainty in the extraction
of the renormalization constants are (i) breaking of the Euclidean
$O(4)$ rotational symmetry to the hypercubic group, because of which
different combinations of $q_\mu$ with the same $q^2$ give different
results in the RI-sMOM scheme; (ii) nonperturbative effects at small
$q^2$; (iii) discretization errors at large $q^2$ other than the
$O(4)$ breaking effects listed above; and (iv) truncation errors in
the perturbative matching to the $\overline{\text{MS}}$ scheme and
running to $2\GeV$.  These systematics are discussed below.

To assess the truncation errors, we compare the conversion factor
using 2-loop matching and 3-loop running versus 1-loop matching and
2-loop running. The 2-loop series for the matching of $Z_T$ is poorly
behaved. For example, the 2-loop series for $Z_T$ when using
horizontal matching between the RI-sMOM and $\overline{\text{MS}}$
schemes is $1 + 0.0052 + 0.0159$ at $q^2=4$~GeV${}^2$ and is $1 +
0.0037 + 0.0078$ at $q^2=25$~GeV${}^2$. In obtaining these estimates
we used the 4-flavor $\alpha_s(\overline{\text{MS}},2\ {\rm
  GeV})=0.3051$ and $\alpha_s(\overline{\text{MS}},5\ {\rm
  GeV})=0.2138$.  On the other hand, the 2-loop series for $Z_S$ at
$q^2=4$~GeV${}^2$, $1 - 0.0157 -0.0039$, is much better behaved.  The
poor convergence of the 2-loop matching factor suggests a systematic
uncertainty of $O(2\%)$ for $Z_T$. Our final error estimates given in
Table~\ref{tab:Zfinal} are larger than this. The series for the factor
describing the running in the continuum is better behaved and the
difference between the 2- and 3-loop estimates between the values of
$q^2$ used in the analysis and 4~GeV${}^2$ is less than $20\%$ of the
quoted errors and starts to become significant only for $q^2 <
1$~GeV${}^2$ where $\alpha_s$ is large. Consequently, we use data with
$q^2 > 4$~GeV${}^2$ when obtaining our final estimates given in
Tables~\ref{tab:Zmethod1and2} and~\ref{tab:resultsrenorm}.

The lattice data in the RI-sMOM scheme have significant $O(4)$
breaking effects.  The size of the spread in $Z_A$, $Z_S$, $Z_T$ and
$Z_V$ and in the ratios $Z_A/Z_V$, $Z_S/Z_V$, and $Z_T/Z_V$ is
illustrated in Fig.~\ref{fig:Zprune} using the $a06m220$ ensemble by
plotting all the data using green circles.  From this full data set,
we pick the points with the smallest values of $\sum_\mu
(p_1)_\mu^4/((p_1)^2)^2 + \sum_\mu (p_2)_\mu^4/((p_2)^2)^2 $, where
$(p_1)_\mu$ and $(p_2)_\mu$ are the momenta in the two external
fermion legs, $q_\mu = (p_1)_\mu - (p_2)_\mu$ and $(p_i)^2 = \sum_\mu
(p_i)_\mu^2$. These points, shown as red diamonds, are expected to
have smaller $O(4)$ breaking effects. Of these red points, the data
between 9.2 and 11.2~GeV${}^2$ used to extract the $Z$'s using method B
discussed below are shown as black diamonds.

The data for the ratios $Z_A/Z_V$, $Z_S/Z_V$, and $Z_T/Z_V$ show much
smaller $O(4)$ breaking, presumably because some of the systematics
cancel. Our final estimates of the renormalized charges are,
therefore, obtained using these ratios as discussed below. Overall, a
better understanding and control over the artifacts due to $O(4)$
symmetry breaking in the RI-sMOM scheme is needed as the contribution
of the uncertainty in the renormalization constants to the errors in
the charges is now larger than the statistical errors and the
excited-state contamination.

After conversion to the $\overline{\text{MS}}$ scheme at 2~GeV, the
selected data points that minimize $\sum_\mu (p_1)_\mu^4/((p_1)^2)^2 +
\sum_\mu (p_2)_\mu^4/((p_2)^2)^2 $ for each $q^2$ are plotted in
Fig.~\ref{fig:Zplots} for the six ensembles as a function of 
$q^2$. The data for $Z_\Gamma$ and $Z_\Gamma/Z_V$ show
nonperturbative effects at low $|q|$ values followed by an
approximate plateau for $Z_A$, $Z_T$, and $Z_V$, whereas the data for
$Z_S$ continue to show a large $q$ dependence. 

From these data in the  $\overline{\text{MS}}$ scheme at 2~GeV, we
extract the estimates in two ways~\cite{Bhattacharya:2015wna}:
\begin{itemize}
\item
Method A: We fit the data with $q^2>0.8$~GeV${}^2$ using the Ansatz
$c/q^2 + Z + d q $, where the first term $c/q^2$ is used to account
for nonperturbative artifacts and the third, $d q$, for the
discretization errors. The errors in the $Z$'s are obtained by using
100 bootstrap samples.
\item
Method B: The estimate for $Z$ is taken to be an average over the data
points in an interval about $q^2 = \Lambda /a $, where the scale
$\Lambda=3$~GeV is chosen to be large enough to avoid nonperturbative
effects and above which perturbation theory is expected to be
reasonably well behaved. This choice satisfies both $q a \to 0$ and
$\Lambda/q \to 0$ in the continuum limit as desired, and $q_\mu a -
\sin(q_\mu a) < 0.05$ to bound discretization effects. In our
simulations, this choice corresponds to $q^2 = 5$, 6.8, and
$10.2$~GeV${}^2$ for the $a=0.12$, 0.09, and 0.06~fm ensembles,
respectively. To estimate the value and the error, we, further choose
an interval of 2 GeV${}^2$ about these $q^2$; i.e., we take the mean
and the standard deviation of the data over the ranges 4--6, 5.8--7.8
and 9.2--11.2 GeV${}^2$.
\end{itemize}
The interval of $q^2$ that contributes to both methods is consistent
with the general requirement that $\eta \Lambda_{\rm QCD} \ll q \ll
\xi \pi/a$, with $\eta$ and $\xi$ of $O(1)$, to avoid both
nonperturbative and discretization artifacts.

\begin{table*}
\centering
\begin{ruledtabular}
\begin{tabular}{c|cccc|ccc}
ID        & $Z_A$      & $Z_S$       &  $Z_T$      & $Z_V$       & $Z_A/Z_V$    & $Z_S/Z_V$   & $Z_T/Z_V$  \\
\hline                                                                                              
a12m310   & 0.929(5)   & 0.912(06)   & 0.899(4)    & 0.904(4)    & 1.035(3)     & 0.975(04)   & 1.000(2) \\
a12m220   & 0.926(7)   & 0.852(20)   & 0.901(6)    & 0.890(6)    & 1.045(4)     & 0.984(13)   & 1.012(3) \\
a09m310   & 0.951(4)   & 0.868(45)   & 0.945(5)    & 0.912(7)    & 1.037(6)     & 0.909(54)   & 1.031(4) \\
a09m220   & 0.899(9)   & 0.867(22)   & 0.931(9)    & 0.889(8)    & 1.018(5)     & 1.007(20)   & 1.050(7) \\
a06m310   & 0.948(5)   & 0.863(11)   & 1.005(6)    & 0.934(4)    & 1.014(3)     & 0.922(10)   & 1.068(5) \\
a06m220   & 0.979(4)   & 0.901(10)   & 1.043(5)    & 0.951(3)    & 1.032(2)     & 0.952(09)   & 1.103(5) \\
\hline                           
a12m310   & 0.979(1)   & 0.925(17)   & 0.989(14)   & 0.933(14)   & 1.0472(21)   & 0.991(17)   & 1.058(2) \\
a12m220   & 0.970(9)   & 0.914(11)   & 0.984(9)    & 0.921(10)   & 1.0526(35)   & 0.994(14)   & 1.068(4) \\
a09m310   & 0.975(10)  & 0.901(13)   & 1.013(11)   & 0.937(11)   & 1.0411(22)   & 0.962(5)    & 1.081(2) \\
a09m220   & 0.972(7)   & 0.878(6)    & 1.018(6)    & 0.934(7)    & 1.0407(14)   & 0.941(8)    & 1.090(3) \\
a06m310   & 0.980(8)   & 0.840(9)    & 1.064(8)    & 0.951(8)    & 1.0300(6)    & 0.883(4)    & 1.118(2) \\
a06m220   & 0.981(7)   & 0.835(3)    & 1.060(5)    & 0.957(6)    & 1.0255(8)    & 0.873(4)    & 1.109(3) \\
\end{tabular}
\end{ruledtabular}
\caption{The renormalization constants $Z_A$, $Z_S$, $Z_T$, $Z_V$ and
  the ratios $Z_A/Z_V$, $Z_S/Z_V$ and $Z_T/Z_V$ in the
  $\overline{\text{MS}}$ scheme at $2\GeV$ obtained on six ensembles
  at the three values of the lattice spacings. (Top) Estimates are
  from method A using the fit $1/q^2 + Z + d q $. (Bottom) Estimates
  from method B obtained by averaging the data over an interval in
  $q^2 $ as described in the text.}
\label{tab:Zmethod1and2}
\end{table*}

\begin{table*}
\centering
\begin{ruledtabular}
\begin{tabular}{c|cccc|ccc}
ID   &   $Z_A$      & $Z_S$     &   $Z_T$    & $Z_V$       & $Z_A/Z_V$ & $Z_S/Z_V$ & $Z_T/Z_V$ \\
\hline                                                      
a12  &  $0.95(3)$   & $0.90(4)$ &  $0.94(4)$ & $0.91(2)$ & $1.045(09)$ & $0.986(09)$ & $1.034(34)$ \\
a09  &  $0.95(4)$   & $0.88(2)$ &  $0.98(4)$ & $0.92(2)$ & $1.034(11)$ & $0.955(49)$ & $1.063(29)$ \\
a06  &  $0.97(3)$   & $0.86(3)$ &  $1.04(3)$ & $0.95(1)$ & $1.025(09)$ & $0.908(40)$ & $1.100(25)$ \\
\end{tabular}
\end{ruledtabular}
\caption{The final mass-independent renormalization constants $Z_A$,
  $Z_S$, $Z_T$, $Z_V$ and the ratios $Z_A/Z_V$, $Z_S/Z_V$ and
  $Z_T/Z_V$ in the $\overline{\text{MS}}$ scheme at $2\GeV$ at the
  three values of the lattice spacings used in our analysis. The
  central value is the average of estimates from the two methods and
  at the two pion masses given in Table~\protect\ref{tab:Zmethod1and2}
  and the errors are taken to be half the spread.}
\label{tab:Zfinal}
\end{table*}

The estimates using method B for all four $Z$'s and at all three
values of the lattice spacing are found to overlap for the two values
of $M_\pi$. As a result, in Ref.~\cite{Bhattacharya:2015wna} we had
neglected possible mass dependence in implementing method A and made a
fit using the Ansatz $c/q^2 + Z + d q $ to the combined data.  In this
work, we analyze the six ensembles separately and present the estimates
from both methods in Table~\ref{tab:Zmethod1and2}.

The estimates from method A show significant dependence on
$M_\pi$ in some cases even though the data at the two values of
$M_\pi$ overlap.  We do not find a uniform trend with $M_\pi$ at the
three different values of $a$, and given the size of the errors in the
data at the two values of $M_\pi$ at fixed $a$ we are not able to make
a reliable extrapolation in $M_\pi$. So we ignore possible dependence
on $M_\pi$ in the data and average the estimates from the two methods
at the two values of $M_\pi$ given in Table~\ref{tab:Zmethod1and2} to
obtain a ``mass-independent'' estimate.  For the error estimate we take
the larger of the two: half the spread or the largest statistical
error.

The data for the ratios $Z_\Gamma/Z_V$ are shown in
Fig.~\ref{fig:Zratios}. As highlighted above, we find that the
systematic effects due to $O(4)$ breaking are smaller in the ratios.
The data again show a rough plateau in $Z_A/Z_V$ and $Z_T/Z_V$ but not
in $Z_S/Z_V$. We also find a significant variation with $M_\pi$ and
between the two methods in some cases, which we include in our error
estimates as explained above.  The final mass-independent
renormalization constants $Z_\Gamma$ and $Z_\Gamma/Z_V$ at the three
lattice spacings are given in Table~\ref{tab:Zfinal}.  The errors in
these renormalization constants are added in quadrature to those in
the extraction of the bare nucleon charges $g_{A,S,T}^{\rm bare}$ and
$g_{A,S,T}^\text{bare}/g_V^\text{bare}$ given in
Table~\ref{tab:resultsbare} to obtain the renormalized charges in two
ways:
\begin{itemize}
\item
Using the product of the ratios,
$(Z_\Gamma /Z_V) \times (g_\Gamma/g_V^{u-d})$, along with the identity $Z_V
g_V^{u-d} = 1$. These are given in Table~\ref{tab:resultsrenorm}.
\item
Using the product $Z_\Gamma g_\Gamma$. These estimates are given in
Table~\ref{tab:resultsgV} for the isovector charges.
\end{itemize}

The two estimates differ by 1\%--3\%. Most of this difference correlates
with the deviation of $Z_V g_V^{u-d}$ from unity as shown in 
Table~\ref{tab:resultsgV}. The deviation is about $1\sigma$:
approximately $3\%$, $ 3\%$ and $2\%$ at the three lattice spacings,
$a=0.12$, $0.09$ and $0.06$~fm, respectively.  The magnitude of these
deviations is consistent with our estimates of errors in the
renormalization constants given in Table~\ref{tab:Zfinal}, which are
now larger than the statistical errors.  A more detailed discussion
of the error budget and the final error estimates are given in
Sec.~\ref{sec:errors}.

The discretization errors in the two estimates of the renormalized
charges are different, so one should only compare the results after
extrapolation.  In Sec.~\ref{sec:results}, we show that even though
the estimates on individual ensembles differ by 1--3\%, the value of the charges
after extrapolation to the continuum limit and the physical pion mass
are consistent. The errors in the data $Z_\Gamma g_\Gamma$ and in the
extrapolated value obtained from them are larger. This is because the
data for both, $Z_\Gamma /Z_V$ and $g_\Gamma/g_V^{u-d}$, have smaller
errors, presumably because some of the systematics cancel in the
ratios.  Our final estimates are, therefore, obtained using $(Z_\Gamma
/Z_V) \times (g_\Gamma/g_V^{u-d})$ with the relation $Z_V g_V^{u-d} =
1$.

\begin{table*}
\centering
\begin{ruledtabular}
\begin{tabular}{c|ccc|ccc|ccc}
ID       & $g_A^{u}$ & $g_A^{d}$ & $g_A^{u-d}$ 
         & $g_S^{u}$ & $g_S^{d}$ & $g_S^{u-d}$ 
         & $g_T^{u}$ & $g_T^{d}$ & $g_T^{u-d}$ \\ 
\hline
a12m310   &  0.923(25) &  -0.326(14)  & 1.248(26)  & 3.43(20)  & 2.61(15)  & 0.82(10) & 0.867(25) & $-$0.218(12) & 1.084(31)  \\   
a12m310*  &  0.932(9)  &  -0.320(5)   & 1.252(9)   & 3.31(6)   & 2.40(4)   & 0.91(4)  & 0.873(9)  & $-$0.215(4)  & 1.087(10)  \\   
a12m220S  &  0.969(37) &  -0.313(21)  & 1.283(39)  & 4.49(52)  & 3.45(37)  & 1.03(32) & 0.858(41) & $-$0.208(21) & 1.067(46)  \\  
a12m220   &  0.942(29) &  -0.332(15)  & 1.273(30)  & 3.87(27)  & 3.21(21)  & 0.67(18) & 0.885(26) & $-$0.218(16) & 1.105(32)  \\ 
a12m220L  &  1.015(75) &  -0.322(22)  & 1.337(83)  & 4.74(72)  & 3.67(63)  & 1.10(17) & 0.855(24) & $-$0.189(18) & 1.043(29)  \\   
a12m220L* &  0.949(11) &  -0.332(6)   & 1.279(12)  & 4.10(18)  & 3.27(14)  & 0.83(6)  & 0.847(16) & $-$0.209(7)  & 1.056(19)  \\   
\hline                                                                                                                           
a09m310   &  0.949(24) &  -0.312(15)  & 1.262(30)  & 3.78(30)  & 2.84(24)  & 0.94(10) & 0.821(25) & $-$0.200(11) & 1.021(31)  \\   
a09m220   &  0.928(23) &  -0.343(16)  & 1.272(30)  & 4.16(25)  & 3.29(19)  & 0.88(13) & 0.815(21) & $-$0.213(9)  & 1.029(22)  \\   
a09m130   &  0.864(42) &  -0.314(25)  & 1.177(44)  & 5.01(58)  & 4.28(37)  & 0.73(37) & 0.764(46) & $-$0.215(21) & 0.980(45)  \\ 
a09m130*  &  0.909(17) &  -0.345(15)  & 1.255(24)  & 5.49(35)  & 4.51(31)  & 0.99(11) & 0.778(23) & $-$0.189(10) & 0.967(28)  \\
\hline                                                                                                                           
a06m310   &  0.873(23) &  -0.299(13)  & 1.172(24)  & 4.31(18)  & 3.03(11)  & 1.28(11) & 0.764(22) & $-$0.187(10) & 0.951(24)  \\   
a06m310*  &  0.895(14) &  -0.317(8)   & 1.212(14)  & 4.12(11)  & 2.94(7)   & 1.18(5)  & 0.782(10) & $-$0.188(4)  & 0.972(12)  \\   
a06m220   &  0.904(48) &  -0.328(25)  & 1.234(51)  & 3.63(40)  & 2.99(20)  & 0.64(29) & 0.749(47) & $-$0.240(18) & 0.990(48)  \\  
a06m220*  &  0.907(16) &  -0.326(9)   & 1.234(17)  & 4.24(9)   & 3.20(5)   & 1.04(6)  & 0.792(9)  & $-$0.192(4)  & 0.984(10)  \\  
\end{tabular}
\end{ruledtabular}
\caption{Estimates of the bare connected charges using the fit ranges
  defined under Case 3 in Table~\protect\ref{tab:4cases}.  The
  isovector charges $g_\Gamma^{u-d} = \langle 0 | \mathcal{O}_\Gamma |
  0 \rangle$ are the same as Case 3 in
  Tables~\protect\ref{tab:3ptgAcompare},~\protect\ref{tab:3ptgScompare}
  and~\protect\ref{tab:3ptgTcompare}.  Ensembles marked with an
  asterisk denote results obtained with the AMA method.}
\label{tab:resultsbare}
\end{table*}

\begin{table*}
\centering
\begin{ruledtabular}
\begin{tabular}{c|ccc|ccc|cccc}
ID       & $g_A^{u}$ & $g_A^{d}$ & $g_A^{u-d}$ 
         & $g_S^{u}$ & $g_S^{d}$ & $g_S^{u-d}$ 
         & $g_T^{u}$ & $g_T^{d}$ & $g_T^{u-d}$ & $g_T^{u+d}$  \\ 
\hline
a12m310   & 0.903(26) & $-$0.319(14)  & 1.221(28) & 3.16(18)  & 2.41(14)  & 0.757(92) & 0.839(37) & $-$0.211(13)  & 1.050(45) & 0.628(32) \\
a12m310*  & 0.914(11) & $-$0.3145(55) & 1.229(14) & 3.066(62) & 2.226(39) & 0.840(36) & 0.848(29) & $-$0.2088(80) & 1.055(36) & 0.640(23) \\
a12m220S  & 0.960(38) & $-$0.310(22)  & 1.270(42) & 4.19(48)  & 3.23(34)  & 0.96(30)  & 0.840(49) & $-$0.204(22)  & 1.046(56) & 0.637(50) \\
a12m220   & 0.917(30) & $-$0.323(16)  & 1.240(32) & 3.56(25)  & 2.95(20)  & 0.62(16)  & 0.853(38) & $-$0.210(17)  & 1.064(47) & 0.641(35) \\
a12m220L  & 0.975(48) & $-$0.309(19)  & 1.284(46) & 4.29(52)  & 3.32(47)  & 1.00(13)  & 0.812(47) & $-$0.180(22)  & 0.992(63) & 0.636(38) \\
a12m220L* & 0.931(13) & $-$0.3254(69) & 1.255(16) & 3.79(16)  & 3.03(14)  & 0.770(54) & 0.822(31) & $-$0.2032(95) & 1.025(39) & 0.618(25) \\
\hline
a09m310   & 0.926(26) & $-$0.304(15)  & 1.231(33) & 3.40(32)  & 2.56(25)  & 0.844(98) & 0.823(33) & $-$0.200(13)  & 1.024(42) & 0.623(29) \\
a09m220   & 0.911(26) & $-$0.337(16)  & 1.249(35) & 3.78(30)  & 2.98(23)  & 0.80(12)  & 0.823(31) & $-$0.215(11)  & 1.039(36) & 0.608(29) \\
a09m130   & 0.868(44) & $-$0.316(26)  & 1.182(48) & 4.65(59)  & 3.97(41)  & 0.67(34)  & 0.789(50) & $-$0.222(23)  & 1.012(52) & 0.567(58) \\
a09m130*  & 0.891(20) & $-$0.338(15)  & 1.230(29) & 4.97(41)  & 4.08(35)  & 0.90(11)  & 0.784(31) & $-$0.191(11)  & 0.975(38) & 0.592(26) \\
\hline
a06m310   & 0.867(24) & $-$0.297(13)  & 1.165(26) & 3.79(23)  & 2.67(15)  & 1.13(11)  & 0.814(29) & $-$0.199(11)  & 1.014(33) & 0.615(28) \\
a06m310*  & 0.888(16) & $-$0.3144(84) & 1.202(18) & 3.62(19)  & 2.58(13)  & 1.038(64) & 0.832(22) & $-$0.2005(65) & 1.034(26) & 0.631(18) \\
a06m220   & 0.913(50) & $-$0.332(27)  & 1.246(56) & 3.25(38)  & 2.67(22)  & 0.57(26)  & 0.812(50) & $-$0.260(21)  & 1.073(53) & 0.553(54) \\
a06m220*  & 0.888(18) & $-$0.3190(90) & 1.208(20) & 3.68(18)  & 2.78(13)  & 0.901(67) & 0.832(21) & $-$0.2016(65) & 1.034(26) & 0.630(18) \\
\end{tabular}
\end{ruledtabular}
\caption{Results for the renormalized charges using the fit
  ranges defined under Case 3 in Table~\protect\ref{tab:4cases}.  
  Estimates of the flavor diagonal charges include only the connected 
  contribution.  The 
  final errors are obtained by adding in quadrature the errors in
  estimates of the ratios $g_\Gamma^{\rm bare}/g_V^{\rm bare}$ 
  to the errors in the ratios of the renormalization constants, $Z_\Gamma/Z_V$ given in
  Table~\protect\ref{tab:Zfinal}. Estimates for $g_T^{u+d}$ neglect
  the disconnected contributions that were shown to be tiny in
  Ref.~\protect\cite{Bhattacharya:2015wna}.  Rest is the same as in
  Table~\protect\ref{tab:resultsbare}.  }
\label{tab:resultsrenorm}
\end{table*}

\begin{table}
\centering
\begin{ruledtabular}
\begin{tabular}{c|ccc|cc}
ID            & $g_A^{u-d}$ & $g_S^{u-d}$ & $g_T^{u-d}$ &  $g_V^{{\rm bare}, u-d}$ & $Z_V g_V^{u-d}$ \\ 
\hline
a12m310       & 1.185(45) & 0.738(97) & 1.019(52) & 1.068(10) & 0.972(23) \\
a12m310*      & 1.189(39) & 0.817(50) & 1.022(44) & 1.065(4)  & 0.969(22) \\
a12m220S      & 1.219(53) & 0.93(29)  & 1.003(61) & 1.055(14) & 0.960(25) \\
a12m220       & 1.209(48) & 0.61(16)  & 1.038(54) & 1.073(12) & 0.977(24) \\
a12m220L      & 1.270(88) & 0.99(16)  & 0.981(50) & 1.088(36) & 0.990(39) \\
a12m220L*     & 1.215(40) & 0.749(62) & 0.993(46) & 1.065(3)  & 0.969(21) \\
\hline                                                                                                                                                                     
a09m310       & 1.199(58) & 0.824(90) & 1.000(51) & 1.060(8)  & 0.975(33) \\
a09m220       & 1.208(58) & 0.78(11)  & 1.008(47) & 1.053(8)  & 0.969(32) \\
a09m130       & 1.118(63) & 0.64(33)  & 0.960(59) & 1.029(18) & 0.947(35) \\
a09m130*      & 1.192(55) & 0.87(10)  & 0.948(47) & 1.055(5)  & 0.971(32) \\
\hline                                                                                                                                                                     
a06m310       & 1.137(42) & 1.10(10)  & 0.989(38) & 1.032(10) & 0.980(23) \\
a06m310*      & 1.176(39) & 1.017(56) & 1.011(32) & 1.034(4)  & 0.982(21) \\
a06m220       & 1.197(62) & 0.55(25)  & 1.030(58) & 1.015(21) & 0.964(29) \\
a06m220*      & 1.197(41) & 0.894(63) & 1.024(31) & 1.047(6)  & 0.995(22) \\
\end{tabular}
\end{ruledtabular}
\caption{Estimates of the renormalized isovector charges obtained
  using $Z_\Gamma \times g_\Gamma^{\rm bare}$ with $Z_\Gamma$ given in
  Table~\protect\ref{tab:Zfinal} and $ g_\Gamma^{\rm bare}$ in
  Table~\protect\ref{tab:resultsbare}. All estimates are obtained
  using the fit ranges defined under Case 3 in
  Table~\protect\ref{tab:4cases}.  We also give results for the bare
  and renormalized $g_V^{u-d}$ in columns five and six. Estimates in
  the sixth column differ from $Z_V g_V^{u-d} = 1$,
  predicted by the conservation of the vector charge, by 2--3\%.
  Ensembles marked with an asterisk denote results obtained with the
  AMA method.}
\label{tab:resultsgV}
\end{table}

%%%%%%%%%%%%%%%%%%%%%%%%%%%%%%%%%%%%%%%%%%%%%%%%%%%%%%%%%%%%%%%%%%%%%
%%%  SECTION                                                      %%%
%%%%%%%%%%%%%%%%%%%%%%%%%%%%%%%%%%%%%%%%%%%%%%%%%%%%%%%%%%%%%%%%%%%%%
\section{Results for the charges $g_A$, $g_S$, $g_T$}
\label{sec:results}

To extrapolate the estimates of the renormalized charges given in
Table~\ref{tab:resultsrenorm} to the continuum limit ($a\rightarrow
0$), the physical pion mass ($M_{\pi^0} = 135$~MeV) and the infinite
volume limit ($L \rightarrow \infty$), we need to use an appropriate
fit Ansatz, motivated by chiral perturbation theory, for the nine data
points. For discussions on predictions of chiral perturbation we
direct the reader to
Refs.~\cite{Bernard:1992qa,Bernard:2006gx,Bernard:2006te,Khan:2006de,Colangelo:2010ba,deVries:2010ah}.
To test these predictions, one ideally requires data at many lattice
spacings and small pion masses. Also, the analysis is simplest if one can 
hold two of the three variables constant to study the variation
with the third.  Our nine data points do not allow such a study.  For
example, estimates at the three different volume points, $a12m220S$,
$a12m220$ and $a12m220L$, at fixed $M_\pi$ and $a$ are consistent
within errors for all the three charges.  In fact, the spread in the
data in most cases is small enough that they can be fit with a linear
Ansatz in each variable. Our goal, given the limited number of data points and
the small variations in each of the three variables, is to make a
simultaneous fit keeping the minimum number of parameters
corresponding to the leading terms in the chiral expansion.

Keeping only the leading corrections in $a$ and $M_\pi L$, we have
studied the following Ansatz motivated by chiral perturbation
theory~\cite{Bernard:1992qa,Bernard:1995dp,Bernard:2006gx,Bernard:2006te,Khan:2006de,Colangelo:2010ba,deVries:2010ah}:
\begin{align}
  g_{A,T} (a,M_\pi,L) &= c_1 + c_2a + c_3 M_\pi^2 + c_3^{\rm log} M_\pi^2 \ln (\frac{M_\pi}{M_\rho})^2  \nonumber \\
&+ c_4 M_\pi^2 \frac{e^{-M_\pi L}}{X(M_\pi L)} \,.
\label{eq:extrapgAT} \\
  g_{S} (a,M_\pi,L) &= c_1 + c_2a + c^{\prime}_3 M_\pi + c_3 M_\pi^2 + \nonumber \\
& c_3^{\rm log} M_\pi^2 \ln (\frac{M_\pi}{M_\rho})^2  +
c^{\prime}_4 M_\pi \frac{e^{-M_\pi L}}{Y(M_\pi L)} \,,
\label{eq:extrapgS}
\end{align}
where $M_\rho$ is the chiral renormalization scale.  Note that the
leading discretization errors are linear in $a$ for our clover-on-HISQ
formalism with unimproved operators, and the leading chiral correction
to $g_S^{u-d}$ starts at $O(M_\pi)$~\cite{Bernard:1992qa}. The finite-volume
correction, in general, consists of a number of terms, each with
different powers of $M_\pi L$ in the denominator and depend on several
low-energy constants (LEC)~\cite{Khan:2006de}. These powers of $M_\pi
L$ are symbolically represented by $X(M_\pi L)$ and $Y(M_\pi
L)$. Since the variation of these factors is small compared to the
exponential over the range of $M_\pi L$ investigated, we set $X(M_\pi
L) = Y(M_\pi L) = {\rm constant}$ and retain only the appropriate
overall factor $M_\pi^n e^{-M_\pi L}$, common to all the terms in the
finite-volume expansion, in our fit Ansatz.

\begin{figure*}[tb]
    \includegraphics[width=0.98\linewidth]{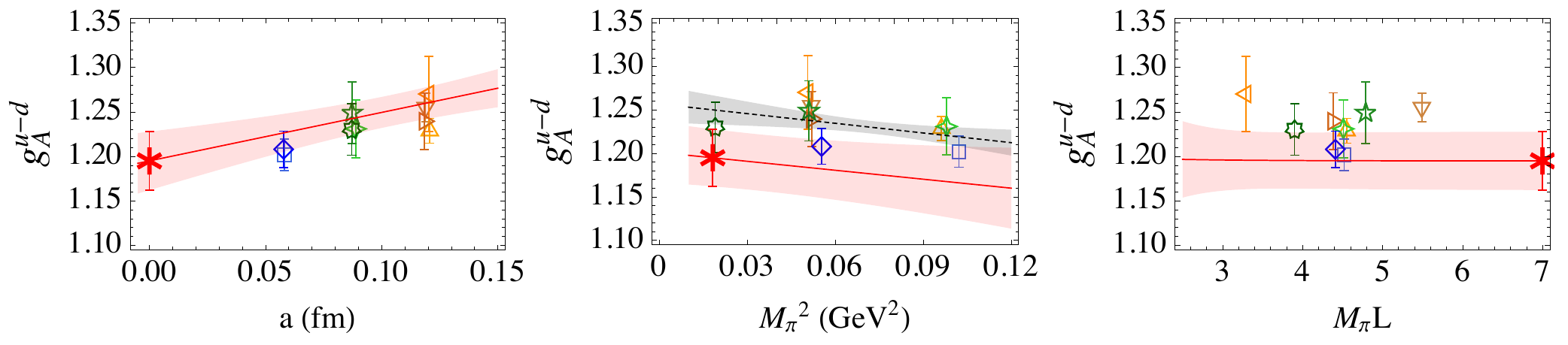}
    \includegraphics[width=0.98\linewidth]{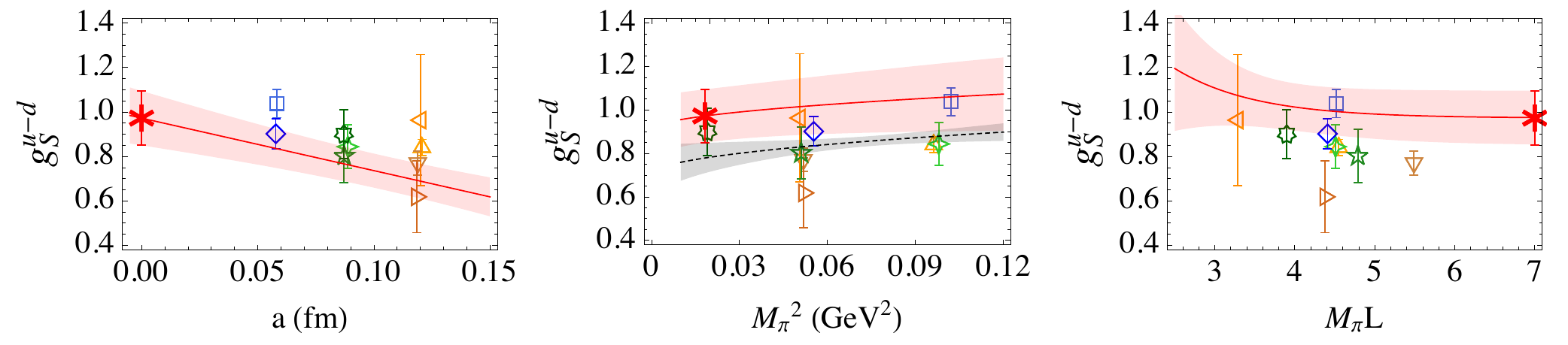}
    \includegraphics[width=0.98\linewidth]{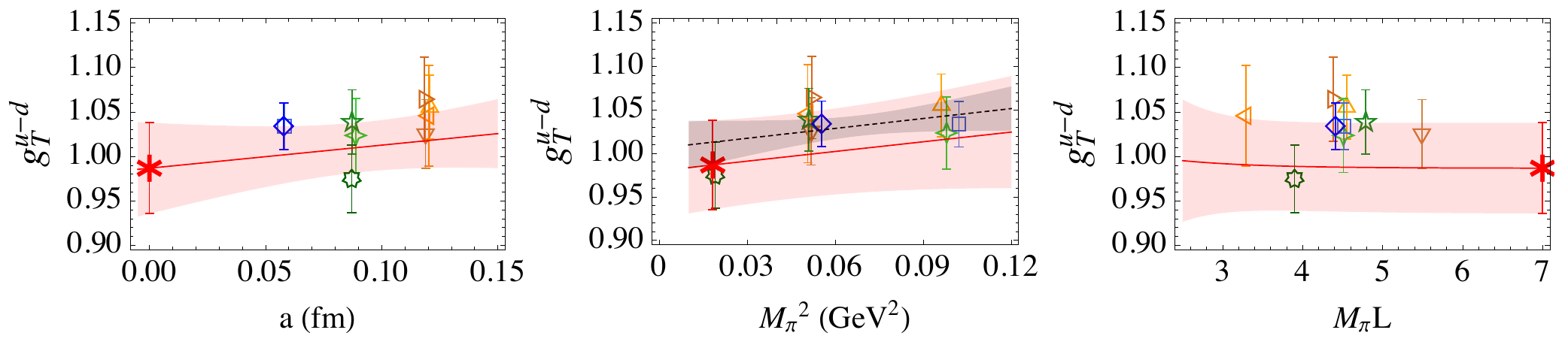}
\caption{The 9-point fit using Eqs.~\protect\eqref{eq:CextrapgAT}
  and~\protect\eqref{eq:CextrapgS} to the data for the renormalized
  isovector charges, $g_A^{u-d}$, $g_S^{u-d}$ and $g_T^{u-d}$, in the
  $\overline{{\rm MS}}$ scheme at $2\GeV$.  The result of the
  simultaneous extrapolation to the physical point defined by
  $a\rightarrow 0$, $M_\pi \rightarrow M_{\pi^0}^{{\rm phys}}=135$~MeV
  and $L \rightarrow \infty$ are marked by a red star.  The error
  bands in each panel show the simultaneous fit as a function of a
  given variable holding the other two at their physical value.  The
  data are shown projected on to each of the three planes.  The
  overlay in the middle figures with the dashed line within the grey
  band, is the fit to the data versus $M_\pi^2$ neglecting dependence
  on the other two variables.  The symbols used for data points from
  the various ensembles are defined in Table~\ref{tab:ens}.
  \label{fig:conUmD_extrap9}}
\end{figure*}

\begin{figure*}[tb]
    \includegraphics[width=0.98\linewidth]{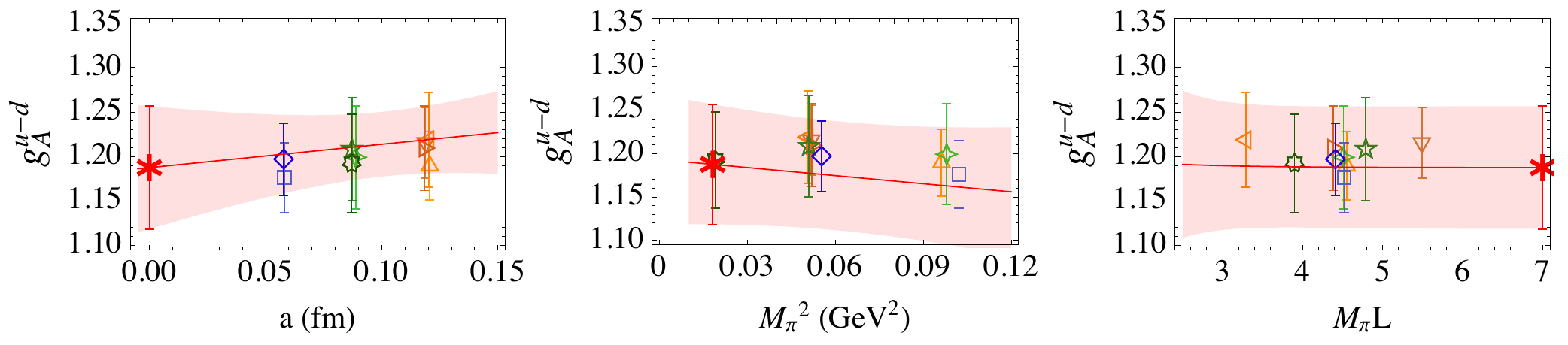}
    \includegraphics[width=0.98\linewidth]{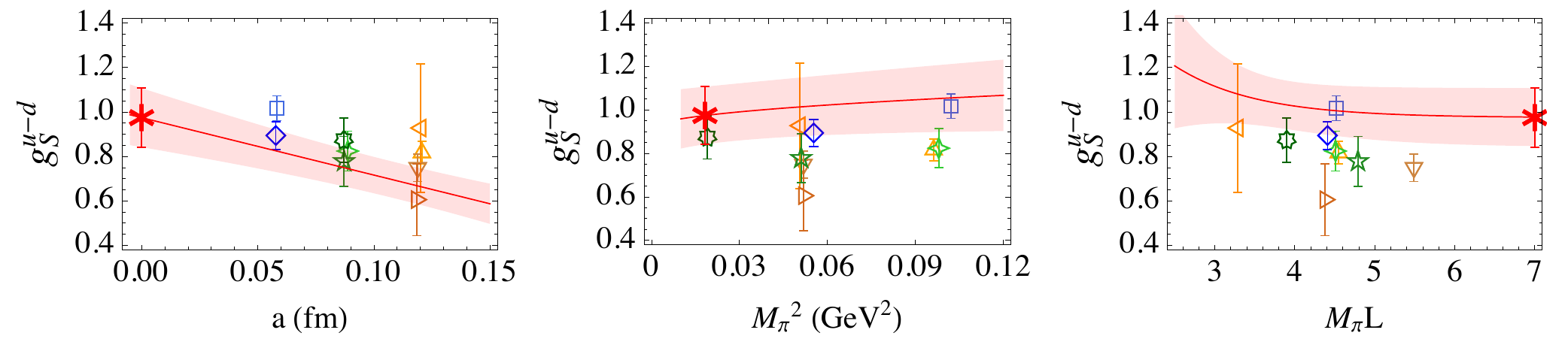}
    \includegraphics[width=0.98\linewidth]{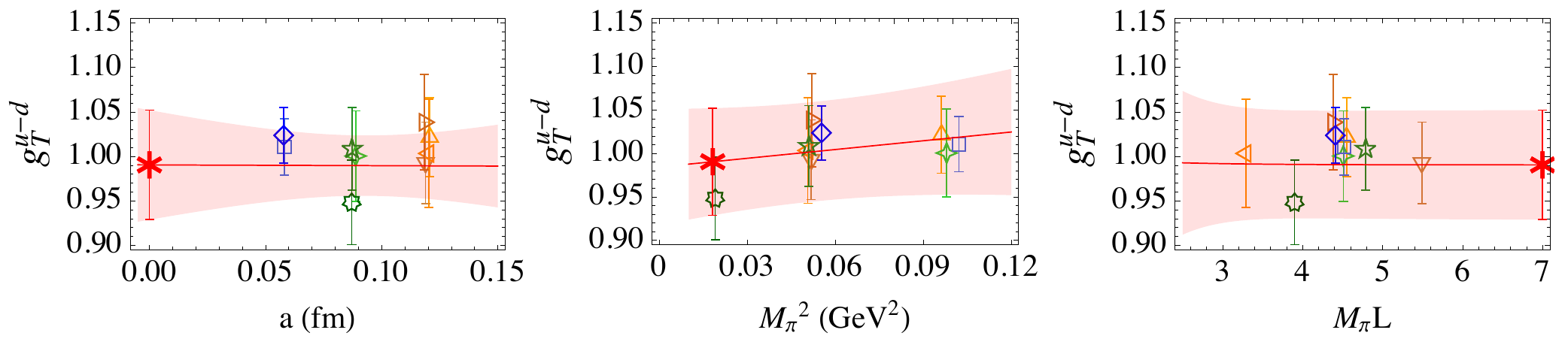}
\caption{The 9-point fit using Eqs.~\protect\eqref{eq:CextrapgAT}
  and~\protect\eqref{eq:CextrapgS} to the data for the renormalized
  isovector charges obtained using $Z_A g_A^{{\rm bare}, u-d}$, $Z_S
  g_S^{{\rm bare}, u-d}$ and $Z_T g_T^{{\rm bare}, u-d}$.  The rest is
  the same as in Fig.~\protect\ref{fig:conUmD_extrap9}.
  \label{fig:UmD_extrap9}}
\end{figure*}

The dependence of the data on the lattice spacing, the pion mass and the
lattice volume is small. We, therefore, investigated fits with the
Ansatz given in Eqs.~\eqref{eq:extrapgAT} and~\eqref{eq:extrapgS} and,
in addition, a number of Ans\"{a}tze with various subsets of terms.  In
each case, the parameters $c_3^{\rm log}$ and $c_4$ (or $c_4^\prime$)
are poorly determined and consistent with zero, reflective of the
small range and small variation in the data with the pion mass and the
lattice volume.  Also, $\chi^2_{\rm d.o.f.} < 1$ in all cases, so it does
not provide a good criterion for deciding the best fit Ansatz. We,
therefore, considered the Ansatz with just the leading term in each of
the three variables:
\begin{align}
  g_{A,T} (a,M_\pi,L) &=& c_1 + c_2a + c_3 M_\pi^2 + 
 c_4 M_\pi^2 {e^{-M_\pi L}} \,,
\label{eq:CextrapgAT} \\
  g_{S} (a,M_\pi,L) &=& c_1 + c_2a + c^{\prime}_3 M_\pi + 
c^{\prime}_4 M_\pi {e^{-M_\pi L}} \,.
\label{eq:CextrapgS}
\end{align}
Using these Ans\"{a}tze, the 9-point fits (9-pt) to the data for the
isovector charges, renormalized using $g_\Gamma^{\rm bare}/g_V^{\rm
  bare} \times Z_\Gamma/Z_V$, are shown in
fig.~\ref{fig:conUmD_extrap9}.  The fits to $g_{A}^{u-d}$ and
$g_{S}^{u-d}$ show a sizable variation with $a$.  We show the same
9-point fit in Fig.~\ref{fig:UmD_extrap9} using the renormalized
charges obtained using $Z_\Gamma \times g_\Gamma$ and find that the
variation of $g_A^{u-d}$ with $a$ is smaller.  The systematics in the
two ways of obtaining the renormalized charges are different but
removed by the extrapolation to the continuum limit as suggested by
the agreement between the results shown in
Table~\ref{tab:9pt-compare}.  The errors in the ratio method,
$g_\Gamma^{\rm bare}/g_V \times Z_\Gamma/Z_V$, are smaller because, as
discussed in Sec.~\ref{sec:renorm}, some of the systematics cancel in
each of the two ratios. We, therefore, use the estimates from the
ratio method in all subsequent analysis.

\begin{table}
\centering
\begin{ruledtabular}
\begin{tabular}{c|ccc}
Method                                                    & $g_A^{u-d}$ & $g_S^{u-d}$ & $g_T^{u-d}$ \\ 
\hline
$g_\Gamma^{\rm bare}/g_V^{\rm bare} \times Z_\Gamma/Z_V$  & 1.195(33)   & 0.97(12)    & 0.987(51)       \\
$\chi^2/$d.o.f.                                           &  0.28       &  0.67       &  0.44           \\           
\hline 
$g_\Gamma^{\rm bare} \times Z_\Gamma$                     & 1.187(69)   & 0.97(13)    & 0.990(62)       \\
$\chi^2/$d.o.f.                                           &  0.05       &  0.65       &  0.39           \\           
\end{tabular}
\end{ruledtabular}
\caption{Results of the 9-point fit to the data for the isovector charges 
  renormalized in two ways: using 
  $g_\Gamma^{\rm bare}/g_V^{\rm bare} \times Z_\Gamma/Z_V$ with $g_V^{\rm bare} Z_V=1$ 
  given in Table~\protect\ref{tab:resultsrenorm}   
  and $g_\Gamma^{\rm bare} \times Z_\Gamma$ given in Table~\protect\ref{tab:resultsgV}. 
  }
\label{tab:9pt-compare}
\end{table}

\begin{figure*}[tb]
    \includegraphics[width=0.98\linewidth]{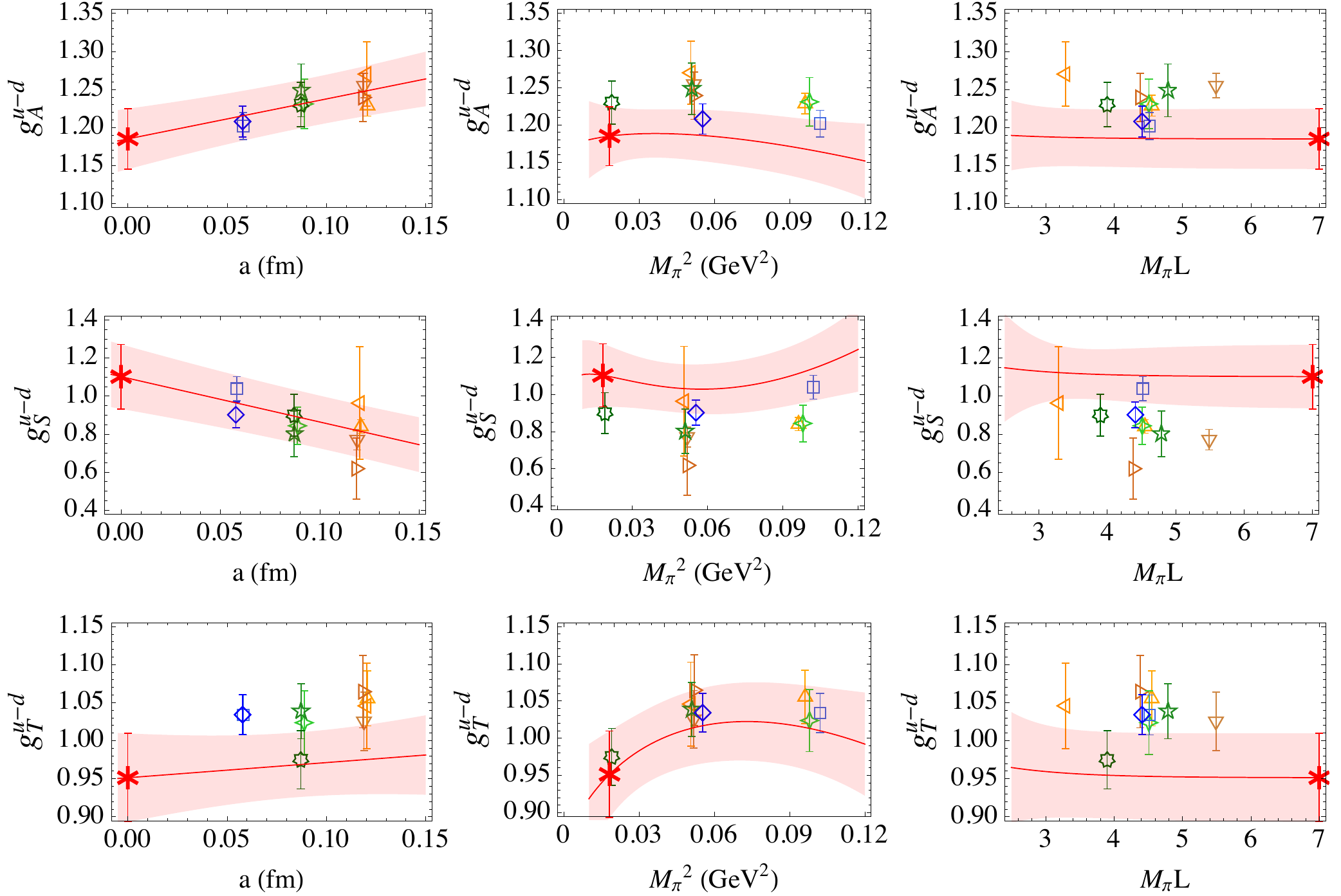}
\caption{The 9-point extrapolation of the isovector charges including
  the the chiral logarithm term, defined in
  Eqs.~\protect\eqref{eq:extrapgAT} and~\protect\eqref{eq:extrapgS},
  in the fit Ansatz.  The rest is the same as in
  Fig.~\protect\ref{fig:conUmD_extrap9}.
  \label{fig:conUmD_extrap9L}}
\end{figure*}

\begin{figure*}[tb]
    \includegraphics[width=0.98\linewidth]{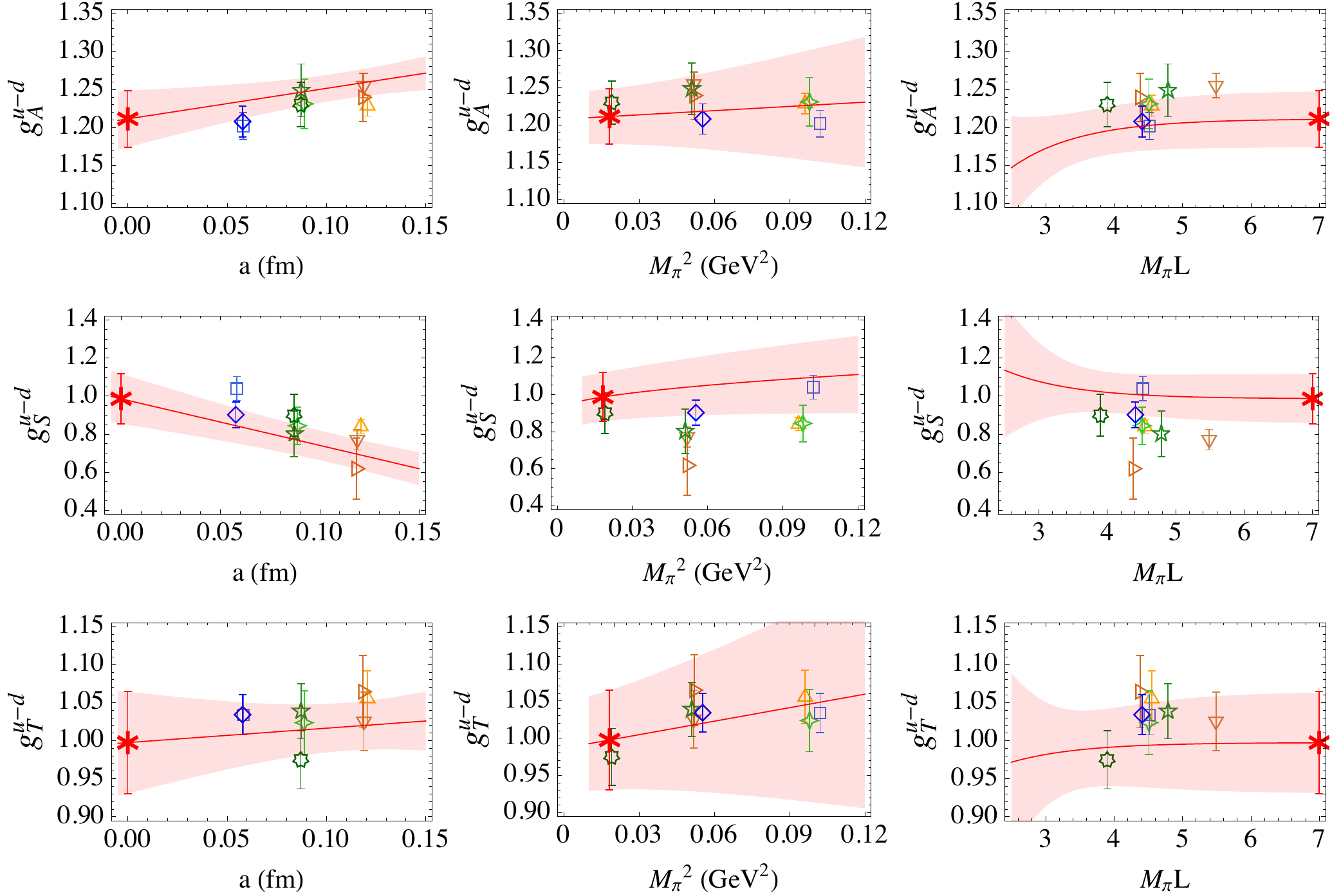}
\caption{The 8-point fit to the isovector charges neglecting
  the $a12m220S$ point.  The rest is the same as in
  Fig.~\protect\ref{fig:conUmD_extrap9}.
  \label{fig:conUmD_extrap8}}
\end{figure*}

We also show two additional fits to illustrate the dependence of the
estimates of the charges on the fit Ansatz: (i) we add the chiral
logarithm term to the 9-point fit and call it the 9-point log fit
(9-ptL) and (ii) we remove the smallest volume point $a12m220S$ and
perform an 8-point fit (8-pt) using Eqs.~\eqref{eq:CextrapgAT}
and~\eqref{eq:CextrapgS}.  These two fits are shown in
Figs.~\ref{fig:conUmD_extrap9L} and~\ref{fig:conUmD_extrap8},
respectively.  The resulting values of the fit parameters and
estimates for the three isovector charges are given in
Table~\ref{tab:9L8fit}. To choose between the 9-point and the 9-point
with chiral logarithm fit, we used the Akaike information criteria
(AIC)~\cite{1100705}.  As can be seen from Table~\ref{tab:9L8fit}, the
$\chi^2$ does not decrease by 2 to justify adding an extra parameter,
the chiral logarithm term. Our final estimates, given in
Sec.~\ref{sec:errors} are, therefore, obtained with the 9-point fit
without the chiral logarithm.

\begin{table*}[hp]
\begin{center}
\renewcommand{\arraystretch}{1.2} % Change horizontal spacing
\begin{ruledtabular}
\begin{tabular}{c|ccccccc|c|c}
             &   $c_1$      &  $c_2$       & $c_3$       & $c_3^{\prime}$ & $c_3^{\rm log}$ & $c_4$        & $c_4^{\prime}$ & $\chi^2/{\rm d.o.f.}$  &  $g_\Gamma$ \\
             &              &  fm${}^{-1}$ & GeV${}^{-2}$& GeV${}^{-1}$   & GeV${}^{-2}$    & GeV${}^{-2}$ &  GeV${}^{-1}$  &                        &             \\
\hline                                                                 
$g_A$(9-pt)  &   1.201(35)  &  0.55(27)    & $-$0.34(38) &                &                 &  1(26)       &                &  0.28                  &  1.195(33)  \\
$g_A$(9-ptL) &   1.166(87)  &  0.53(27)    & $-$1.1(1.8) &                & $-$0.62(1.4)    &  $-$3(26)    &                &  0.30                  &  1.185(40)  \\
$g_A$(8-pt)  &   1.208(35)  &  0.40(31)    & 0.19(68)    &                &                 &  $-$43(53)   &                &  0.12                  &  1.211(37)  \\
\hline                                                                                                                                             
$g_S$(9-pt)  &   0.91(15)   &  $-$2.4(0.9) &             &    0.48(59)    &                 &              &  20(29)        &  0.67                  &  0.97(12)   \\
$g_S$(9-ptL) &   0.62(30)   &  $-$2.4(0.9) &             &    13(12)      & 21(19)          &              &  4(32)         &  0.53                  &  1.10(17)   \\
$g_S$(8-pt)  &   0.91(15)   &  $-$2.5(1.0) &             &    0.57(68)    &                 &              &  14(38)        &  0.82                  &  0.99(13)   \\
\hline                                                                                                                                             
$g_T$(9-pt)  &   0.980(55)  &  0.26(46)    & 0.37(57)    &                &                 &  6(39)       &                &  0.44                  &  0.987(51)  \\
$g_T$(9-ptL) &   0.85(12)   &  0.20(47)    & $-$2.7(2.5) &                & $-$2.4(1.9)     &  $-$9(39)    &                &  0.15                  &  0.951(58)  \\
$g_T$(8-pt)  &   0.986(61)  &  0.19(54)    & 0.6(1.1)    &                &                 &  $-$20(10)   &                &  0.54                  &  0.997(67)  \\
\end{tabular}
\end{ruledtabular}
\caption{Values of the fit parameters defined in
  Eqs.~\protect\eqref{eq:extrapgAT}
  and~\protect\eqref{eq:extrapgS}. The 9-point fit to the isovector
  charges includes terms with $c_1,\ c_2,\ c_3$ (or $c_3^{\prime}$ for $g_S^{u-d}$)
  and $c_4$ (or $c_4^{\prime}$ for $g_S^{u-d}$). The 9-ptL fit includes the chiral
  logarithm term $c_3^{\rm log}$. The 8-pt fit uses the same Ansatz as
  the 9-pt fit but neglects the $a12m220S$ data point.   The
  last column gives the value of the charge at the physical point.}
\label{tab:9L8fit}
\end{center}
\end{table*}


We find that the smallest volume point, $a12m220S$, provides a 
large lever arm in the determination of the shape of the
finite-volume correction, as can be seen by comparing the 9-pt
(Fig.~\ref{fig:conUmD_extrap9}) versus the 8-pt
(Fig.~\ref{fig:conUmD_extrap8}) fits.  Nevertheless, the fits do not
show a significant difference for $M_\pi L \, \gsim 4$.

To further explore the sensitivity of the data for the isovector
charges $g_{A,S,T}^{u-d}$ to chiral logarithms, we make 8-point fits 
(neglecting the $a12m220S$ point and 
assuming that the finite-volume corrections can be ignored for $M_\pi
L \, \gsim 4$) with and without a chiral logarithm term:
\begin{align}
g_{A,T}(a, M_\pi) &= c_1 + c_2 a + c_3 M_\pi^2 + c_3^{\rm log} M_\pi^2 \ln \frac{M_\pi^2}{M_\rho^2} \, . \nonumber \\
g_{S}(a, M_\pi) &= c_1 + c_2 a + c_3^{\prime} M_\pi + c_3^{\rm log} M_\pi^2 \ln \frac{M_\pi^2}{M_\rho^2} \, .
\label{eq:gAchiralsimple}
\end{align}
The chiral renormalization scale is taken to be
$M_\rho=0.775$~GeV. Results for the parameters $c_i$ are summarized in
Table~\ref{tab:chiralfit} with and without the $c_3^{\rm log}$
term. The parameter $c_3$ (or $c_3^{\prime}$) is again poorly
constrained by the data since the variation in the estimates between $M_\pi
= 135$ and $320$~MeV is comparable to the errors in the
individual points for all three charges as can be seen in
Figs.~\ref{fig:conUmD_extrap9},~\ref{fig:conUmD_extrap9L}
and~\ref{fig:conUmD_extrap8}. The main effect of adding the $c_3^{\rm
  log}$ term is an adjustment with $c_3$: there is a large
cancellation between the contributions of these two terms. Also, the
errors in both $c_3$ and $c_3^{\rm log}$ are $O(1)$ in all three
charges. To test whether the chiral logarithm term improves the predictive
power of the fit, we again use the Akaike information
criterion~\cite{1100705}. The $\chi^2$ changes by much less than two
units, indicating that adding the chiral logarithm is not justified.
Also, for the scalar charge, we find little difference in the fits
between using $c_3 M_\pi^2$ versus $c_3^\prime M_\pi$ as the leading
chiral term.

The bottom line is that changing the fit Ansatz from
Eqs.~\eqref{eq:extrapgAT} and~\eqref{eq:extrapgS} to
Eqs.~\eqref{eq:CextrapgAT} and~\eqref{eq:CextrapgS} to that given in
Eq.~\eqref{eq:gAchiralsimple} does not significantly change the
estimates for the charges as can be seen by comparing the last column
of Tables~\ref{tab:9L8fit} and~\ref{tab:chiralfit}.

Our final results for the isovector charges using the 9-point fit are
presented in Table~\ref{tab:IVresults} in Sec.~\ref{sec:errors} after
we revisit our error analysis and assign an additional systematic
error to account for the uncertainty due to the various fit Ansatz
discussed above.

\begin{table*}[hp]
\begin{center}
\renewcommand{\arraystretch}{1.2} % Change horizontal spacing
\begin{ruledtabular}
\begin{tabular}{c|ccc|cc|c|c}
             &   $c_1$      &  $c_2$       &   $c_3$      & $c_3^\prime$ & $c_3^{\rm log}$  & $\chi^2/$d.o.f.    & $g_\Gamma$   \\
             &              &  fm${}^{-1}$ & GeV${}^{-2}$ & GeV${}^{-1}$ & GeV${}^{-2}$     &                    &              \\
\hline                                                                                                                          
$g_A^{u-d}$  &   1.171(85)  &  0.51(27)    & $-$1.0(1.7)  &              &  $-$0.5(1.4)     &  0.24              &  1.188(38)   \\
$g_A^{u-d}$  &   1.201(35)  &  0.52(27)    & $-$0.31(28)  &              &                  &  0.23              &  1.195(32)   \\
\hline                                                                                                                          
$g_S^{u-d}$  &   0.60(28)   &  $-$2.46(88) &              & 14(10)       &  22(17)          &  0.41              &  1.12(13)    \\
$g_S^{u-d}$  &   0.91(15)   &  $-$2.61(87) &              & 0.75(47)     &                  &  0.68              &  1.01(11)    \\
\hline                                                                                                                          
$g_T^{u-d}$  &   0.85(12)   &  0.20(48)    & $-$2.5(2.4)  &              &  $-$2.4(1.9)     &  0.16              &  0.955(58)   \\
$g_T^{u-d}$  &   0.982(55)  &  0.23(48)    & 0.43(44)     &              &                  &  0.43              &  0.990(50)   \\
\end{tabular}
\end{ruledtabular}
\caption{Values of the fit parameters for the 8-point fit defined in
  Eq.~\eqref{eq:gAchiralsimple} are given in the top row for each
  charge. The bottom row uses the same fit Ansatz but with $c_3^{\rm
    log}$ set to 0.  The last column gives the value of the charge at
  the physical point.}
\label{tab:chiralfit}
\end{center}
\end{table*}

To obtain the flavor-diagonal charges $g_\Gamma^u$ and $g_\Gamma^d$
and the isoscalar combination $g_\Gamma^{u+d}$ requires calculation of
the disconnected contributions. We have not carried out any new
simulations to update estimates of the disconnected contributions to
the tensor charges that were shown to be $O(1\%)$ of the connected
contribution and consistent with zero in
Ref.~\cite{Bhattacharya:2015wna}.  We, therefore, consider the
connected part to be a good approximation to the full result and
present updated results for $g_T^u$, $g_T^d$ and $g_T^{u+d}$ in
Table~\ref{tab:FD9L8fit}.  The disconnected contributions to the axial
and scalar charges are larger, about $0.1$ and $O(1)$, respectively,
and we are working on a more detailed analysis of these.  We,
therefore, do not present results for the isoscalar combinations
$g_A^{u+d}$ and $g_S^{u+d}$ but give only the connected contributions
to the flavor-diagonal charges $g_{A,S}^{u,d}$ using the same three fits,
9-point, 9-point with log, and 8-point, in
Table~\ref{tab:FD9L8fit}.

\begin{table*}[htbp]
\centering
\begin{ruledtabular}
\begin{tabular}{c|ccc|ccc|cccc}
ID        & $g_A^u$  &  $g_A^d$  &  $g_A^{u-d}$ & $g_S^u$   & $g_S^{d}$ & $g_S^{u-d}$  & $g_T^{u}$  & $g_T^{d}$    & $g_T^{u-d}$ & $g_T^{u+d}$ \\ 
\hline
9-pt  &  0.856(27) & $-$0.335(15) & 1.195(33)  & 4.94(30)   & 4.00(22)   & 0.97(12)   & 0.792(42)  & $-$0.194(14) & 0.987(51)  & 0.598(36) \\
9-ptL &  0.841(31) & $-$0.342(20) & 1.185(40)  & 5.59(48)   & 4.59(39)   & 1.10(17)   & 0.766(48)  & $-$0.183(17) & 0.951(58)  & 0.579(41) \\
8-pt  &  0.877(31) & $-$0.333(16) & 1.211(37)  & 5.09(33)   & 4.13(25)   & 0.99(13)   & 0.796(55)  & $-$0.197(17) & 0.997(67)  & 0.599(46) \\
\end{tabular}
\end{ruledtabular}
\caption{Estimates of the connected parts of the flavor diagonal
  charges and $g_T^{u+d}$ with the three different fits described in
  Table~\protect\eqref{tab:9L8fit}. The estimates for the isovector
  charges $g_{A,S,T}^{u-d}$ are reproduced from
  Table~\protect\eqref{tab:9L8fit} to make comparison easier.}
\label{tab:FD9L8fit}
\end{table*}

\begin{figure*}[tb]
  \subfigure{
    \includegraphics[width=0.95\linewidth]{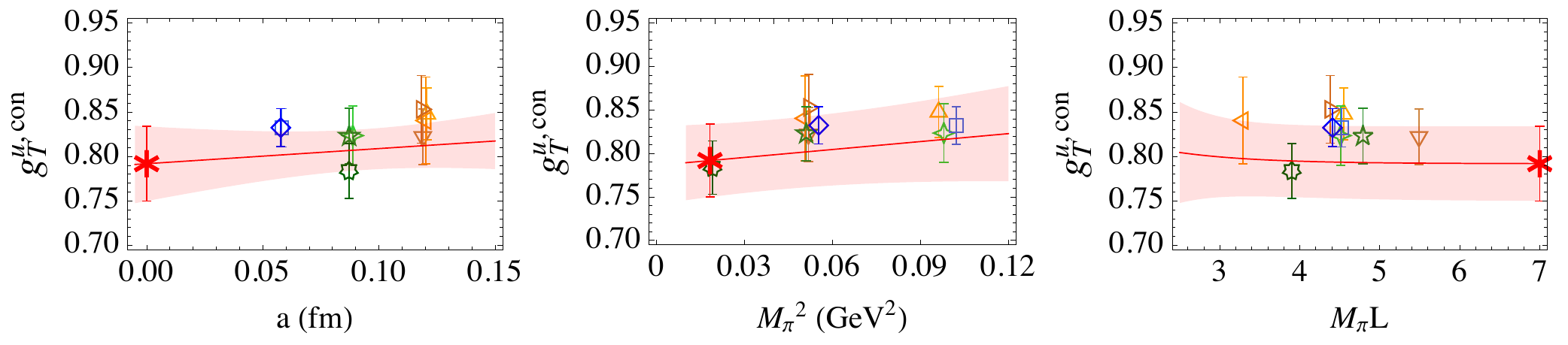} 
}
  \hspace{0.04\linewidth}
  \subfigure{
    \includegraphics[width=0.95\linewidth]{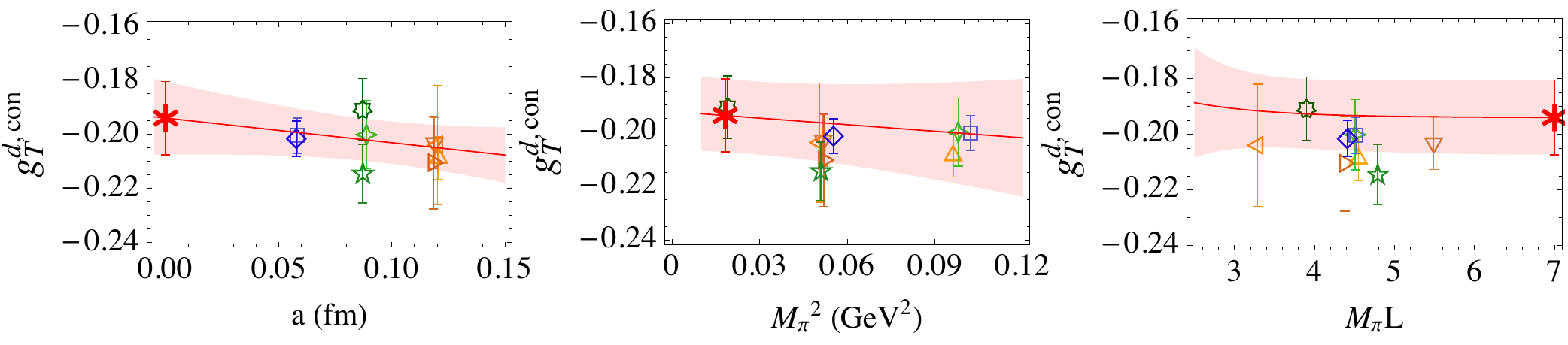}
  }
\caption{9-point simultaneous fits to the connected contributions to 
  $g_T^u$ and $g_T^d$ versus $a$, $M_\pi^2$ and $M_\pi L$ using
  Eq.~\protect\eqref{eq:CextrapgAT}. 
  The dependence on the three variables 
  $M_\pi^2$, $a$ or $M_\pi L$ is small. The data symbols are defined in
  Table~\protect\ref{tab:ens}.
  \label{fig:extrap-gT-diagonal}}
\end{figure*}

\begin{figure*}[tb]
  \subfigure{
    \includegraphics[width=0.95\linewidth]{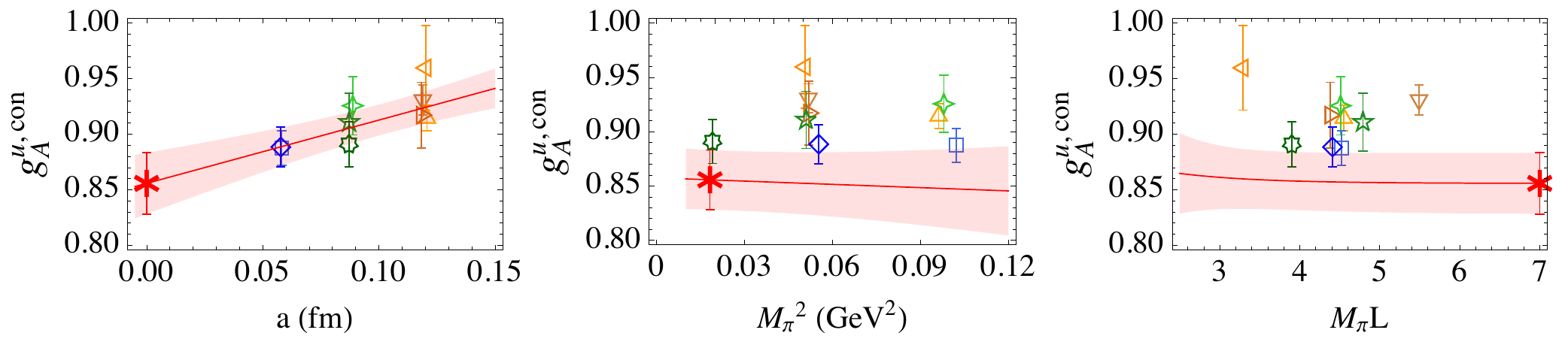}
  }
  \hspace{0.04\linewidth}
  \subfigure{
    \includegraphics[width=0.95\linewidth]{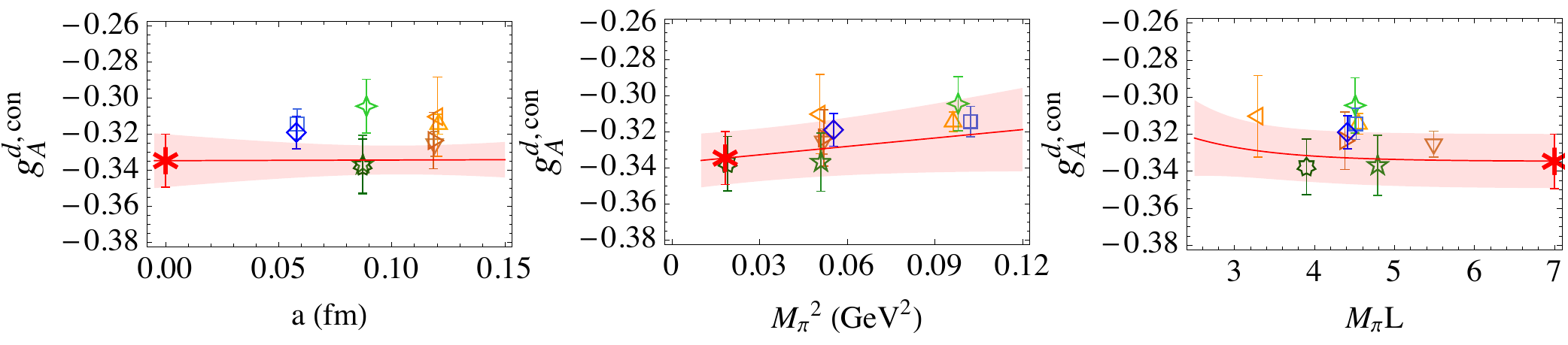}
  }
\caption{9-point simultaneous fits to the connected contributions to
  $g_A^u$ and $g_A^d$ versus $a$, $M_\pi^2$ and $M_\pi L$ using
  Eq.~\protect\eqref{eq:CextrapgAT}.
  The data for $g_A^u$ show a notable dependence on the lattice spacing $a$. The 
  rest of the panels show 
  little dependence on either $M_\pi^2$ or $M_\pi L$. The data symbols
  are defined in Table~\protect\ref{tab:ens}.
  \label{fig:extrap-gA-diagonal}}
\end{figure*}

\begin{figure*}[tb]
  \subfigure{
    \includegraphics[width=0.95\linewidth]{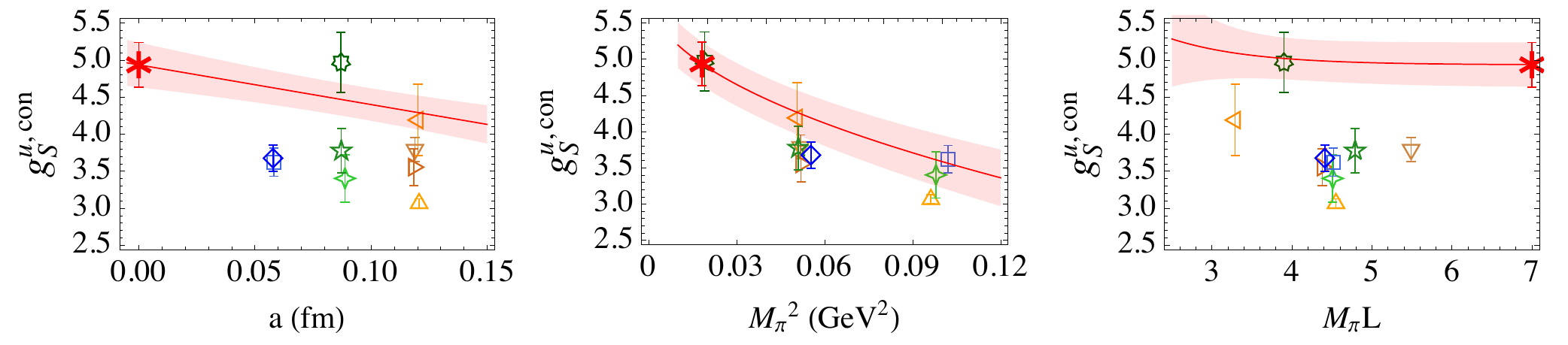}
  }
  \hspace{0.04\linewidth}
  \subfigure{
    \includegraphics[width=0.95\linewidth]{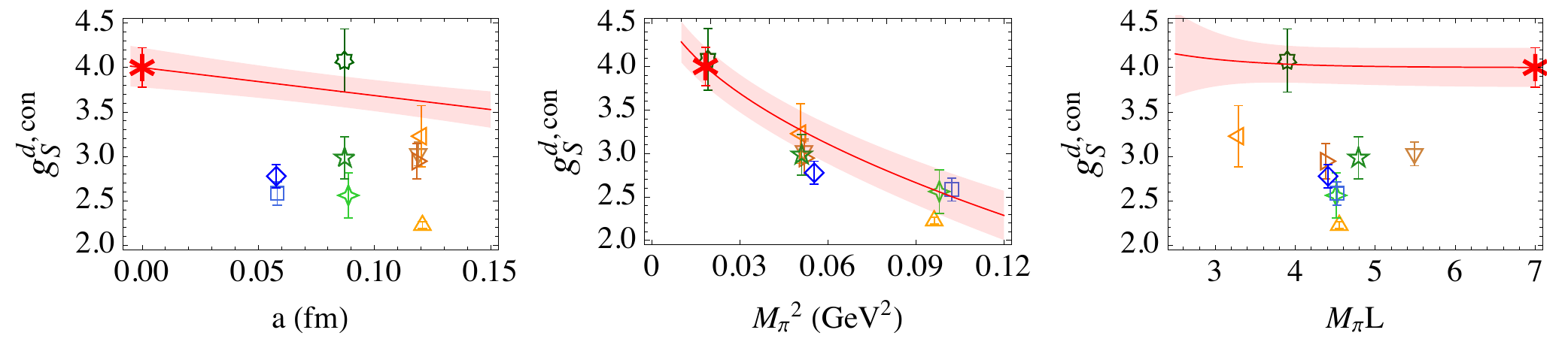}
  }
\caption{9-point simultaneous fits to the connected contributions to 
  $g_S^u$ and $g_S^d$ versus $a$, $M_\pi^2$ and $M_\pi L$ using
  Eq.~\protect\eqref{eq:CextrapgS} plus a $c_3 M_\pi^2$ term. 
  The data show significant dependence on $M_\pi^2$ and are much larger in 
  magnitude compared to $g_{A,T}^u$ and $g_{A,T}^d$. The data symbols are defined in
  Table~\protect\ref{tab:ens}.
  \label{fig:extrap-gS-diagonal}}
\end{figure*}

We show the behavior of the connected parts of the flavor-diagonal
charges $g_{A,S,T}^{u,d}$ versus the lattice spacing and the pion mass
in Figs.~\ref{fig:extrap-gT-diagonal},~\ref{fig:extrap-gA-diagonal}
and~\ref{fig:extrap-gS-diagonal} using 9-point fits with $c_3^{\rm
  log}=0$ in the Ansatz given in Eq.~\eqref{eq:gAchiralsimple}.  The
plots show that estimates of $g^u_T$ are about 4 times larger than
of $g_T^d$, and both are essentially flat with respect to the pion mass,
the lattice spacing and the lattice volume. The behavior of
$g_A^{u,d}$, shown in Fig.~\ref{fig:extrap-gA-diagonal}, is similar in
magnitude and sign to that in $g_T^{u,d}$. Again, data show little
dependence on the pion mass, however, there is a notable increase of
$g_A^{u}$ with $a$ that carries over to $g_A^{u-d}$ plotted in
Fig.~\ref{fig:conUmD_extrap9}.  The 9-point fits with the
$c_3^\prime$ term for $g_S^u$ and $g_S^d$ are shown in
Fig.~\ref{fig:extrap-gS-diagonal}. These data are much larger in
magnitude and show a significant dependence on the quark mass.

The final results for the connected parts of the flavor-diagonal
charges $g_\Gamma^u$ and $g_\Gamma^d$ from the 9-point fit are
presented in Table~\ref{tab:FDresults} in Sec.~\ref{sec:errors}.  The
new estimates of $g_{T}^{u,d}$, needed to analyze the contribution of
the EDM of the quarks to the neutron EDM, supersede the values
presented in Refs.~\cite{Bhattacharya:2015wna,Bhattacharya:2015esa}.

%%%%%%%%%%%%%%%%%%%%%%%%%%%%%%%%%%%%
\section{Confirmation of The 2-state Analysis}
\label{sec:confirmation}
%%%%%%%%%%%%%%%%%%%%%%%%%%%%%%%%%%%%

%
\begin{table*}
\begin{center}
\renewcommand{\arraystretch}{1.2} % Change horizontal spacing
\begin{ruledtabular}
\begin{tabular}{l|c|cc|c|ccc}
Ensemble ID &  Type  & $\sigma$   &  $N_{\rm GS}$ & $t_\text{sep}/a$ & $N_\text{conf}$  & $N_{\rm meas}^{\rm HP}$  & $N_{\rm meas}^{\rm AMA}$  \\
\hline
a06m310 & AMA  & 6.5   &  70   &  $\{16,20,22,24\}$ & 1000 & 4000  &  64,000   \\
a06m310 & AMA2 & 12    &  250  &  $\{18,20,22,24\}$ & 500  & 2000  &  64,000   \\
a06m220 & AMA  & 5.5   &  70   &  $\{16,20,22,24\}$ & 650  & 2600  &  41,600   \\
a06m220 & AMA2 & 11    &  230  &  $\{18,20,22,24\}$ & 650  & 2600  &  41,600   \\
\end{tabular}
\end{ruledtabular}
\caption{The smearing parameters, the values of $t_{\rm sep}$ and the
  statistics used in the two AMA simulations on the $a06m310$ and
  $a06m220$ ensembles to test the efficacy of the 2-state fit in
  controlling excited-state contamination. The second set of AMA measurements
  with the larger smearing size is labeled AMA2. The number of HP
  measurements used to correct the bias in the AMA method is listed under 
  $N_{\rm meas}^{\rm HP}$.  }
\label{tab:NewEns}
\end{center}
\end{table*}

The analysis in the previous sections was predicated on the assumption
that the 2-state Ansatz given in Eqs.~\eqref{eq:2pt} and~\eqref{eq:3pt} 
resolves the excited-state contamination in the 2- and 3-point
functions. In this section, we provide further confirmation of this
assumption using additional high-statistics AMA simulations on the
$a06m310$ and $a06m220$ ensembles with different smearing parameters. 
These are listed in Table~\ref{tab:NewEns} under the label AMA2.

\begin{table*}[h]
\centering
\begin{ruledtabular}
\begin{tabular}{c|c|ccccc|c}
ID       & Type & Fit Range  & $aM_0$      & $aM_1$     & ${\cal A}_0^2 \times 10^{11}$  & ${\cal A}_1^2 \times 10^{11}$   & ${\cal A}_1^2/{\cal A}_0^2$   \\
\hline
a06m310 & AMA  &  8---30 & 0.3268(23) & 0.56(3) & 0.58(3)                  & 0.95(10)                 & 1.66(11) \\
a06m310 & AMA2 &  5---25 & 0.3282(17) & 0.68(5) & 1.32(4)$\times 10^{-11}$ & 1.5(2)$\times 10^{-11}$  & 1.16(14) \\
a06m220 & AMA  &  8---30 & 0.3069(18) & 0.63(3) & 11.3(4)                  & 39.2(5.0)                & 3.47(35) \\
a06m220 & AMA2 &  4---30 & 0.3037(13) & 0.64(2) & 2.66(6)$\times 10^{-9}$  & 3.8(2)$\times 10^{-9}$   & 1.41(4)  \\
\end{tabular}
\end{ruledtabular}
\caption{Comparison of the masses $aM_0$ and, $aM_1$ and the
  amplitudes ${\cal A}_0$ and ${\cal A}_1$ obtained using the AMA
  method with different smearing parameters on the $a06m310$ and
  $a06m220$ ensembles.  The AMA estimates are the same as in
  Table~\protect\ref{tab:2ptfits1}. The smearing parameters and the
  statistics for the two runs are given in
  Table~\protect\ref{tab:NewEns}.}
\label{tab:2ptfitsNew}
\end{table*}

\begin{table*}[h]
\centering
\begin{ruledtabular}
\begin{tabular}{c|c|ccc|ccc|ccc}
         &      &  \multicolumn{3}{c|}{Axial}  &  \multicolumn{3}{c|}{Scalar}  &  \multicolumn{3}{c}{Tensor}  \\
ID       & Type & $\langle 0 | \mathcal{O}_\Gamma | 0 \rangle$ & $\langle 0 | \mathcal{O}_\Gamma | 1 \rangle$ & $\langle 1 | \mathcal{O}_\Gamma | 1 \rangle$ 
                & $\langle 0 | \mathcal{O}_\Gamma | 0 \rangle$ & $\langle 0 | \mathcal{O}_\Gamma | 1 \rangle$ & $\langle 1 | \mathcal{O}_\Gamma | 1 \rangle$ 
                & $\langle 0 | \mathcal{O}_\Gamma | 0 \rangle$ & $\langle 0 | \mathcal{O}_\Gamma | 1 \rangle$ & $\langle 1 | \mathcal{O}_\Gamma | 1 \rangle$ \\
\hline
a06m310 & AMA  & 1.212(14) & $-$0.060(17) & $-$1.2(1.4) & 1.18(5) & $-$0.40(4) & $-$0.5(1.0) & 0.972(12) & 0.128(10) &  0.50(22) \\
a06m310 & AMA2 & 1.210(13) & $-$0.042(26) & $-$2.6(6.2) & 1.17(7) & $-$0.43(8) & $-$9(25)    & 0.987(12) & 0.219(18) &  0.6(4.0) \\
a06m220 & AMA  & 1.234(17) & $-$0.121(18) & $-$6.3(3.9) & 1.04(6) & $-$0.30(4) & $-$0.7(2.6) & 0.984(10) & 0.103(8)  & $-$0.53(59) \\
a06m220 & AMA2 & 1.222(13) & $-$0.063(21) & $-$4.2(3.9) & 0.79(8) & $-$0.17(7) &  38(25)     & 0.969(11) & 0.203(16) &  1.6(2.4) \\
\end{tabular}
\end{ruledtabular}
\caption{Comparison of results with the two different smearing parameters defined in Table~\protect\ref{tab:NewEns}
  for the three unrenormalized matrix elements $\langle 0 |
  \mathcal{O}_\Gamma | 0 \rangle$, $ \langle 1 | \mathcal{O}_\Gamma |
  0 \rangle$ and $ \langle 1 | \mathcal{O}_\Gamma | 1 \rangle$ for the
  isovector axial, scalar and tensor operators using the 2-state
  Ansatz given in Eqs~\eqref{eq:2pt} and~\eqref{eq:3pt}. The fit ranges for the 3-point functions are
  defined under Case 3 in Table~\protect\ref{tab:4cases}.  The AMA estimates are the same as given in 
  Tables~\protect\ref{tab:3ptgAcompare},~\protect\ref{tab:3ptgScompare} and~\protect\ref{tab:3ptgTcompare}.}
  \label{tab:3ptbareME3New}
\end{table*}

\begin{figure}[h]
    \includegraphics[width=0.94\linewidth]{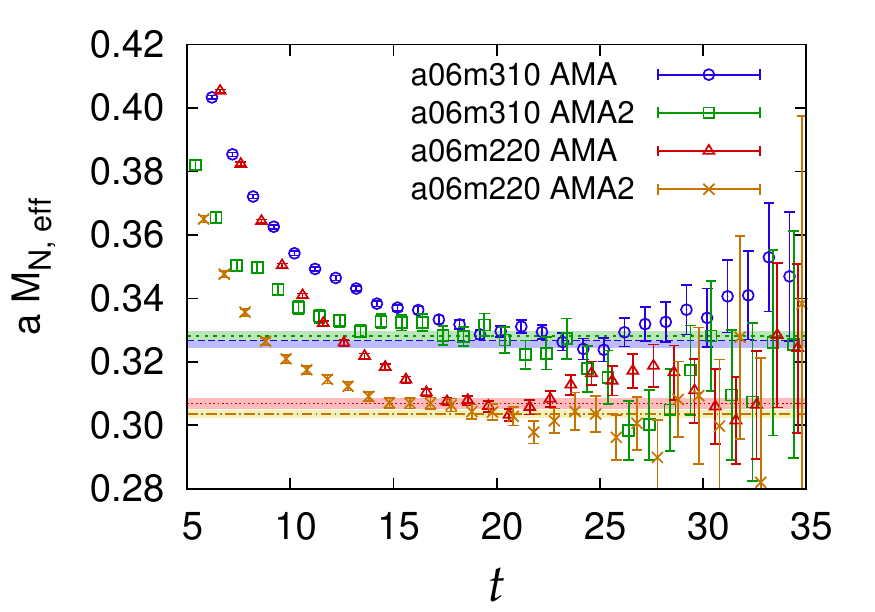}
\caption{Comparison of the nucleon effective mass obtained on the
  $a06m310$ and $a06m220$ ensembles with different smearing sizes. The
  lattice parameters are summarized in Table~\protect\ref{tab:NewEns}.  }
  \label{fig:Mna06m310Comp}
\end{figure}

\begin{figure*}[tbp]
  \subfigure{
    \includegraphics[height=1.85in,trim={0.095cm 1.22cm 0 0},clip]{figs/gA_a06m310_AMA_xtrap}
    \includegraphics[height=1.85in,trim={0.9cm   1.22cm 0 0},clip]{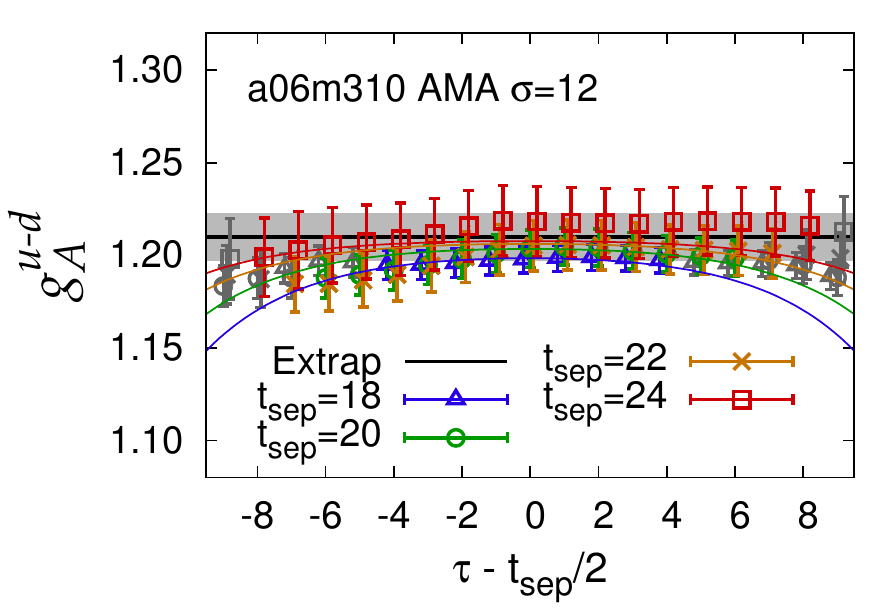}
}
  \subfigure{
    \includegraphics[height=1.85in,trim={0.095cm 1.22cm 0 0},clip]{figs/gS_a06m310_AMA_xtrap}
    \includegraphics[height=1.85in,trim={0.9cm   1.22cm 0 0},clip]{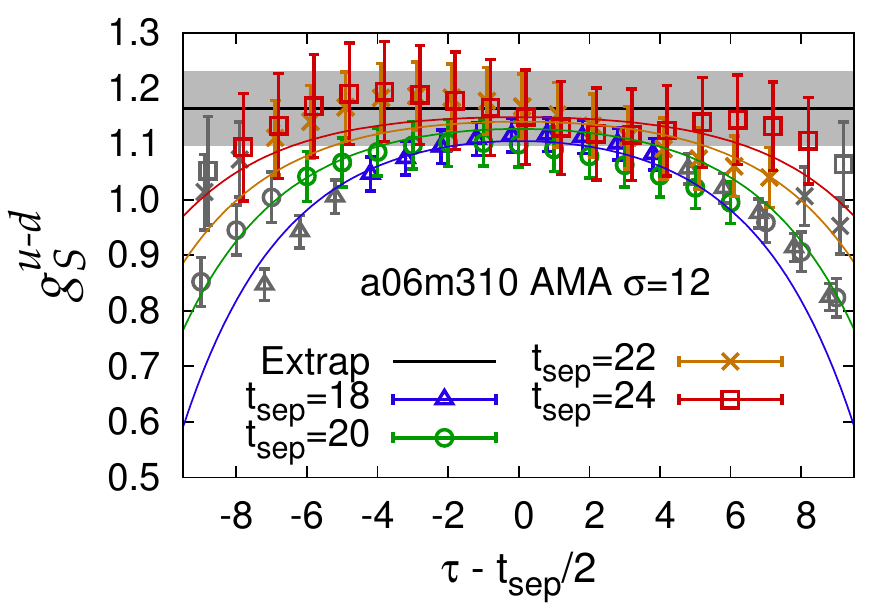}
}
  \subfigure{
    \includegraphics[height=1.85in,trim={0.095cm 1.22cm 0 0},clip]{figs/gT_a06m310_AMA_xtrap}
    \includegraphics[height=1.85in,trim={0.9cm   1.22cm 0 0},clip]{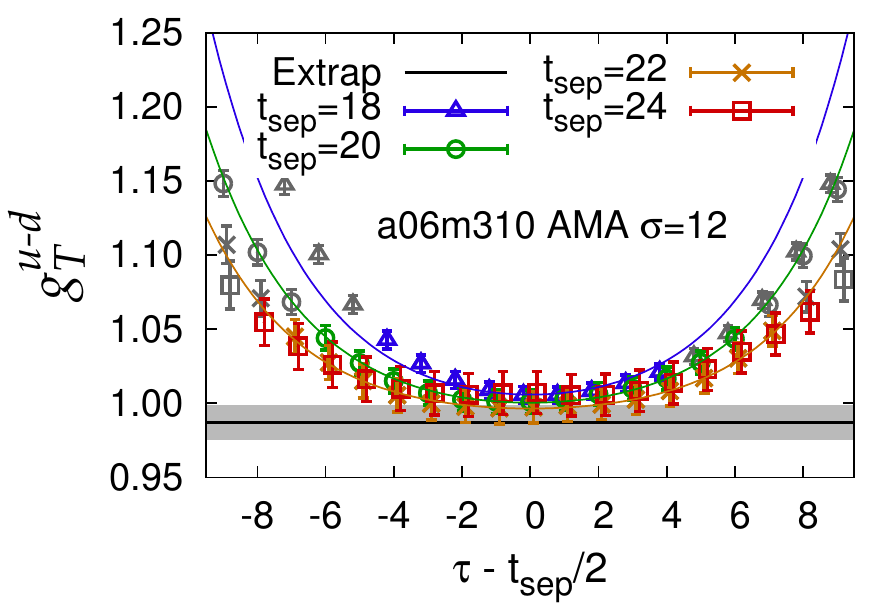}
}
  \subfigure{
    \includegraphics[height=2.26in,trim={0.095cm 0.11cm 0 0},clip]{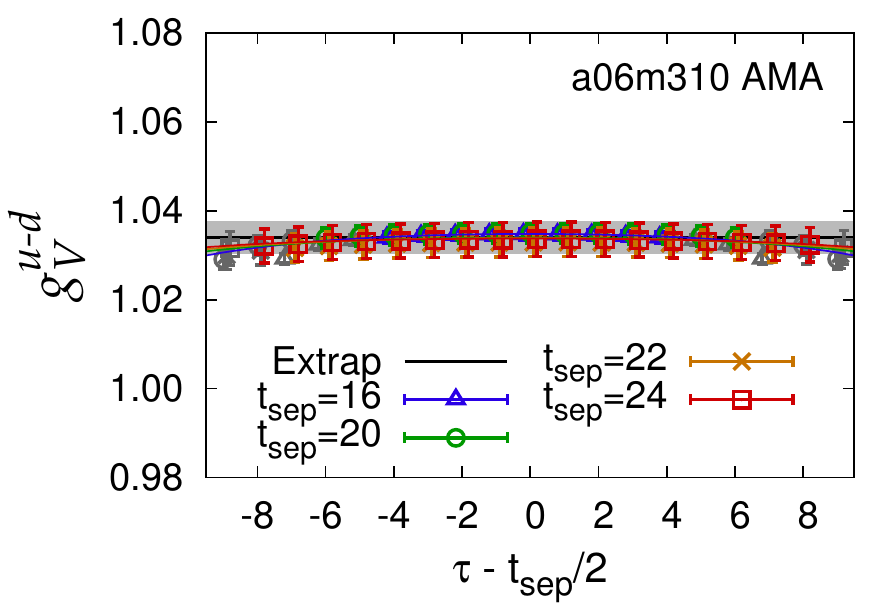}
    \includegraphics[height=2.26in,trim={0.9cm   0.11cm 0 0},clip]{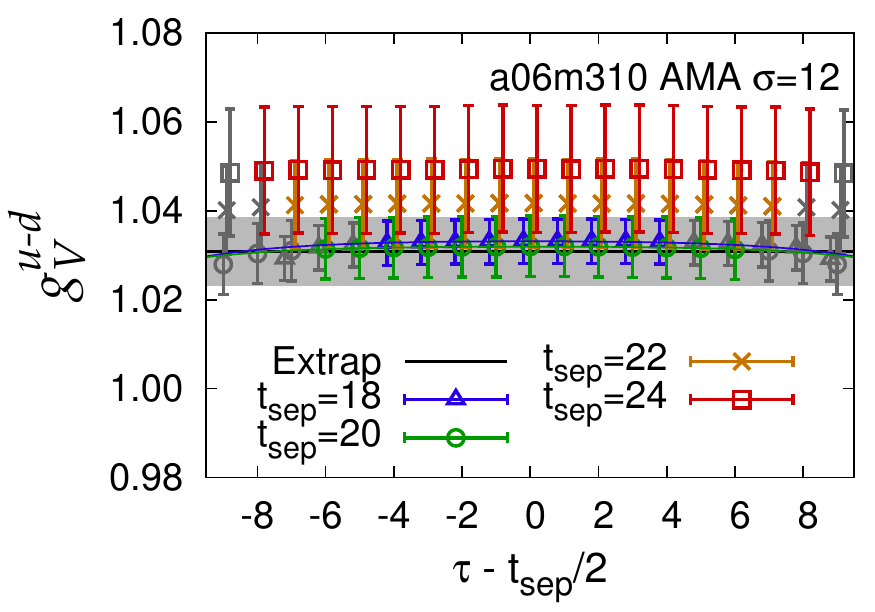}
}
\caption{Comparison of the excited-state contamination in the
  extraction of unrenormalized isovector charges $g_{A,S,T,V}^{u-d}$
  from the $a06m310$ ensemble. The plots on the left are with smearing
  parameters $\{\sigma=6.5, {N_{\rm GS}=70} \}$ and on the right with
  $\{\sigma=12, {N_{\rm GS}=250} \}$. Both calculations were done
  using the AMA method with parameters summarized in
  Table~\protect\ref{tab:NewEns}.}
  \label{fig:a06m310Comp}
\end{figure*}

\begin{figure*}[tbp]
  \subfigure{
    \includegraphics[height=1.85in,trim={0.095cm 1.22cm 0 0},clip]{figs/gA_a06m220_AMA_xtrap}
    \includegraphics[height=1.85in,trim={0.9cm   1.22cm 0 0},clip]{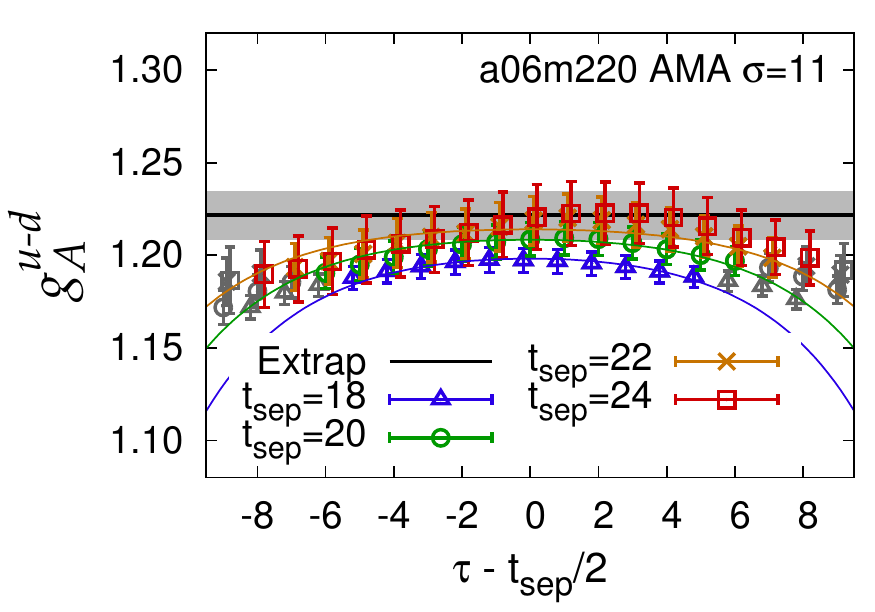}
}
  \subfigure{
    \includegraphics[height=1.85in,trim={0.095cm 1.22cm 0 0},clip]{figs/gS_a06m220_AMA_xtrap}
    \includegraphics[height=1.85in,trim={0.9cm   1.22cm 0 0},clip]{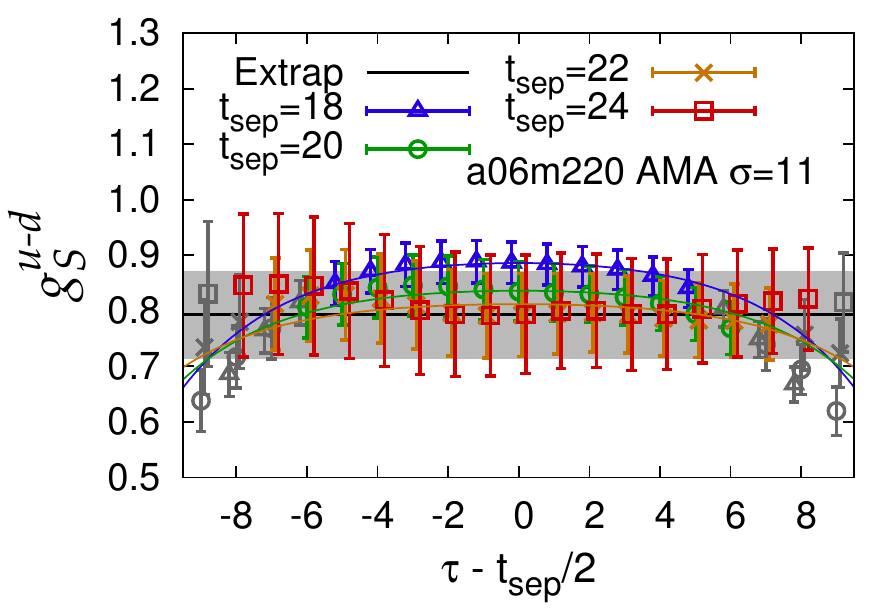}
}
  \subfigure{
    \includegraphics[height=1.85in,trim={0.095cm 1.22cm 0 0},clip]{figs/gT_a06m220_AMA_xtrap}
    \includegraphics[height=1.85in,trim={0.9cm   1.22cm 0 0},clip]{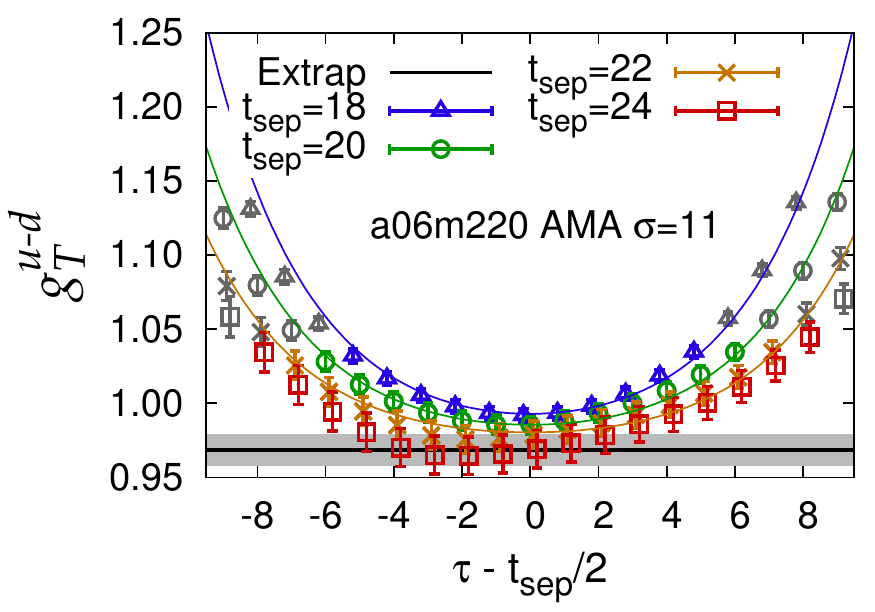}
}
  \subfigure{
    \includegraphics[height=2.26in,trim={0.095cm 0.11cm 0 0},clip]{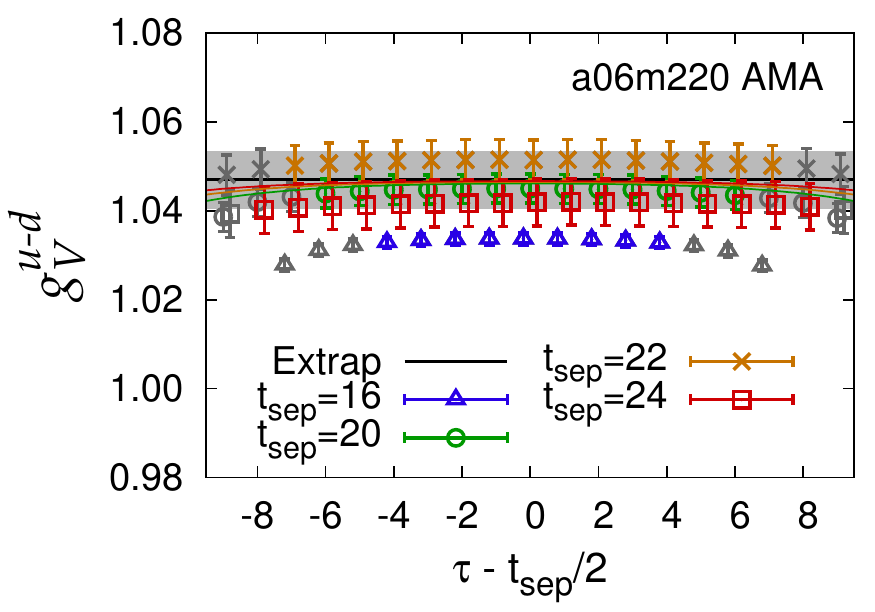}
    \includegraphics[height=2.26in,trim={0.9cm   0.11cm 0 0},clip]{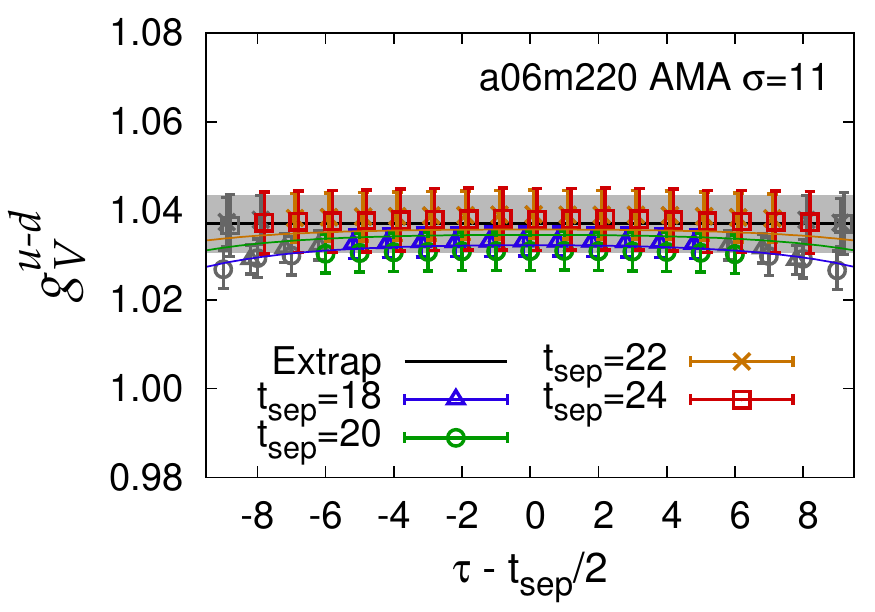}
}
\caption{Comparison of the excited-state contamination in the
  extraction of unrenormalized isovector charges $g_{A,S,T,V}^{u-d}$ 
  from the $a06m220$ ensemble. The plots on the left are with smearing
  parameters $\{\sigma=5.5, {N_{\rm GS}=70} \}$ and on the right with
  $\{\sigma=11, {N_{\rm GS}=230} \}$. Both calculations were done
  using the AMA method with parameters summarized in
  Table~\protect\ref{tab:NewEns}.}
  \label{fig:a06m220Comp}
\end{figure*}

A comparison of the effective mass for the two smearings is shown in
Fig.~\ref{fig:Mna06m310Comp}, and of the excited-state contamination
using Eqs.~\eqref{eq:2pt} and~\eqref{eq:3pt} using fits to the 3-point
functions data is shown in Figs.~\ref{fig:a06m310Comp} and
\ref{fig:a06m220Comp}. The results of the fits to the 2-point function are 
given in Table~\ref{tab:2ptfitsNew}, and for the three matrix elements
$\langle 0 | \mathcal{O}_\Gamma | 0 \rangle$, $ \langle 1 |
\mathcal{O}_\Gamma | 0 \rangle$ and $ \langle 1 | \mathcal{O}_\Gamma |
1 \rangle$ are given in Table~\ref{tab:3ptbareME3New}. A summary of
the notable features is as follows:
\begin{itemize}
\item
Increasing the smearing size $\sigma$ reduces the ratio ${\cal
  A}_1/{\cal A}_0$ and the relative contribution of the
excited states. Estimates of $M_0$ and the isovector charges $g_A^{u-d}$,
$g_T^{u-d}$ and $g_V^{u-d}$ (given by $\langle 0 | \mathcal{O}_\Gamma | 0
\rangle$) with the two different smearing sizes agree within
$1\sigma$. The one exception is $g_S^{u-d}$ from the $a06m220$ ensemble.
\item
The excited-state contamination in $g_A^{u-d}$ is significantly
reduced with the larger smearing size from $ 15\%$ to $ 5\%$. This can
be seen by comparing the data at the central values of $\tau$. The
2-state fit to both sets of data gives consistent estimates of the
$t_\text{sep} \to \infty$ value.
\item
The excited-state effect in $g_T^{u-d}$ at the central values of $\tau$ is
less than $5\%$ in all cases as shown in
Figs.~\ref{fig:gT7},~\ref{fig:a06m310Comp} and~\ref{fig:a06m220Comp}.
Comparing the data with the two smearing sizes, we find that the
2-state fit gives consistent and stable estimates of the
$t_\text{sep} \to \infty$ value for both $a06m310$ and $a06m220$
ensembles.
\item
The data for $g_S^{u-d}$ at different $t_\text{sep}$ with the larger
smearing size overlap and are not well resolved as shown in
Figs.~\ref{fig:a06m310Comp} and~\ref{fig:a06m220Comp}.  The pattern of
variation of the $a06m220$ AMA2 data versus $t_\text{sep}$ is opposite
to that seen in the other three ensembles listed in
Table~\ref{tab:NewEns}. The 2-state fit takes this into account with
an unreasonably large value of $ \langle 1 | \mathcal{O}_\Gamma | 1
\rangle$ and with the opposite sign. As a result, the two estimates of
$g_S^{u-d}$ from the $a06m220$ ensemble differ by roughly $3\sigma$.  We
consider this inverted pattern of the $ \langle 1 | \mathcal{O}_\Gamma
| 1 \rangle$ contribution in the AMA2 data to be a statistical
fluctuation and regard the AMA estimate to be more reliable.
\item
Estimates of $\langle 0 |\mathcal{O}_\Gamma | 1 \rangle$ agree in sign and 
roughly in magnitude for all the charges. 
\item
Estimates of $\langle 1 |\mathcal{O}_\Gamma | 1 \rangle$ for the three
charges are poorly determined with either smearing size.
\item
The differences in $M_1$, $\langle 0 |\mathcal{O}_\Gamma | 1 \rangle$
and $\langle 1 | \mathcal{O}_\Gamma | 1 \rangle$ are large in some
cases.  These differences reflect the fact that the 2-state fit lumps
the contribution of all the excited states into one ``effective''
excited state, and their contributions vary with the smearing size and
the fit ranges.  Much higher-statistics data with better interpolating
operators that enable a three-state fit will be needed to obtain
reliable estimates of these first excited-state parameters.
\item
The errors in the two matrix elements $ \langle 0 | \mathcal{O}_\Gamma
| 1 \rangle$ and $ \langle 1 | \mathcal{O}_\Gamma | 1 \rangle$ given
by the 2-state fit are larger in the AMA2 analysis even though the
total number of measurements in the two cases is the same. With
reduced excited-state contamination, the data at different $\tsep$
overlap and the 2-state fit becomes less stable with respect to values
of $\tsep$ used in the fit. Consequently, higher statistics are needed
to provide reliable $t_\text{sep} \to \infty$ estimates.
\end{itemize}

The bottom line is that these two additional simulations validate the
results based on 2-state fits to data at multiple values of $\tsep$
presented in Sec.~\ref{sec:results}. They also show that the large
excited-state contamination in $g_A^{u-d}$ and $g_S^{u-d}$ on the $a=0.06$~fm
lattices with smearing parameter $\sigma \approx 6$ is significantly
reduced with $\sigma \approx 11$. On the other hand, estimates at
different $t_\text{sep}$ start to overlap with reduced excited-state
contamination, and the 2-state fit becomes less stable. This is most
obvious in the $g_S^{u-d}$ data on the $a06m220$ ensemble. In retrospect, a
more effective compromise balancing the two effects would have been
achieved with an intermediate value for the smearing parameter,
$\sigma=9$ ($\approx 0.55$~fm in physical units), on the $a=0.06$~fm
ensembles.

%                                                                                                                                                                                                                  

%%%%%%%%%%%%%%%%%%%%%%%%%%%%%%%%%%%%
\section{Assessing the estimation of errors}
\label{sec:errors}
%%%%%%%%%%%%%%%%%%%%%%%%%%%%%%%%%%%%

The estimation of errors in our analysis has four main components:
\begin{itemize}
\item
Statistical and excited-state contamination (SESC): Errors from these
two sources are estimated together in the combined 2-state fit.
\item
Uncertainty in the determination of the renormalization constants
$Z_\Gamma$: The $Z$'s are estimated as a function of $q^2$ using the
RI-sMOM scheme on the lattice and then converted to the 
$\overline{\text{MS}}$ scheme at $2\GeV$ using perturbation theory. As
discussed in Sec.~\ref{sec:renorm}, there is significant spread in the
data due to breaking of rotational symmetry on the lattice. Second,
the 2-loop series for the matching factor for $g_T^{u-d}$ does not
show a convergent behavior.  Third, the data for the four charges do
not uniformly show a large enough interval, $\eta \Lambda_{\rm QCD}
\ll q \ll \xi \pi/a$, over which they are independent of $q^2$, so we are
not able to extract a unique estimate of $Z_\Gamma$.  The two methods
we use give significantly different estimates. We use this difference,
which is larger than the individual statistical uncertainty, to assign the
error.  These error estimates are added in quadrature to those in the
bare charges obtained from the 2-state fit to get the total error in
the renormalized charges on each ensemble.
\item
Discretization errors due to nonzero lattice spacing $a$,
finite-volume effects characterized by $M_\pi L$ and uncertainty in
the chiral behavior: The errors due to these three systematics are
obtained using a simultaneous fit. The Ansatz used for the final
results, keeping the lowest-order corrections in each variable, is
given in Eqs.~\eqref{eq:CextrapgAT} and~\eqref{eq:CextrapgS}.
\item
Uncertainty due to the number of terms retained in the combined fit
Ansatz as discussed in Sec.~\ref{sec:results}.
\end{itemize}

A recapitulation of the important features observed in the data for the
isovector charges is as follows:
\begin{itemize}
\item
Estimates of the isovector axial charge, $g^{u-d}_A$, converge from
below with respect to excited-state contamination on all the
ensembles. The data plotted in Fig.~\ref{fig:conUmD_extrap9} show no
significant dependence on the pion mass or the lattice volume. The
largest variation is the increase with the lattice spacing and the
size of the slope is dictated by the smaller estimates on the
$a=0.06$~fm ensembles. Also, as discussed in Sec.~\ref{sec:results},
even though this slope is different in the fits to renormalized
charges calculated in the two ways, the extrapolated results are
consistent. Overall, the spread in the estimates under changes in the
fit Ansatz is small, about $0.03$.
\item
The isovector scalar charge, $g^{u-d}_S$, also converges from below
with respect to excited-state contamination. The data show a decrease
with the lattice spacing but no significant variation with the pion
mass.  Our error estimate accounts for (i) the large excited-state
contamination on the two finest ensembles, $a06m310$ and $a06m220$;
(ii) the uncertainty in the determination of the renormalization
constant $Z_S$; and (iii) the larger, by a factor of 3--5, statistical
errors compared to those in $g_A^{u-d}$ and $g_T^{u-d}$ on the various
ensembles as summarized in Table~\ref{tab:resultsbare}. The fits
capture the variation of $g_S^{u-d}(a, M_\pi, M_\pi L)$ with respect
to both the lattice spacing $a$ and the pion mass $M_\pi$ as shown in
Figs.~\ref{fig:conUmD_extrap9} and~\ref{fig:conUmD_extrap8}. The
largest sensitivity to the fit Ansatz comes from adding a chiral
logarithm term, which tends to increase the estimate by about $0.12$
as shown in Table~\ref{tab:9L8fit}. However, as shown in
Fig.~\ref{fig:extrap-gS-diagonal}, keeping just the leading polynomial correction, 
$O(M_\pi)$ gives very good fits to the data, and the AIC indicates that adding 
the chiral logarithm term does not improve the fits significantly. Nevertheless, 
we will take half the change, 0.06, as an estimate of the systematic error 
due to the choice of the fit Ansatz.
\item
Our analysis of the isovector tensor charge, $g_T^{u-d}$, shows that
(i) estimates converge from above for each ensemble and the fit using
the 2-state Ansatz given in Eqs.~\eqref{eq:2pt} and~\eqref{eq:3pt} accounts for the
excited-state contamination within the quoted errors. (ii) The data
for the renormalization constant in the RI-sMOM scheme show a window
in $q^2$ over which the estimates in the $\overline{\text{MS}}$ scheme
at $2\GeV$ are constant within errors as discussed in
Sec.~\ref{sec:renorm} and Ref.~\cite{Bhattacharya:2015wna}. On the
other hand, the poorly behaved 2-loop series for the matching factor
suggests that this systematic uncertainty could be as large as $0.02$. We have
taken this systematic into account when estimating the error in $Z_T$ given in
Table~\ref{tab:Zfinal}.  (iii) The estimates from the nine ensembles
display little dependence on the lattice spacing, pion mass or the
lattice volume, as shown in Fig.~\ref{fig:conUmD_extrap9}. (iv)
The largest change in estimates, about 0.04, is again due to adding
a chiral logarithm term to the fit Ansatz.
\end{itemize}

Based on the data, fits, and trends observed, we propose at an error
budget from each of the sources that is summarized in
Table~\ref{tab:errors}.  The entries are constructed as follows: For
the statistical uncertainty and excited-state contamination, we
consider the data shown in
Figs.~\ref{fig:gA7},~\ref{fig:gS7},~\ref{fig:gT7},~\ref{fig:a06m310Comp}
and~\ref{fig:a06m220Comp} and the errors in the 3-point function data,
the efficacy of the 2-state fit, and the difference between the two AMA
estimates on the $a06m310$ and $a06m220$ ensembles. For estimating the
uncertainty in the renormalization constants, we use the errors given
in Table~\ref{tab:Zfinal}, which are consistent with the size of the
deviations of $Z_V g_V^{u-d}$ from unity given in
Table~\ref{tab:resultsgV}. For assessing the error associated with the
extrapolation in the lattice spacing, we take half of the total spread
in the central values at the three lattice spacings $a=0.12$, $0.09$, 
and $0.06$~fm. Similarly, for the dependence on the quark mass, we
take half of the spread in the central values at the three masses,
$M_\pi \approx 135$, $220$, and $310$~MeV.  The finite-volume
correction is observed to be small in all cases and the error budget
is assigned to be half the change on going from the 9-point to the 8-point
fit estimates shown in Table~\ref{tab:9L8fit}; i.e., the direction is given by
the change on removing the smallest $M_\pi L$ point $a12m220S$.  Our
combined fits include these systematics, and the error estimates 
given in Table~\ref{tab:9L8fit} are consistent with
the error budget summarized in Table~\ref{tab:errors}.
 
An error estimate due to the choice of the fit Ansatz is the least
straightforward to assess as discussed next.  With nine data points,
we can explore only a limited space of lowest order corrections given
in Eqs.~\eqref{eq:extrapgAT} and~\eqref{eq:extrapgS}. Within this
subspace, the largest variation with respect to the fit Ansatz is in
$g_S^{u-d}$ and $g_T^{u-d}$ obtained with and without the chiral
logarithm term as shown in Figs.~\ref{fig:conUmD_extrap9}
and~\ref{fig:conUmD_extrap9L} and quantified in
Table~\ref{tab:9L8fit}. As pointed out in Sec.~\ref{sec:results}, in
fits with chiral logarithms, the two terms proportional to $c_3$ (or
$c_3^\prime$) and $c_3^{\rm log}$ compete and the errors in them are
large. The Akaike information criterion indicates that including the
chiral logarithm term does not improve the fit sufficiently to warrant
it. Nevertheless, we take half the spread (0.06 for $g_S^{u-d}$ and
0.02 for $g_T^{u-d}$) between the two 9-point estimates given in
Table~\ref{tab:9L8fit} as a conservative estimate of the uncertainty
due to the fit Ansatz.  Estimates of $g_A^{u-d}$ are much more stable
under changes in the fit Ansatz, with the central value varying
between 1.18 and 1.21. Again, we take a conservative value, $0.02$, as
an estimate of this systematic uncertainty.

In Table~\ref{tab:errors}, we also indicate the direction in which our
analysis changes the estimate as a result of taking into account a
given systematic. For example, the slope of the fit to $g_A^{u-d}$
versus $a$ shown in Fig.~\ref{fig:conUmD_extrap9} is positive. As a
result, the central value after extrapolation to $a=0$ is lower than the data
points. We indicate this by attaching a $\Downarrow$ to the estimate
shown in Table~\ref{tab:errors}. For the ratios of renormalization
constants, we compare the estimates from methods A and B given in
Table~\ref{tab:Zmethod1and2}. Taking estimates using method B as the
baseline, we assign a $\Downarrow$ if taking the average with method A
lowers the final estimate as shown in Table~\ref{tab:Zfinal}. The direction 
of the SESC estimate is taken to be the direction of convergence with $\tsep$.

\begin{table}
\centering
\begin{ruledtabular}
\begin{tabular}{c|ccc}
Error From          &  $g_A^{u-d}$ & $g_S^{u-d}$ &   $g_T^{u-d}$     \\
\hline                                                      
SESC                &  $0.02$  $\Uparrow$    & $0.05$   $\Uparrow$     &  $0.02$   $\Downarrow$  \\
$Z$                 &  $0.01$  $\Downarrow$  & $0.04$   $\Uparrow$     &  $0.04$   $\Downarrow$  \\
$a$                 &  $0.02$  $\Downarrow$  & $0.04$    $\Uparrow$    &  $0.01$   $\Downarrow$  \\
Chiral              &  $0.02$  $\Uparrow$    & $0.03$  $\Downarrow$    &  $0.02$   $\Downarrow$  \\
Finite volume       &  $0.01$  $\Uparrow$    & $0.01$  $\Uparrow$      &  $0.01$   $\Uparrow$  \\
\hline                                                      
Error quoted        &  $0.033$               & $0.12$                  &  $0.046$                \\
\hline
Fit Ansatz          &  $0.02$                & $0.06$                  &  $0.02$                 \\
\end{tabular}
\end{ruledtabular}
\caption{Estimates of the error budget for the three isovector charges
  due to each of the five systematic effects described in the text.
  The symbols $\Uparrow$ and $\Downarrow$ indicate the direction in
  which a given systematic is observed to drive the central value
  obtained from the 9-point fit. The second last row gives the errors
  in our best estimate given in Table~\protect\ref{tab:9L8fit} using
  the 9-point fit.  The last row gives the additional systematic
  uncertainty that accounts for the variations due to the
  choice of the fit Ansatz.  }
\label{tab:errors}
\end{table}

Our final estimates for the isovector charges, including a second 
error to account for the variation in the estimates 
with the fit Ansatz, are given in
Table~\ref{tab:IVresults}.  Estimates of the flavor-diagonal charges
given in Table~\ref{tab:FDresults} and of $g_T^{u+d}$ given in
Table~\ref{tab:IVresults} are based on only the connected
diagrams. We, therefore, consider it premature to assign an additional
systematic uncertainty due to the fit Ansatz.

\begin{table}[ht]
\begin{center}
\renewcommand{\arraystretch}{1.2} % Change horizontal spacing
\begin{ruledtabular}
\begin{tabular}{l|cc|cc}
           & $g_A^{u-d}$    &   $g_S^{u-d}$  &  $g_T^{u-d}$   & $g_T^{u+d,{\rm con}}$     \\
\hline
$g_\Gamma$ & 1.195(33)(20)  & 0.97(12)(6)    & 0.987(51)(20)  & 0.598(33)  \\
\hline
$\chi^2/\text{d.o.f.}$ & 0.28           & 0.67           & 0.44           & 0.30      \\
\end{tabular}
\end{ruledtabular}
\caption{Final estimates of the renormalized isovector charges for the
  proton from the 9-point fit described in
  Sec.~\protect\ref{sec:results}.  The first error includes
  statistical and all systematic uncertainties except that due to the
  extrapolation Ansatz, which is given by the second error estimate.
  As explained in the text, in the connected estimate of $g_T^{u+d}$
  we only give the first error.  Estimates for the neutron are
  obtained by the $u \leftrightarrow d$ interchange. }
\label{tab:IVresults}
\end{center}
\end{table}

\begin{table*}[t]
\begin{center}
\renewcommand{\arraystretch}{1.2} % Change horizontal spacing
\begin{ruledtabular}
\begin{tabular}{l|cc|cc|cc}
                        & $g_A^u$    &  $g_A^d$       &   $g_S^u$  &  $g_S^d$  &  $g_T^u$   & $g_T^d$       \\
\hline
$g_\Gamma$              & 0.856(27)  & $-$0.335(15)   & 4.94(30)   & 4.00(22)  & 0.792(42)  & $-$0.194(14)  \\
$\chi^2/\text{d.o.f.}$  & 0.59       &  0.28          & 1.6        & 2.1       & 0.38       &  0.48         \\
%% \hline
%% 0.856(26)  & $-$0.331(14)   & 4.75(32)   & 3.78(24)  & 0.800(38)  & $-$0.192(12)  \\
%% 0.52       &  0.32          & 1.47       & 2.07      & 0.38       &  0.62         \\
\end{tabular}
\end{ruledtabular}
\caption{Final estimates of the connected part of the renormalized
  flavor-diagonal charges of the proton. The $\chi^2/\text{d.o.f.}$
  for the 9-point fit are given in the second row.
  %% estimates from the 7-point fits
  Estimates for the neutron are
  obtained by the $u \leftrightarrow d$ interchange. }
\label{tab:FDresults}
\end{center}
\end{table*}

To summarize, the first error quoted in $g_T^{u-d}$ is dominated by that in $Z_T$
and, as shown in Table~\ref{tab:errors}, all five systematic effects
are in the same direction. Since the other four systematics are small,
we are confident that our error estimate, $0.046$, covers 
these five systematics. Our new result, $g_T^{u-d}=0.987(51)(20)$,
confirms the conclusion reached in Ref.~\cite{Bhattacharya:2015wna}
that all the systematics are under control at a few percent level in
this calculation.

The extraction of the scalar charge $g_S^{u-d} = 0.97(12)(6)$ is less
precise since the statistical errors in $g_S^{u-d}$ on each ensemble
are still 10\%--15\%, that is, a factor of 3--5 larger than those in
$g_T^{u-d}$. Also, all sources of systematic uncertainty are at the $
5\%$ level. The dominant systematic is from including a chiral
logarithm in the fit Ansatz---the shift in the estimate is much larger
than in $g_A^{u-d}$ or $g_T^{u-d}$.  Considering the size and sign of
the first five systematic uncertainties, we again conclude that the
first error, $0.12$, and the additional $0.06$ due to the fit Ansatz
are conservative.

Our estimate $g_A^{u-d} = 1.195(33)(20)$ is smaller than the
experimental result $1.276(3)$ by about $ 7\%$. This difference could
be due to a combination of the observed few percent effect in the
various systematics. Thus, a higher-statistics study on ensembles at
smaller $a$ and closer to the physical $M_\pi$ is needed to improve 
the estimate. 

%%%%%%%%%%%%%%%%%%%%%%%%%%%%%%%%%%%%
\section{Comparison with Previous Works}
\label{sec:comparison}
%%%%%%%%%%%%%%%%%%%%%%%%%%%%%%%%%%%%

In this section we compare our results with previous determinations of
the isovector charges $g_A^{u-d}$, $g_S^{u-d}$ and $g_T^{u-d}$. In
this comparison, it is important to note that we have, for the first
time, taken all the systematics into account by uniformly using the
2-state fit with multiple values of $\tsep$ to address excited-state
contamination and by making a combined fit in the three variables
$a$, $M_\pi^2$ and $M_\pi L$ using Eqs.~\eqref{eq:CextrapgAT}
and~\eqref{eq:CextrapgS}. Our error estimates from this combined fit are
larger (for example, the fits versus only $M_\pi^2$, shown in
Fig.~\ref{fig:conUmD_extrap9} as grey overlays, have a much narrower error
band and give $g_A^{u-d}=1.25(2)$); however, we claim they are realistic as discussed in
Sec.~\ref{sec:errors}. Also, our final results include a second error
estimate: the first error includes statistical and all systematic
uncertainties except that due to the extrapolation Ansatz, which is
given by the second error estimate.

\subsection{$g_A^{u-d}$}
\label{sec:gAcompare}

\begin{figure}
\begin{center}
\includegraphics[width=0.46\textwidth]{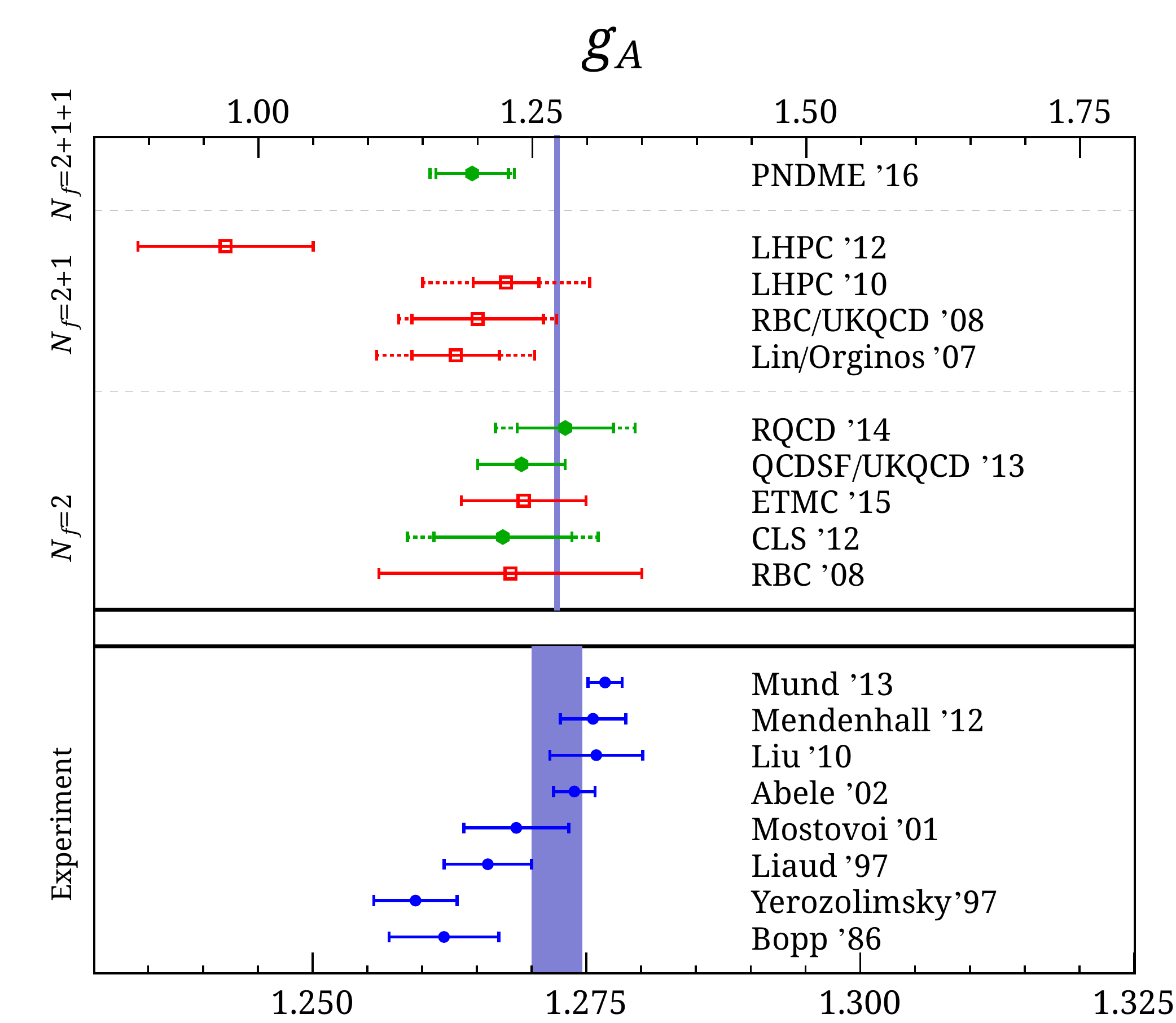}
\end{center}
\vspace{-0.5cm}
\caption{ (Top) A summary plot showing the current estimates of $g_A^{u-d}$
  from lattice-QCD calculations with 2+1+1, 2+1 and 2 flavors. The
  data are taken from the following sources: 
  PNDME'16 (this work);
  LHPC'12~\protect\cite{Green:2012ud};
  LHPC'10~\protect\cite{Bratt:2010jn};
  RBC/UKQCD'08~\protect\cite{Lin:2007ap};
  RQCD'14~\protect\cite{Bali:2014nma};
  QCDSF/UKQCD'13~\protect\cite{Horsley:2013ayv};
  ETMC'15~\protect\cite{Abdel-Rehim:2015owa};
  CLS'12~\protect\cite{Capitani:2012gj};
  RBC'08~\protect\cite{Yamazaki:2008py}.  (Bottom) The experimental
  results have been taken from the following sources: Mund'13~\protect\cite{Mund:2012fq};
  Mendenhall'12~\protect\cite{Mendenhall:2012tz};
  Liu'10~\protect\cite{Liu:2010ms};
  Abele'02~\protect\cite{Abele:2002wc};
  Mostovoi'01~\protect\cite{Mostovoi:2001ye};
  Liaud'97~\protect\cite{Liaud:1997vu};
  Yerozolimsky'97~\protect\cite{Erozolimsky:1997wi};
  Bopp'86~\protect\cite{Bopp:1986rt}.  The blue band highlights the
  2014 PDG average value 1.2723(23)~\protect\cite{Agashe:2014kda}.
  Note the change in scale between the upper (lattice QCD) and the
  lower (experimental) panels. The lattice-QCD estimates in red
  indicate that estimates of excited-state contamination, or
  discretization errors, or chiral extrapolation were not
  presented. When available, systematic errors have been added to
  statistical ones as outer error bars marked with dashed lines. }
\label{fig:gA}
\end{figure}

Calculations of $g_A^{u-d}$ are considered a test of the lattice-QCD
method to provide accurate estimates of the properties of the
nucleon. A summary of experimental and lattice-QCD determination of
$g_A^{u-d}$ is given in Fig.~\ref{fig:gA}. We note that over time the
experimental estimate has increased steadily to its present value
$1.276(3)$. On the other hand, most previous lattice-QCD estimates
have been in the range 1.1--1.2~\cite{Constantinou:2014tga}. A
significant reason for the lattice-QCD estimates being low has been
excited-state contamination, since it can make a large negative
contribution depending on the nucleon interpolating operator (the
smearing parameter $\sigma$ in our study) as shown in
Figs.~\ref{fig:gA7},~\ref{fig:a06m310Comp} and
\ref{fig:a06m220Comp}. In this work, we have shown that analyses using
a combination of well-tuned smeared sources for generating quark
propagators, performing simulations at multiple values of $t_{\rm
  sep}$, and a simultaneous 2-state fit to the data at a number of
values of $t_{\rm sep}$ reduces this contamination to the size of the
statistical errors which are about $2\%$. With $O(50,000)$
measurements on ensembles with $M_\pi L\, \gsim 4$ and spanning
0.12--0.06~fm in lattice spacing and 135--320~MeV in the pion mass,
the uncertainty from the chiral fit and continuum extrapolation is
also reduced to about $2\%$ from each of these systematics.  Our
analysis of the full error budget is presented in
Sec.~\ref{sec:errors}.

\begin{figure*}
\begin{center}
\includegraphics[width=0.46\textwidth]{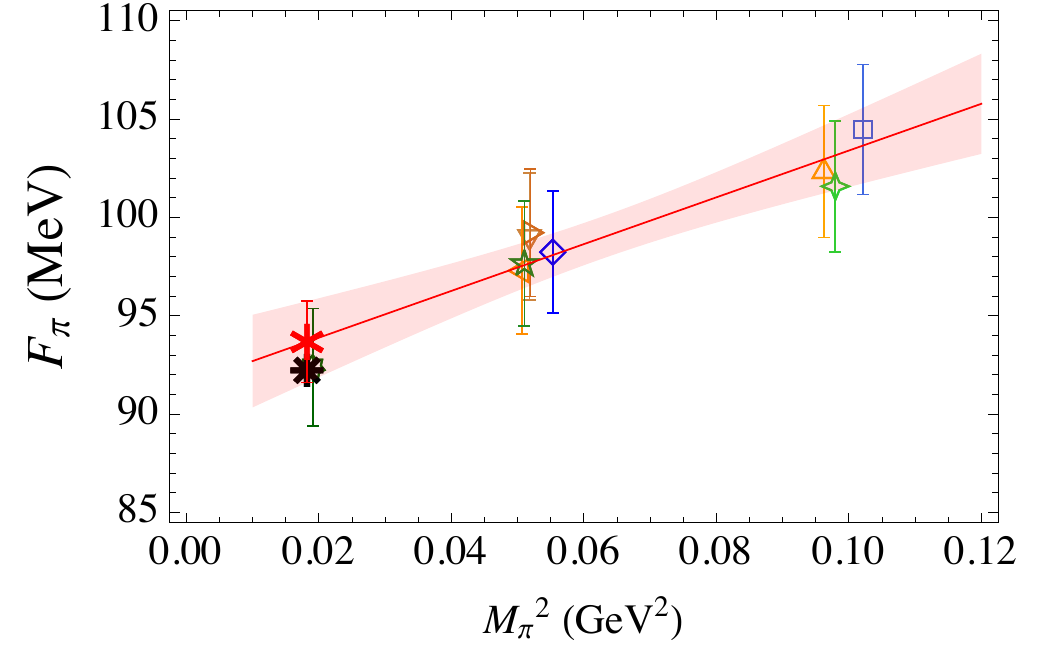}
\includegraphics[width=0.46\textwidth]{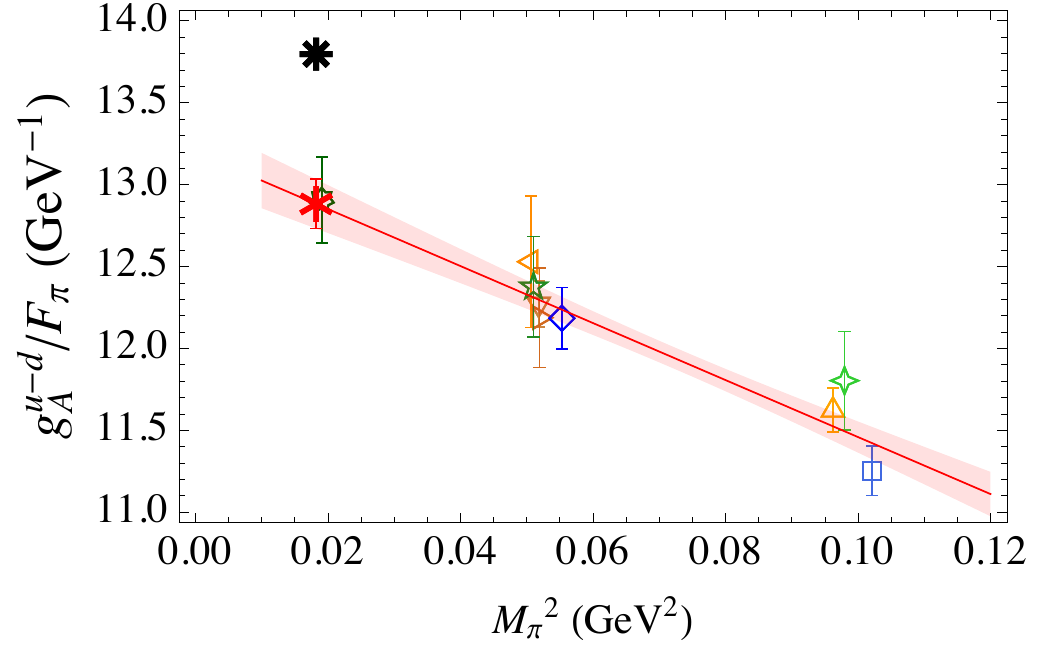}
\end{center}
\vspace{-0.5cm}
\caption{Fits to $F_\pi$ (left panel) and $g_A/F_\pi$ (right panel)
  data given in Table~\protect\ref{tab:resultsFpi} assuming a linear
  dependence on $M_\pi^2$ for both.  The lattice scale of these HISQ
  ensembles used to convert $M_\pi$ and $F_\pi$ to physical units was
  determined using $r_1$~\cite{Bazavov:2012xda}.  Since the
  extrapolated $F_\pi$, red star, matches the experimental value at
  the physical point, the value of the renormalization factor
  independent ratio $g_A/F_\pi$ remains about $7\%$ below the
  experimental result shown as a black star. }
\label{fig:gAfpi}
\end{figure*}

All but one previous lattice-QCD results underestimate $g_A^{u-d}$ as
shown in Fig.~\ref{fig:gA}.  The exceptional result is from the RQCD
Collaboration~\cite{Bali:2014nma} that finds a large slope with
$M_\pi^2$ in the renormalization factor independent ratio
$g_A/F_\pi$. Extrapolating this ratio to the physical pion mass gave
the estimate $g_A = 1.280(44)(46)$ even though the majority of their
data for $g_A^{u-d}$ are $\lsim 1.2$. Our data for $F_\pi$ and
$g_A/F_\pi$, given in Table~\ref{tab:resultsFpi}, show little
dependence on the lattice spacing and lattice volume. We, therefore,
analyze them in Fig.~\ref{fig:gAfpi} versus just $M_\pi^2$ and plot
the result of a fit linear in $M_\pi^2$. The fit shows that our
clover-on-HISQ estimate, $g_A/F_\pi=12.88(15)$, at the physical pion
mass remains low by about $7\%$ compared to the experimental value
13.80. This is because our $F_\pi r_1$ data, where $r_1$ is used to
set the scale of the HISQ lattices~\cite{Bazavov:2012xda}, extrapolate
to the physical value. On the other hand, the error in the ratio
$g_A/F_\pi$ is much smaller than in $g_A$, and therefore the deviation is 
more significant.  The difference between the two calculations
suggests that further analysis is needed to quantify the various
systematics in $g_A^{u-d}$.

\begin{table}
\centering
\begin{ruledtabular}
\begin{tabular}{c|c|cc|c}
ID              & $g_{A,{\rm bare}}^{u-d}$ & $F_\pi^{\rm bare}$   & $F_\pi$ & $g_A^{u-d}/F_\pi$ \\ 
                &                          & (MeV)                & (MeV)   & (GeV${}^{-1}$) \\ 
\hline
a12m310*        & 1.252(9)   & 107.7(1.0)   & 102.3(3.4)   & 11.62(14) \\
a12m220S        & 1.283(39)  & 102.4(1.0)   &  97.3(3.2)   & 12.53(40) \\
a12m220         & 1.273(30)  & 104.5(1.0)   &  99.2(3.2)   & 12.19(31) \\
a12m220L*       & 1.279(12)  & 104.2(0.9)   &  99.0(3.2)   & 12.27(15) \\
\hline                                                                                                                             
a09m310         & 1.262(30)  & 106.9(1.0)   & 101.6(3.3)   & 11.80(30) \\
a09m220         & 1.272(30)  & 102.8(0.9)   &  97.6(3.2)   & 12.38(31) \\
a09m130*        & 1.255(24)  &  97.2(0.8)   &  92.4(3.0)   & 12.91(26) \\
\hline                                                                                                          
a06m310*        & 1.212(14)  & 107.7(0.7)   & 104.5(3.3)   & 11.25(15) \\
a06m220*        & 1.234(17)  & 101.3(0.7)   &  98.2(3.1)   & 12.19(19) \\
\end{tabular}
\end{ruledtabular}
\caption{Estimates of the unrenormalized $g_A^{u-d}$ (reproduced from
  Table~\protect\ref{tab:resultsbare}) and the bare and renormalized
  $F_\pi$ obtained from the HP measurements. The dominant source of
  error in the renormalized $F_\pi$ is the uncertainty in the $Z_A$
  given in Table~\protect\ref{tab:Zfinal}. We also give the
  renormalization factor independent ratio $g_A^{u-d}/F_\pi$,
  calculated as a ratio of the bare quantities, and which is plotted
  in Fig.~\protect\ref{fig:gAfpi}.  }
\label{tab:resultsFpi}
\end{table}

To reconcile the roughly $2\sigma$ difference between our result
$g_A^{u-d}=1.195(33)(20)$ and the experimental value
$g_A^{u-d}=1.276(3)$ would require all five $O(2\%)$ systematics given
in Table~\ref{tab:errors} to eventually move the result in the same
direction or one or more of the systematic effects have been grossly
underestimated.  To gain a better understanding of how the various
sources of errors contribute and to reduce the overall uncertainty to
$O(2\%)$ will require at least $O(200,000)$ measurements on the seven
ensembles at different $a$ and $M_\pi$ used in this study and the
analysis of one additional ensemble at $a=0.06$~fm and
$M_\pi=135$~MeV.  Increasing the statistics by a factor of four will
reduce the errors in the data with the largest $\tsep$ we have
analyzed and thus improve the $\tsep \to \infty$ estimates. Adding the
point at the physical quark mass and the smallest lattice spacing
$a=0.06$~fm, will further constrain the chiral fit.  This level of
precision is achievable with the next generation of leadership-class
computing resources.

\subsection{$g_S^{u-d}$}
\label{sec:gScompare}

There are few estimates of $g_S^{u-d}$ using lattice QCD.  The RQCD
Collaboration reported $1.02(18)(30)$~\cite{Bali:2014nma}, and the
LHPC Collaboration obtained
$1.08(28)$~\cite{Green:2012ej}.\footnote{The recent estimates from the
  European Twisted Mass Collaboration
  (ETMC)~\cite{Abdel-Rehim:2015owa} are not included in our comparison
  since they have not been extrapolated to the continuum limit. Their
  results are $g_S^{u-d}=1.23(10)$ from the $2+1+1$-flavor calculation
  at $M_\pi=373$~MeV and $a=0.083$~fm and $2.16(34)$ from their
  physical-mass $2$-flavor calculation at $a=0.093$~fm. } While these
estimates are consistent with our result $g_S^{u-d} = 0.97(12)(6)$, it
is clear that the errors are still large in all lattice-QCD estimates.
Our work suggests that increasing the measurements to $O(200,000)$ on
all the ensembles (the same program needed to reduce the overall
uncertainty in $g_A^{u-d}$ to $2\%$) will also reduce the uncertainty
in $g_S^{u-d}$ to less than $10\%$.

\begin{figure}
\begin{center}
\includegraphics[width=0.46\textwidth]{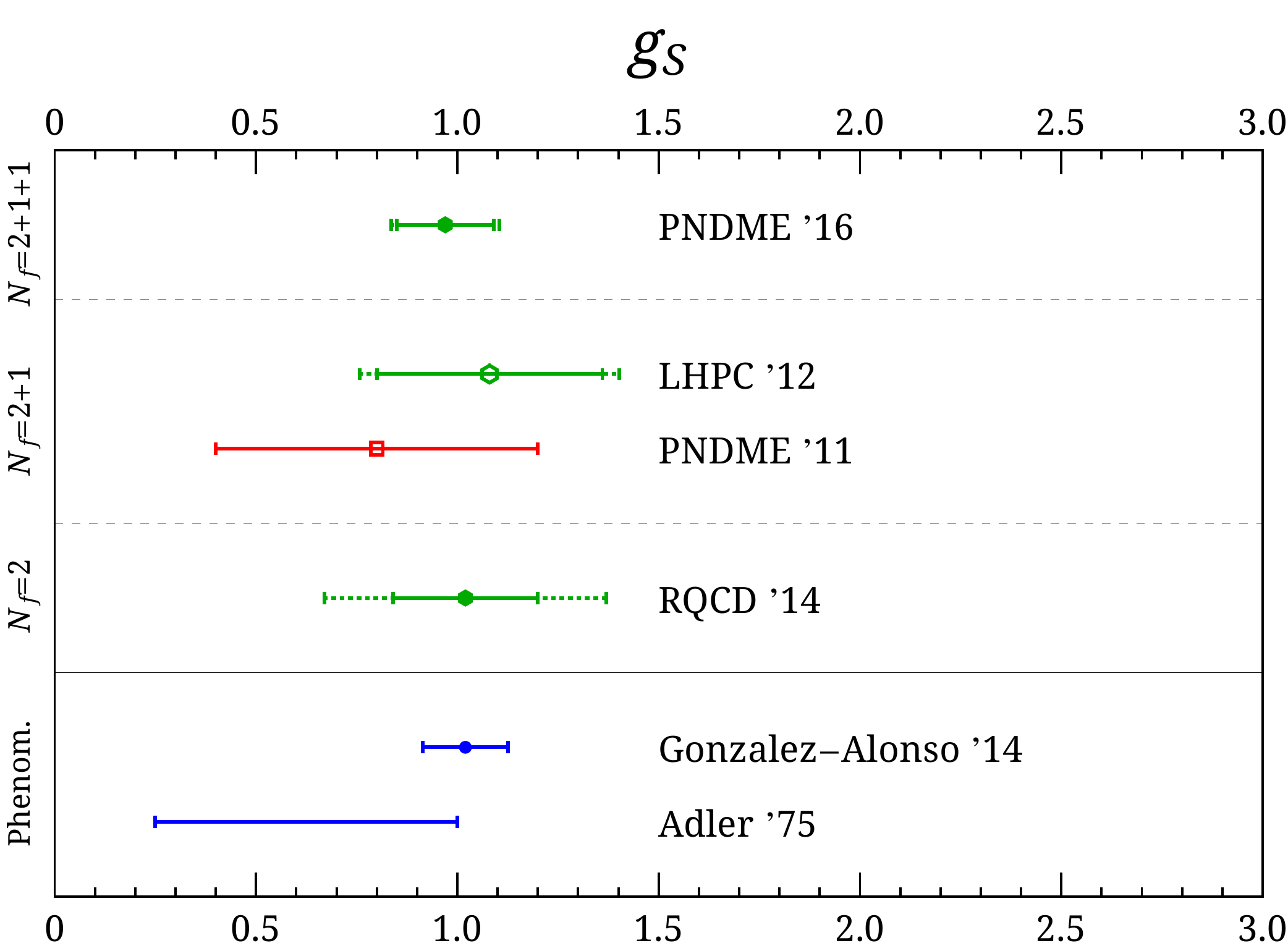}
\end{center}
\vspace{-0.5cm}
\caption{A summary plot showing estimates for $g_S^{u-d}$ from lattice
  QCD and phenomenology. The data are taken from the following sources: PNDME'16 (this
  work); LHPC'12~\protect\cite{Green:2012ej};
  PNDME'11~\protect\cite{Bhattacharya:2011qm};
  RQCD'14~\protect\cite{Bali:2014nma}.  The estimates based on the
  conserved vector current and phenomenology are taken from 
  Gonzalez-Alonso~\protect\cite{Gonzalez-Alonso:2013ura} and
  Adler~\protect\cite{Adler:1975he}.  The rest is the same as in
  Fig.~\protect\ref{fig:gA}. }
\label{fig:gS}
\end{figure}

Gonzalez-Alonso {\it et al.}~\cite{Gonzalez-Alonso:2013ura} used the
conserved vector current (CVC) relation $g_S/g_V = (M_N-M_P)^{\rm QCD}/
(m_d-m_u)^{\rm QCD}$ to obtain $g_S^{u-d}$. In their analysis, the
estimates of the two mass differences on the right-hand side were
obtained using the global lattice-QCD data. Their result,
$g_S^{u-d}=1.02(8)(7)$, is consistent with our estimate discussed
above and shown in Fig.~\ref{fig:gS}. These two lattice-QCD estimates,
using CVC versus our direct calculation of the charge, have very
different systematic uncertainties, so their consistency is a
nontrival check. We, therefore, consider our result,
$g_S^{u-d}=0.97(12)(6)$, from a direct calculation as having achieved
the target accuracy of about 10\%--15\% needed to put bounds on scalar
and tensor interactions at the TeV scale when combined with
experimental measurements of $b$ and $b_\nu$ parameters in neutron
decay experiments with $10^{-3}$
sensitivity~\cite{Bhattacharya:2011qm}.  The current and prospective
status of these bounds is given in Sec.~\ref{sec:est}.

We can use CVC in reverse to combine our result $g_S=0.97(12)(6)$ with
the lattice QCD determination of $(m_d-m_u)^{\rm QCD}$ given by the
Flavor Lattice Average Group (FLAG)~\cite{FLAG:2016qm} to predict
$(M_N-M_P)^{\rm QCD}$.  Using the 2+1-flavor estimates
$m_d=4.68(14)(7)$~MeV and $m_u=2.16(9)(7)$~MeV from FLAG, we get
$(M_N-M_P)^{\rm QCD} = 2.44(38)$~MeV, and using the 2+1+1-flavor FLAG
estimates $m_d=5.03(26)$~MeV and $m_u=2.36(24)$~MeV gives 
$(M_N-M_P)^{\rm QCD} = 2.59(49)$~MeV.  These two estimates update our
old ``PNDME'' value quoted by Gonzalez-Alonso {\it et al.}~\cite{Gonzalez-Alonso:2013ura} 
and are now competitive with other lattice QCD estimates given by them.

\subsection{$g_T^{u-d}$}
\label{sec:gTcompare}

\begin{figure}
\begin{center}
\includegraphics[width=0.46\textwidth]{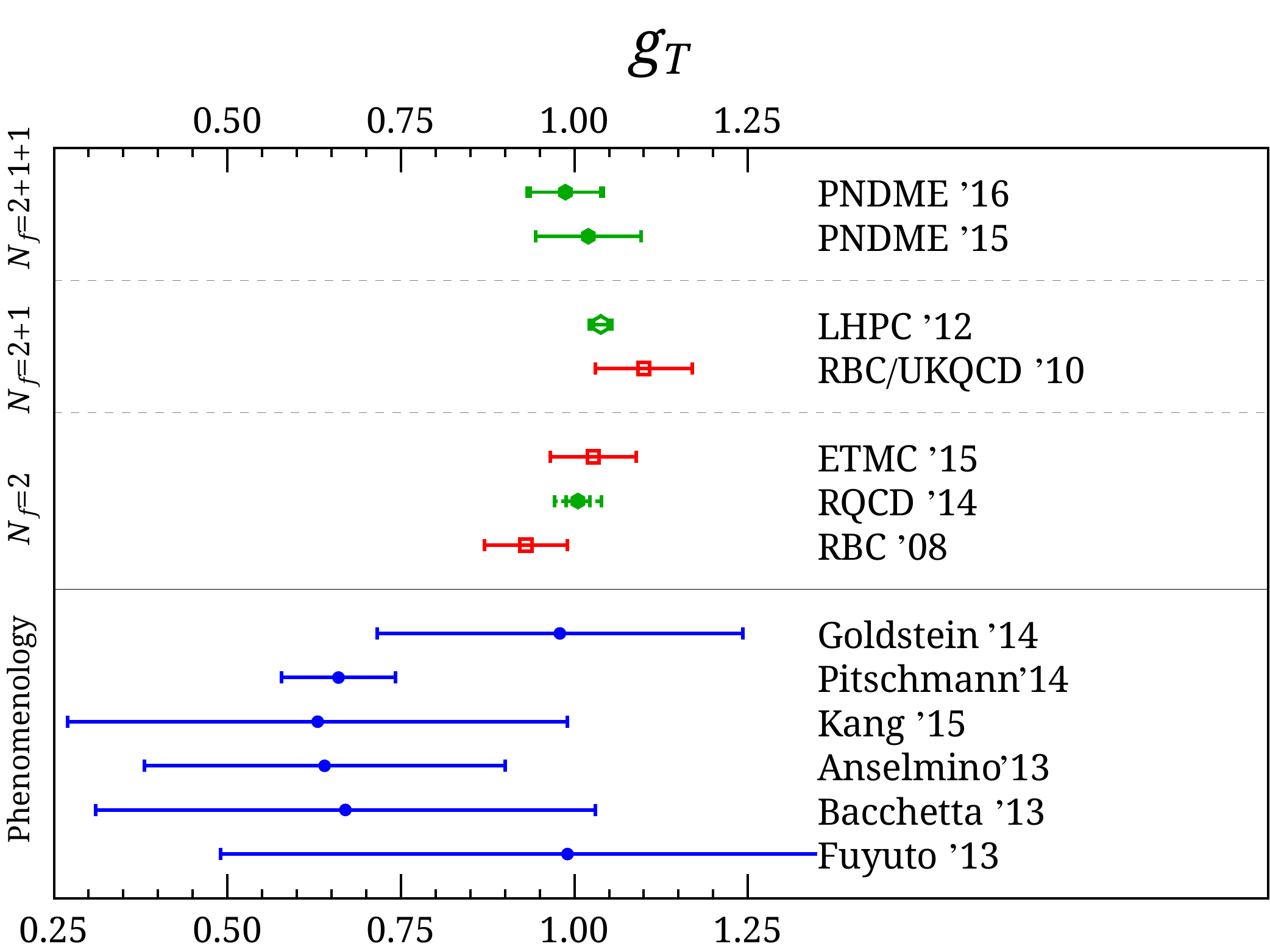}
\end{center}
\vspace{-0.5cm}
\caption{A summary plot showing estimates for $g_S^{u-d}$ from lattice
  QCD and phenomenology. The data are taken from the following sources: 
  PNDME'16 (this work); 
  PNDME'15~\protect\cite{Bhattacharya:2015wna}
  LHPC'12~\protect\cite{Green:2012ej};
  RBC/UKQCD'10~\protect\cite{Aoki:2010xg}
  ETMC'15~\protect\cite{Abdel-Rehim:2015owa};
  RQCD'14~\protect\cite{Bali:2014nma}; and 
  RBC'08~\protect\cite{Yamazaki:2008py}. 
  The phenomenological estimates are taken from the following sources: 
  Kang'15~\protect\cite{Kang:2015msa}; 
  Goldstein'14~\protect\cite{Goldstein:2014aja}; 
  Pitschmann'14~\protect\cite{Pitschmann:2014jxa}; 
  Anselmino~\protect\cite{Anselmino:2013vqa}; 
  Bacchetta'13~\protect\cite{Bacchetta:2012ty}; and 
  Fuyuto~\protect\cite{Fuyuto:2013gla}.  
  The rest is the same as in Fig.~\protect\ref{fig:gA}.
   }
\label{fig:gT}
\end{figure}

In Ref.~\cite{Bhattacharya:2015wna}, we presented a detailed
discussion of all the errors in the determination of the isovector
charge $ g_T^{u-d}$ and compared our results with those obtained by
other collaborations. Our new estimate, $ g_T^{u-d} = 0.987(51)(20)$,
improves on the value $1.020(76)$ reported in
Ref.~\cite{Bhattacharya:2015wna}.\footnote{The small decrease can be
  traced back to the better control over excited-state contamination
  in this study with the AMA method and thus better determination of
  the $\tsep \to \infty$ value that converges from above.} It is also
consistent with the estimate $g_T^{u-d} = 1.038(11)(12)$ from the LHPC
Collaboration ($N_f=2+1$ hypercubically nested EXP (HEX) smeared
clover action~\cite{Capitani:2006ni}, domain-wall action, and 
domain-wall-on-asqtad actions)~\cite{Green:2012ej}, and $g_T^{u-d} =
1.005(17)(29)$ by the RQCD Collaboration ($N_f=2$ $O(a)$-improved
clover fermions)~\cite{Bali:2014nma}. The conclusion reached in
Ref.~\cite{Bhattacharya:2015wna}, that the error estimates from the
various sources are estimated reliably, is validated by this
higher-statistics study.  In Fig.~\ref{fig:gT}, we update the
lattice-QCD results and show that they are more accurate than the
sum-rule, Dyson-Schwinger, and phenomenological estimates (integral
over the longitudinal momentum fraction of the experimentally measured
quark transversity distributions). Given the consistency of the
lattice-QCD estimates and our better understanding of the
excited-state contamination and other systematic effects, we consider
our error estimate to be conservative and $ g_T^{u-d} = 0.987(51)(20)$
a reliable value to use in phenomenology.

\begin{figure*}
\begin{center}
\includegraphics[width=.85\textwidth]{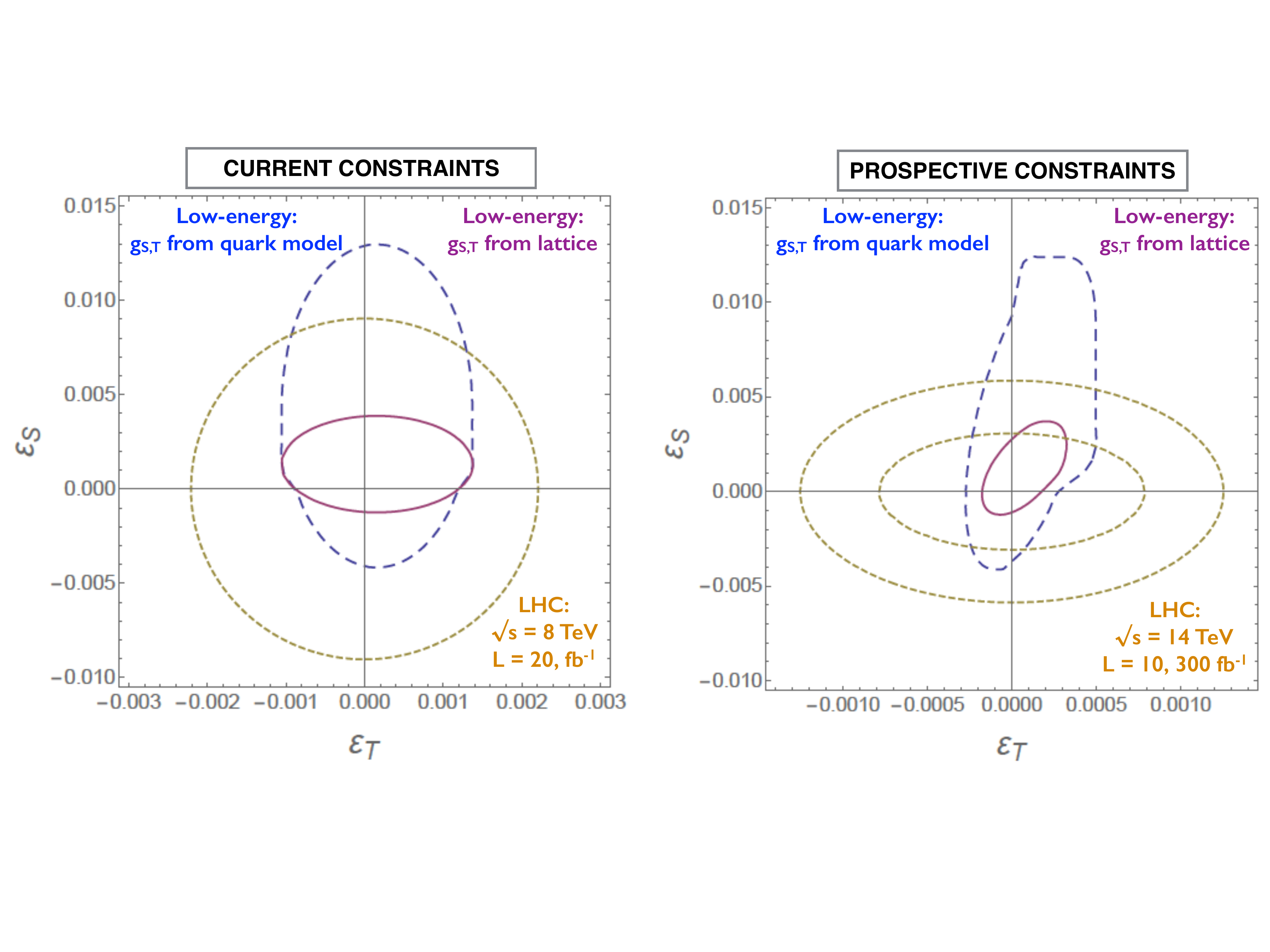}
\end{center}
\vspace{-2.5cm}
\caption{Left panel: current $90 \%$ C.L. constraints on $(\epsilon_S)$
  and $(\epsilon_T)$ from beta decays ($\pi \to e \nu \gamma$ and $0^+
  \to 0^+$) and the LHC ($p p \to e \nu + X$) at $\sqrt{s}=8$ TeV.
  Right panel: prospective $90 \%$ C.L. constraints on $(\epsilon_S)$
  and $(\epsilon_T)$ from beta decays and the LHC ($p p \to e \nu +
  X$) at $\sqrt{s}=14$ TeV.  The low-energy constraints correspond to
  0.1\% measurements of $B,b$ in neutron decay and $b$ in $^6$He
  decay.  In both panels we present the low-energy constraints under
  two different scenarios for the scalar and tensor charges $g_{S,T}$:
  quark model~\cite{Herczeg:2001vk} (large dashed contour) and lattice
  QCD results given in Table~\protect\ref{tab:IVresults} (smaller
  solid contour).  }
\label{fig:eSeT}
\end{figure*}

%%%%%%%%%%%%%%%%%%%%%%%%%%%%%%%%%%%%%%%%%%%%%%%%%%%%%%%%%%%%%%%%%%%%%
%%%  SECTION                                                      %%%
\section{Constraining new physics using precision beta decay measurements}
\label{sec:est}
%%%%%%%%%%%%%%%%%%%%%%%%%%%%%%%%%%%%%%%%%%%%%%%%%%%%%%%%%%%%%%%%%%%%%%%%%%%%%%%%

Our improved results for the isovector tensor charges $g_{S}^{\rm u-d}$ and
$g_T^{\rm u-d}$ enable more stringent tests of nonstandard scalar and tensor
charged-current interactions, parametrized by the dimensionless
couplings $\epsilon_{S,T}$~\cite{Bhattacharya:2011qm,
  Cirigliano:2012ab}:
\begin{eqnarray}
{\cal L}_{\rm CC}  &=&
- \frac{G_F^{(0)} V_{ud}}{\sqrt{2}} \  \Big[ \
 \epsilon_S  \  \bar{e}  (1 - \gamma_5) \nu_{\ell}  \cdot  \bar{u} d 
 \nonumber \\
&+  & 
\epsilon_T \   \bar{e}   \sigma_{\mu \nu} (1 - \gamma_5) \nu_{\ell}    \cdot  \bar{u}   \sigma^{\mu \nu} (1 - \gamma_5) d
\Big] ~. 
\end{eqnarray}
These nonstandard couplings, $\epsilon_{S,T}$ can be constrained at
low energy by precision beta-decay measurements (of the pion, neutron,
and nuclei) as well at the Large Hadron Collider (LHC) through the
reaction $pp \to e \nu + X$. The LHC constraint is 
valid provided the mediator of the new interaction is
heavier than a few TeV.

In Fig.~\ref{fig:eSeT} (left panel), we update the analysis of
constraints on $\epsilon_T$ and $\epsilon_S$ presented in
Refs.~\cite{Bhattacharya:2011qm, Cirigliano:2012ab} by using our improved
estimate of the tensor charge and the LHC data from the 2012 run at
$\sqrt{s} = 8$~TeV, for an integrated luminosity of approximately
$20~{\rm fb}^{-1}$~\cite{Khachatryan:2014tva,ATLAS:2014wra}.  The
current bound on $\epsilon_T$ is dominated by the radiative pion decay
$\pi \to e \nu \gamma$~\cite{Bychkov:2008ws,Mateu:2007tr}, while the
bound on $\epsilon_S$ is dominated by the Fierz interference term in
$0^+ \to 0^+$ nuclear beta decays~\cite{Hardy:2008gy}.  

In the prospective bounds shown in Fig.~\ref{fig:eSeT} (right panel),
the change in low-energy constraints is due to improvements in
experiments: We assume future measurements of the Fierz interference
term and the neutrino asymmetry parameter ($b$ and $b_\nu$) in neutron
decay will be at the level of
$10^{-3}$~\cite{Dubbers:2007st,WilburnUCNB,abBA,Pocanic:2008pu}.  A
similar constraining power on $\epsilon_T$ can be achieved with a
$0.1\%$-level measurement of the Fierz term in the pure Gamow-Teller
$^6$He decay~\cite{Knecht:2012jb}.

Between the current and prospective constraints, the improvement in
the LHC bounds comes from both the increase in center-of-mass energy
$\sqrt{s}$ and the integrated luminosity.  In both panels, the impact
of using our estimates of $g_{S,T}$ given in Table~\ref{tab:IVresults}
over the quark model estimates $0.25 < gS < 1.0$ and $ 0.6 < gT <
2.3$~\cite{Herczeg:2001vk} is large. Furthermore, the lattice-QCD
reduction of uncertainties in $g_{S,T}$ implies that the constraining
power of beta decays can actually be stronger than of the LHC for this
class of nonstandard interactions.

The current upper bounds on the effective couplings $\epsilon_{S,T} =
(v/\Lambda_{S,T})^2$ correspond to lower bounds for the effective
scales $\Lambda_{S} > 7$~TeV and $\Lambda_T >13$~TeV.  
If future low-energy experiments find a nonzero signal, then
combined with precision calculations of $g_{S,T}$, we will be able to
predict the scale of new particles to be probed at the LHC or future
colliders.  Even without a signal, with improved precision, we will be
able to tighten the lower bounds on the scale of new physics in these
channels, which will help rule out certain classes of BSM models.

%%%%%%%%%%%%%%%%%%%%%%%%%%%%%%%%%%%%%%%%%%%%%%%%%%%%%%%%%%%%%%%%%%%%%
%%%  SECTION                                                      %%%
%%%%%%%%%%%%%%%%%%%%%%%%%%%%%%%%%%%%%%%%%%%%%%%%%%%%%%%%%%%%%%%%%%%%%
\section{Conclusions}
\label{sec:conclusions}

We have presented a high-statistics study of the isovector and
flavor-diagonal charges of the nucleon using clover-on-HISQ lattice
QCD. By using the AMA error-reduction technique we show that the
statistical precision of the data can be improved significantly.
Also, keeping one excited state in the analysis of data at multiple
values of $\tsep$ allows us to isolate and mitigate excited-state
contamination. Together, these two improvements allow us to
demonstrate that the excited-state contamination in the axial and the
tensor channels has been reduced to the $ 2\%$ level.  The
high-statistics analysis of nine ensembles covering the range
0.12--0.06~fm in the lattice spacing, $M_\pi =$ 135--320~MeV in the
pion mass, and $M_\pi L =$ 3.3--5.5 in the lattice size allowed us to
analyze the systematic errors due to lattice discretization,
dependence on the quark mass and finite volume. As a result, this is
the first work that is able to include the effect of these three
systematic uncertainties by making a simultaneous fit in all the three
variables $a$, $M_\pi^2$ and $M_\pi L$. In the case of the isovector
charges, we also assign a second error estimate to account for the
variation in the results due to the choice of the extrapolation Ansatz
as discussed in Sec.~\ref{sec:errors}.  Our final estimates are given
in Tables~\ref{tab:IVresults} and~\ref{tab:FDresults}.

One of the largest sources of uncertainty comes from the calculation
of the renormalization constants for the quark bilinear operators.
These are calculated nonperturbatively in the RI-sMOM scheme over a
range of values of $q^2$. As discussed in Sec.~\ref{sec:renorm}, the
breaking of the rotational symmetry gives rise to a large spread in
the data. After conversion to $\overline{\text{MS}}$ scheme at $2\GeV$
using perturbation theory, which shows poor convergence for $Z_T$,
the data, especially for $g_S^{u-d}$, do not show a large scaling window in
which they are independent of $q^2$. Our estimates of errors take into
account these uncertainties, and are therefore larger than those
obtained by other collaborations. 

Our estimate $g_A^{u-d}=1.195(33)(20)$ is about $2 \sigma$ (about
$7\%$) below the experimental value $g_A/g_V = 1.276(3)$.  In
Sec.~\ref{sec:errors} we analyze five systematic effects and find that
each contributes at the 1\%--2\% level, with roughly equal distribution
in sign. Improvement in our understanding of these five systematic
factors: residual excited-state contamination, uncertainty in the
determination of the renormalization constants, the Ansatz used for
the chiral fit, error in the chiral fit and the continuum
extrapolation, requires a still higher-statistics calculation.  Based
on the various systematic errors discussed in Sec.~\ref{sec:errors},
we claim in Sec.~\ref{sec:gAcompare} that the overall uncertainty can
be reduced to $ 2\%$ by increasing the statistics to $O(200,000)$
measurements on $O(2000)$ configurations on the seven ensembles shown
in Fig.~\ref{fig:gA7} and an additional physical mass HISQ ensemble at
$a=0.06$~fm.  Last, to address possible systematic effects due to
using the clover-on-HISQ formulation versus a unitary lattice
formulation, we have started calculations with similar statistics and
methodology using the clover-on-clover
formulation~\cite{Yoon:2016dij}.

For the tensor charges, we find that the dependence on the lattice
volume, lattice spacing and the light-quark mass is small, and a
simultaneous fit in these variables, keeping just the lowest-order
corrections, gives reliable estimates of the physical value.  Our
final estimate for the isovector tensor charge, $g_T^{u-d} =
0.987(51)(20)$, is in good agreement with the previously reported
estimate~\cite{Bhattacharya:2015wna} and is more accurate than
phenomenological estimates as shown in Fig.~\ref{fig:gT}.

We have also updated our estimates for the connected parts of the
flavor-diagonal charges. New estimates of the tensor charges of the
proton, needed for the analysis of the contribution of the
quark EDM to the neutron
EDM~\cite{Bhattacharya:2015wna,Bhattacharya:2015esa}, are
$g_T^{u}=0.792(42)$ and $g_T^{d}=-0.194(14)$.

The extraction of the scalar charge of the proton has larger
uncertainty.  The statistical errors in the lattice data for
$g_S^{u-d}(a, M_\pi, M_\pi L)$ are 3--5 times those in
$g_T^{u-d}(a,M_\pi,M_\pi L)$, and the data show significant dependence
on the lattice spacing $a$ and a weaker dependence on the pion mass
$M_\pi$.  Our estimate, $g_S^{u-d}=0.97(12)(6)$, is in very good
agreement with the estimate $g_S^{u-d}=1.02(8)(7)$ obtained using the
conserved vector current relation in
Ref.~\cite{Gonzalez-Alonso:2013ura}.  Previous lattice-QCD estimates
summarized in Fig.~\ref{fig:gS} have larger errors but are consistent
with these two estimates as discussed in Sec.~\ref{sec:comparison}.
Combining our estimate $g_S^{u-d}=0.97(12)(6)$ with the 2+1+1-flavor
estimate of the difference of light quarks masses $(m_d-m_u)^{\rm
  QCD}=2.67(35)$~MeV given by the FLAG~\cite{FLAG:2016qm}, we obtain 
$(M_N-M_P)^{\rm QCD} = 2.59(49)$~MeV

Finally, our results, $g_S^{u-d}=0.97(12)(6)$ and $g_T^{u-d} =
0.987(51)(20)$, meet the target uncertainty of $15\%$ required to
maximize the impact of future measurements of the helicity-flip parts
of the neutron decay distribution with $10^{-3}$
accuracy~\cite{Bhattacharya:2011qm}.  The status of current and
prospective constraints on novel scalar and tensor interactions,
$\epsilon_{S,T}$, using our improved estimates of $g_{S,T}$ are given
in Sec.~\ref{sec:est}.  Our goal for the near future is to further
understand and reduce all the systematic uncertainties in the estimate
of $g_A^{u-d}$, a benchmark for evaluating the accuracy achievable in
lattice-QCD calculations of the matrix elements of quark bilinear
operators within nucleon states.

\begin{acknowledgments}
We thank the MILC Collaboration for providing the 2+1+1-flavor HISQ
lattices used in our calculations.  We also thank Gunnar Bali, 
Jeremy Green and Martin Gonzalez-Alonso for discussions. Simulations were carried out on
computer facilities of (i) the USQCD Collaboration, which are funded
by the Office of Science of the U.S. Department of Energy, (ii) the
Extreme Science and Engineering Discovery Environment (XSEDE), which
is supported by National Science Foundation Grant No. ACI-1053575,
(iii) the National Energy Research Scientific Computing Center, a DOE
Office of Science User Facility supported by the Office of Science of
the U.S. Department of Energy under Contract No. DE-AC02-05CH11231;
and (iv) Institutional Computing at Los Alamos National Laboratory.
The calculations used the Chroma software
suite~\cite{Edwards:2004sx}. This material is based upon work
supported by the U.S. Department of Energy, Office of Science of High
Energy Physics under Contract No.~DE-KA-1401020 and the LANL LDRD
program. The work of H-W.L. and S.D.C. was supported by DOE Grant
No.~DE-FG02-97ER4014. H-W.L. is supported in part by the M. Hildred
Blewett Fellowship of the American Physical Society. T.B., V.C., H-W.L. and
B.Y. thank the Institute for Nuclear Theory at the University of
Washington for its hospitality during the completion of this work.
\end{acknowledgments}

\clearpage
%
%-----------
% reference
%-----------
%\bibliographystyle{abbrv} %%% physical review
%% \makeatletter
%% \ifx\@bibitemShut\undefined\let\@bibitemShut\relax\fi
%% \makeatother
\bibliography{ref} %%% ref.bib file

\end{document}